\documentclass[twoside,11pt]{PhDthesisPSnPDF}
\pdfoutput=1 

\newcommand{\figuremacro}[3]{
	\begin{figure}[!t]
		\centering
		\includegraphics[width=1\textwidth]{#1}
		\caption[#2]{#3}   
		\label{#1}
	\end{figure}
}

\newcommand{\figuremacroW}[4]{
	\begin{figure}[!t]
		\centering
		\includegraphics[width=#4\textwidth]{#1}
		\caption[#2]{#3}  
		\label{#1}
	\end{figure}
}

\newcommand{\figuremacroH}[4]{
	\begin{figure}[!t]
		\centering
		\includegraphics[height=#4\textheight]{#1}
		\caption[#2]{#3}  
		\label{#1}
	\end{figure}
}

\newcommand{\figuremacroFP}[4]{
	\begin{figure}[!p]
		\centering
		\includegraphics[height=#4\textheight]{#1}
		\caption[#2]{#3}  
		\label{#1}
	\end{figure}
}

\newcommand{\figuremacroSW}[4]{
	\begin{sidewaysfigure}[!p]
		\centering
		\includegraphics[height=#4\textheight]{#1}
		\caption[#2]{#3}  
		\label{#1}
	\end{sidewaysfigure}
}

\newcommand{\figref}[1]{Figure~\ref{#1}}

\newcommand{\sref}[1]{Section~\ref{#1}}

\newcommand{\nucl}[3]{
	{^{#1}_{#2}\textrm{#3}}
}

\newcommand{\rng}{\,--\,}

\long\def\symbolfootnote[#1]#2{\begingroup%
\def\thefootnote{\fnsymbol{footnote}}\footnote[#1]{#2}\endgroup}

\usepackage{rotating}
\usepackage{multirow}
\usepackage{nomencl}

\ifpdf  
    \pdfcatalog { /PageMode (/UseOutlines)
                  /OpenAction (fitbh)  }
\fi

\title{{\huge Propagation of Coronal Mass Ejections in the Inner Heliosphere}}

\ifpdf
  \author{\href{mailto:maloneys@tcd.ie}{Shane A. Maloney, B.Sc. (Hons.)}}
  \collegeordept{\href{http://www.tcd.ie/physics}{School of Physics}}
  \university{\href{http://www.tcd.ie}{Trinity College Dublin}}

  \crest{\includegraphics[width=4cm]{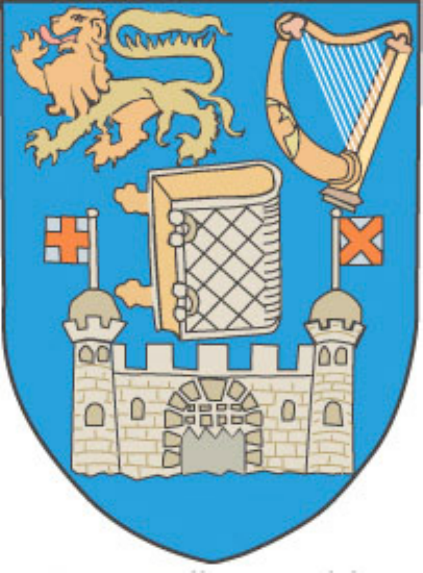}}
  
\else
  \author{Shane Maloney}
  \collegeordept{School of Physics}
  \university{Trinity College Dublin}
  \crest{\includegraphics[width=4cm]{logo}}
\fi

\degree{Philosophi\ae Doctor (PhD)}
\degreedate{September 2011}

\begin{document}


\renewcommand\baselinestretch{1.2}
\baselineskip=18pt plus1pt


\maketitle  

\onehalfspace


%
%
%





%
\frontmatter
\setcounter{page}{3}






\begin{declaration}        

I declare that this thesis has not been submitted as an exercise for a degree at this or any other university and it is entirely my own work. \\

I agree to deposit this thesis in the University's open access institutional repository or
allow the library to do so on my behalf, subject to Irish Copyright Legislation and
Trinity College Library conditions of use and acknowledgement.

\vspace{20mm}

\centering
{\bf Signature: .......................................... Date: .........................}

\end{declaration}



\begin{abstracts}        

Solar Coronal mass ejections (CMEs) are large-scale ejections of plasma and magnetic field from the corona, which propagate through interplanetary space. CMEs are the most significant drivers of adverse space weather on Earth, but the physics governing their propagation through the Heliosphere is not well understood. This is mainly due to the limited fields-of-view and plane-of-sky projected nature of previous observations. The Solar Terrestrial Relations Observatory (STEREO) mission launched in October 2006, was designed to overcome these limitations.

In this thesis, a method for the full three dimensional (3D) reconstruction of the trajectories of CMEs using STEREO was developed. Observations of CMEs close to the Sun  ($<$\,15\,$R_{\odot}$)  were used to derive the CMEs trajectories in 3D. These reconstructions supported a pseudo-radial propagation model. Assuming pseudo-radial propagation, the CME trajectories were extrapolated to large distances from the Sun (15\rng240\,$R_{\odot}$). It was found that CMEs slower than the solar wind were accelerated, while CMEs faster than the solar wind were decelerated, with both tending to the solar wind velocity.

Using the 3D trajectories, the true kinematics were derived, which were free from projection effects. Evidence for solar wind (SW) drag forces acting in interplanetary space were found, with a fast CME decelerated and a slow CME accelerated toward typical SW velocities. It was also found that the fast CME showed a linear dependence on the velocity difference between the CME and the SW, while the slow CME showed a quadratic dependence. The differing forms of drag for the two CMEs indicated the forces responsible for their acceleration may have been different. Also, using a new elliptical tie-pointing technique the entire front of a CME  was reconstructed  in 3D. This enabled the quantification of its  deflected trajectory, increasing angular width, and propagation from 2 to 46\,$R_{\odot}$ (0.2 AU). Beyond 7\,$R_{\odot}$, its motion was shown to be determined by aerodynamic drag. Using the reconstruction as an input for a 3D magnetohydrodynamic simulation, an accurate arrival time at the L1  Lagrangian point near Earth was determined.

CMEs are known to generate bow shocks as they propagate through the corona and SW. Although CME-driven shocks have previously been detected indirectly via their emission at radio frequencies, direct imaging has remained elusive due to their low contrast at optical wavelengths. Using STEREO observations, the first images of a CME-driven shock as it propagates through interplanetary space from 8\,$R_{\odot}$ to 120\,$R_{\odot}$ (0.5 AU) were captured. The CME was measured to have a velocity of $\sim$\,1000\,km\,s$^{-1}$ and a Mach number of $4.1\pm 1.2$, while the shock front standoff distance ($\Delta$) was found to increase linearly to $\sim$\,20\,$R_{\odot}$ at 0.5 AU. The normalised standoff distance ($\Delta/D_{O}$) showed reasonable agreement with semi-empirical relations, where $D_{O}$ is the CME radius. However, when normalised using the radius of curvature ($\Delta/R_{O}$), the standoff distance did not agree well with theory, implying that $R_{O}$ was underestimated by a factor of $\sim$\,3\rng8. This is most likely due to the difficulty in estimating the larger radius of curvature along the CME axis from the observations, which provide only a cross-sectional view of the CME. The radius of curvature of the CME at 1\,AU was estimated to be $\sim$\,0.95\,AU

\end{abstracts}




\begin{dedication} 

To my parents, \\ who have supported and encouraged me throughout

\end{dedication}



\begin{acknowledgements}      

I am extremely grateful to my supervisor Peter Gallagher, whose experience, advice, and enthusiasm has been vital on many occasions. It has been a pleasure working with you these past four years.

I must also thank James McAteer and Shaun Bloomfield whose guidance and discussions have been most helpful.

I must also express my appreciation to all the staff in the School of Physics without your efforts the school would grind to a halt. 

During my times at Goddard a huge thank you must go to Alex, Jack and Dominic. Whether it was helping to sort out my financial situation, or ideas and thoughts on my research or outside of work, you helped make my time at Goddard all that is was.
 
A huge thank you also goes out to my fellow comrades in arms in the office, past and present; Aidan, Claire, David, Eoin, Eamon, Joseph, Jason, Larisza, Paul C, Paul H, Peter, Sophie. Whether it was discussing the finer points of IDL, general banter in the office, morning coffee, or after work pints you have made my time here more enjoyable then I imagined it could be. My gratitude also goes to all the guys from ACTTT and BOXMASTERS who kept me from leading a completely sedentary lifestyle.

Also a general thank you to my friends and family, you have done more than you know and are all extremely important to me. Special thanks to my girlfriend, Lucy, for putting up with me, particularly over the last few weeks and months.

\end{acknowledgements}



\setcounter{secnumdepth}{3} 
\setcounter{tocdepth}{3}    
\tableofcontents            



\chapter{List of Publications}
\begin{enumerate}
	\item Gallagher, P. T., \textbf{Maloney, S.~A.} et al. \\ 
	``A comprehensive overview of the 2011 June 7 event'',  \\
	Astronomy \& Astrophysics, {\it in prep}
	
	\item P\'{e}rez-Su\'{a}rez, D., \textbf{Maloney, S~ A.}, Higgins, P.~A. et al. \\
	``Studying Sun-planet connections using the Heliophysics Integrated Observatory (HELIO)''
	Solar Physics {\it in review}
		
	 \item \textbf{Maloney, S.~A.} and Gallagher, P.~T. \\ 
	``STEREO Direct imaging of a coronal mass ejection driven shock to 0.5\,AU'', \\
	Astrophysical Journal Letters, 736, L5, 2011

	\item Byrne, J.~P., \textbf{Maloney, S.~A.}, McAteer, R.~T.~J., Refojo, J.~M. and Gallagher, P.~T. \\ 
	``Propagation of an Earth-directed coronal mass ejection in three dimension'', \\
	Nature Communications, 1, 74, 2010
	
	\item Mierla, M., \textbf{Maloney, S.~A.} et al. \\ 
	``On the 3-D reconstruction of Coronal Mass Ejections using coronagraph data'', \\
	 Annales Geophysicae, 28, 203, 2010	

	\item \textbf{Maloney, S.~A.} and Gallagher, P.~T. \\ 
	``Solar Wind Drag and the Kinematics of Interplanetary Coronal Mass Ejections'', \\
	Astrophysical Journal Letters, 724, L127, 2010
	
	\item \textbf{Maloney, S.~A.}, Gallagher, P.~T. \& McAteer, R.~T.~J. \\ 
	``Reconstructing the 3-D trajectories of CMEs in the inner Heliosphere'', \\
	Solar Physics, 256, 149, 2009 
\end{enumerate}


\listoffigures	

\listoftables  







\mainmatter





\onehalfspace

\pagestyle{fancy}




\chapter{Introduction}
\label{chap:intro}
\ifpdf
    \graphicspath{/}
\else
    \graphicspath{{1_introduction/figures/EPS/}{1_introduction/figures/}}
\fi


\hrule height 1mm
\vspace{0.5mm}
\hrule height 0.4mm
\noindent
\\	{\it In this chapter I will introduce the fundamental physics and concepts that are discussed in this thesis, beginning with a short introduction to the Sun, its various layers, atmosphere, and activity. This is followed by an introduction to coronal mass ejections (CMEs) which consists of a historical account of CME observations and their interpretation and a brief review of the modern perspective on CMEs. Finally, I discuss some of the open questions surrounding CMEs.} \\
\hrule height 0.4mm
\vspace{0.5mm}
\hrule height 1mm

\newpage

 \begin{quote}
``Physicists say we are made of stardust. Intergalactic debris and far-flung atoms, shards of carbon nanomatter rounded up by gravity to circle the sun. As atoms pass through an eternal revolving door of possible form, energy and mass dance in fluid relationship. We are stardust, we are man, we are thought. We are story.''\\
Glenda Burgess (The Geography of Love: A Memoir)
 \end{quote}
 
Since the dawn of mankind the Sun has been a source of great wonder and fascination. Some early civilisation worshipped the Sun as deity, indeed some of our oldest antiquities such as Newgrange and Stonehenge are believed to be solar observatories of a form. Building these monuments required detailed knowledge of the Sun's motion relative to the Earth, and ever since then we have been increasing our knowledge of the Sun. As science and technology advanced we began to probe the Sun's structure, composition and energy source. This led to a picture of the Sun, in terms of stellar equations, as a hot ball of gas in equilibrium, with gravitational contraction being balanced by the energy released during nuclear fusion. The modern era of space borne observations has revealed the dynamic and active nature of the Sun in the form of sunspots, filaments, prominences, flares and coronal mass ejections (CMEs). The Sun is the power source for all life on Earth and will also ultimately be the cause of its death.
 
 \section{The Sun}
 \label{sec:sun}
 The Sun, our nearest star, is a main sequence star of spectral type G2V. The Sun has a total luminosity $L_{\odot}=(3.84\pm0.04)\times10^{26}$\,W, mass $M_{\odot}=(1.9889\pm0.0003)\times10^{30}$\,kg and a radius $R_{\odot}=(6.959\pm0.007)\times10^8$\,m \citep{foukal2004solar}. As all stars, the Sun was born from a giant molecular cloud of approximate mass $10^4-10^6\,M_{\odot}$ which began to gravitationally collapse and fragment. The process of collapse and fragmentation continued until one of these fragments attained a central temperature large enough to start hydrogen fusion, about $4.6\times10^{9}$ years ago \citep{prialnik2009introduction}. At this point fusion taking place in the core produced enough energy to counterbalance the gravitational collapse. Currently the Sun is in a stable configuration, on the Main Sequence, where it is in hydrostatic equilibrium ($\nabla P = -\rho g$). The Sun will continue to maintain this stable state for about another $5 \times 10^9$ years before entering the red giant phase. At this point the Sun will expand to about 100 times its current size and begin shedding its outer layers, due to successive nuclear burning in ever more distant shells. This ultimately leads to the total loss of the outer envelope exposing a degenerate core, in which all nuclear burning has ceased, called a white dwarf \citep{phillips1995guide}.

\subsection{Solar Interior}
\label{sec:solinterior}
\figuremacroFP{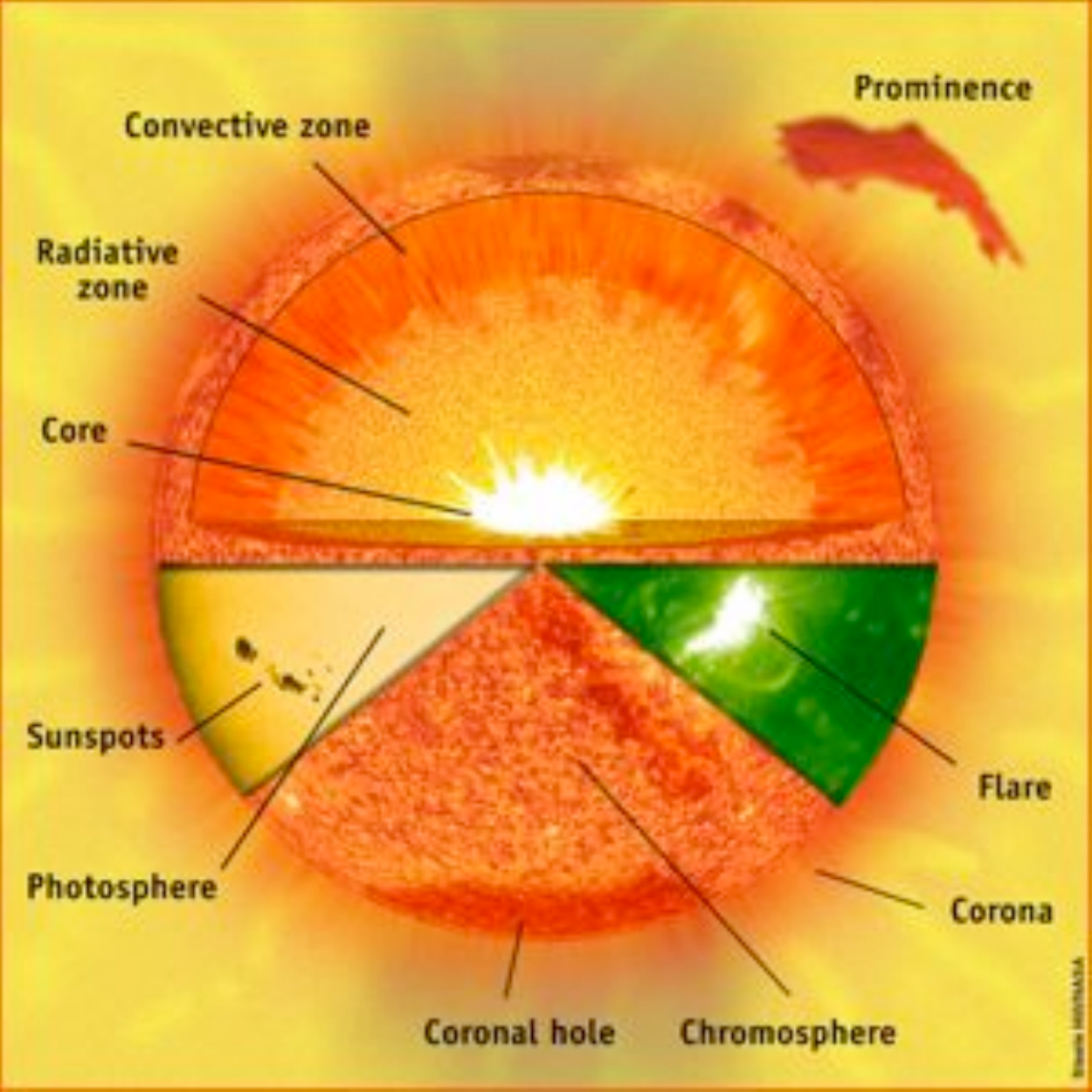}{Structure of the Sun}{At the centre of the Sun is the core ($\leq$0.25\,$R_{\odot}$) where temperatures reach $\sim$1.5$\times10^7$\,K, high enough for fusion to take place. The energy generated at the core from the fusion process is transported towards surface via thermal radiation in the radiative zone (0.25\rng0.70\,$R_{\odot}$). At this point the solar plasma is cool enough to from highly ionised atoms and becomes optically thick. As a result it is convectively unstable and energy is transported through mass motions in the convection zone (0.7\rng1.0\,$R_{\odot}$). The visible surface of the Sun, the photosphere, is a thin layer in the atmosphere where the bulk of the Sun's energy is radiated, its spectra is well matched to a blackbody with peak temperature of 5,600\,K. Above the Sun's visible surface lies the chromosphere and finally the corona where the temperature soars back up to 1\rng$2\times10^6$\,K. Image courtesy of Steele Hill \href{http://sohowww.nascom.nasa.gov/explore/images/layers.gif}{NASA/GSFC}.}{0.60}
 
The modern picture of the Sun's interior structure has been built up over time, the three most important contributions to this have been the `standard solar model' (SSM; \citealt{bahcall1982p767}), helioseismology, and solar neutrino observations. As we cannot directly observe the interior of the Sun we have to model its structure and then compare the model to observed properties, iteratively changing the model parameters until they match the observations. The SSM is essentially several differential equations derived from fundamental physics principles, which are constrained by boundary conditions (mass, radius, and luminosity). The SSM treats the Sun as a spherically symmetric, quasi-static system which is powered by nuclear reactions at its hottest part, the core. The system is assumed to start out as a cloud of primordial gas which collapses under gravity. The composition of this cloud and hence the Sun can only be altered by the process of `nuclear burning' in the core. All the energy generated by the nuclear burning is transported by radiation except where convection is a more efficient process.

The process of `nuclear burning' or nuclear fusion occurring in the core at about $1.5 \times 10^7$\,K, is the power source of the Sun. The proton-proton (p-p) chain describes the basic process which occurs:
\begin{align}
	\nucl{1}{1}{H} + \nucl{1}{1}{H} & \rightarrow \nucl{2}{1}{H} + e^{+} +\nu_{e} \\
	\nucl{2}{1}{H} + \nucl{1}{1}{H}  & \rightarrow \nucl{3}{2}{He} + \gamma \label{eq:ppsplit} \\
	\nucl{3}{2}{He} + \nucl{3}{2}{He} & \rightarrow \nucl{4}{2}{He} + 2 \nucl{1}{1}{H}
 \end{align}
 where $\nucl{1}{1}{H}$ is hydrogen, $\nucl{2}{1}{H}$ is an isotope of hydrogen (deuterium), $\nucl{4}{2}{He}$ is helium, $\nucl{3}{2}{He}$ is an isotope of helium with one neutron, $e^{+}$ a positron, $\nu_{e}$ an electron neutrino and $\gamma$ a photon. The p-p chain splits off into three branches at \eqref{eq:ppsplit} which operate simultaneously, their reaction rates are determined by the density, temperature, and elemental abundances in the core. The net result of the p-p chain is the fusion of 4 hydrogen nuclei to from a Helium nuclei or $\alpha$-particle:
\begin{equation}
	4 \: \nucl{1}{1}{H} \rightarrow \nucl{4}{2}{He}
\end{equation}
with the products weighing less than sum of the initial masses by 0.02866\,amu or $4.8 \times 10^{-29}$\,kg. Using Einstein's mass energy relation $E=mc^{2}$ this results in a energy output of $4.28 \times 10^{-12}$\,J (26.73\,MeV) of which varyingly small proportions are carried away by the neutrinos \citep{phillips1995guide, foukal2004solar}. Comparing the total energy output of the Sun ($L_{\odot}$) to the energy released during one cycle of the p-p chain we can estimate the reaction rate to be $8.97 \times 10^{37}$\,s$^{-1}$. The region where the temperatures and densities are high enough for fusion to occur is known as the core, and extends out to about 0.25\,$R_{\odot}$ (\figref{sunpartsfull}). Outside of the core is the radiative zone (0.25\rng0.70\,$R_{\odot}$) where thermal radiation is the most efficient means of transporting the intense energy generated in the core, in the form of high energy photons, outward. The temperature drops from about $7 \times 10^{6}$\,K at the bottom of the radiative zone to $2 \times 10^{6}$\,K just below the convection zone. As the radiation is thermal it is described by the Planck (blackbody) equation, the specific radiative intensity is given by:
\begin{equation}
	B_{\lambda}(T)=\frac{2hc^{2}}{\lambda^5}\frac{1}{\exp \left (\frac{hc}{\lambda k_{B} T} \right )-1} \quad \left[ \frac{\mathrm{W}}{\mathrm{m}^2\,\mathrm{sr}\,\mathrm{m}}\right]
	\label{eq:plankwav}
\end{equation}
where $h$ is the Planck constant, $c$ is the speed of light, $k$ is the Boltzmann constant. The peak of this function is given by Wein's displacement law $\lambda_{max} = 2.8977 \times 10^{-3}\,T^{-1}$\,m\,K thus the radiative zone is dominated by x-ray and gamma-ray photons. Due to the still high densities ($2\times10^{4}$\,--\,$2\times 10^{2}$\,kg\,m$^{-2}$) in the radiative zone the mean free path of the photons is very small ($\sim 9.0\times10^{-2}$\,cm) hence it can take tens to hundreds of thousands of years to escape \citep{mitalas1992p759}. As the temperature continues to fall, highly ionised atoms begin to form once an appreciable number of atoms have formed the plasma becomes optically thick, and as a result, becomes unstable as indicated by the upper limit of  the Eddington luminosity:
\begin{equation}
	\kappa F < 4 \pi c\,G m
	\label{eq:eddlim}
\end{equation}
where $\kappa$ is the opacity, $F$ the radiation flux, $G$ the gravitational constant, $c$ the speed of light and $m$ the mass coordinate \citep{prialnik2009introduction}. The sudden increase in opacity above the radiative zone causes the plasma to become convectively unstable. In this region called the convection zone (0.7--1.0\,$R_{\odot}$), convection is the dominant form of energy transport. If the mass motions are rapid enough to assume they are adiabatic, and radiation pressure negligible, the Schwarzchild criterion for stability against convection may be used:
\begin{equation}
	\left| \frac{dT}{dr} \right|_{star}\!\! = \left( \frac{\gamma -1}{\gamma} \right) \left| \frac{dP}{dr} \right|_{star} 
	\label{eq:schwarzchildcrit}
\end{equation}
where $\gamma=C_{P}/C_{V}$ is the ration of specific heats \citep{prialnik2009introduction}. This gives a lower limit on the conditions necessary for convection to occur. In other words for convection to occur the temperature gradient must be larger than the adiabatic gradient.

\figuremacro{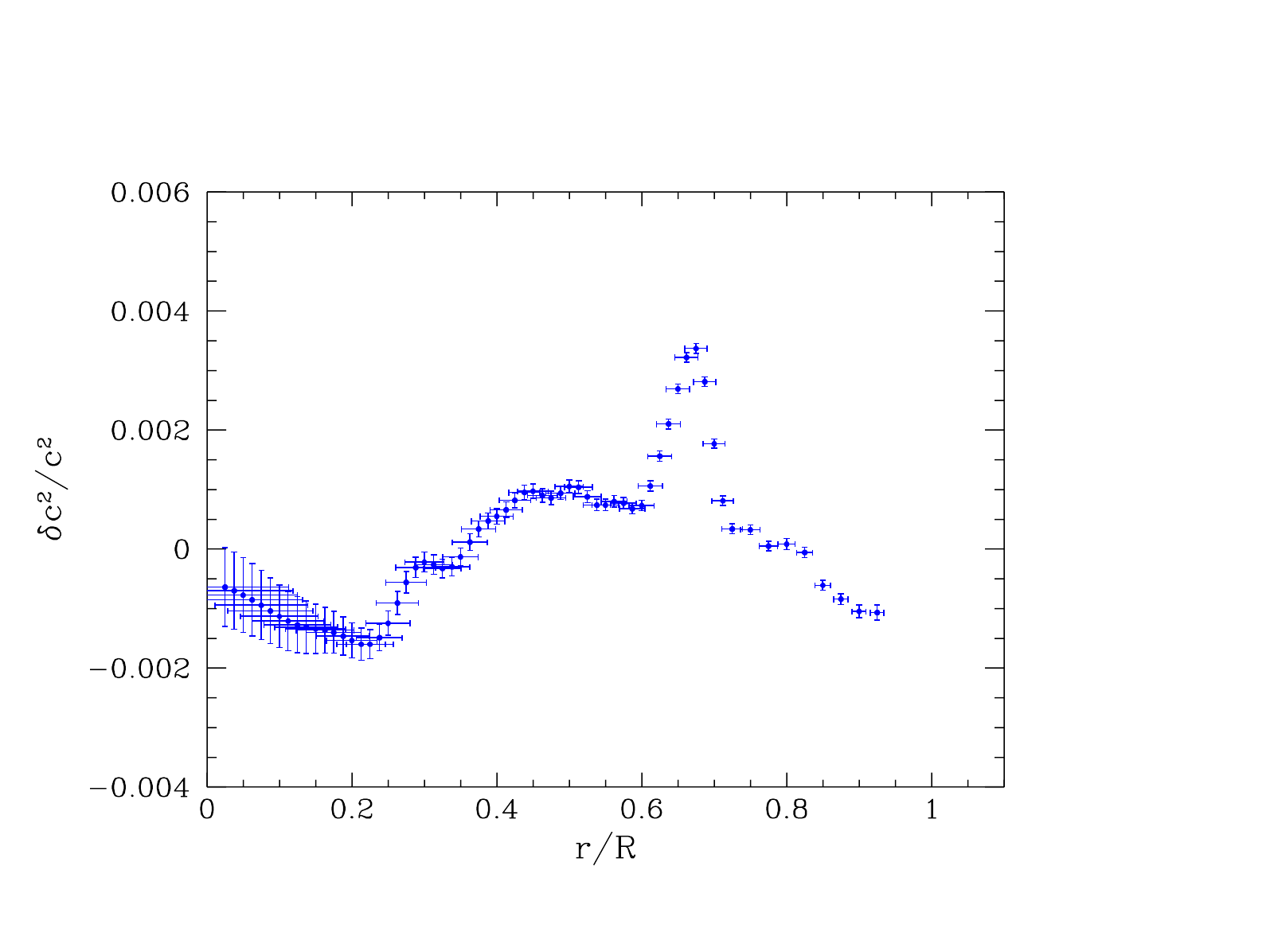}{Sound Speed in the Interior of the Sun}{This plot shows the relative differences between the squared sound speed in the Sun inferred from two months of Michelson Doppler Imager (MDI) data and from the SSM. The feature at about 0.7\,r/R corresponds to the tachocline. The horizontal bars show the spatial resolution, and the vertical bars are the error estimates. Image courtesy of \href{http://soi.stanford.edu/results/sspeed.html}{Stanford Solar Oscillations Investigation (SOI)}.}

Helioseismology allows us to probe the solar interior by studying the propagation of sound waves in the Sun. Solar pressure waves (p-modes) are believed to be generated by the turbulence in the convection zone near the surface of the sun. Only certain allowed modes (spherical harmonics) can persist, as a result the Sun `rings' like a bell. As these acoustic waves travel, they are refracted due to the waves' speed dependance on temperature, hence depth. This means that for all allowed modes the waves will be propagating normal to the solar surface when they reach it. This motion of the wave can be detected as Doppler shifts at the surface. Different modes penetrate to different depths and by combining a large number, the entire solar interior can be studied. The comparison of the properties derived from helioseismology, and that of the SSM can be used to change the model parameters to better fit the data, and the modern comparisons are extremely good as shown in \figref{helio_res}

During the fusion process a large number of neutrinos are produced and escape the Sun. These neutrinos can be detected on Earth and the flux compared to predictions from the SSM. When this was first done for the rare $\nucl{8}{}{B}$ neutrino flux \citep[p-p III chain;][]{prialnik2009introduction}, there was a large discrepancy between the predicted flux and the observed flux, which was only $\sim 0.4$ of the predicted value and led to the so-called `solar neutrino problem' \citep{bahcall2003p15}.  Results from helioseismology strongly suggested that the SSM was correct, but the experimental neutrino results were rigorously tested and found to be correct, leaving a serious issue with either the standard model for particle physics, or with the SSM. The problem was resolved when it was theorised that neutrinos could oscillate, that is an electron neutrino ($\nu_{e}$) could become a muon neutrino ($\nu_{\mu}$) as it propagated from the Sun to the Earth. This was a large step as it required the neutrino to have a small but finite mass. It became clear that the early experiment to measure the neutrino flux was only sensitive to electron neutrino ($\nu_{e}$) and so could have `missed' some of the flux. Today the $\nucl{8}{}{B}$ measured neutrino flux (from the 3 flavours, electron, muon, and tau) is $(5.44\pm0.99)\times10^6$\,neutrinos/cm$^2$\,s in good agreement with the SSM prediction of $(5.05^{+1.0}_{-0.8})\times10^6$\,neutrinos/cm$^2$\,s \citep{bahcall2003p15}.

\figuremacro{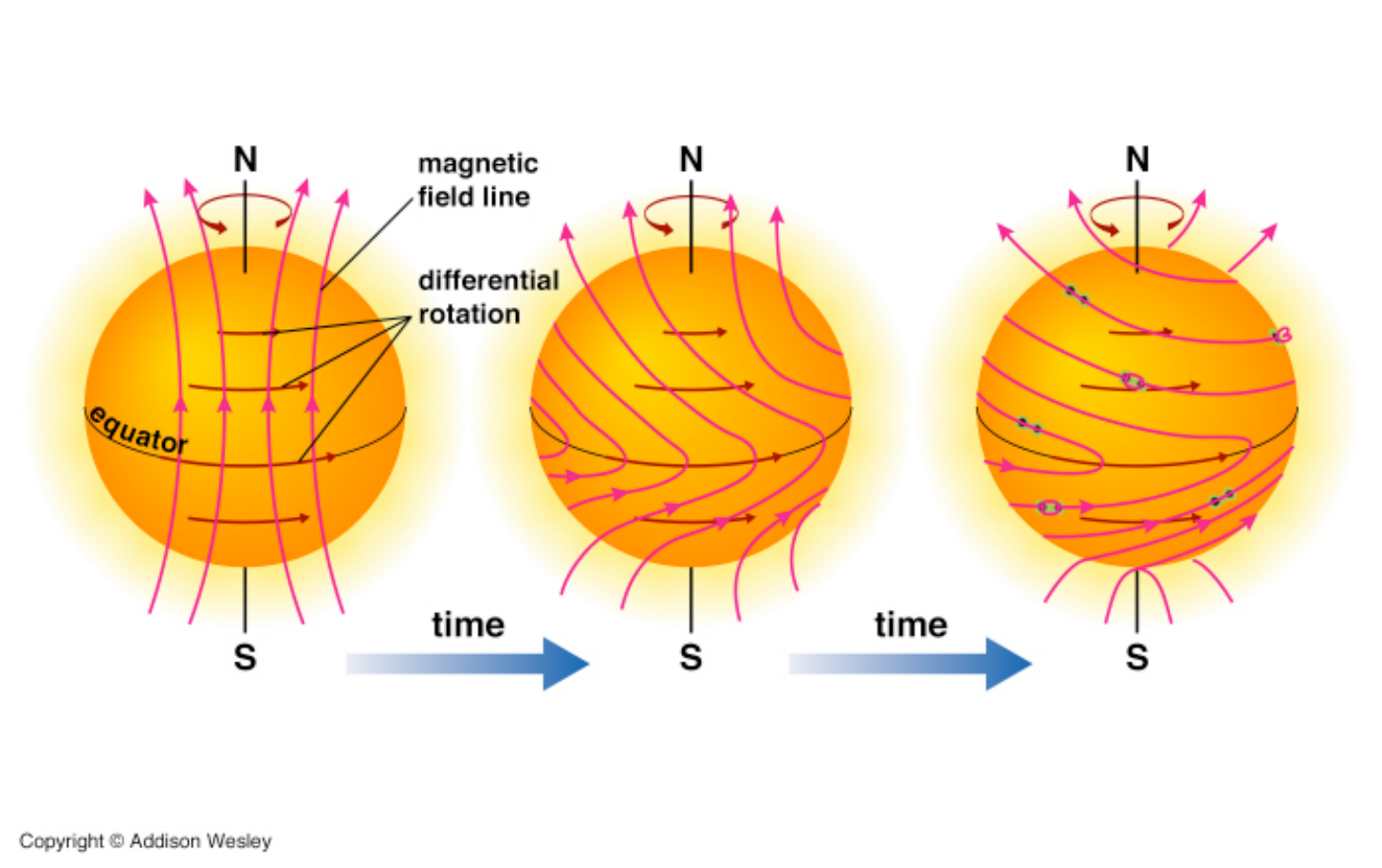}{$\alpha \Omega$-Dynamo}{{\it Left:} Shows the Sun's bipolar field. {\it Middle:} The magnetic field is being twisted by differential rotation. {\it Right:} Loops of magnetic field begin to break the surface forming sunspots. (From The Essential Cosmic Perspective, by Bennett et al)}

The core and the radiative zones of the Sun rotate rigidly (as a solid body) but the convection zone rotates differentially, there is a thin interface between the two regions known as the tachocline. Due to the meeting of the two bodies rotating at different rates this region is subjected to large shear flows. These flows are believed to be the mechanism that generates the Sun's large-scale magnetic field and powers the solar dynamo. The Sun's magnetic field is mainly dipolar and aligned to the rotation axis, thus each hemisphere has an opposite dominant polarity (\figref{sun_magnetic} left). The differential rotation of the convection zone winds-up this field. This large scale twisting which transforms poloidal field to toroidal field is know as the $\Omega$-effect (\figref{sun_magnetic} middle). As the field is twisted up the magnetic pressure increases and bundles of magnetic field lines (flux ropes) can become unstable and rise up in the from of loops. Due to solar rotation, the Coriolis effect twists these loop back towards north-south orientation reinforcing the original poloidal field, this is known as the $\alpha$-effect (\figref{sun_magnetic} right) and completes the $\alpha \Omega$-dynamo. When magnetic loops become buoyant and rise up through the surface they are visible as sunspots on-disk and mark the footprints of large loops which extend into the solar atmosphere. In a given hemisphere the leading sunspot and trailing sunspot will have opposite polarities, this order is reversed in the other hemisphere (Hale's Law). Also the tilt angle of the sunspots pairs have a mean value of 5.6$^{\circ}$ relative to the solar equator (Joy's Law). Sunspots are known to migrate from high latitudes towards the equator over an 11 year cycle (Sporer's Law; see \figref{ButterflyDiagram}). The net affect is an increase in opposite polarity field at the poles, ultimately the majority of the field will be oppositely oriented and the dipole will flip. This occurs every 11 years, thus a complete cycle takes 22 years (N to S to N). The activity of the Sun, in the form of active regions, flares, transient events, and other associated phenomenon, is modulated by this cycle (see \figref{ButterflyDiagram} lower).

\figuremacro{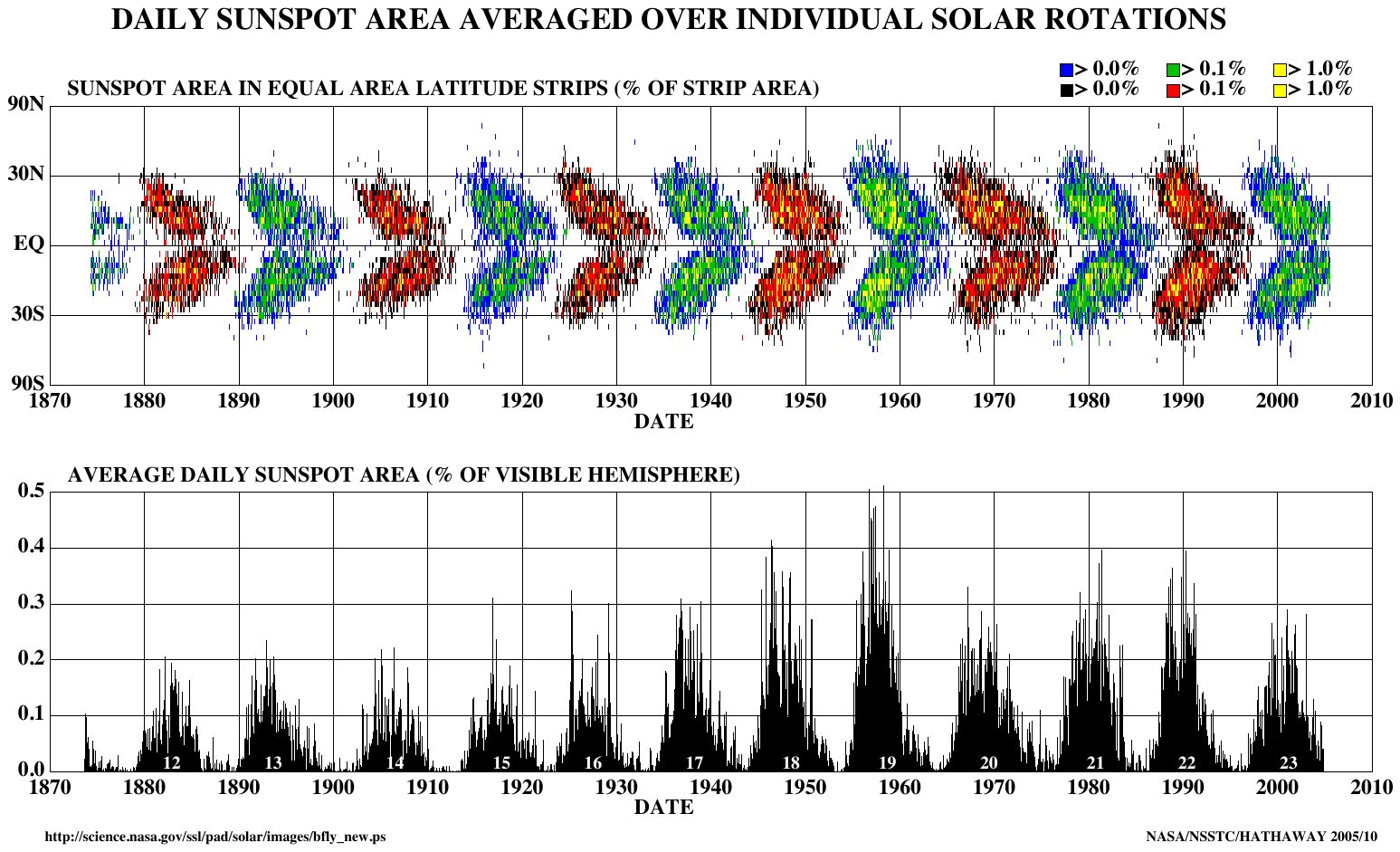}{Butterfly Diagram}{The position of the sunspots in equal area latitude strips, averaged over a solar rotation with respect to time (top). The butterfly pattern is clear as the decrease in the upper limit of sunspot latitudes with time. (bottom) The average sunspot area as a function of time. The 11 year modulation is clear in both of these plots. Image courtesy of \href{http://solarscience.msfc.nasa.gov/images/l}{NASA MSFC}.}

\subsection{Solar Atmosphere}
\label{sec:solatmosphere}
\figuremacroW{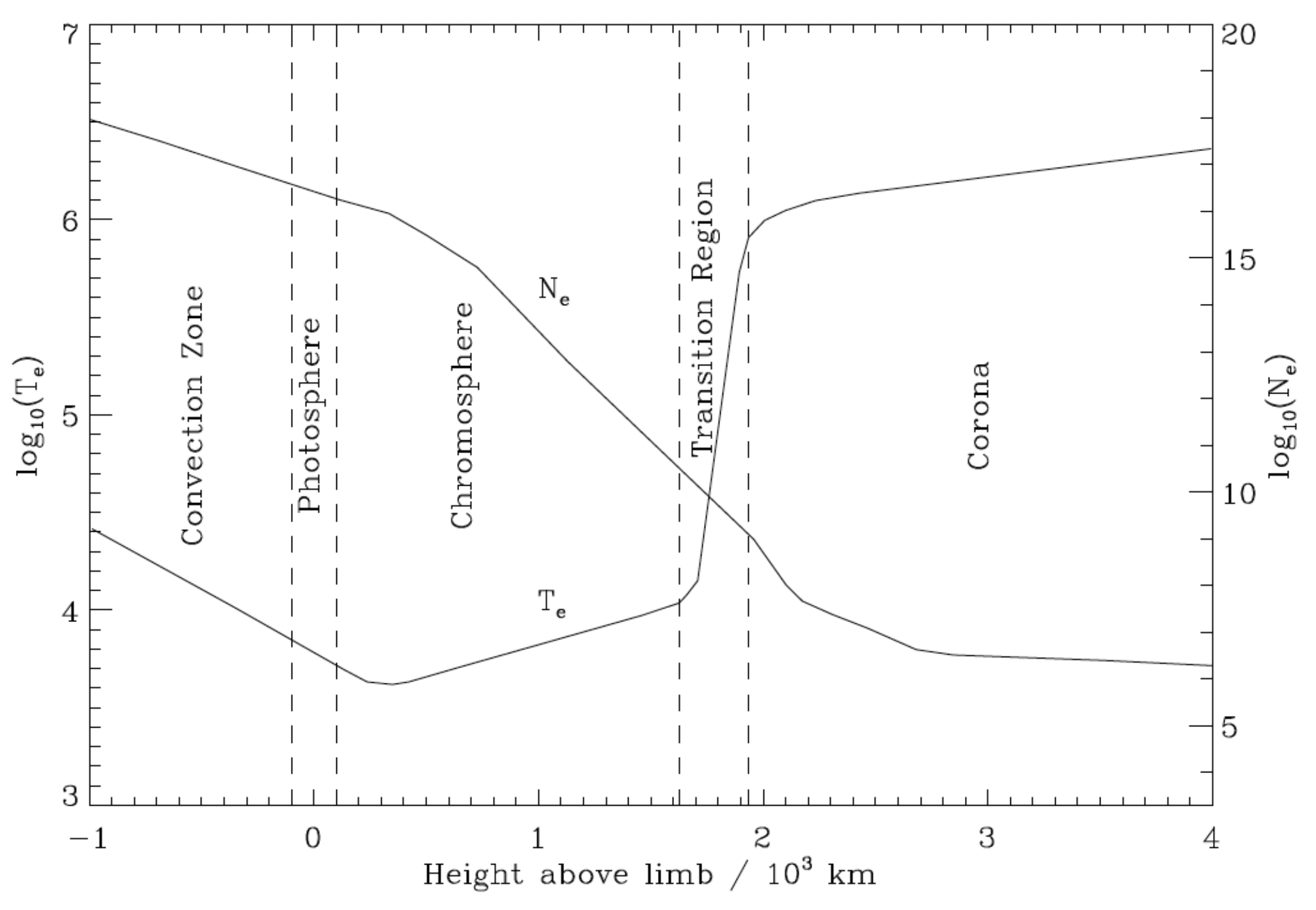}{Temperature and Electron Density in the Solar Atmosphere}{A 1D model of the electron density $N_{e}$ [cm$^{-3}$] and temperature $T_{e}$ [K] profile thought the solar atmosphere from \citet{gallagher2000PhD} after \citet{gabriel1982p245}. Neutral atoms are present in the photosphere and chromopshere but the plasma is fully ionised in the corona due to the higher temperature.}{0.8}

The Sun's atmosphere is composed of all the regions above the photosphere. Until now we have referred to the photosphere as the visible surface of the Sun, it is in fact a very thin layer of the solar atmosphere. The solar atmosphere is usually separated into three regions, the photosphere, chromosphere and corona based on their density, temperature, and composition as shown in \figref{atmos_model}. However, this separation is a simplification as the atmosphere is an in-homogenous mix of different plasma properties due to up-flows, down-flows, heating, cooling and other dynamic processes.     The density of the plasma generally decreases through these regions with increasing height. The temperature decreases, reaching a minimum in the chromosphere, then slowly rises until there is a rapid increase at the transition region which continues into the corona. This rapid increase in temperature leads to the so-called `coronal heating problem'.

An important parameter in describing the solar atmosphere is the plasma-$\beta$ term, the ratio of the thermal to magnetic pressures:
\begin{equation}
	\beta = \frac{p_{th}}{p_{mg}}=\frac{nk_{B}T}{B^{2}/2 \mu_{0}},
\end{equation}
where $n$ is the number density and $\mu_{0}$ the permeability of free space. In the photosphere the plasma-$\beta$ is large and the plasma motions carry the field with them (\figref{plasmabeta}). Ascending into the chromosphere and corona the plasma-$\beta$ becomes small and the plasma is constrained to follow the magnetic fields. Continuing upwards the plasma-$\beta$ drops again and the magnetic field is advected out with the solar wind plasma flow and ultimately forms the Parker spiral (\figref{plasmabeta}).

\figuremacroW{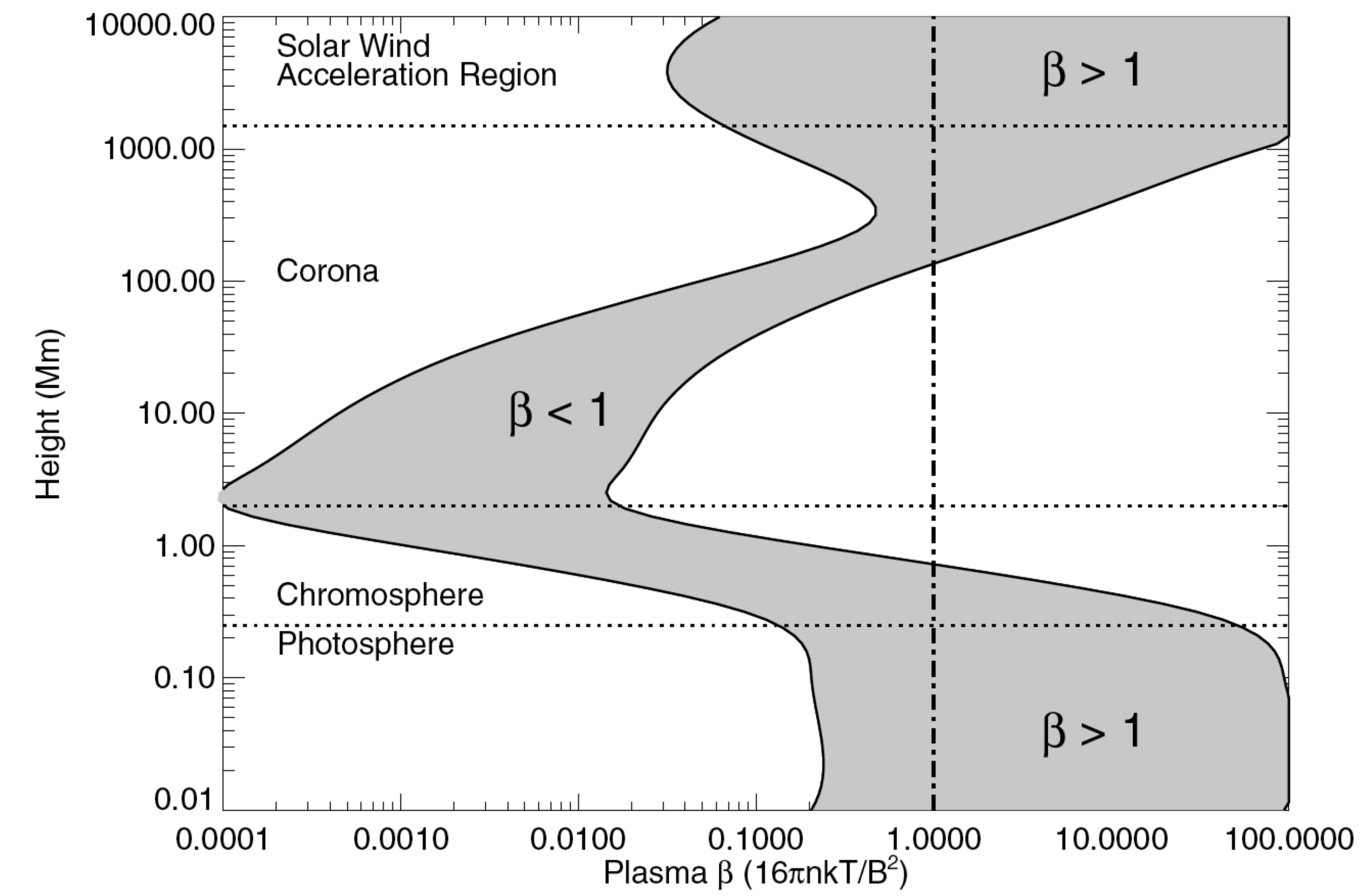}{Plasma-$\beta$ in the Solar Atmosphere}{Plasma-$\beta$ in the solar atmosphere as a function of height for two magnetic field strengths of 100\,G and 2500\,G. The layers of the atmosphere are segregated by the dotted lines. The corona is the only region in which the magnetic pressure dominates over the thermal pressure, a low $\beta$ plasma \citep{aschwanden2006physics}.}{0.9}

\subsubsection{Photosphere}
The photosphere is the visible surface of the Sun and is defined as the height where the optical depth, at visible wavelengths, equals 2/3 ($\tau_{5000} \approx 2/3$, $I = I_{0}e^{-\tau}$). This is the mean optical depth at which the photospheric radiation is emitted or where the effective temperature ($T_{eff}=5776$) and blackbody temperature of the photosphere match. This can be seen substituting $B(T)=\sigma / \pi T^4$ and $F = \sigma T^{4}_{eff}$, where $F$ is the total radiative flux, in the general solution of the radiative transfer equation:
\begin{align}
B(\tau) &= \frac{3}{4}( \tau + 2/3) \frac{F}{ \pi}
\end{align} which gives, \begin{align}
\sigma T^4 &= \frac{3}{4}(\tau + 2/3) \sigma T^{4}_{eff}
\end{align}
implying that $\tau=2/3$ \citep{foukal2004solar}. The temperature drops from 6,400\,K at the base of the photosphere to 4,400K at the top. The spectrum of photospheric radiation is that of a blackbody with a large number of absorption features, Fraunhofer lines, due the upper layers of the atmosphere superimposed on it. The photospheric number density ranges from $\sim$10$^{19}$\rng$10^{21}$\,m$^{-3}$ over the depth of the photosphere (500\,km).

One of the main observable features in the photosphere is granulation due to the convective motions. Granules are small-scale features made up of brighter regions isolated by darker lanes, this interpreted as the upflow of hot, bright material to the surface which then flows horizontally and cools, flowing back down in the dark lanes. The plasma-$\beta$ is much larger than one throughout the photosphere which means the magnetic fields are tied to the flows. Typical granules are of the order of 1,000\,km in diameter and have lifetimes of 5\rng10 minutes with vertical flow velocities of hundreds to thousands km\,s$^{-1}$. There are also larger scale flow patterns know as mesogranulation and supergranulation. Mesogranules are typically 7000\,km in diameter, have lifetime of  hours with vertical flows of the order of tens of m\,s$^{-1}$. Supergranules are larger still at diameters of $3 \times 10^{4}$\,km, and consequently have longer lifetimes of days, they have large horizontal flows and smaller vertical flows of the order of 0.5\,km\,s$^{-1}$. Sunspots are also found in the the photosphere they appear as darker regions due to their lower temperature (4,000\,K) as convection is suppressed by the strong magnetic fields (kG). Sunspots play an important role in the activity of the Sun as they are the source of solar flares and many CMEs.
 
\subsubsection{Chromosphere}
The chromosphere lies above the photosphere, the temperature initially decreases to a minimum of $\sim$\,4,500\,K before increasing to $\sim$\,20,000\,K with increasing height. It occupies a region approximately 2,000\,km thick and with a density of about 10$^{16}$\,m$^{-3}$ but the mass density decreases by a factor of 10$^6$. The structure of the chromosphere is split between the hot bright magnetic network and the cooler darker internetwork \citep{Gallagher1999p251}. Jet-like structures, called spicules, with diameters of hundreds of kilometers and attaining heights of tens of thousand of kilometers, with flows of the order of 30\,km\,s$^{-1}$ lasting 5\rng10 minutes are ubiquitous.

The source of the chromosphere's increasing temperature is not fully understood, but looking at the details we can infer some physical properties. The initial decrease in temperature is due to the decrease in the density of H$^{-}$ ions, decreasing the plasmas' ability to absorb radiation from below, thus the temperature falls. Further out, some non-radiative form of energy is deposited, energy which ionises the hydrogen. The free electrons produced excite atoms which de-excite by line emission such as {H-$\alpha$}, Ca~{\sc ii} and Mg~{\sc ii}. There is a balance between the energy input and radiative losses, forming a broad plateau in temperature at about 6,000\,K. This balance is limited by the supply of neutral hydrogen, as this decreases the number of neutral or partially ionised atoms decreases, and the temperature rapidly rises. At about 20,000\,K there is thought to be another plateau due to Lyman-$\alpha$, but as height increases the ionisation of hydrogen increases and Lyman-$\alpha$ emission can no longer balance the energy input and the temperature rises rapidly. This transition marks the edge of the chromosphere and the start of the transition region.

While the nature of the heating mechanism is unclear, from the observations it is clear there must be some form of energy deposition occurring. Neither radiation nor conduction can not be the source as the temperature is lower at the base of the lower chromosphere and photosphere than in chromosphere proper (and would thus violate the laws of thermodynamics). Mass motions are neither observed nor applicable since the chromosphere is in hydrostatic equilibrium. The most likely source of the energy (heat flux) is the dissipation of compressional or sound waves as proposed by \citet{biermann1946p118} and \citet{schwarzschild1948}. In this paradigm, the convective plasma motions of the photosphere, launches sound waves into the chromosphere which travel upwards with little dissipation. As the density drops, the waves steepen and form shocks which rapidly dissipate energy, heating the chromosphere. This type of acoustic heating is not appropriate in the network regions where the strong magnetic field suppress the convective motions which drive the waves. This led to the idea of Alfv\'{e}n wave heating, first introduced by \citep{Osterbrock1961p347}. Aflv\'{e}n waves are magneto-hydrodynamic waves which propagate along magnetic fields, the restoring force is provided by magnetic tension and the ion mass provides the inertia. Here the magnetic field itself is responsible for transporting and depositing the energy from the photospheric motions. This type of heating matches well with observations of plage and emerging flux regions, which both show strong heating, implying the heat flux is related to the magnetic field strength.

Filaments are seen as dark channels in on-disk H$\alpha$ observations often over active regions or as prominences when observed on the limb as bright features. Spicules which are jets of plasma are also observed on the limb, typically reaching heights of $\sim$3,000\rng10,000\,km above the solar surface and lasting only $\sim$5\rng15 minutes.
The transition region lies between the chromosphere and corona, here the temperature rapidly jumps (over 100\,km) to above 1\,MK. Above the transition region the magnetic field dominates and determines the structures. The high temperatures result in prominent emission from carbon, oxygen and silicon ions in the UV and EUV.

\subsubsection{Corona}
The tenuous, hot, outer layer of the atmosphere is known as the corona. The electron density of the corona ranges from $\sim$10$^{14}$\,m$^{-3}$ at its base, 2,500\,km above the photosphere, to $\lesssim$10$^{12}$\,m$^{-3}$ for heights $\gtrsim$1\,$R_{\odot}$ \citep{aschwanden2006physics}. The density varies depending on the feature, the open magnetic structures of coronal holes can have densities in the region of (0.5\rng1.0)$\times10^{14}$\,m$^{-3}$, streamers (3\rng5)$\times10^{14}$\,m$^{-3}$ while active regions have densities in the region of $2\times10^{14}$\rng$10^{15}$\,m$^{-3}$. The temperature in the corona is generally above $1\times10^{6}$\,K but again varies across different coronal features. Coronal holes have the lowest temperature (less than $1\times10^{6}$\,K) followed by quiet Sun regions at 1\rng2$\times10^{6}$\,K,  and active regions are the hottest at  2\rng6$\times10^{6}$\,K with flaring loops reaching even higher temperatures. The high temperatures reached in the corona give rise to EUV and X-ray emission with highly ionised iron lines being a prominent feature. The visible corona during eclipses is due to Thomson scattering of photospheric light from free electrons in the coronal plasma. The corona has a number of components:
\begin{itemize} 
	\item K-corona (kontinuierliches spektrum) is composed of Thomson-scattered photospheric radiation and dominates below $\sim$2\,$R_{\odot}$. The scattered light is strongly polarised parallel to the solar limb as a result of the Thomson scattering mechanism. The high temperatures mean the electrons have high thermal velocities which wash out (due to thermal broadening) the Fraunhofer lines, producing a white-light continuum. The intensity of the K-corona is proportional to the density summed along the line-of-sight.
	
	\item F-corona (Fraunhofer corona) is composed of photospheric radiation Rayleigh-scattered off dust particles, and dominates above $\sim$2\,$R_{\odot}$. It forms a continuous spectrum with the Fraunhofer absorption lines superimposed. The radiation has a very low degree of polarisation. The F-corona is also know as Zodiacal light and can be seen with the naked eye at dawn or dusk under favourable conditions.
	
	\item E-corona (Emission) is composed of line emission from visible to EUV due to various atoms and ions in the corona, containing many forbidden line transitions, thus it contains many polarisation states. Some of the strongest lines are Fe~{\sc xiv} 530.3\,nm (green-line; visible), H-$\alpha$ at 656.3\,nm (visible), and Lyman-$\alpha$ 121.6\,nm (UV).
	
	\item T-corona (Thermal) is composed of thermal radiation from heated dust particles. It is a continuous spectrum according to the temperature and colour of the dust particles.
\end{itemize}
 
 The earliest descriptions of the corona were based on a static model with a heat input at some level $r_{0}$ which is described by $T_{0}$, in a spherically symmetrical corona \citep{Chapman1957}. The goal then was to derive the density, pressure, and temperature with respect to this reference level. The equation of hydrostatic equilibrium is:
\begin{align}
 	\frac{dp}{dr} = -\rho\frac{GM_{\odot}}{r^{2}}
	\label{hydro}
\end{align}
where the density of the plasma is $\rho = nm_{p}$ and the pressure due to electrons and protons is $p=2nk_{B}T$. Rewriting the pressure in terms of density, and substituting into \eqref{hydro} gives:
\begin{align}
	\frac{dp}{p} = -\frac{GM_{\odot}m_{p}}{2 k_{B}T}\frac{dr}{r^{2}}
\end{align}
and integrating:
\begin{align}
	\label{hydropressure}
	p(r) = p_{0}\exp \left ( -\frac{GM_{\odot}m_{p}}{2 k_{B}}\int_{r_{0}}^{r} \frac{dr}{r^{2}T}  \right ).
\end{align}

Due to the high coronal temperatures, conduction should play an important role. The temperature distribution is determined by the conservation of conductive flux $q = k \nabla T$ where $k$ is the thermal conductivity which in the absence of sources or sinks reduces to:
\begin{align}
	\nabla \cdot q = 0.
\end{align}
In the case of spherical symmetry, this can be written:
\begin{align}
	\frac{1}{r^{2}} \frac{d}{dr} \left ( r^{2} k \frac{dT}{dr} \right ) = 0. 
\end{align}
this implies that:
 \begin{align}
 	\label{eq:hcon}
	 r^{2} k \frac{dT}{dr}=\textrm{constant}.
\end{align}
For a fully ionised hydrogen plasma to a good approximation we have $k(T)=k_{0}T^{5/2}$ \citep{Spitzer1962} thus we can write \eqref{eq:hcon}:
 \begin{align}
 	r^{2}T^{5/2}\frac{dT}{dr}=\textrm{constant}
 \end{align}
 integrating yields:
 \begin{align}
	T(r) = T_{0}\left ( \frac{r_{0}}{r} \right )^{2/7}.
\end{align}
The integral from \eqref{hydropressure} may be evaluated:
\begin{align}
\int_{r_{0}}^{r} \frac{dr}{r^{2}T_{0}\left ( \frac{r_{0}}{r} \right )^{2/7}}  = \frac{7}{5} \frac{1}{T_{0}r_{0}} \left ( 1 - \left ( \frac{r_{0}}{r} \right )^{5/7} \right )
\end{align}
 and from \ref{hydropressure} pressure is given by:
 \begin{align}
 	\label{staticp}
 	p(r) = p_{0}\exp \left ( - \frac{GM_{\odot}m_{p}}{2 k_{B}}  \frac{7}{5} \frac{1}{T_{0}r_{0}} \left ( 1 - \left ( \frac{r_{0}}{r} \right )^{5/7} \right )  \right ),
 \end{align}
 and similarly for density:
 \begin{align}
 	\label{staticrho}
  	\rho(r) = \rho_{0} \left( \frac{r_{0}}{r} \right )^{2/7}   \exp \left ( - \frac{GM_{\odot}m_{p}}{2 k_{B}}  \frac{7}{5} \frac{1}{T_{0}r_{0}} \left ( 1 - \left ( \frac{r_{0}}{r} \right )^{5/7} \right )  \right ).
 \end{align}
 
Inspecting \eqref{staticrho} and \eqref{staticp}, as $r \rightarrow \infty$, the pressure tends to a constant value while the density goes to infinity which is clearly unphysical. Even ignoring the density issue, the pressure value the static model tends to is some seven orders of magnitude greater than pressure in the interstellar medium (ISM) which must provide a boundary condition at large radii. 
 
\subsection{Solar Wind and Heliosphere}
\label{ssec:solwin}
Eugene Parker was one of the next scientists to tackle the problem and ultimately solve it in the form of the solar wind. To explain the observation that comet tails always point away from the Sun, both when approaching it and receding from it, Biermann, in 1953 suggested that there must be a continuous outflow from the Sun \citep[][and references therein]{Biermann1957p109} . \citet{1958parker664} was the first person to make the connection between Biermann's and Chapman's work: that the heat flow in Chapman's static corona could drive the stream of particles (solar wind) Biermann speculated must exist.

The Parker model is a spherically symmetric, static, isothermal model of the corona with only radial flows. The conservation of mass equation ($\nabla \cdot (\rho v) =0$) for this system is:
\begin{align}
 	\frac{d}{dr}(r^{2} \rho \mathbf{v}) = 0,
 \end{align}
 therefore
 \begin{align}	
	r^{2} \rho v = \textrm{constant}.
 \end{align}
 The radial component  of the momentum equation ($\rho( \mathbf{v}\cdot \nabla \mathbf{v}) = -\nabla P + \rho  \mathbf{g} $) can then be written:
 \begin{align}
 	\rho v \frac{dv}{dr} = - \frac{dp}{dr} - \frac{GM_{\odot}\rho}{r^{2}} \nonumber \\
	v \frac{dv}{dr} = - \frac{1}{\rho}\frac{dp}{dr} - \frac{GM_{\odot}}{r^{2}}.
	\label{parkerform}
 \end{align}
As the model is isothermal we can write the equation of state $p=2 \rho k_{B} T /m_{p}$ so \mbox{\eqref{parkerform} can} be written by eliminating $\rho$ using the equation of state:
\begin{align}
	\left ( v - \frac{2 k_{B}T}{m_{p}} \frac{1}{v} \right ) \frac{dv}{dr} = \frac{4 k_{B} T}{r m_{p}} - \frac{GM_{\odot}}{r^{2}}.
	\label{parkerform2}
\end{align}
A critical point occurs when $dv/dr \rightarrow 0$ hence we define:
\begin{align}
	v_{c} = \sqrt{\frac{2k_{B}T}{m_{p}}} \qquad r_{c} = \frac{GM_{\odot}}{2v_{c}^{2}} 
\end{align}
and rewrite \eqref{parkerform2}:
\begin{align}
	\label{parkerdiff}
	\left ( v^{2} - v_{c}^{2} \right ) \frac{1}{v} \frac{dv}{dr} = 2\frac{v_{c}^{2}}{r^{2}} \left ( r - r_{c} \right ).
\end{align}
Integrating \eqref{parkerdiff} equation yields:
\begin{align}
	\left ( \frac{v}{v_{c}} \right )^{2} - \ln 	\left ( \frac{v}{v_{c}} \right )^{2} = 4 \ln 	\left ( \frac{r}{r_{c}} \right ) + 4 \frac{r_{c}}{r} + C,
	\label{parker_sols}
\end{align}
Parker's equation.
\figuremacroW{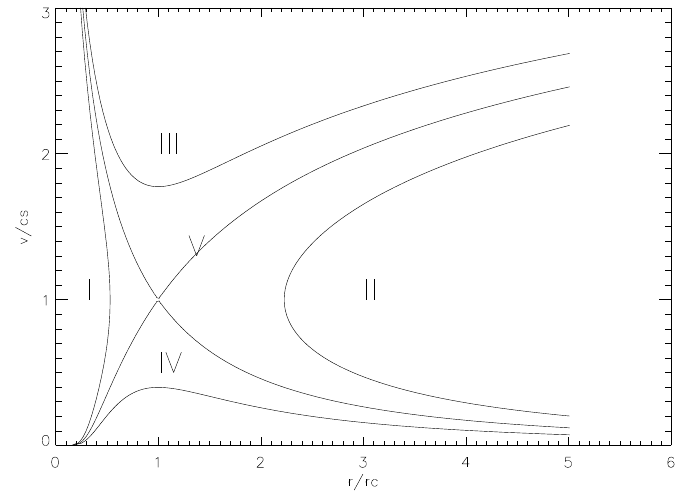}{Parker's Solar Wind Solutions}{The solar wind velocity $v$ as a function of radius $r$ for various values of the constant $C$. The five different classes of solution are indicated by the labels I-V. cs is the sound speed $v_{c}$ and rc is the sonic radius $r_{c}$}{0.8}

The various possible solutions to \eqref{parker_sols} depend on the constant $C$ and are shown in \figref{parker_sw_sols}. While they are all mathematically valid most have unphysical properties. Solution I is double-valued which is unphysical as it implies the SW leaves the solar surface with sub-sonic velocities, reaches a maximum radius and returns to the surface at super-sonic velocities. Solution II is double-valued and never reaches the solar surface so is clearly unphysical. Solution III starts at super-sonic velocities at the solar surface which is not observed so this solution is neglected. Solution IV know as the `solar breeze' because inspecting \figref{parker_sw_sols} we see that as $r \rightarrow 	\infty$ that $v \rightarrow 0$ thus at large distances we may approximate \eqref{parker_sols} with 
\begin{align}
	-\ln \left ( \frac{v}{v_{c}} \right )^{2} \approx 4 \ln \left ( \frac{r}{r_{c}} \right ) \implies \frac{v}{v_{c}} \approx \left ( \frac{r_{c}}{r} \right )^{2} \implies r^{2} v \approx r_{c}^{2} v_{v} = \text{constant}.
\end{align}
 Inserting this into the mass continuity equation $\rho = \text{const}/r^{2}v=\text{const}/r_{c}^{2}v_{c}$ the density tends to a constant and as the plasma is isothermal the pressure must also be a finite constant. This is not physical as the extremely small ISM pressure cannot balance the SW pressure. Solution V crosses through a critical point at $r = r_{c}$, $v=v_{c}$ called the sonic point. This point can be used to constrain the integration constant $C=-3$. Again inspecting \figref{parker_sw_sols} we see as $r \rightarrow \infty$ that $v \gg v_{c}$ so \eqref{parker_sols} is approximated by
\begin{align}
 	\left ( \frac{v}{v_{c}} \right )^{2} \approx 4 \ln \left ( \frac{r}{r_{c}} \right ) \implies  \frac{v}{v_{c}} \approx 2 \sqrt{\ln \left( \frac{r}{r_{c}}\right )}.
 \end{align}
 From the mass continuity equation the density is given by
 \begin{align}
 	\rho = \frac{\text{const}}{r^{2}v} \approx \frac{\text{const}'}{r^{2} \sqrt{\ln \left( \frac{r}{r_{c}}\right ) }}
 \end{align}
so the density will tend to zero, and as the plasma is isothermal so will the pressure thus this solution can match the ISM pressure. So the Parker solar wind solution is given by solution V which starts off sub-sonically at the solar surface, the velocity increases monotonically with height, reaching the sound speed at the critical point (or sonic point), and propagates super-sonically thereafter. We can estimate the critical radius and velocity by assuming a typical temperature for the solar wind of $T \approx 10^{6}$\,K, deriving $v_{c}\approx 120$\,km\,s$^{-1}$ and $r_{c} = 12$\,R$_{\odot}$ which corresponds to a solar wind velocity at 1\,AU of 430\,km\,s$^{-1}$ very close to the measured velocity of $\sim$400\,km\,s$^{-1}$. In 1959 the first indications of the presence of the solar wind came from the Russian Lunik III and Venus I spacecraft and was confirmed in 1962, just four years after its prediction, by Mariner II spacecraft measurements analysed by \citet{1962Neugebauerp1095}.

As the plasma-$\beta$ of the solar wind is much larger than one, (see \figref{plasmabeta}) the magnetic field will be advected outward with the solar wind as the field lines are `frozen in' (see Section \ref{sec:mhdeqs}). The combination of this radial advection with the rotation of the Sun forms what is known as the Parker spiral. If we assume that solar gravitation and solar wind acceleration can be neglected beyond some distance $r_{0}$, then the radial outflow velocity ($v_{r}$) can be approximated by a constant $v$. In a spherical coordinate system which rotates with the Sun we can write the velocity components as:
\begin{align}
	v_{r}=v, \qquad v_{\theta}=0, \qquad v_{\phi} = \omega \left ( r - r_{0} \right ) \sin \theta
\end{align}
where $\omega$ is the angular velocity of the Sun ($\omega = 2.7 \times 10 ^{-6}$\,rad\,s$^{-1}$). A differential equation for the velocity stream lines can be obtained from $\mathbf{v} \times d\mathbf{S}=0$:
\begin{align}
	\frac{dr}{v_{r}} =  \frac{r d \theta}{v_{\theta}} = \frac{r \sin \theta d \phi}{v_{\phi}}
\end{align} 
Integrating this equation from $r_{0}$ to $r$ and from $\phi_{0}$ to $\phi$ gives:
\begin{align}
	\frac{r}{r_{0}} - 1 - \ln \left ( \frac{r}{r_{0}} \right ) = \frac{v}{r_{0} \omega} \left ( \phi - \phi_{0} \right ),
\end{align} when $r>r_{0}$ this equation can be approximated by:
\begin{align}
	( r - r_{0}) \approx \frac{v}{\omega} (\phi - \phi_{0}) 
\end{align}
which is in the form of an Archimedean spiral shown in \figref{parkerspiral}.

\figuremacroW{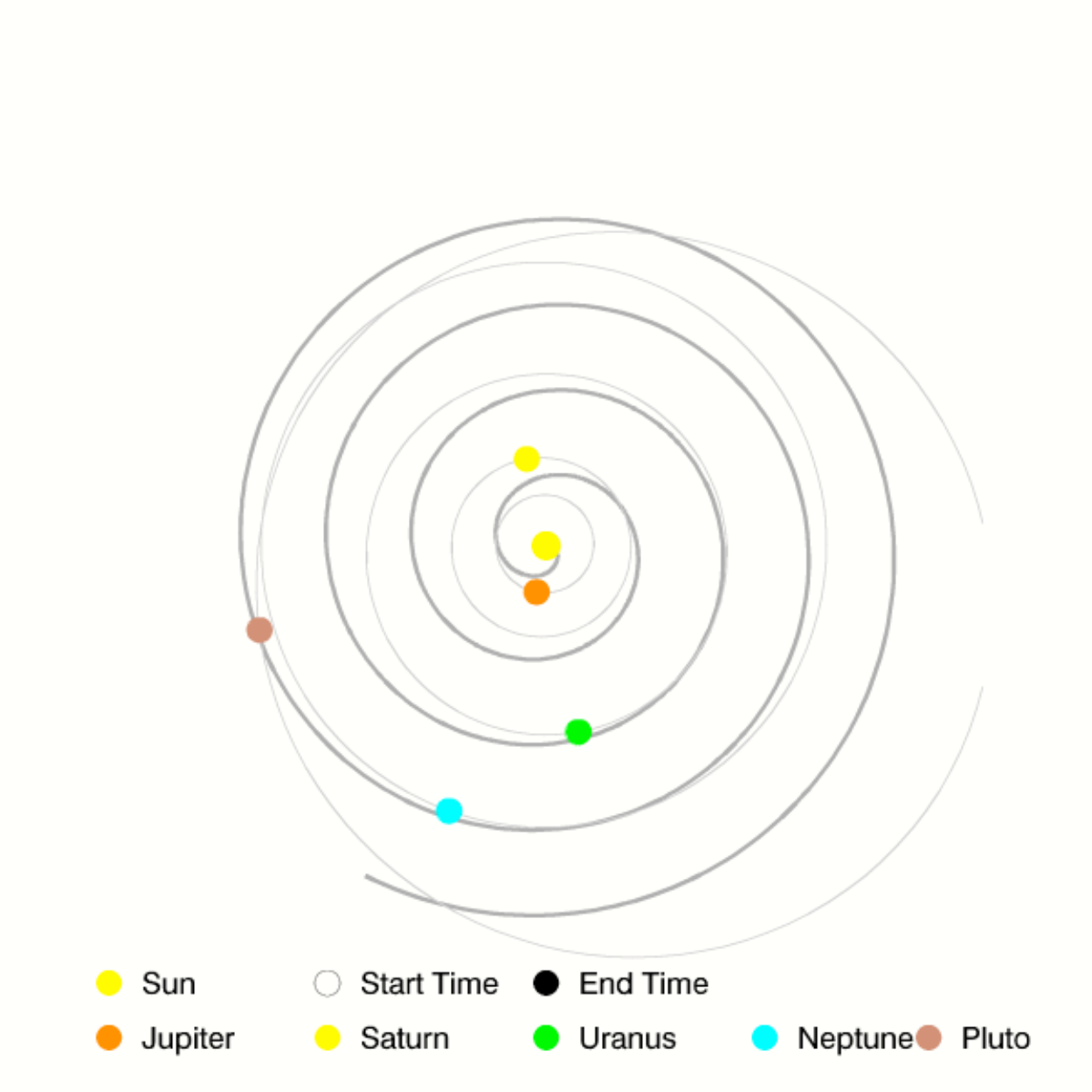}{Parker Spiral}{The interplanetary magnetic field showing the Parker spiral geometry for a solar wind speed of 600\,km\,s$^{-1}$ and a field line starting at a Carrington longitude of 0$^{{\circ}}$.}{0.75}

As the magnetic field `frozen in' to the flow and considering that $\nabla \cdot \mathbf{B}=0$ we may write:
\begin{align}
	B_{r}(r, \theta, \phi) &= B(\theta, \phi_{0}) \left ( \frac{r_{0}}{r} \right )^{2} \nonumber \\
	B_{\theta}(r, \theta, \phi) &= 0 \nonumber \\
	B_{\phi}(r, \theta, \phi) &= B(\theta, \phi_{0}) \left ( \frac{\omega}{v} \right ) (r-r_{0}) \left ( \frac{r_{0}}{r} \right )^{2} \sin \theta  \nonumber \\
\end{align} where $B(\theta, \phi_{0})$ is the magnetic field at $r = r_{0}$. The angle of the magnetic field with respect to the radial direction can be obtained from:
\begin{align}
	\tan \psi = \frac{B_{\phi}}{B_{r}} = \left ( \frac{\omega}{v} \right ) (r-r_{0}) \sin \theta
\end{align} and when $r$ is large this is approximated by:
\begin{align}
	\tan \psi = \frac{B_{\phi}}{B_{r}} =\left  ( \frac{\omega r}{v} \right ) \sin \theta. 
\end{align}
Inserting typical values of the solar wind at Earth of $v=400$\,km\,s$^{-1}$, $\omega=2.7\times10^{-6}$\, rad\,s$^{-1}$ and using $r=1$\,AU\,$=1.5\times10^{8}$\,km and $\sin(90^{\circ}) = 1$ we find $\psi \sim 45^{\circ}$ close to the measured value \citep{goossens2003}.


\figuremacroW{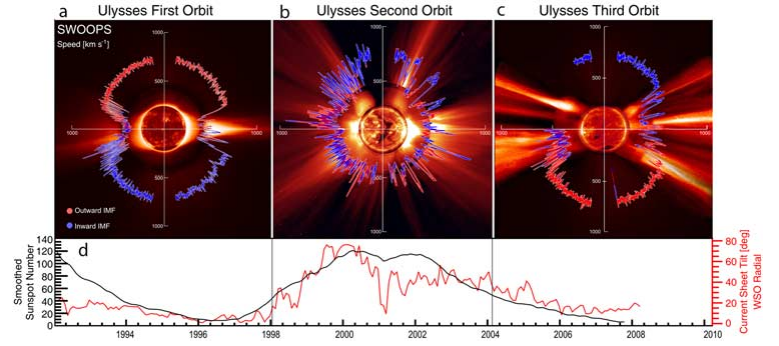}{Solar Wind Speed Measured by Ulysses}{(a-c) Polar plots of solar wind speed with the magnetic field polarity indicated by colour for three Ulysses orbits. The background image are blended composites from SOHO/EIT 195\,{\AA}, the Mauna Load K coronameter and SOHO/LASCO C2. (d) Show smooth sunspot number and heliospheric current sheet tilt angle \citep{McComas2008p8103}.}{1.0}

\figuremacroFP{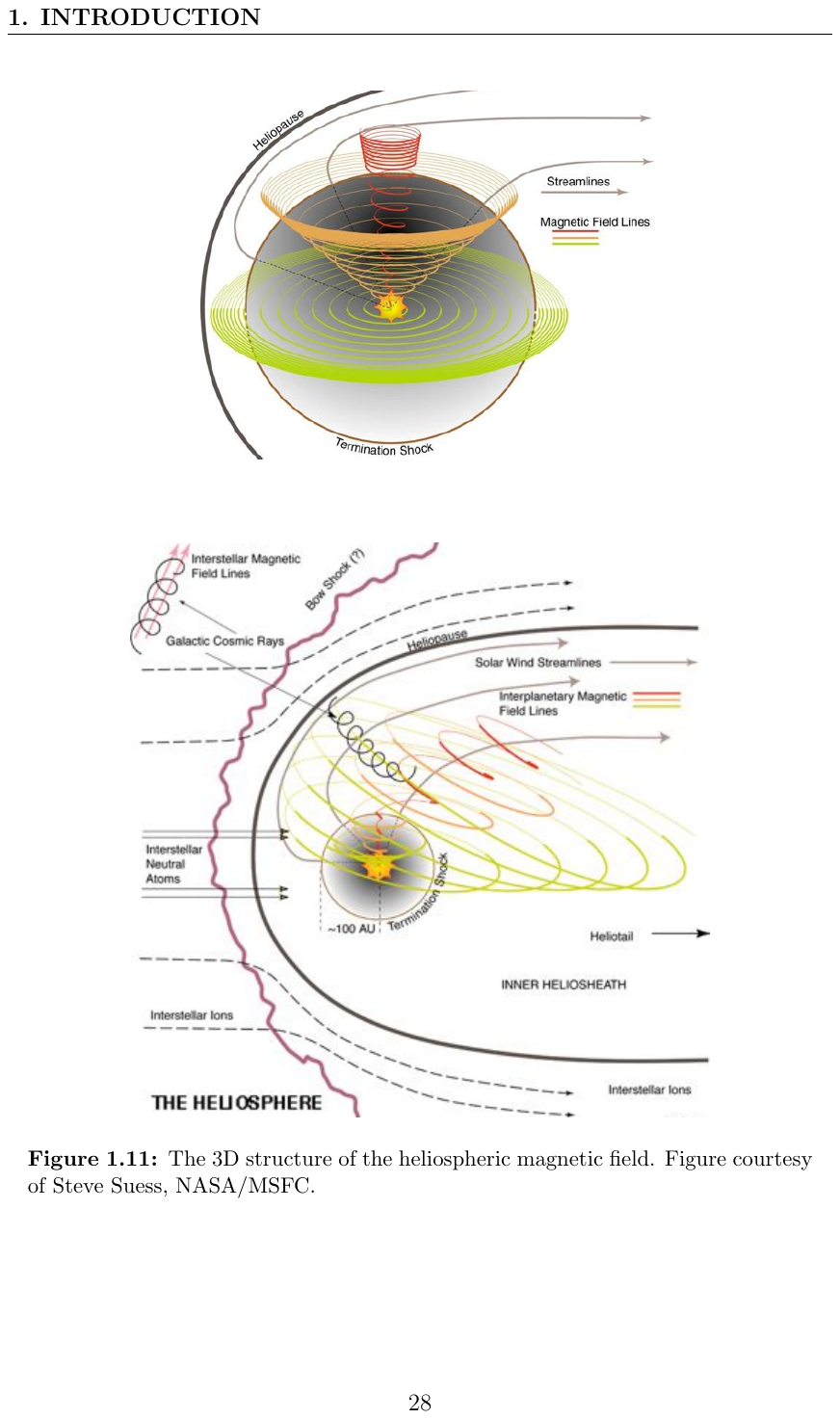}{The Heliosphere}{Depiction of the Heliosphere and its prominent features. Shown are the spiralling magnetic field lines, termination shock, heliopause, and bow shock. Image courtesy of Steve Suess, NASA/MSFC.}{0.9}

It is now known that the solar wind consists of two components: the slow solar wind with typical 1\,AU values for velocity, density, and temperature of $\sim$400\,km\,s$^{-1}$, $\sim$10\,cm$^{-3}$, $\sim$1.4$\times$10$^{5}$\,K respectively; and the fast solar wind with typical 1\,AU values for velocity, density, and temperature of $~$800\,km\,s$^{-1}$, $\sim$3\,cm$^{-3}$, and $\sim$1.0$\times$10$^{5}$\,K respectively originating from open magnetic field regions (see \figref{swoops}). The two different speed streams can interact to form co-rotating interaction region (CIRs) where the fast wind ploughs into the slow solar wind and can form shocks. Another feature of the solar wind is the heliospheric current sheet (HCS) which separates the two opposite polarities of the Sun's magnetic, field forming what is know as the `ballerina skirt'. 

The solar wind cannot expand forever and eventually runs into the ISM at what is known as the termination shock, where the solar wind transitions back to sub-sonic velocities (\figref{heliosphere}). The Voyager II spacecraft recently (and previously Voyager I) passed though this region at some 70\rng90\,AU, details of these shock crossings can be found in \citet{Richardson2008p63}, \citet{Burlaga2005p2027} and \cite{Decker2008p67}. Outside of the termination shock lies the heliosheath where the ISM and SW are in pressure balance, its outer boundary is the heliopause which also marks the edge of the Heliosphere. In this region the interaction between the ISM and SW causes turbulence and heating of plasma. As the Sun travels around our galaxy and encounters the ISM a bow shock is believed to from ahead of the heliopause (\figref{heliosphere}).

\section{Coronal Mass Ejections}
\label{sec:CMEs}
\begin{quotation}
	{\it ``We define a coronal mass ejection to be an observable change in coronal structure that (1) occurs on a time scale between a few minutes and several hours and (2) involves the appearance (and outward motion) of a new, discrete, bright, white-light feature in the coronagraph field of view.''}
	
	\raggedleft	-\cite{Hundhausen1984p2639}
\end{quotation}	 
\figuremacroW{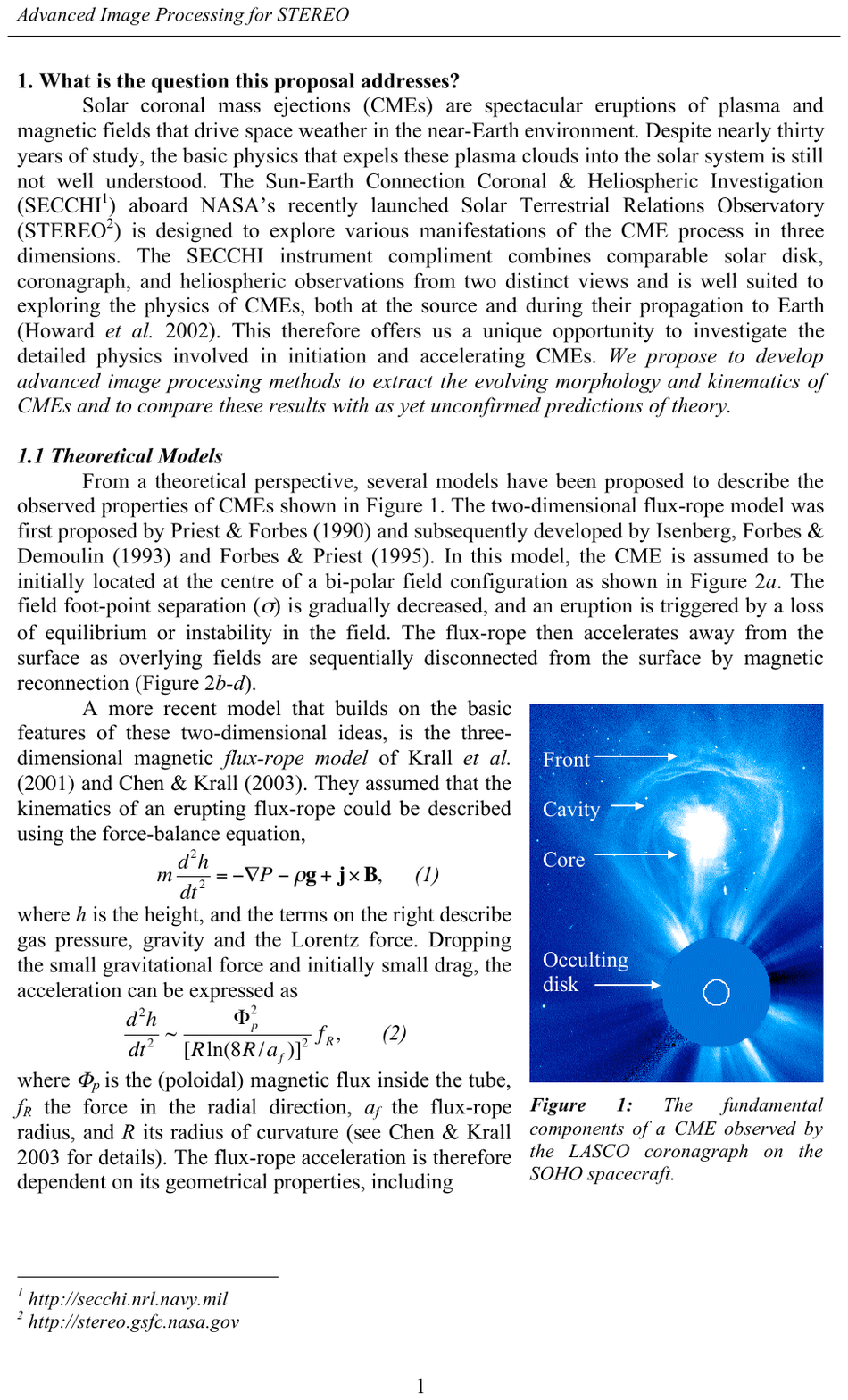}{Typical `lightbulb' CME}{The typical components of a CME observed by the LASCO coronagraph on SOHO spacecraft.}{0.5}

Coronal Mass Ejections (CMEs) are large scale eruptions of plasma and magnetic field which propagate from the Sun into the Heliosphere. A typical CME has a magnetic field strength of tens of nT, a mass in the range of $10^{13}$--$10^{16}$\,g \citep{vourlidas2002p91} and velocity between $\sim$10\rng2,000\,km\,s$^{-1}$ some times even reaching 3,500\,km\,s$^{-1}$ close to the \mbox{Sun \citep{yashiro2004p7105}.} At 1\,AU, CME velocities (300\rng1,000\,km\,s$^{-1}$) tend to be closer to the solar wind speed \citep{Lindsay1999p12515,Wang2005p1230,Gopalswamy2006p145}. The energies associated with CMEs are of the order of 10$^{24}$--10$^{25}$\,J making CMEs the most energetic events on the Sun \citep{vourlidas2002p91}. Although CMEs often exhibit a three-part structure which consists of a bright front followed by a dark cavity and bright core (see \figref{cmefeat}), they may also exhibit more complex structures \citep{Pick2006p341}. In fact less than $\sim$30\% of CME events possess all the three parts \citep{webb1987p383}.

\figuremacroFP{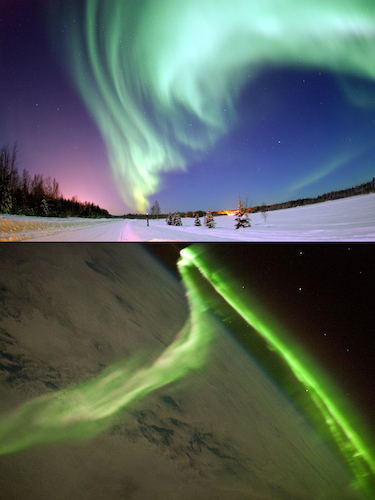}{Images of Aurora from the ground and ISS}{Two images of the aurora from Eielson Air Force Base at Bear Lake, Alaska (top) and from onboard the ISS (bottom). Images courtesy of Wikimedia Commons.}{0.8}

CMEs are know to be the most important driver of adverse space weather on Earth and in the near-Earth environment as well as on other planets \citep{Schwenn2005p1033,Prange2004p78}. The most famous phenomena associated with space weather is the Aurora Borealis or Northern Lights (see \figref{auroa1}). The Aurora is caused by energetic particles traveling along the Earth's magnetic field lines interacting with atoms (mainly nitrogen and oxygen) in the upper atmosphere producing emission (resonance or recombination). One of the most extreme space weather events in recent history occurred on 2 September 1859 \citep{Carrington1859p13}. This event was associated with a white light flare observed by Carrington.  About 18 hours after the flare a severe geomagnetic storm occurred causing widespread sightings of the Aurora down to latitudes as low as 18$^{\circ}$ North and the loss of a significant portion of the telegraph service for many hours \citep{Green2006p145}. As society has become more reliant on technology it has also become more susceptible to the adverse affects of space weather. On the 13 March 1989 a geomagnetic storm caused large geomagnetically induced currents (GIC) which caused the failure of a transformer that ultimately led to collapse of the Hydro-Qu\'{e}bec network. This left some six million people without power for nine hours causing substantial economic losses upwards of \$13.2 million  \citep{Bolduc2002p179}. The storm was caused by a CME ejected from the Sun on 10 March 1989 impacting the Earth some $\sim 50$ hours later. Other CMEs have knocked-out or caused damage to satellites, most recently on 5 April 2010, when Galaxy~15 or `Zombiesat' (Intelsat) stopped responding to ground commands. The satellite continued to broadcast while it drifted forcing other satellites to take evasive action to avoid interference from `Zombiesat'. Galaxy 15 was subsequently recovered and placed into a safe mode. More generally, the increased radiation due to space weather poses hazards to astronauts and passengers on long distance flights, especially those over the poles. Also, polar flights may be restricted due to communications blackouts caused by space weather. In the modern era, society's dependence on GPS, communication satellites and inter-connected power grids mean space weather is an increasing concern. The cost of a severe space weather event was estimated to be up to \$2 trillion in a recent report by the National Research Council, \citep{spcweatherEcoImp2008}. As such ,the monitoring and prediction of space weather is of the utmost importance to society. Understanding the formation, acceleration and propagation of CMEs is vital to this goal as they are the main drivers of adverse space weather.

\subsection{Historical Observations}
\label{sec:hist}
\figuremacro{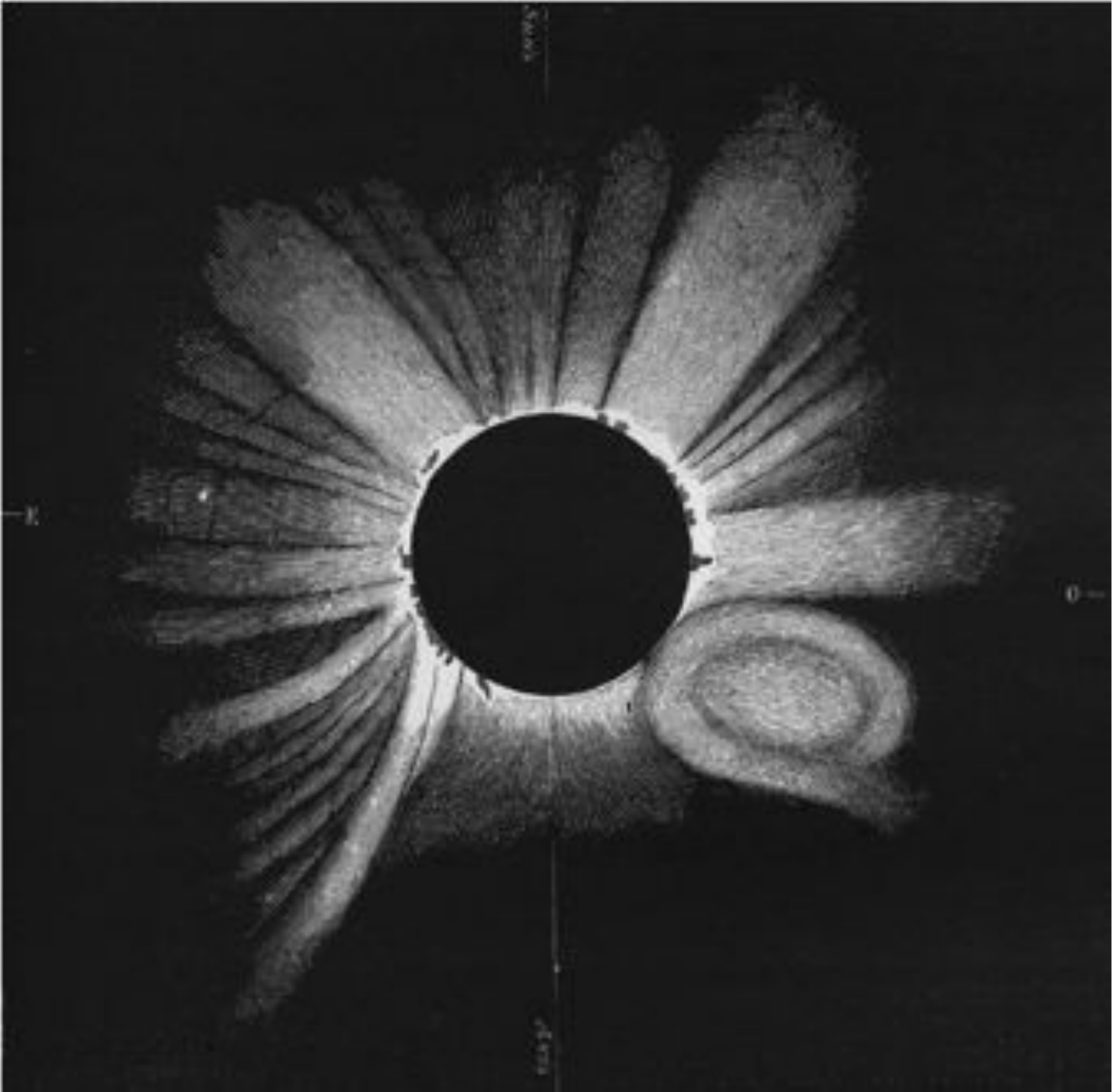}{1860 Eclipse Sketch}{Drawing of the 1860 eclipse recorded by G. Temple from Torreblanca , Spain with what is probably a CME \citep{Eddy1974p235}.}
The earliest observation of a CME probably dates back to the eclipse of 1860 in a drawing recorded by G. Temple shown in \figref{firstcmeobs}. It took 113 years for the CME to be formally discovered, the first definitive observation being made by \citet{tousey1973p713} using the coronagraph on-board the seventh Orbiting Solar Observatory (OSO-7). Following this a number of space-borne coronagraphs recored numerous CMEs. While the first imaging observation of a CME was in 1973, it is now apparent that CMEs and their effects had been observed much earlier, in a number of different types of observations. For example, geomagnetic disturbances caused by CMEs had been recorded as early as 1724. CMEs were also observed in radio observations via interplanetary scintillation (IPS) from the 1960s, however it was not until 1980s that the IPS could be directly related to the CMEs. Also fast CMEs produce shock waves and these shock waves then produce radio emissions at the local plasma frequency. These are known as Type II bursts and lead indirectly to the first height-time plot of a CME or at least the shock it drove as shown in \figref{cmeheighttime}. These shock waves can also be detected in measurements of the solar wind plasma {\it in situ}.

It became clear, soon after the sunspot cycle was first discovered, that there was a link between the sunspot cycle and geomagnetic activity. One of the first people to suggest that transient ejections of plasma from the Sun could be the source of geomagnetic storms was \citet{Lindemann1919p669}, and later Chapman and Ferro also suggested that coronal transients could account for the geomagnetic activity \citep{Chapman1931p77,Chapman1931p171,Chapman1932p147}. At this time there were two main interpretations of these transients: flare induced transient ejections of plasma from formerly closed regions which drag the magnetic field outward with them \citep{Piddington1958p589,gold1962p100}; and shocks formed in open regions by the rapidly increasing expansion speed of the solar wind caused by the flare which would not drag any magnetic field with them \citep[e.g.,][]{parker1963}.

\figuremacroW{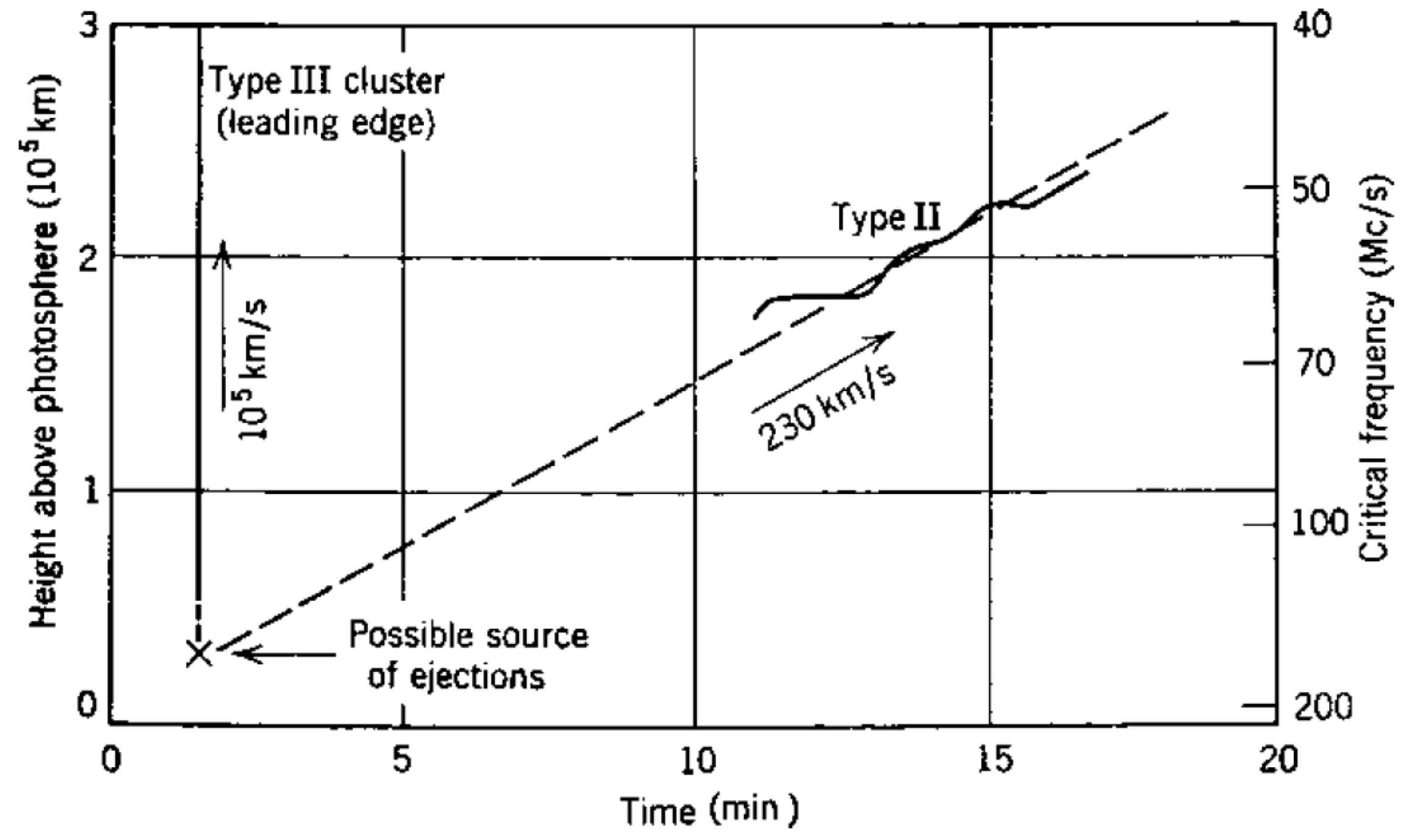}{Type II and III Radio Burst Height-Time Plot.}{A height-time plot derived from Type II and III radio burst from \citet{wild1954p532}.}{0.8}

There were many signatures of these interplanetary transients such as solar energetic particle (SEP) events detected at Earth \citep{Forbush1946p771}. \citet{wild1963p291} suggested a two stage mechanism was required to accelerate the particles up to the observed energies: an initial flare acceleration followed by further acceleration by an outwardly-moving fast magnetohydrodynamic shock. The radio signatures of such moving shock fronts were commonly detected as Type II radio bursts.

IPS is caused by the diffraction of radio-waves, from distant pulsars, by inhomogeneities in the solar wind plasma: as the solar wind is moving it produces a time-varying intensity signal which can be detected at Earth. This technique led to the detection and study of many transient events from the 1960s onward \citep[e.g.][]{Houminer1973p136,Houminer1972p1703}. Correlations were found between these events and {\it in situ} measurement at Earth \citep{Houminer1974p1041,Rickett1975p237} and in geomagnetic activity indexes \citep{Vlasov1982p536}. The exact nature of the transient density variations and their relationship, or lack thereof, to coronal transients was unclear. Attempts were made to relate the IPS observations with solar source features \citep{Tappin1983p1171,Hewish1988p195} but with limited success.

Entry into the space age allowed direct measurements of the properties of these coronal transients. The {\it in situ} signatures of the outwardly propagating shocks were fist discovered in Mariner 2 observations \citep{Sonett1964p153}, and later in Vela 3 observations \citep{Gosling1968p61}. It was found that the plasma driving these shocks had different properties compared to that of the normal solar wind. The material often showed enhanced helium content \citep{Hirshberg1972p467}, and low proton and electron temperatures \citep{Gosling1973p2001}. These early {\it in situ} observations indicated the Parker model was incorrect, and that the disturbances were associated with previously closed field regions. In some cases, streaming super-thermal electrons were found, suggesting the field lines threading the plasma were either connected to the Sun at both ends, or disconnected from it entirely. \citet{Hundhausen1970p4631} used observations of shock disturbances to estimate the mass and energy associated with  a large shock to be $\sim10^{13}$\,kg and $\sim10^{25}$\,J\symbolfootnote[1]{These values are remarkably close to the modern measurements of CME mass and energy.}.

The magnetic signature of these transient events was first observed by \citet{Burlaga1981p6673} in observations from five different spacecraft. The signature was a smooth rotation of the magnetic field vector following the shock, which they called a `magnetic cloud' (MC) citing earlier theoretical work. However, the link between this signature and the coronal transients was not made until a following paper \mbox{\citep{Klein1982p613}}, which also identified the basic combination of properties that defines a MC which are still used today namley; low temperatures, high magnetic field strength, and a smoothly-rotating magnetic field vector.

Around this time, the first space-borne coronagraphs were appearing. This led to the first remote imaging observation of these coronal transients \citep{tousey1973p713}. Soon after, a direct link between these transients and radio Type II bursts was made \citep{Stewart1974p203,Stewart1974p219}. The near continuous monitoring of the solar corona by various coronagraphs in the 1970s provided some unexpected results. Far more ejections were found than would have been expected on the frequency of occurrence of shock wave disturbances, and the ejections were far more commonly associated with eruptive prominences than impulsive flares \citep{Gosling1974p4581,Munro1979p201}. However many were followed by long-duration soft X-ray events lasting hours \citep{Sheeley1975p377,Webb1976p159}. Also, surprisingly, most of the material in the ejection was of coronal origin rather than flare or prominence ejecta \citep{Hildner1975p363}. The ejections also showed a wide range of speeds from hundreds of kilometres a second to greater than 1,200\,km\,s$^{-1}$ \citep{gosling1976p389}, morphologies, and energetics \citep{howard1985p8173}. The unexpected results led to the so-called `Solar Flare Myth'. The first use of the term `coronal mass ejection' appears to be from  \cite{gosling1976p389}, and likely arose from the more tentative `mass ejection coronal transient' which first appeared in \citep{Hildner1975p363}. 

\subsection{The `Solar Flare Myth'}
\label{sec:myth}
Up until the early 1990s, despite all the unexpected results, all the activity discussed above was primarily attributed to solar flares, and CMEs were believed to be the result of flare-driven shock waves. However the evidence against such an interpretation had been growing throughout the 1970s and 1980s. For example, CMEs and geomagnetic storms were often not associated with flares, and the energy required to launch the CME was greater than the flare \citep{MacQueen1980p605}. Also, coronagraphic observations demonstrated that the flanks of the CMEa did not move laterally as the loop top moved outward through the corona which was not consistent with the shock interpretation \citep{Sime1984p2113}. Further, in a number of papers  which back-projected CMEs to find their onset times, none were found to be coincident with flares and typically, the flare occurred some time after the CME onset \citep{Harrison1990p917,Harrison1989p2333}. The debate continued and intensified with publication of Gosling's \citeyear{Gosling1993p18937} paper entitled `The solar flare myth' in which the author demonstrated that the source of interplanetary shocks and of most geomagnetic storms were CMEs not flares, and that the CME flare relationship was secondary at best. In this paper Gosling proposed a ``modern paradigm'' describing the relationship between flares, CMEs, and geomagnetic activity shown in \figref{flaremyth}

\figuremacroW{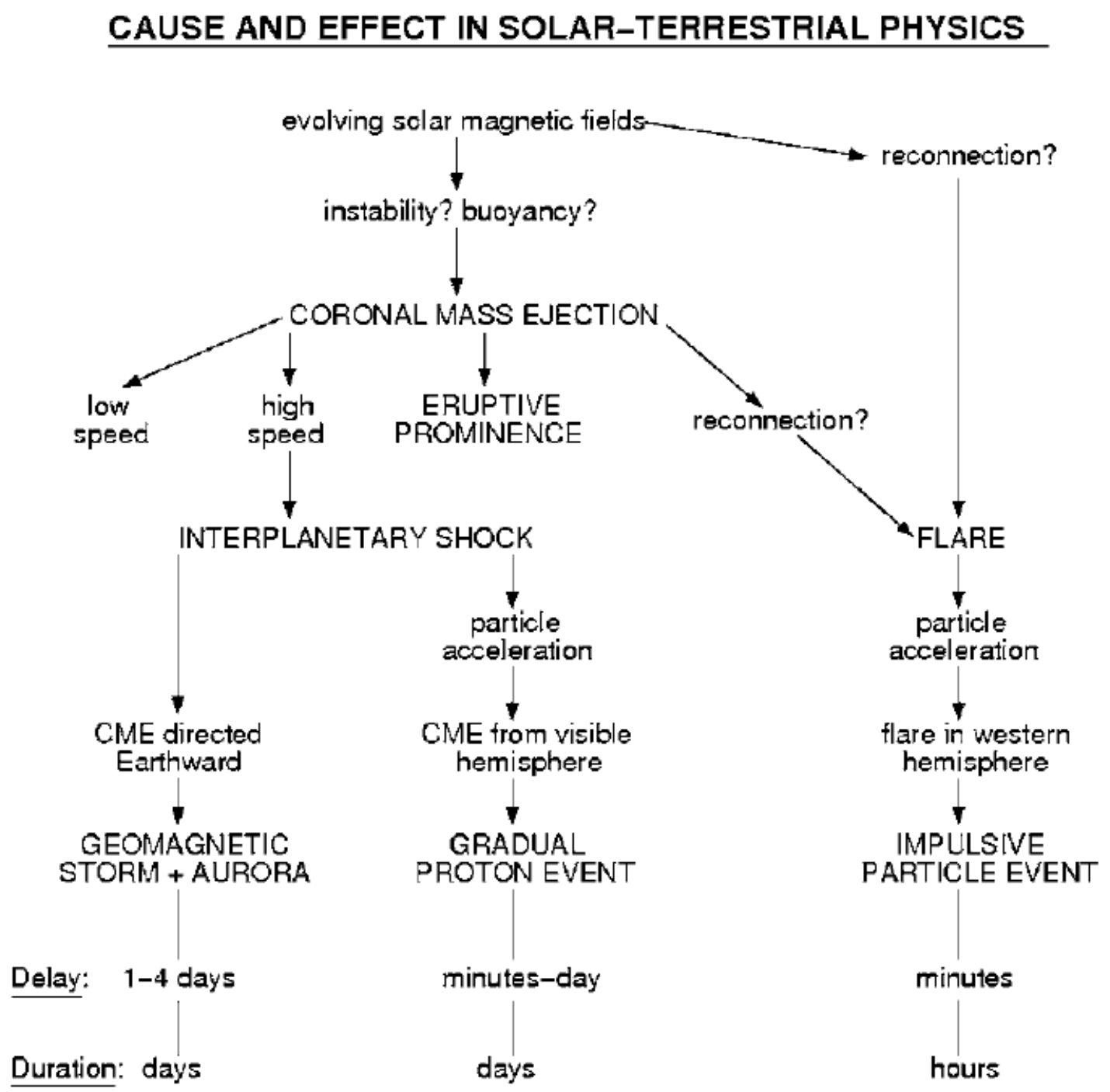}{Gosling's CME Flare Paradigm}{Gosling's interpretation of the relationship between flares, CME and geomagnetic activity. Note the flare is at best from a common process ``evolving solar magnetic fields'' (reconnection) or secondary to the CME launch process \cite{Gosling1993p18937}.}{0.8}

\subsection{Current Understanding}
\label{sec:currentview}
CMEs originate wherever flares and prominence eruptions occur. Both flares and prominences are associated with active regions (ARs), which are regions of high magnetic field with or without sunspots. The most energetic CMEs seem to come from ARs which contain sunspots of opposite polarity. Cool prominences suspended above ARs also often produce CMEs, prominences can also form above neutral lines between sunspots of opposite polarities in ARs. Studies of {\it in situ} measurements of CMEs  at 1\,AU indicate CMEs probably remain attached at the Sun. The counter-streaming (bi-directional) particles in CMEs indicate that the `legs' remain anchored on either side of the neutral line \citep{Kahler1991p9419,Farrugia1993p7621,Farrugia1993p15497}. So closed magnetic structures seem to be the basic characteristic of CME-producing regions on the Sun. Thus the energy required to power the CME must ultimately come from the magnetic field itself.

\figuremacroW{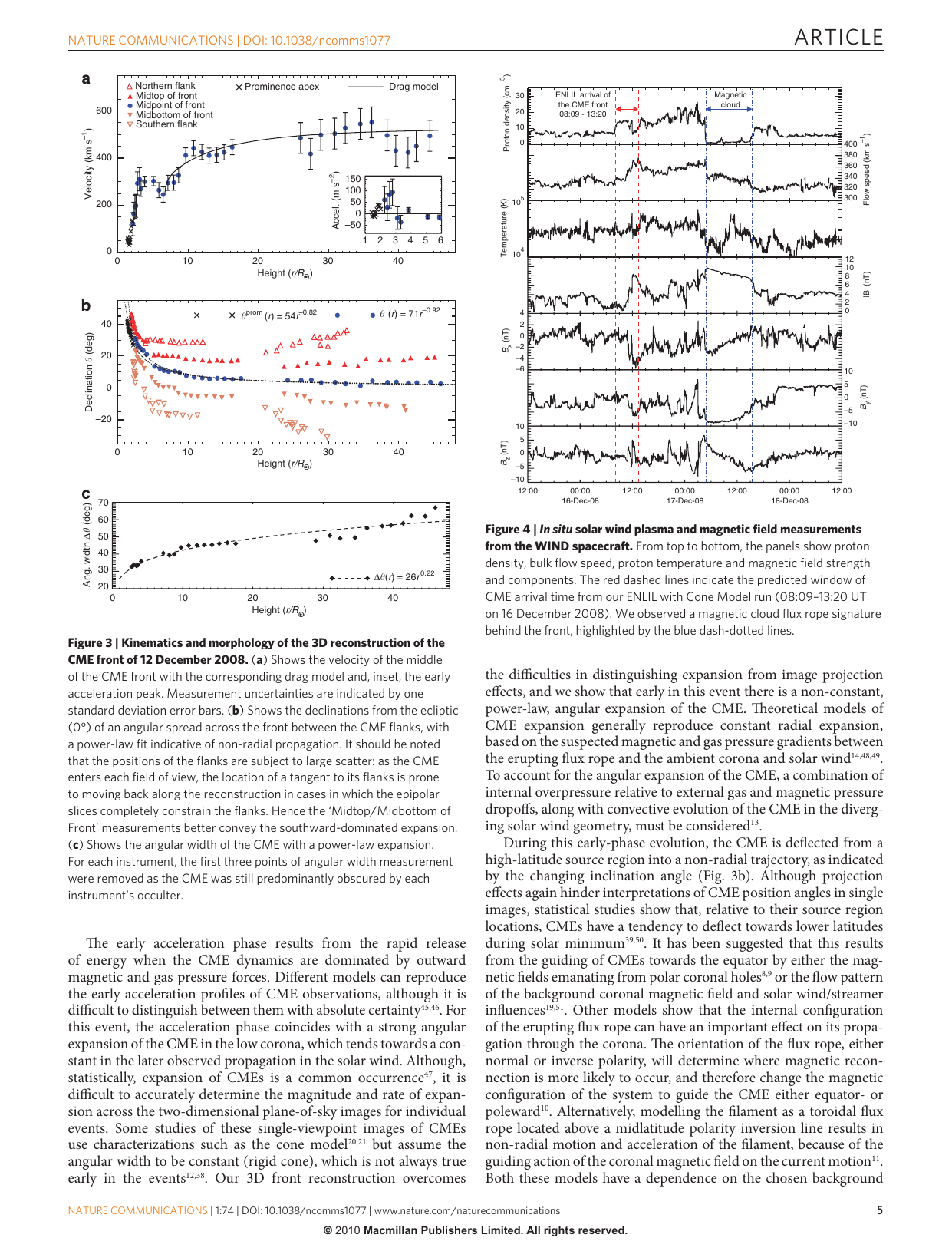}{Magnetic Cloud {\it in situ} Measurements}{Top to bottom, the panels show proton
density, bulk flow speed, proton temperature and magnetic field strength
and components. The red dashed lines indicate a predicted window of
CME arrival time. The magnetic cloud flux rope signature is
behind the front, highlighted by the blue dash-dotted lines \citep{Byrne2010p74}.}{0.8}

The general appearance of a CME was shown previously in \figref{cmefeat}, but not all CMEs share this appearance. CMEs are also associated with on-disk `dimming'  regions, typically on either side of the neutral line \citep[see review by Sterling in][]{wilsion2003}. Dimming is the reduction in the intensity of radiation due to physical changes in the plasma (mass motion, density, temperature) typically observed in X-rays \citep{bastian1999} or EUV \citep{Gopalswamy2000p1457}. The long term observations of LASCO have also discovered a number of unclassified morphologies, flux ropes \citep{Chen2000p481,Plunkett2000p371}, prominence-less CMEs \citep{Gopalswamy2001p149,Yashiro2003p2631}, and jet-like CMEs \citep{Yashiro2003p2631} which contain no aspect of the three-part structure.

Much of the material contained in CMEs is already present in the corona so it is expected to have coronal temperatures and densities. This is not true of the cool (400\rng8,000\,K), dense ($10^{10}$\rng$10^{11}$\,cm$^{-3}$) prominence material of the core. The cavity is believed to be of coronal temperature but lower density than the front and core. This is often interpreted as a flux rope geometry often seen in 1\,AU {\it in situ} measurements, \figref{mc} shows an example. The decreased density, temperature, increased magnetic field strength and smooth rotation of the magnetic field are the classic signatures of a MC. Not all CMEs measured  at 1\,AU have MC structures, one possible explanation of this is that the spacecraft trajectory though the CME cloud only samples one part and misses the MC as shown in \figref{cmestruct}.

\figuremacroW{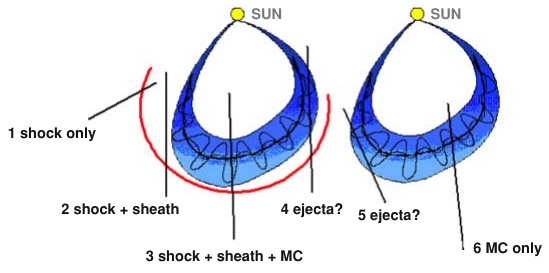}{Possible Spacecraft track through a CME}{Possible spacecraft tracks through a fast shock driving CME, slow CME and corresponding features which would be sampled \cite{Gopalswamy2006p145}.}{0.7}

CMEs have a wide range of speeds, accelerations and widths \citep[see][for an overview]{Gopalswamy2004p201}. The kinematic evolution of CMEs is complex, consisting of multiple phases. \cite{Sheeley1999p4739} first classified CMEs into two classes: (1) slow gradual CMEs associated with prominences; and (2) impulsive fast CMEs often associated with flares. Studies based on larger samples indicate that slow CMEs arise from prominence lift-offs and streamer blowouts, and fast CMEs from ARs and flares \citep{Gonzalez2003p1039,Gopalswamy2000p1457}. \citet{Zhang2001p452,Zhang2006p1100} suggest a three phase model of initation, acceleration and propagation and relate this to other observable properties such as soft X-rays. Numerous studies have attempted to relate the observed projected kinematics to the true kinematics: \mbox{\citet{dalLago2003p2637}} found $v_{rad}=0.88v_{exp}$; while \citet{Schwenn2005p1033} found the 1\,AU travel time followed $T_{ar}=203.0 - 20.77 \ln v_{exp}$. These type of statistical studies only provide a rough estimate of the corrections \citep{Vrsnak2007p339}, which are not accurate enough for detailed comparisons to models. Detailed studies of single events are necessary to make such comparisons.

\figuremacroW{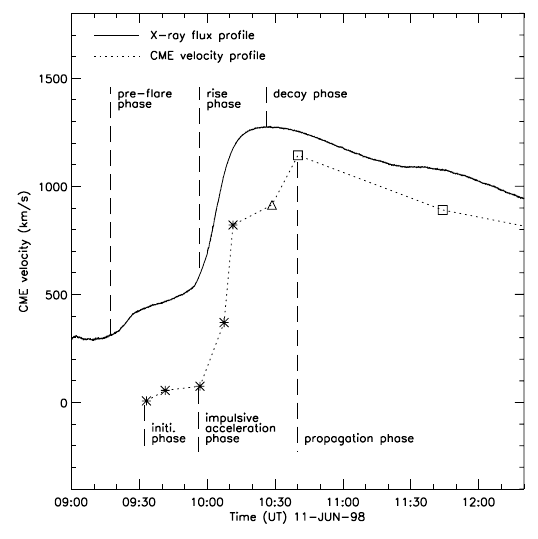}{CME Velocity Evolution with soft X-ray light Curve}{Evolution of a CME velocity (LASCO) in terms of a three phases profile (dashed line) and the flare (GOES) soft X-ray flux temporal profile (solid line) for the event on 1998 June 11 \citep{Zhang2001p452}.}{0.8}

\figuremacroW{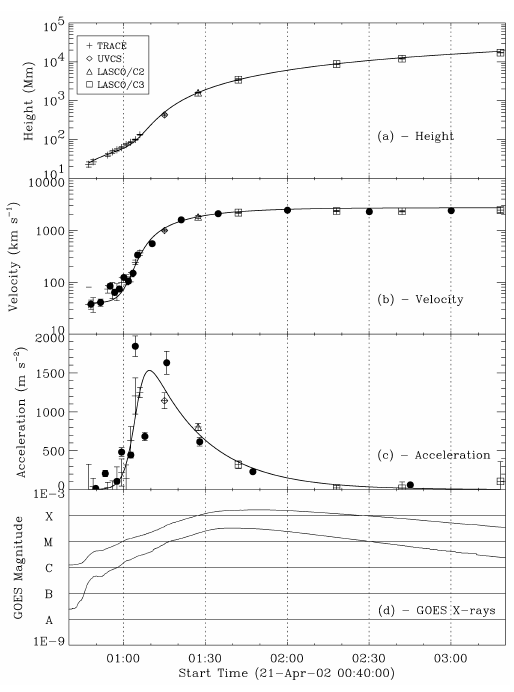}{CME Kinematics Fit with double exponential and corresponding soft X-ray Flux}{The (a) height, (b) velocity, (c) acceleration profiles and (d) the GOES soft X-ray flux \citep{Gallagher2003p53}. The velocity was obtained by taking the first and second numerical derivatives: the first-difference values are also shown as filled circles. The solid line gives the best fit to the data using a double exponential acceleration profile \eqref{galo2003}.}{0.8}

One such study was conducted by \citet{Zhang2001p452} who found a strong correlation between the soft X-ray flux and CME accleration. \figref{zhang2001} shows the velocity evolution for a CME, and the x-ray light curve: notice that the two vary nearly in step. \citet{Gallagher2003p53} conducted a similar study, but extended the data set back towards the Sun using UCVS and TRACE. They used a double exponential profile:
\begin{align}
	\label{galo2003}
	a(t) = \left [  \frac{1}{a_{r} \exp (t/\tau_{r})} + \frac{1}{a_{d} \exp (t/\tau_{d})}  \right ]
\end{align}
to fit the acceleration where $a_{r}$, $a_{d}$ are the initial accelerations and $\tau_{r}$, $\tau_{d}$ are the e-folding times for the rise and decay phase. Using this they found an acceleration peak of $\sim$1,500\,km\,s$^{-2}$ which corresponded to the soft X-ray rise phase (\figref{gallagher2003}) and supported a relationship between the X-ray flux and CME acceleration. This relationship, between X-ray flux and CME acceleration, was fully realised in a paper by \citet{Temmer2008p95}. They compared the CME kinematics to the hard X-ray flux and found the peak acceleration corresponded to the peak X-ray flux \figref{temmer08}. This is consistent with the previous observation as the derivative of the soft X-ray flux is often used as a proxy for hard X-ray flux. The relationship between the peak acceleration and hard X-ray flux indicates a strong relationship between magnetic reconnection and acceleration. This could be interpreted as reconnection occurring in a current sheet behind the flux rope as suggested by a number of models (Section \ref{sec:cmeforminit}). \citet{Lin2010p44} tried specifically to test the CME kinematics against a number of models. They found peak accelerations of $\sim$1,500\,km\,s$^{-2}$ and $\sim$600\,km\,s$^{-2}$ close to the Sun ($<3$\,$R_{\odot}$). The authors show it is difficult to give merit to one model above another, as within the inherent scatter they all reproduce the kinematics to a good degree.

\figuremacro{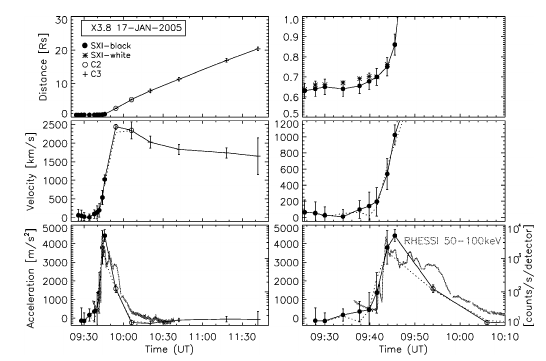}{CME Kinematics with corresponding hard X-ray Flux}{Event kinematics from top to bottom, distance-time profile $d(t)$, velocity $v(t)$, and acceleration $a(t)$ of the CME as observed by GOES SXI and LASCO and (bottom) the RHESSI 50\rng100\,keV hard X-ray flux of the associated flare \citep{Temmer2008p95}. The left panels show the full CME height range covered by the LASCO FOV the right panels zoom into the early acceleration phase of the CME.}

The above studies are subject to large uncertainties due to projection effects. A number of efforts have been made to overcome this limitation, such as forward modelling, tomographic, and polarimetric techniques. Forward modelling uses a pre-assumed geometry such as the cylindrical model \citep{Cremades2004p307}, or cone model \citep{Xie2004p3109,Xue2005p8103,Zhao2002p1223}, and varies the model parameters to best match the 2D observations. The kinematics can then be derived from these best-fit models, however they are subject to the large and unknown errors about the pre-supposed geometry.

The polarimetric technique of \citet{Moran2004p66} uses the ratio of unpolarised to polarised brightness of the Thomson-scattered K-corona to estimate the average line-of-sight distance from the instrument plane of sky. An example of this reconstruction technique is shown in \figref{moron2004}. This technique is only applicable up to 5\,$R_{\odot}$ as beyond this the F-corona can no longer be considered unpolarised. Yet other studies attempt to tie on-disk or {\it in situ} signatures of CMEs together: \cite{Demoulin2008p347,Howard2008p1104}. These studies have made progress but the fundamental problem of uncertainties in the derived kinematics still remain a problem.

\figuremacroW{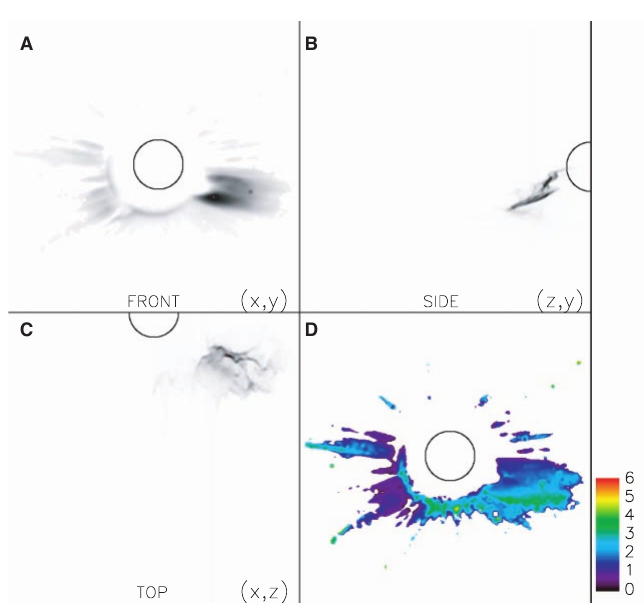}{Polarimetric CME Reconstruction}{(A) Total brightness a CME occurring on 31 October 1998 at 4:56UT (front view), (B) a reconstructed side view of the CME in the (z, y) plane, (C) a reconstructed top view of the CME in the (x, z) plane, and (D) a topographical map of the CME displaying distance from the (x, y) plane \citep{Moran2004p66}. The colour bar indicates distance from the sky plane in $R_{\odot}$. The solar disk is outlined.}{0.6}

From comparing CME speeds close to the Sun to those at 1\,AU (see \figref{icmeaccl}), and by studying the derived speed from radio observations (see \figref{cmetoicmespeed}), it is clear that CMEs are accelerated as they propagate. This has been studied in terms of a `drag' force \citep{Cargill1996p4855,Vrsnak2002p1019,Tappin2006p233} and an extended Lorentz force \citep{Chen1996p27499}. The limited spatial nature of coronagraphic and {\it in situ} measurements, and both temporal and spatial limitation of radio and heliospheric imagers, mean that little can really be said about the exact nature of the forces governing CME propagation. However,  it is clear there is some force at play which tends to equalise the CME velocity to that of the background solar wind speed.

\figuremacroW{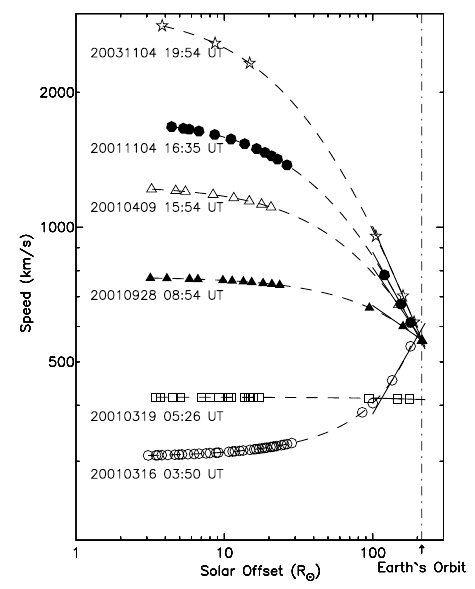}{CME Acceleration far from the Sun}{Speed-distance profiles plotted on a log-log scale \citep{Manoharan2006p345}. Average initial speed of the CME increases from bottom ($\sim$300\,km\,s$^{-1}$) to top ($\sim$2500\,km\,s$^{-1}$). The date and the start time of each CME are shown. These profiles converge toward the ambient solar wind speed at the Earth's orbit. The data points at $R \le 30\,R_{\odot}$ and $R \ge 80\,R_{\odot}$ are from LASCO and IPS measurements, respectively.}{0.6}

\figuremacro{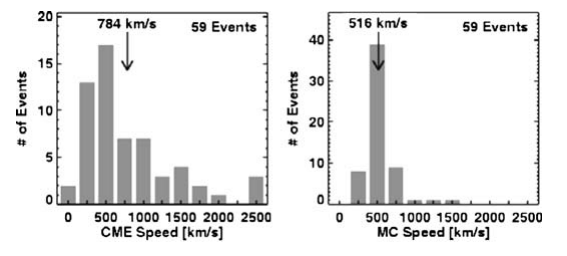}{Comparison of CMEs speeds close to the Sun to those at 1\,AU}{The speed distribution of CMEs (left) and ICMEs or MCs (right) for a set of 59 CME-ICME pairs \citep{Gopalswamy2006p145}. The average of the distribution is indicated on each panel.}

The STEREO mission was proposed to remove or diminish many of the issues outlined above. The STEREO mission aims to achieve this by providing dual perspective views along the entire Sun-Earth line. It will facilitate the tracking of CME signatures close to the Sun all the way to Earth, and their association to the {\it in situ} measurements. The high resolution and cadence images from dual perspectives will allow the 3D trajectory of CMEs to be reconstructed over an extended range, allowing accurate kinematics to be extracted and compared to theory in an attempt to answer some of the fundamental questions outlined above. 

\subsection{Open Questions}
\label{sec:openqs}
While our knowledge of CMEs has greatly expanded with improving observations and theoretical interpretations, new questions have been raised. Some of the most fundamental questions about CMEs still remain unanswered. How is the energy required to launch a CME built up and stored? While a number of models have been developed to answer this (see Section \ref{sec:cmeforminit}), there is still no consensus on the matter. Another key question is what leads to the release of this energy and the eruption of a CME? A related question is: are CMEs pre-existing structures, or are  they formed during the eruption? Again a number of models attempt to answer these questions but, as of yet, still fall short. Some authors claim \citep{Zhang2001p452} that there are two (or more) types of CME: slow, gradual CMEs which are accelerated slowly over large distances; and impulsive  fast CMEs accelerated low in the corona, often associated with flares. It is not clear if these two distinct types of CME exist, and are due to different processes, or if they are the extreme ends of a continuous spectrum of CME properties.

Once CMEs leave the Sun they are accelerated by an interaction with the solar wind; very little is known about this interaction other than that it tends to accelerate most CME towards the solar wind speed while others seem to be unaffected. Attempts have been made to model this interaction as a form of drag, or by the action of an extended Lorentz force. A combination of limited observations, and uncertainties in what observations there are, have left this particular question wide open. Also at 1\,AU why do some CMEs contain MC structures, and others not, and if not, are they still connected to the Sun? This could be an observational effect, but it could also be a manifestation of different underlying structures and mechanisms at play in some CMEs.
 
\section{Thesis Outline}
The work undertaken for this thesis enhances the understanding of the kinematics and morphologies of CMEs as they propagate through the inner Heliosphere. Until now, studies of these properties have been hampered for a number of reasons. The most significant of these are projection effects which affect not only the morphology but also the kinematics. Also, the lack of contiguous high resolution and high cadence CME observation through the inner heliosphere has lead to significant uncertainties in the interplanetary CME kinematics.

Chapter 2 discusses CME theory and related phenomena. Chapter 3 details CME observations,  instrumentation, as well as the data reduction, and conversion to physical units used in this work. Chapter 4 details the methods used to analyse the observations. Crucial to this work are the 3D reconstruction techniques which are outlined as well as the drag modelling and fitting. The 3D reconstructions are the basis for the work undertaken in Chapter 5, where the trajectories of a number of CMEs were reconstructed and are presented.  Chapter 6 presents results from drag modelling efforts on a number of CMEs. Also presented are the results from a new 3D reconstruction method called `elliptical tie-pointing' and the drag modelling results from this. In Chapter 7, a CME-driven shock is studied and compared with semi-empirical shock relations. Finally, Chapter 8 presents the main conclusions of the thesis and details possible future work that could follow on from these new developments.



\chapter{Coronal Mass Ejections and Related Theory} 
\label{chap:theory}


\ifpdf
    \graphicspath{{/}}
\else
    \graphicspath{{2/figures/EPS/}{2/figures/}}
\fi


\hrule height 1mm
\vspace{0.5mm}
\hrule height 0.4mm
\noindent
\\	{\it In this chapter the details of the theoretical framework used to describe CMEs and CME related phenomena are presented. We have seen in the previous chapter that magnetised plasmas are ubiquitous on the Sun and throughout the Heliosphere. The complex interaction of these plasmas may be understood in terms of magnetohydrodynamics (MHD). The application of this theory to the various aspects of CME evolution is presented. Some of the current CME models associated with initiation, acceleration, and propagation are reviewed. The relevant theoretical description of shocks and its relation to CME-driven shocks is also presented.} \\
\hrule height 0.4mm
\vspace{0.5mm}
\hrule height 1mm

\newpage

\section{Magnetohydrodynamic (MHD) Theory}
\label{sec:mdh}
Magnetohydrodynamics combines Maxwell's equations with the theory of fluid mechanics in an attempt to describe the interplay between the plasma flow's effect on the magnetic field, and the altered magnetic fields effect on the plasma flow. Simply put, moving charges (plasma flow) will generate a magnetic field, this field will affect any other changes, altering their motion and thus a feedback between the flow and magnetic field is set up. In MHD this feedback is described by the induction equation.

\subsection{Maxwell's Equations}
\label{sub:maxwell}
Maxwell's equations form a closed set of equations and describe the interaction of magnetic ($\mathbf{B}$) and electric ($\mathbf{E}$) fields. They can be written in a derivative form in vacuum as follows:
\begin{align}
	\label{eq:gauss}
	\nabla \times \mathbf{B} & = \mu _{0} \mathbf{j} + \frac{1}{c^{2}}\frac{\partial \mathbf{E}}{\partial t} \\
	\label{eq:solin}	
	\nabla \cdot \mathbf{B} & = 0 \\
	\label{eq:maxwell}
	\nabla \times \mathbf{E} & = - \frac{\partial \mathbf{B}}{\partial t} \\
       \label{eq:ampere}
       \nabla \cdot \mathbf{E} & = \frac{1}{\epsilon_{0}} \rho
\end{align}
where $\textbf{j}$ is the current density, $\rho$ is the charge density, $\mu_{0}$ is the magnetic permeability of vacuum, $\epsilon_{0}$ is the permittivity of free space and $c$ is the speed of light. If the typical plasma velocities are much less than the speed of light, the second term in Amp\`ere's law \eqref{eq:gauss} may be neglected giving:
\begin{align}
	\label{eq:amperem}
	\nabla \times \mathbf{B} & = \mu_{0} \mathbf{j}.
\end{align}

\subsection{Fluid Equations}
\label{sub:fluid}
The fluid equations, or Navier-Stokes equations, arise from applying Newton's Second Law ($F=ma$) to a fluid (continuum), and can be written in the most general form (Cauchy momentum equation; momentum conservation): 
\begin{align}
	\label{eq:cauchy}
	\rho \left ( \frac{\partial \mathbf{v}}{ \partial t} + \mathbf{v} \cdot \nabla \mathbf{v} \right ) = - \nabla p + \nabla 			\cdot \mathbb{T} + \mathbf{f}
\end{align}
where $\mathbf{v}$ is the velocity vector, $p$ is the pressure, $\mathbb{T}$ is the stress tensor and $\mathbf{f}$ represents other body forces. Using the convective derivative ($D/Dt=\partial / \partial t + \mathbf{v} \cdot \nabla$), makes it more clear that it is just a statement of Newton's  Second Law:
\begin{align}
	\label{eq:cauchconv}
	\rho \frac{D \mathbf{v}}{Dt} = - \nabla p + \nabla \cdot \mathbb{T} + \mathbf{f}.
\end{align}
Generally, the mass continuity equation (mass conservation):
\begin{align}
	\frac{\partial \rho}{ \partial t}+\nabla \cdot \left ( \rho \mathbf{v} \right ) = 0,
	\label{eq:masscton}
\end{align}
is used in conjunction with the fluid equations. This is an incomplete description, the stress tensor $\mathbb{T}$ is still unknown, however using knowledge of the viscous behaviour of the fluid, we may make some assumptions on the form and properties of $\mathbb{T}$. For an incompressible Newtonian fluid the equations may be written:
\begin{align}
	\rho \left ( \frac{\partial \mathbf{v}}{ \partial t} + \mathbf{v} \cdot \nabla \mathbf{v} \right ) = - \nabla p + \mu 				\nabla^{2}\mathbf{v} + \mathbf{f}
 	\label{eq:ns}
\end{align}
where $\mu$ is the (constant) dynamic viscosity, and the mass continuity equation reduces to $\nabla \cdot \mathbf{v}=0$. The Navier-Stokes equations are strictly an expression of conservation of momentum and do not fully describe the system: additional information such as an energy equation, or an equation of state is necessary. An energy equation which is often used is:
\begin{equation}
	\frac{D}{Dt} \left ( \frac{p}{\rho^{\gamma}} \right ) = \mathcal{L}
	\label{eq:energy}
\end{equation}
where $\mathcal{L}$ is the total loss function. Assuming the fluid is described by an ideal gas $p={\rho R T}$, then for adiabatic processes $\mathcal{L}=0$, thus \eqref{eq:energy} becomes:
\begin{equation}
	\frac{\partial \rho}{\partial t}  + \mathbf{v} \cdot \nabla p = - \gamma p \nabla \cdot \mathbf{v}
	\label{eq:energy2}
\end{equation}
The Navier-Stokes equations can be non-dimensionalised by scaling each of the quantities by a characteristic measure:
\begin{equation*}
	x'=\frac{x}{L},\quad v'=\frac{\mathbf{v}}{V}, \quad t'=\frac{V}{L}t, \quad \nabla'=\frac{1}{L}\nabla
\end{equation*}
where $L$ is characteristic length, $V$ characteristic velocity, $L/V$ characteristic time scale. Then, depending on the regime under study, the equation can be normalised by:
\begin{align}
	p' &= \frac{1}{\rho V^{2}}p\text{, if viscous effects are small}, \\
	\text{or by }p' &= \frac{L}{\mu V}p\text{, if viscous effects are large.}
\end{align}
The Navier-Stokes equations thus become:
\begin{align}
	\label{eq:nshigh}
	\frac{\partial \mathbf{v'}}{ \partial t'} + \mathbf{v'} \cdot \nabla' \mathbf{v'} &= - \nabla p' + \frac{1}{\Re} \nabla'^			{2}\mathbf{v'} + \mathbf{f'}
\end{align}
and:
\begin{align}
		\label{eq:nslow}
	\Re \left ( \frac{\partial \mathbf{v'}}{ \partial t'} + \mathbf{v'} \cdot \nabla' \mathbf{v'} \right ) &= - \nabla p' + 				\nabla'^{2}\mathbf{v'} + \mathbf{f'}
\end{align}
where $\Re$ is the Reynolds number ($\Re=\rho V L/\mu$). In the low Reynolds number regime where viscous effects dominate, the pressure is provided by the viscosity of the fluid ($p \sim \mu L/V$); in the high Reynolds number regime where inertial effects dominate,  the pressure is provided by dynamic or ram form ($p \sim \rho V^{2}$). Equations \eqref{eq:nshigh} and \eqref{eq:nslow} can be solved to give the force acting on a body in a hight $\Re$ flow:
\begin{align}
	\label{aerodrag}
	F &= \frac{1}{2} C_{D} A \rho V^{2}
\end{align}where $C_{D}$ is the drag coefficient, $A$ is the cross-sectional area of the object perpendicular to the flow direction and a low $\Re$ flow or Stokes flow:
\begin{align}
	\label{stokesdrag}
	F &= 6\pi \mu a V
\end{align}
where $a$ is the radius of the object. The drag coefficient ($C_{D}$) is defined as the drag force normalised by $\rho U^{2}$ and the area $A$.


\subsection{MHD equations}
\label{sec:mhdeqs}
The ideal and resistive MHD equations are obtained by taking the adiabatic, inviscid (viscosity neglected) fluid equations, assuming that the only body forces are due to gravity and the Lorentz force, namely:
\begin{align}
	\rho \left ( \frac{\partial \mathbf{v}}{ \partial t} + \mathbf{v} \cdot \nabla \mathbf{v} \right ) = - \nabla p + 						\mathbf{j} \times \mathbf{B}  + \rho \mathbf{g} \\
	\frac{\partial \rho}{\partial t}  + \mathbf{v} \cdot \nabla p = - \gamma p \nabla \cdot \mathbf{v} \\
	\frac{\partial \rho}{ \partial t}+\nabla \cdot \left ( \rho \mathbf{v} \right ) = 0
\end{align}
these may be combined with Maxwell's equations through Ohm's law:
\begin{align}
	\mathbf{j}=\sigma (\mathbf{E} + \mathbf{v} \times \mathbf{B})	
	\label{eq:ohms}
\end{align}
where $\sigma$ is the conductivity of the plasma. Ohm's law can be rewritten using Ampere's law \eqref{eq:ampere} to obtain:
\begin{align}
	\mathbf{E}=- \mathbf{v} \times \mathbf{B} + \frac{1}{\mu_{0} \sigma } \nabla \times \mathbf{B},
\end{align}
and so Faraday's Law \eqref{eq:maxwell} becomes: 
\begin{align}
	\frac{\partial  \mathbf{B}}{\partial t} = \nabla \times \left ( \mathbf{v} \times \mathbf{B} \right ) - \nabla \times 			\left ( \eta \nabla \times \mathbf{B} \right ),
	\label{eq:indpart}
\end{align}
where $\eta=1/ \mu_{0} \sigma$ is the magnetic diffusivity. Assuming that $\eta$ is constant \eqref{eq:indpart} becomes:
\begin{align}
	\frac{\partial  \mathbf{B}}{\partial t} &= \nabla \times \left ( \mathbf{v} \times \mathbf{B} \right ) - \eta \nabla 					\times \left ( \nabla \times \mathbf{B} \right )
\end{align}
using a vector identity this can be written:
\begin{align}	
	\frac{\partial  \mathbf{B}}{\partial t} &= \nabla \times \left ( \mathbf{v} \times \mathbf{B} \right ) - \eta \left 						[ \nabla^{2} \mathbf{B} - \nabla ( \nabla \cdot \mathbf{B}) \right ].
\end{align}
Using the solenoid constraint $\nabla \cdot \mathbf{B}=0$ we obtain the induction equation:
\begin{align}	
	\frac{\partial  \mathbf{B}}{\partial t} &= \nabla \times \left ( \mathbf{v} \times \mathbf{B} \right ) - \eta \nabla^{2} 				\mathbf{B}.
	\label{eq:induct}
\end{align}
This equation is vital for any model that considers a magnetised plasma, as it describes how a magnetic configuration will respond to fluid motions and vice versa -- which is more dominant depends on the ratio of the terms. In a similar fashion to the fluid equations, we may define a dimensionless quantity $R_{M}$, called the magnetic Reynolds number to be:
\begin{align}
	R_{M}&=\frac{\nabla \times \left ( \mathbf{v} \times \mathbf{B} \right )}{\eta \nabla^{2} \mathbf{B}}, \\
	   	& = \frac{VL}{\eta},
\end{align}
This can thought of as providing two different timescales (1) changes due to fluid motion dominating when $R_{M}$ is large: 
\begin{align}
	\frac{B}{\tau _{motion}} \approx \frac{V}{L}B \implies \tau_{motion} \approx \frac{L}{V}
\end{align}
where $t = V/L$ is the characteristic time scale, and (2) changes due to diffusion of the field when $R_{M}$ is small:
\begin{align}
	\frac{B}{\tau _{diffusion}} \approx \eta \frac{B_{0}}{L^{2}} \implies \tau_{diffusion} \approx \frac{L^{2}}				{\eta}
\end{align}

Let us consider a typical sunspot with $L=10^{7}$\,m, $\eta=1$\,m$^{2}$\,s$^{-1}$ and $V=10^{6}$\,m\,s$^{-1}$, so the magnetic Reynolds number will be:
\begin{align}
	R_{M} = \frac{10^{7}10^{6}}{1}=10^{13}  \gg 1,
\end{align}
so the sunspot is dominated by plasma motions and we can estimate the diffusion time scale to be:
\begin{align}
	\tau_{diffusion} = \frac{L^{2}}{\eta}= \frac{({10^{7})}^{2}}{1}\approx10^{14} \approx 31,000 \text{ years}
	\label{eq:splifetime}
\end{align}

In the case of a perfectly conducting fluid ($R_{M} \rightarrow \infty$) the magnetic field lines must move with the plasma, or are `frozen in' to the plasma \citep{Alfvn1943} and the induction equation reduces to:
\begin{align}
	\frac{\partial  \mathbf{B}}{\partial t} &= \nabla \times \left ( \mathbf{v} \times \mathbf{B} \right )
\end{align}
which forms the basis of ideal MHD. In this case, the magnetic field is tied to the plasma motions.


 \subsection{Magnetic Reconnection}
 \label{sec:recon}
 \figuremacroW{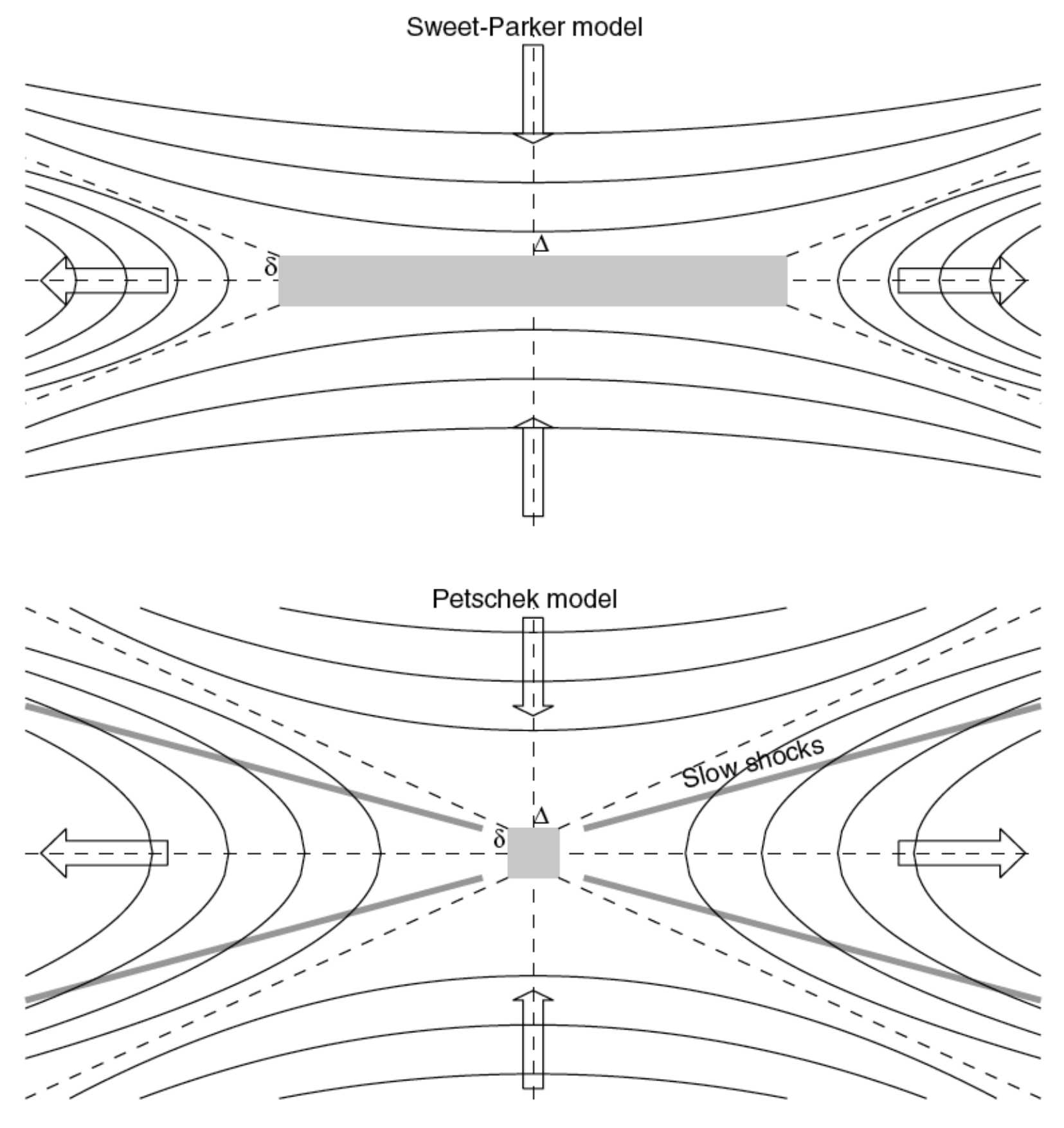}{Sweet-Parker and Petshek Reconnection Geometries}{Geometry of the Sweet-Parker (top) and Petshek (bottom) reconnection models \citep{aschwanden2006physics}. The diffusion region has a length $\Delta$ and width $\delta$.}{0.95}
 
It is clear from a comparison of the lifetimes of sunspots of days to weeks, and our estimation of the lifetime of a sunspot, thousands of years \eqref{eq:splifetime} that some other process must play a role in the dynamic evolution of magnetised plasmas. That process is magnetic reconnection. This is also indicated from the rapid energy release associated with CMEs and flares. Magnetic reconnection is generally defined as a change in the connectivity of field lines with time, where the energy stored in the magnetic field is converted into particle, thermal, and kinetic energies. In an ideal plasma, the field lines are `forzen-in' and thus coupled to the plasma motion. However when regions of opposite polarity flux are in close proximity a boundary layer with large currents must form to separate the two regions (this can be seen from \eqref{eq:amperem}). The small but non-zero resistivity (or large but not infinite conductivity) in the boundary layer opposes these currents. At the very centre of the boundary layer the neutral layer the magnetic field must go to zero, so the field smoothly and continuously changes across the region. The total pressure in this region must balance on both sides, $B_{1}+p_{1} = p_{nl}=B_{2}+p_{2}$ where $p$ is thermal pressure and $B$ the magnetic pressure. Thus, in the neutral layer ,the plasma-$\beta$ becomes large and non-ideal effects, such as diffusion can take place. This region is known as the diffusion region, grey area in \figref{reconmodels}. This formalism was first developed by \citet{sweet1958p123}, who showed diffusion could occur in the boundary layer, and then \citet{parker1957p509} who derived the scaling laws and together is known as Sweet-Parker reconnection.

\subsubsection{Sweet-Parker Reconnection}
\label{sub:sweetparker}
The same MHD equations cannot be applied simultaneously to describe inside and outside the diffusion region. However, using some simplifications, separate solutions for the two regions can be found and matched. The Sweet-Parker model is a steady-state 2D model where the diffusion region is assumed to be much longer than it is width ($\Delta \gg \delta$; \figref{reconmodels}~(top)). For a steady state incompressible flow ($\nabla \cdot \mathbf{v}=0$), we can see that the inflow and outflow must balance $v_{in}\Delta=v_{out}\delta$. Outside of the diffusion region, ideal MHD applies so there are no currents and the z-component of Ohm's law \eqref{eq:ohms} is $E_{z}+v_{in}B_{x}=0$. Inside the diffusion region there are large currents and the flow has stagnated ($v=0$), thus the z-component of Ohm's law is $E_{z}=\eta J_{z}$. For a steady state flow, this implies that the electric field $E_{z}$ inside and outside the diffusion zone must be the same by Faraday's Law \eqref{eq:maxwell}. Thus we can write $v_{in}B_{x} = \eta J_{z}$ or $v_{in}=\eta J_{z}/B_{x}$ and integrate Ampere's Law \eqref{eq:amperem} around the region to obtain $B_{x}=\mu J_{z}\delta$, and thus we arrive at:
\begin{align}
	v_{in} = \frac{\eta}{\delta}.
\end{align}
Now, if we assume the x-component of the magnetic field is completely destroyed, then by conservation of energy $B_{x}/2\mu=1/2 \rho v_{out}^{2}$ and thus $v_{out}= B_{x}/\sqrt{\mu \rho} \equiv v_{A}$ and we can write in inflow velocity in terms of the outflow velocity (or Alfv{\'e}n velocity):
\begin{align}
	v_{in}^{2} = \left( v_{in}\frac{\delta}{\Delta} \right )\left ( \frac{\eta}{ \delta} \right ) \text{ or } v_{in}^{2} = v_{out}^{2} \left( \frac{\eta}{v_{A} \Delta} \right).
\end{align}
The rate of reconnection is given by the ratio of the inflow to outflow velocities:
\begin{align}
	M =  \left ( \frac{v_{in}}{v_{out}} \right )^{1/2} = \left (\frac{\eta}{v_{A} \Delta} \right )^{1/2} = \frac{1}{\sqrt{S}}
\end{align}
where $S=L v_{A}/\eta$ is the Lundquist number (or magnetic Reynolds number). The Lundquist number can be estimated from typical values of $L=10^{7}$\,m and $v_{A}=10^{6}$\,m\,s$^{-1}$ to obtain $S \approx 10^{13}$ and hence $M\approx10^{-7}$.

For a typical solar flare/CME some $\sim$\,10$^{23}$\,J are released over a time scale of about $10^{2}$\,s, which means an energy release rate of $\sim10^{21}$\,J\,s$^{-1}$. Assuming all this energy comes from magnetic reconnection we can write:
\begin{align}
	\frac{dE_{M}}{dt} = \frac{B^{2}}{2 \mu} \frac{dV}{dt} \approx \frac{B^{2}}{2 \mu} L^{2}v_{out} \approx  \frac{B^{2}}{2 \mu} \frac{L^{2}v_{A}}{\sqrt{S}}.
\end{align}
where $E_{M}$ the magnetic energy. Taking a field strength of $10^{-2}$\,T we get an energy release rate of $\sim10^{15}$\,J\,s$^{-1}$, which corresponds to a time of tens of days for a flare rather then minutes much too long compared to observations. Sweet-Parker reconnection is thus too slow, with the limiting factors being the ratio of the width to length of the diffusion region and the Alfv{\'e}n velocity.

\subsubsection{Petschek Reconnection}
\label{sub:petshek}
\citet{Petschek1964p425} put forward an altered version of the Sweet-Parker model, this altered configuration is shown in  \figref{reconmodels}~(bottom). In this configuration the size of the diffusion region is very compact ($\delta \approx \Delta$) compared to Sweet-Parker ($\Delta >> \delta$). Due to the smaller size, the propagation time through the diffusion region is reduced and reconnection can proceed at a faster rate. However, as the size is reduced so is the amount of plasma that can flow through the diffusion region. The result of this is that much of the inflowing plasma turns around outside of the diffusion region and two slow mode shocks form due to the abrupt changes from $v_{in}$ to $v_{out}$. These shock waves are the main site where inflowing magnetic energy is converted to heat and kinetic energy. By assuming a potential field in the inflow region \citet{Petschek1964p425} found the external field scales logarithmically with distance $L$.  Using this he estimated the reconnection rate, at a distance $L$, where the magnetic field dropped to half its original value, to be:
\begin{align}
	M = \frac{\pi}{8 \ln (L/\Delta)} \approx \frac{\pi}{8 \ln (S)}.
\end{align}
The reconnection rate only logarithmically depends on the Lundquist number (or $R_{M}$) using our previous estimate of $S$ this gives $M \approx 0.01$ which is four orders of magnitude faster than Sweet-Parker.

\figuremacroW{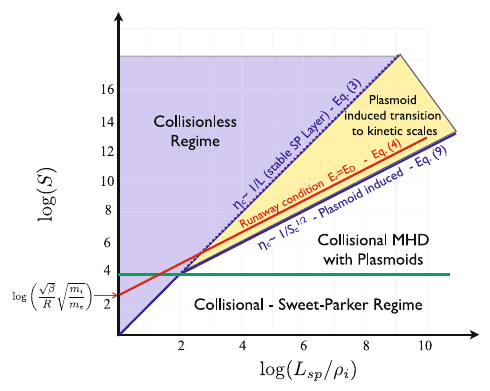}{Magnetic Reconnection Regimes}{ The emerging parameter space for the various reconnection regimes. S is the Lunquist number, and L$_{sp}$ is the length of the reconnection region normalised by density \citep{Daughton2011p52}.}{0.75}

Both of the models discussed have no external driver, such as an inflow. They also require a potential field in the inflow region. Relaxation of these simplifications has led to a family of solutions. These generalisation of 2D reconnection can be summarised in terms of internal Alfv\'{e}n Mach number at the entrance to the diffusion region $M_{Ai}$, and the Alfv\'{e}n Mach number at the exterior inflow $M_{Ae}$ ,as:
\begin{align}
	 \left ( \frac{M_{Ai}}{M_{Ae}}\right )^{1/2} = 1 - \frac{4}{\pi} (1-b) \left [ 0.834 - \ln \tan \left (\frac{\pi}{4}  S^{-1} M_{Ae}^{-1/2} M_{Ai}^{-3/2} \right ) \right ]
\end{align}
This contains the Sweet-Parker, Petschek ($b=0$) and other hybrid solutions \citep{Priest2000,Schrijver2009,aschwanden2006physics}. \figref{reconregime} shows how as the length scales ($L_{sp}$) and Lundquist number ($S$) vary, different regimes become important. 3D reconnection is significantly different from 2D reconnection in many regards for example in the 3d case x-point less reconnection can occur, as well as reconnection where the magnetic field lines have the same topology. Reconnection plays a vital role in many CME models and in the interpretation of observations, although not all models require magnetic reconnection.

\section{Coronal Mass Ejection Theory}
\label{sec:cmetheory}
CMEs are known to be associated with flare and filament eruptions, but the exact mechanism of their driver is unknown. Additionally, there is a debate about whether the flux rope  (which becomes the CME) is a pre-existing structure, or is formed during the eruption. A number of theoretical models have been put forth to describe the forces and mechanisms responsible for the initiation and acceleration of CMEs. In nearly all models some instability is the trigger for the eruption, these instabilities may be understood using the cartoon mechanical analogies shown in \figref{cmetriggers}.

\begin{itemize}
	\item {\bf Thermal Blast:} Early models proposed that the rapid increase in thermal pressure caused by a flare cannot be constrained by the magnetic field and thus drives the CME outward. Observations suggest this is not correct as many CMEs have no associated flare, or the flare occurs after the CME (see Section \ref{sec:hist} and \ref{sec:myth})  

	\item {\bf Dynamo:} These models require the rapid generation of magnetic flux by stressing of the magnetic field \citep{Klimchuk1990p745}. This driver, known as flux injection \citep{Krall2000p964}, corresponds to one of the following cases; (1) pre-existing field lines become twisted; (2) new ring shaped field lines rise up and become detached from the photosphere; or (3) arch shaped field lines rise up while remaining rooted in the photosphere. The first case requires footpoint motion some two orders of magnitude faster then observed \citep{Krall2000p964}. The second case has not been observed and there is no obvious force to lift the mass. The third case is the most plausible for this model.

\figuremacroFP{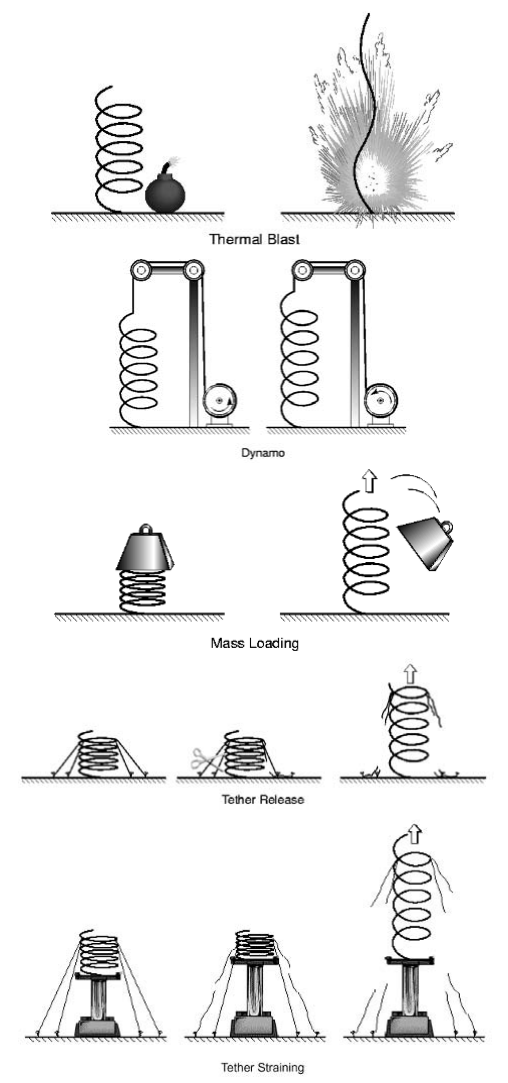}{Cartoon Showing Mechanical Analogies for CME Initiation Mechanisms}{Cartoons showing the mechanical analogues for different CME eruption mechanisms \citep{Klimchuk2001p143}.}{0.9}
	
	\item {\bf Mass Loading:} The slow build up of mass which is then removed causes an eruption. In terms of observations it is consistent with growing quiescent and eruptive filaments. Theoretically, the magnetic energy pre- and post-eruption are compared to show the plausibly of the transition from a higher to lower energy state \citep[e.g.][]{Low1999p109}. Two possible forms of mass loading are: (1) by prominences with dense cool material contained in a compact volume; (2) by relatively less dense material spread over a large volume, which is susceptible to the Rayleigh-Taylor instability. The first concept is supported by numerous observations of coincident filament and CMEs, where the derived mass of the filament is a critical constraint \citep{Zhang2004p1043}. The second concept is supported by low density cavities observed in some helmet streamers, but there are also many which do not contain low density cavities \citep{Hundhausen1999p143}.
	
	\item {\bf Tether Release:} Magnetically dominated loops, such as those in the corona, remain stable due to a balance between the upward force of magnetic pressure ($\nabla \mathbf{B}^{2}/2\mu_{0}$) and the inward force of magnetic tension ($(\mathbf{B}\cdot \nabla)\mathbf{B}/\mu_{0}$). The field lines which provide the tension are sometimes called tethers. In the mechanical analogy, once a tether is released then the tension on the remaining tethers increases, this can continue until the stain becomes so large that the remaining tethers fail and the spring (CME) is launched upwards. A 2D model recreates this type of behaviour by driving the footpoints towards each other, until a catastrophic loss of equilibrium occurs and the x-point jumps discontinuously to a new height \citep[e.g.][]{Isenberg1993p368}. In non-ideal MHD, reconnection at the x-point after the tethers are cut would launch the CME \citep[e.g.][]{Lin2000p2375,Amari2000p49,Mikic1999p918}. 
	
	\item {\bf Tether Straining:} This is very similar to the tether release model except the strain is increased by some external force rather than a constant strain being redistributed among fewer and fewer tethers. A physical model of this is the `magnetic breakout' model \citep{Antiochos1999p485,Aulanier2000p447}. This type of behaviour can also found using mass loading \citep{Zhang2002pD4}. The equilibrium loss model of \citet{Forbes199p377} is also of the tether straining type, as are the sheared arcade model of \citet{Mikic1994p898}; \citet{Linker1995p45} and flux rope models of \citet{Wu1995p325,Wu2000p1101}.
\end{itemize}

The last three models are known as storage and release models where magnetic stress is slowly built up over time before the eruption, some of which is then rapidly released during the eruption. The tether straining and release models seem the most likely scenarios, as they are able to reproduce a number of observable features through 2D and 3D modelling. The full dynamical evolution of CMEs includes three phases: initiation, acceleration and propagation \citep{Zhang2001p452}. Below I will outline a representative number of models which describe the initiation and acceleration (Section \ref{sec:cmeforminit}) of CMEs and, separately, their propagation (Section \ref{sec:cmepropagation}). 

\subsection{Initiation and Acceleration}
\label{sec:cmeforminit}
One of the initial problems facing storage and release CME models was the apparent paradox that the pre-eruption closed field system was of lower magnetic energy than post-eruption open field system \citep{Sturrock1984p341}. It was argued that the relaxation of the pre-existing field releases more energy than it takes to open the field or the free magnetic energy stored in the field should be greater than the energy required to open the field \citep{Barnes1972p659}. Following this \citet{Kopp1976p85} proposed a three stage model for eruptive flares: (1) energy is stored in a force free arcade or flux rope; (2) the eruption occurs, fully opening the field; and (3) the open field is closed via reconnection to a nearly force-free state. The transition from 1) to 2) would be ideal and 2) to 3) non-ideal. \citet{Aly1984p349} conjectured that this scenario was energetically unfeasible. He argued that the open field configurations must always have higher magnetic energy than corresponding force-free configurations, as long as the field is simply connected. In 1991 both \citet{Aly1991p61} and \citet{Sturrock1991p655} developed proofs of Aly's conjecture. These results seemed to imply that CMEs were energetically impossible, however there are a number of ways to avoid this problem: the field may not be simply connected but contain x- and o-points, the field lines are not opened up to infinity or only a portion of the closed field lines are opened.

\subsubsection{Catastrophe Model}
\figuremacroW{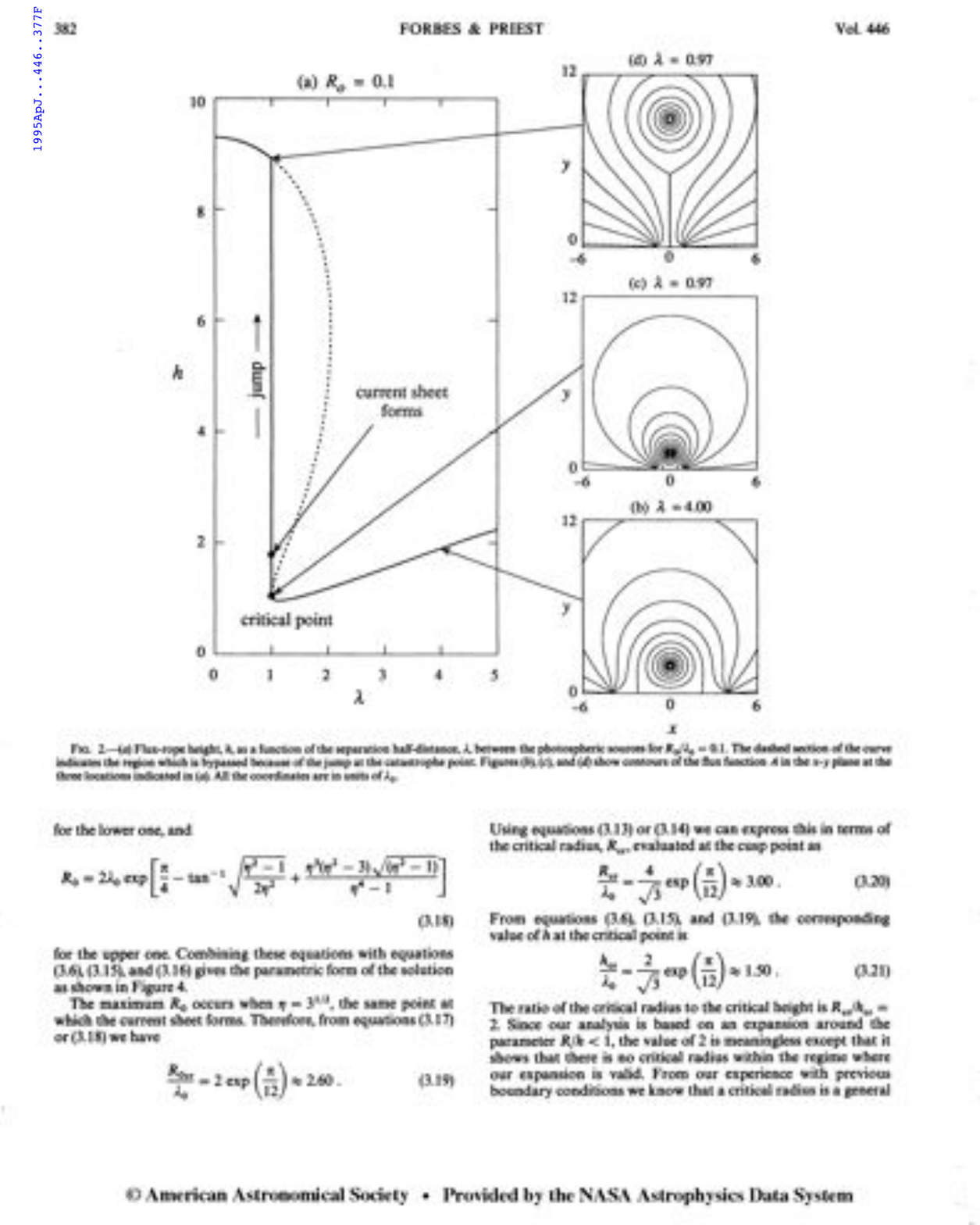}{Theoretical Evolution of 2D Flux Rope}{The theoretical evolution of a 2-D flux rope of radius $R_{0}/\lambda_{0}=0.1$ \citep{Forbes1995p377}. (b-c) The footpoints are slowly moved towards each other and flux rope height decreases. (d) At some critical distance $\lambda_{0}$ the flux rope loses equilibrium and abruptly jumps in height. (a) Shows the resulting height evolution as a function of the footpoint separation $\lambda$. All coordinates are in units of $\lambda_{0}$.}{0.8}

In these models a 2D flux rope catastrophically loses equilibrium and erupts due to footpoint motions in the photosphere \citep{Priest1990p319,Forbes1991p294,Isenberg1993p368,Forbes1995p377}. \figref{2dfluxrope} illustrates one such model where a flux rope is suspended above the photosphere due to an equilibrium between the upward magnetic pressure gradient force and the downward magnetic tension force. As the footpoint separation distance ($\lambda$) is reduced, the flux rope moves downward due to increasing magnetic tension, and as the magnetic pressure grows the magnetic energy of the system also increases. When the critical footpoint separation is reached and equilibrium is lost, the pressure gradient force becomes larger than the tension force and the flux rope is propelled upwards. If all the kinetic energy is dissipated then the flux rope will stabilise at the upper equilibrium, as shown in \figref{2dfluxrope}. If not, the flux rope will oscillate between the critical height and some point above the upper equilibrium height. If reconnection occurs in the current sheet which forms below the flux rope, it will increase in height without limit. An analytical expression for the kinematics of the flux rope, provided it is thin ($R \ll h$), and before the formation of the current sheet $h/\lambda_{0} \ll 2$ \citep{Priest2000}, is:
\begin{align}
	\dot{h} \approx \sqrt{\frac{8}{\pi}} v_{A0} \left [ \ln \left ( \frac{h}{\lambda_{0}}\right )+ \frac{\pi}{2} - \tan^{-1}\frac{h}{\lambda_{0}} \right ]^{-1/2} + \dot{h}_{0}
\end{align}
where $h$ is the flux rope height, $\dot{h}$ is the velocity of the flux rope, $\dot{h}_{0}$ is an initial perturbation velocity, $\lambda_{0}$ is the critical footpoint separation, and $v_{A0}$ is the Alfv\'{e}n speed at $h=\lambda_{0}$. Simplified expressions may be obtained by separating the evolution into an `early' ($t \ll \lambda_{0}/v_{A0}$, $h/\lambda_{0} \ll 1$) phase:
\begin{align}
	\dot{h} \simeq \dot{h}_{0} + \frac{2 v_{A0}}{\sqrt{3 \pi}} \left ( \frac{\dot{h}_{0}}{\lambda_{0}} \right )^{3/2} t^{3/2}
\end{align}
 and a `late'' ($h/\lambda_{0} \gg 1$, $\ln h \ll \ln R$) phase:
 \begin{align}
 		\dot{h} \approx \sqrt{\frac{8}{\pi}} v_{A0} \left [ \ln \left ( \frac{h}{\lambda_{0}} \right ) - \frac{\pi}{2} \right ]^{1/2}.
 \end{align}
Once the current sheet forms, analytical solutions of the CME evolution can no longer be derived. 

\subsubsection{Toroidal Instability Model}

\figuremacroFP{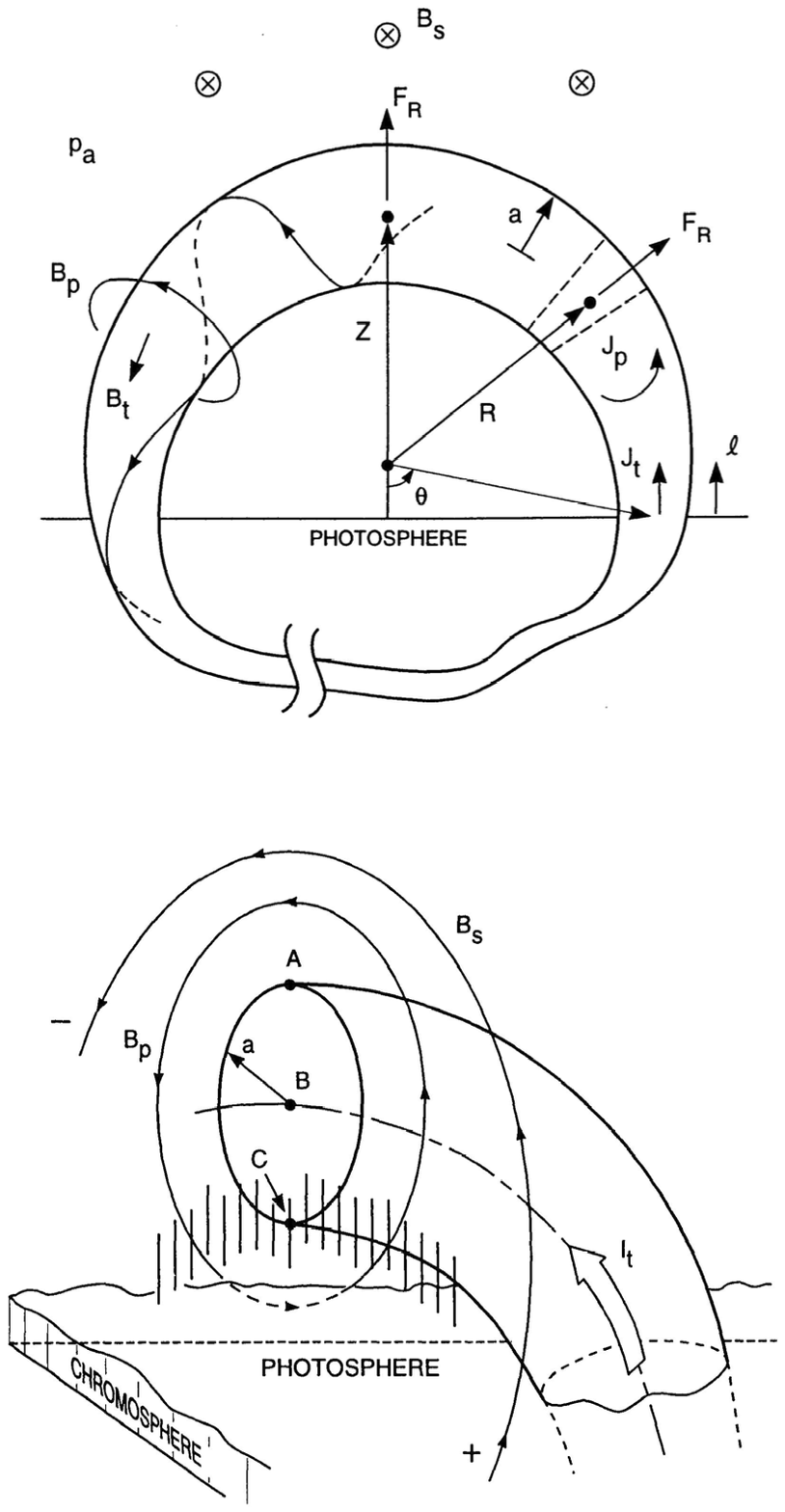}{3D Flux Rope Model}{(top) A 3D flux rope modelled as a current loop with subscripts ``t'' and ``p'' referring to toroidal and poloidal directions. (bottom) Geometry of the flux rope as a model for a prominence as indicated by vertical hash lines. The ambient field $B_{S}$ marks the boundary of the flux rope \citep{Chen1996p27499}.}{0.8}

This model consists of an extension of the 2D flux rope to  a 3D magnetic flux rope suspended above the photosphere, which is anchored in the photosphere at both ends \figref{chenmodel} \citep{Chen1989p453,Chen1996p27499,Chen2000p481,Krall2001p1045,Krall2000p964}. It is initially in equilibrium, and erupts as a result of poloidal magnetic flux being injected into the flux rope, triggering the toroidal instability (TI). There are two dominant forces acting on the flux rope: the outward Lorentz self-force (hoop force), and the inward Lorentz force due to the background field. The flux rope is susceptible to an eruptive instability if the background magnetic field ($B_{S}$) decreases sufficiently quickly and vanishes at infinity. A sudden increase in the poloidal flux would then lead the flux rope to rapidly and continually expand, triggering a CME eruption. The condition on the background magnetic field for this to occur is:
\begin{align}
	-R \frac{d ln B_{S}}{dr} > 3/2
\end{align} where $R$ is the major radius of the flux rope see \figref{chenmodel}~(top). An equation of motion for this system may be written:
\begin{align}
M\frac{dZ}{dt} = F_{R}+F_{g}+F_{d}
\end{align} where $F_{R}$, $F_{g}$ and $F_{d}$ are radial, gravitational and drag forces respectively. The radial component can be written:
\begin{align}
	\label{rq:chenflxrp}
F_{R} = \frac{I_{t}^{2}}{c^{2}R} \left [ \ln \left ( \frac{8R}{a} \right ) + \frac{1}{2} \beta_{p} - \frac{1}{2}\frac{\overline{B_{t}}^{2}}{{B_{pa}}^{2}} + 2 \left ( \frac{R}{a} \right ) \frac{B_{s}}{B_{p}} - 1 + \frac{\xi_{i}}{2} \right ] + F_{g} + F_{d}
\end{align} where $\xi_{i}$ is the internal inductance, $\beta_{p}$ is the ratio of the average gas to magnetic pressure at the flux rope boundary. This equation is too complex to find analytical solutions, but has been numerically solved an example is shown in \figref{cheneg}. 

 \figuremacroW{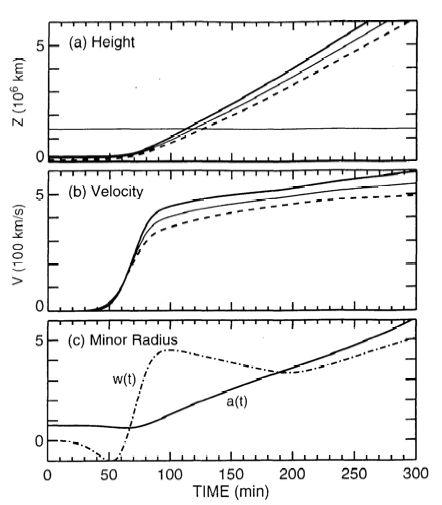}{Toroidial Instability Simulation}{Toroidal instability simulation. (a) The height of the apex (thick), centre of mass (thin), and trailing edge (dashed). (b) Corresponding speed. (c) The minor radius $a(t)$ in $10^{5}$\,km and $w(t)$ in 10\,km\,s$^{-1}$ \citep{Chen1996p27499}.}{0.5}

\figuremacroFP{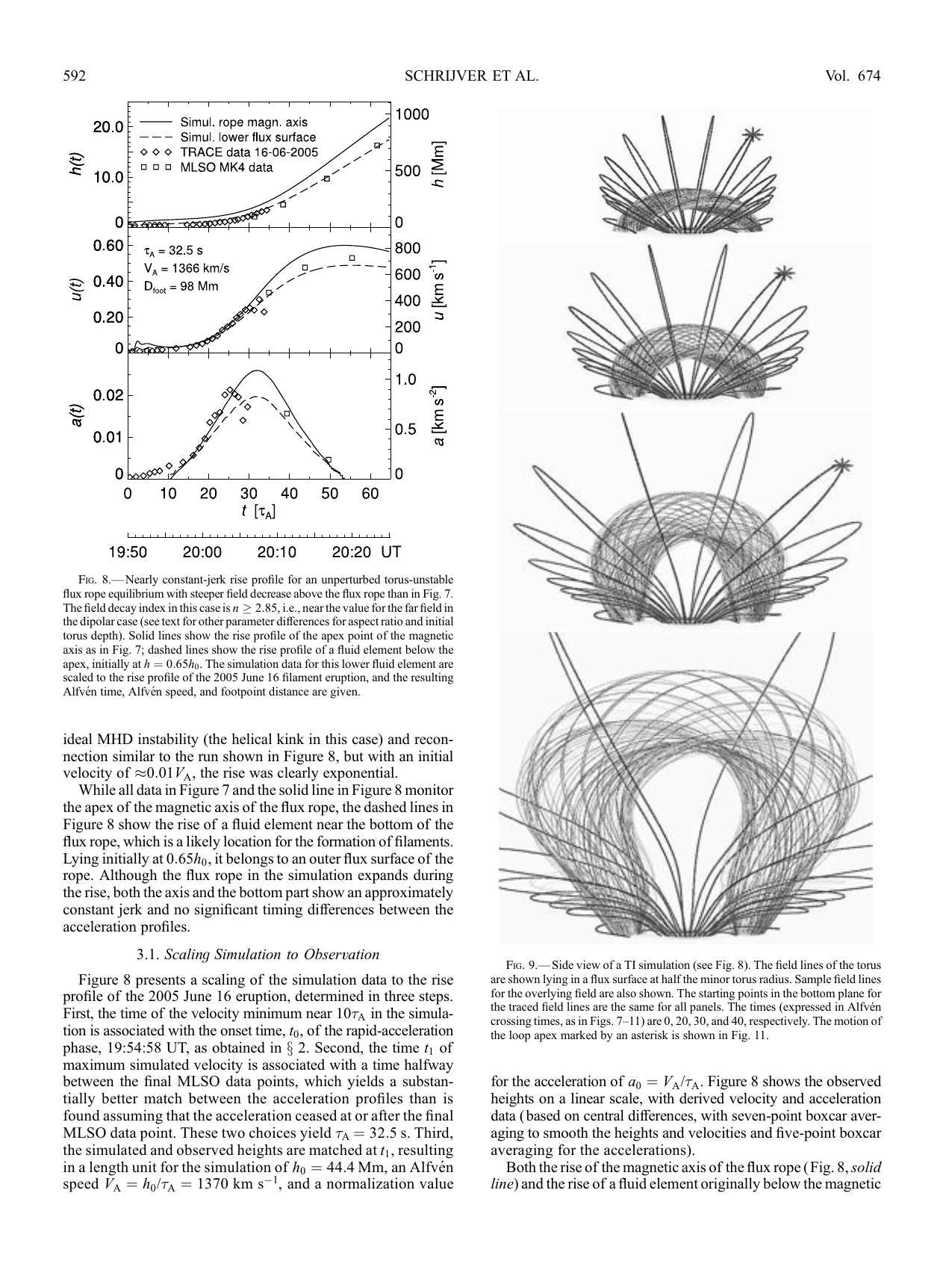}{3D Toroidal Instability Simulation}{Side view of a 3D toroidal instability \citep{Schrijver2008p586}. The field lines of the torus are shown lying in a flux surface at half the minor torus radius. Sample field lines for the overlying field are also shown. The starting points in the bottom plane for the traced field lines are the same for all panels. The times (expressed in Alfv\'{e}n crossing times) are 0, 20, 30, and 40, respectively.}{0.85}

\citet{Kliem2006p5002} were able to derive an analytical solution based on a simplified model. They assumed only two counteracting Lorentz forces acted on the flux rope, ignored changes in the external field its effects on the flux rope, and assumed a simple profile for the external field ($B_{S}(H)=B_{0}H^{-n}$). They found that at the beginning the instability can be approximated by:
\begin{align}
h(\tau) = \frac{P_{0}}{P_{1}} \sinh (P_{1}\tau)\text{, }h \equiv \frac{H}{H_{0}} -1 \ll 1 
\end{align}where $H$ is the height of the flux rope, $H_{0}$ is the height of the flux rope at on-set of the instability, $\tau$ is time normalised by Alfv\'{e}n time, 
$P_{0}$ comprises initial parameters of the flux rope and $P_{1}$ is associated with the external magnetic field profile. A distinctive feature of their expression and simulation was that the acceleration shows a fast rise and a more gradual decay. However, \citet{Schrijver2008p586} was able to demonstrate that the evolution of the instability can be changed by tuning the initial conditions. The hyperbolic height dependance can change to a polynomial form, which changes the acceleration profile to a more gradual one. One issue with these models is that they contain no twist as is often observed in CME eruptions (see \figref{3dchenflr}). Still, a number of studies have shown good agreement between simulations and observations \citep{Krall2000p964,Krall2001p1045}. Recent attempts have tried to establish the amount of energy available from the photosphere to create the instability. Using an ideal flux rope event \citet{Schuck2010p68} found there was insufficient energy to launch the CME.

\subsubsection{Kink Instability Model}

\figuremacroW{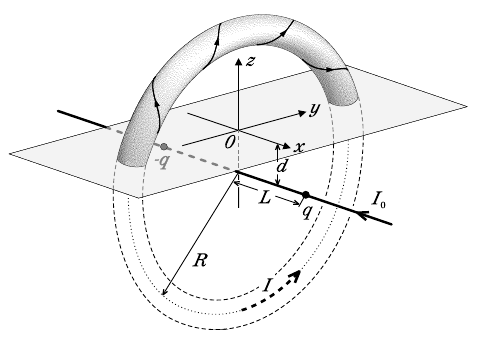}{3D Potential Magnetic Field Bundle}{Model for Kink instability the magnetic field under study is modeled by a force-free circular flux tube with the total current $I$, a pair of magnetic charges $-q$, $q$ and a line current $I_{0}$ \citep{Titov1999p707}.}{0.6}

The kink instability (KI) aries from a similar set up as the TI, except it involves the twisting of the photospheric footpoints (increasing the toroidal flux) of an initially potential flux bundle see
\figref{titovflxrope} \citep{Torok2003p1043,Torok2004p27,Torok2005p97,Fan2003p105,Rachmeler2009p1431}. The twisting motion serves to both form the flux rope, and also to move it into a region of possible instability. The continual twisting of a flux rope of finite length must lead to a KI.
\figuremacroFP{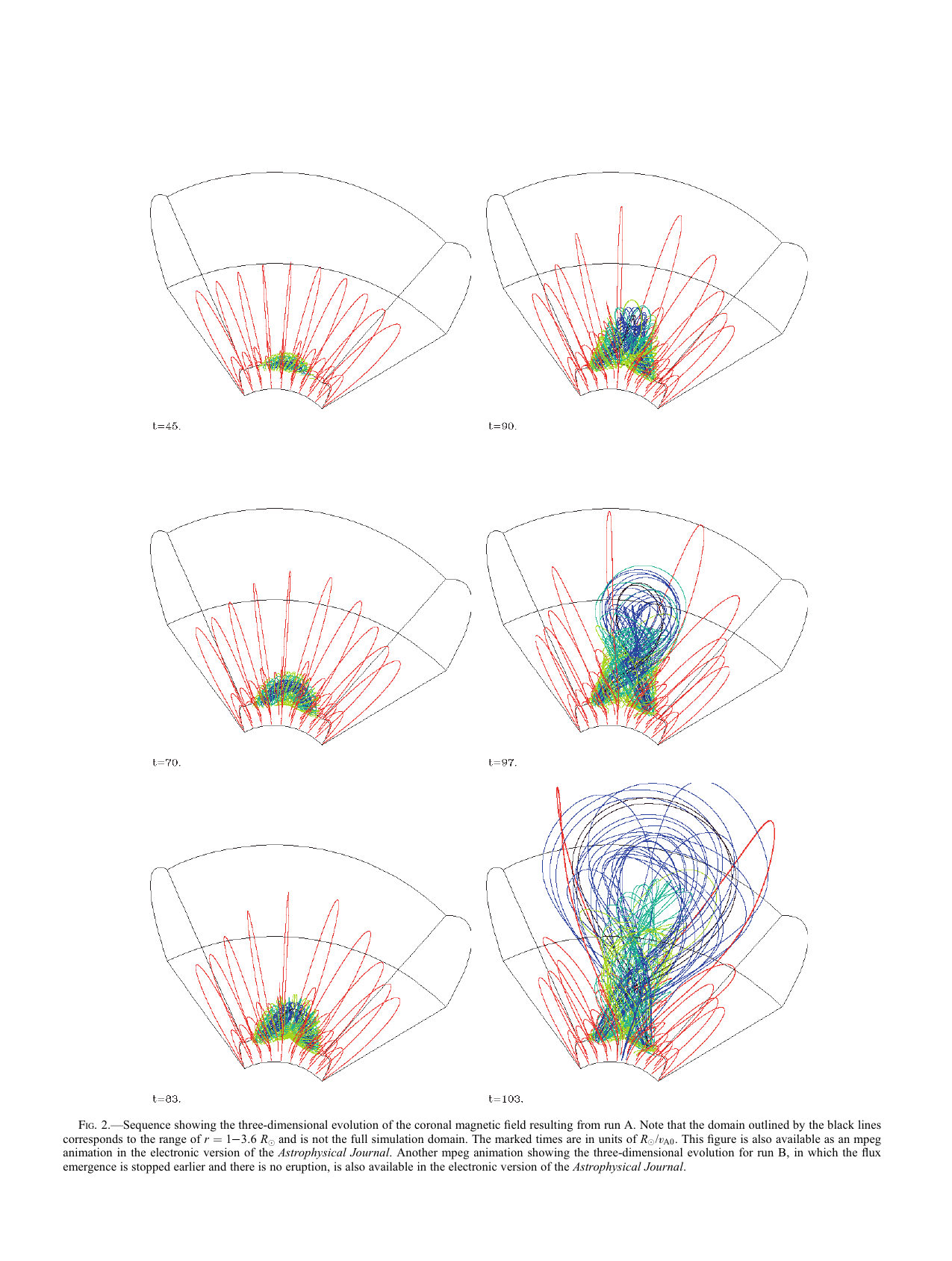}{3D Kink Instability Simulation}{Simulation of the kink instability \citep{Fan2005p543}. The black lines corresponds to the range of $r=1$\rng$3.6\,R_{\odot}$ and is not the full simulation domain. The marked times are in units of $R_{\odot}/v_{A0}$.}{0.8}
The threshold for such an instability depends on the ratio of the azimuthal to axial field components ($B_{\phi}/B_{z}$), the length-to-width ratio of the flux rope ($L/r$), and the radial profile of the flux rope. It can be expressed in terms of the flux rope twist:
\begin{align}
	\Phi = \frac{L B_{\phi}(r_{0})}{r_{0}B_{z}(r_{0})}
\end{align} where $r_{0}$ is a characteristic radius of the flux rope.
For a straight cylindrically symmetric flux rope with fixed ends of uniform twist (Gold-Hoyle equilibrium), the threshold was numerically found to be $\Phi_{GH}=2.49 \pi$ \citep{Hood1981p297}. There are no analytical solutions for the kinematics of the KI model, but it has been studied extensively using numerical simulations. Once the flux rope achieves ``supercritical twist'' two current sheets form, the first helical wrapping the rising kinked flux rope, and the other a vertical sheet below the flux rope \citep{Titov1999p707,Fan2003p105,Fan2004p1123}. The formation of the vertical current sheet enables the eruption to proceed with, or without, reconnection. The KI model recreates some of the phenomena observed in CME-related eruptions including soft x-ray sigmoids \citep{Kliem2004p23} and the twisted structures which often appear in CMEs and prominences (see \figref{3dkinksim}). Significantly it also enbles the build up of substantial mass and can release up to 25\% of the stored energy \citep{Torok2005p97}. However not all simulations of the KI form current sheets and so remain confined i.e., with no CME eruption \citep{Fan2005p543}, and the amount of twist required to produce an explosive eruption may be non-physical.

\subsubsection{Magnetic Breakout Model}
\figuremacroFP{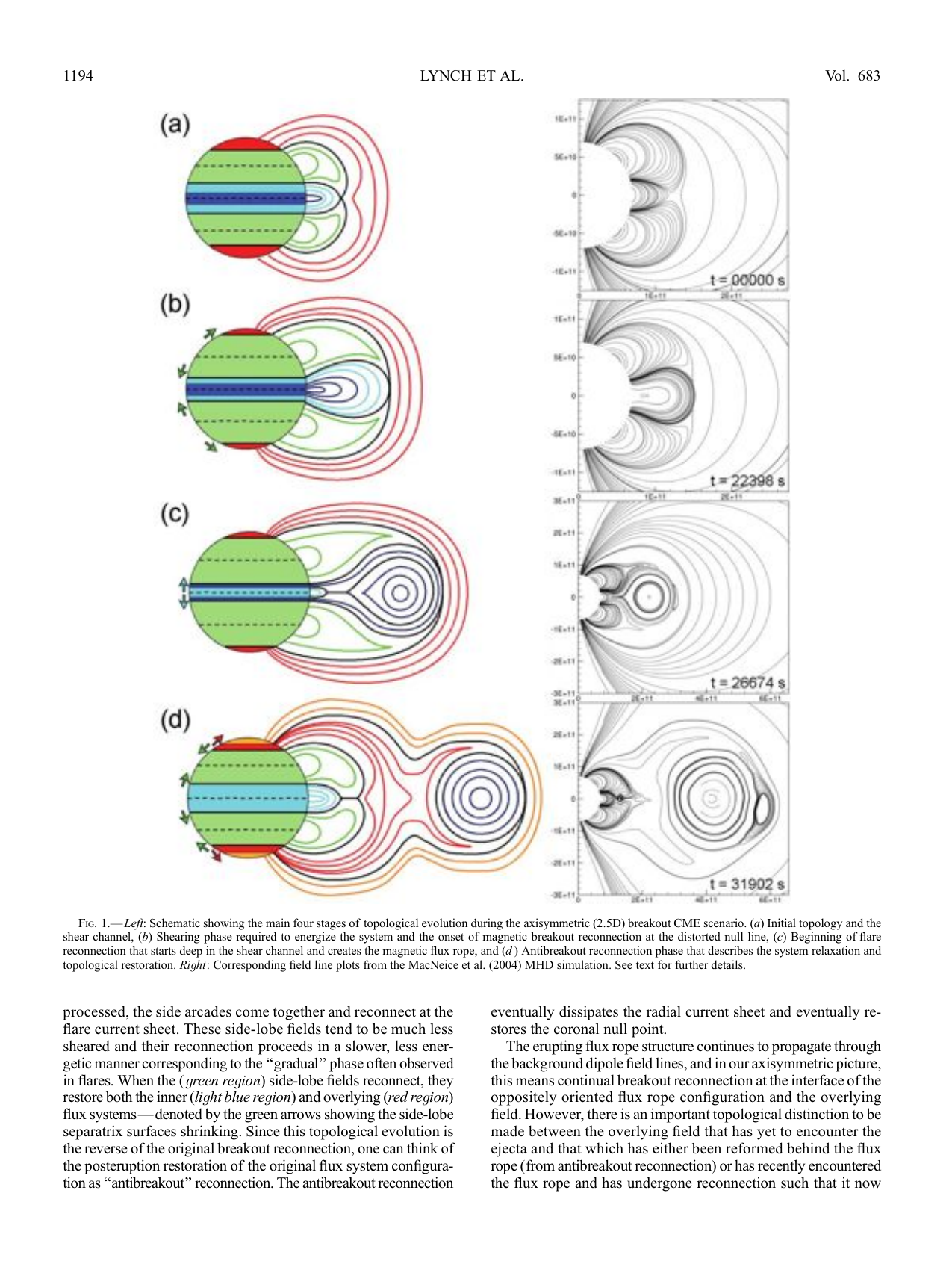}{Breakout Schematic and Simulation}{Schematic showing topological layout (left) and evolution of breakout model \citep{Lynch2008p1192}. (a) Initial multipolar topology, (b) shearing phase which energises the system causes magnetic breakout reconnection at the distorted null line, (c) flare reconnection occurs low down and forms the flux rope, and (d) the system relaxing after the eruption.}{0.7}

In the magnetic breakout (BO) model the CME eruption is triggered by shearing which causes reconnection between the overlying field and the multipolar field below it, as shown in \figref{bomodel} \citep{Antiochos1999p485,MacNeice2004p1028,Lynch2004p589,Lynch2008p1192,DeVore2005p1031}. This configuration has four distinct flux systems: a central low-lying arcade straddling the equator (\figref{bomodel}~a-blue); two low-lying side arcades (one on each side of the central arcade \figref{bomodel}~a-green); and a large scale (polar) arcade overlying the three low-lying arcades (\figref{bomodel}~a-red). There is a null point above the central arcade. Shearing concentrated at the equatorial neutral line causes the central arcade to rise, distorting the null point into an x-point (\figref{bomodel}~b). Continued shearing causes the central arcade to rise even more, stretching the x-point to form a current sheet. Reconnection occurring in the current sheet transfers flux from the overlying field and the un-sheared field to the side arcades (\figref{bomodel}~c), thus creating a passage for the CME without opening the field. As the central arcade continues to rise a current sheet forms behind it (\figref{bomodel}~d). A disconnected flux rope is created due to reconnection in this current sheet which will result in flare activity. The feedback between the outward expansion drives faster breakout reconnection which, in turn, causes more expansion and leads to an explosive eruption. After the secondary reconnection cutsoff the current sheet, the side arcades will move inward and a third reconnection phase begins to restore and reform the magnetic fields (flare ribbons). There are no analytical expressions for the kinematics of the BO model though a number of simulations have produced kinematic profiles. \citet{Lynch2004p589} used a 2.5D simulation to show that the kinematics could be well represented with a constant acceleration profile:
\begin{align}
	h(t) = h_{0} + v_{0}(t) + \frac{1}{2} a_{0} t^{2}.
\end{align}
\figuremacroW{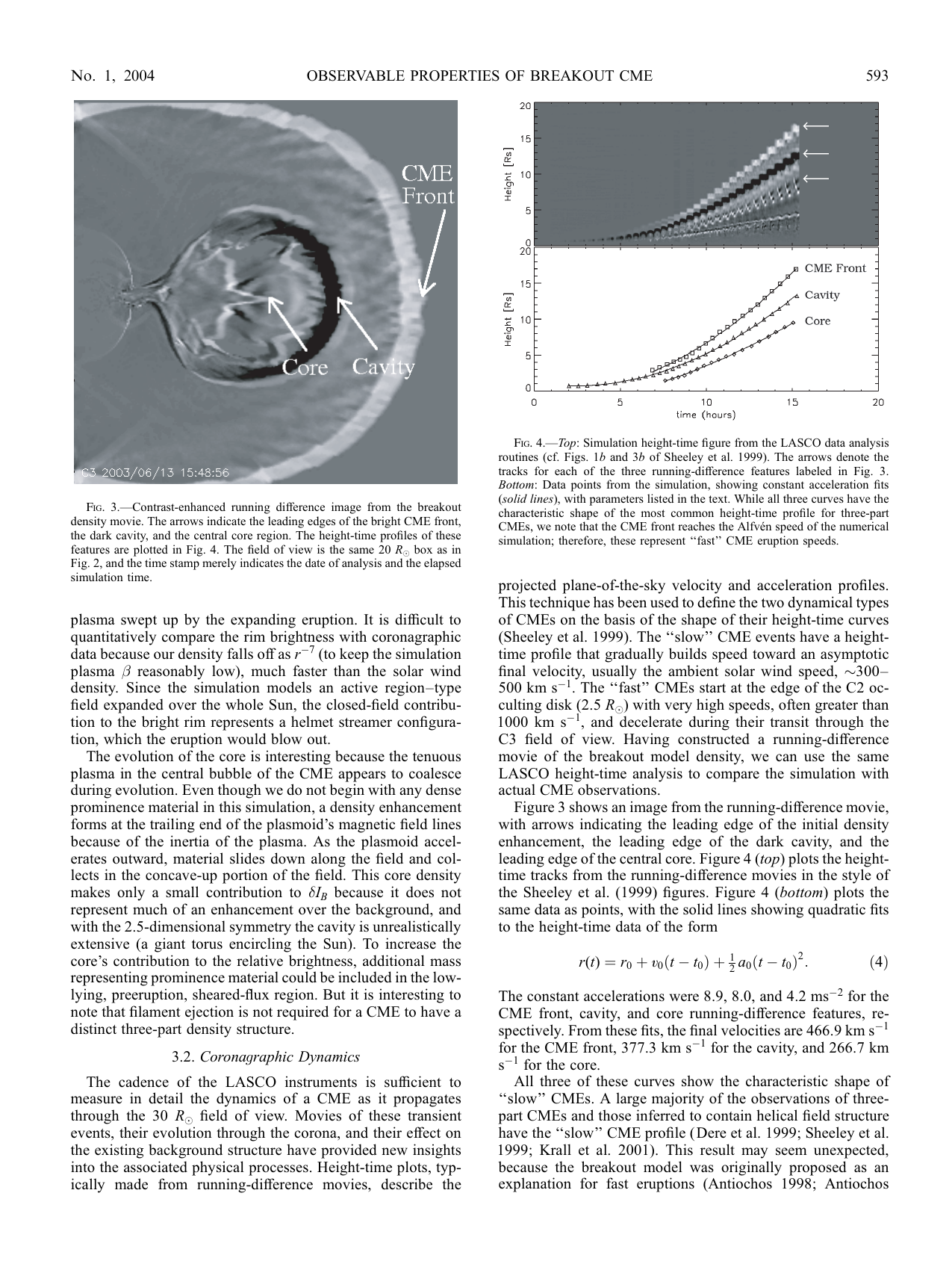}{Breakout CME Simulation}{Running difference image from the breakout model density data \citep{Lynch2004p589}. The arrows indicate the leading edges of the bright CME front, the dark cavity, and the central core region. The time stamp indicates the date of analysis and the elapsed simulation time.}{0.5} 3D simulations by \citet{Lynch2008p1192} resulted in more complex kinematics which consisted of three phases of constant acceleration: 1) extended low acceleration during shearing and breakout; 2) followed by a short period of high acceleration due to the secondary reconnections; 3) and the final phase of zero acceleration or constant velocity. \citet{DeVore2008p740} found a similar but not identical profile with three phases of constant acceleration:  1) extended low acceleration during shearing; 2) short fast acceleration with breakout and secondary reconnections; 3) and a fast deceleration during the restoration phase.

The \citet{DeVore2008p740} results may not be applicable to CMEs as they led to confined events. The BO model reproduces some observed CME and CME related phenomena \citep{vanderHolst2007p77} and flares occur during the secondary (flare reconnection) and also during the third phase (ribbon flares).  It also reproduces the classic three-part CME structure \citep{Lynch2004p589} as shown in \figref{bocme}. It avoids the Aly-Sturrock problem (\sref{sec:cmeforminit}) by opening only a portion of the field and not the entire field. The BO model can also proceed without reconnection using an ideal instability as shown by \citet{Rachmeler2009p1431}.

\subsection{Propagation}
\label{sec:cmepropagation}
During the propagation phase the dynamics of CMEs are not very well understood and the forces causing the residual acceleration which is sometimes observed are not known. Some models have been developed in an effort to understand it in the context of a ``drag'' force -- for example, the `snow plough' model \citep{Tappin2006p233}, or the aerodynamic drag model \citep{Cargill2004p135}, and others model it in terms of a Lorentz force ``flux rope'' model \citep{Chen1996p27499}. Examining the physical properties of the CME and the environment into which it is launched can give some insight into the mechanisms that might be at play. The solar wind is a high-$\beta$ plasma which means that magnetic effects such as reconnection and magnetic pressure are less of an influence than the hydrodynamics of the plasma. This also means that Alfv\'{e}n waves transporting energy can more easily be converted to fast mode waves which efficiently dissipate their energy in the corona. However, CMEs are typically low-$\beta$ structures even at 1\,AU, with plasma-$\beta$ of 0.1 common. This means treating CMEs as solid bodies moving through a flow is not a bad approximation. CMEs often drive shocks which build-up a compressed plasma region ahead of them, called the sheath. Many CMEs also contain the structured magnetic field of a flux rope or, magnetic clouds.


\subsubsection{Aerodynamic Drag}
The aerodynamic drag model invokes the drag caused by fluid flow over a fixed body. If we define the Lorentz, gravitational, pressure gradient and drag forces as $F_{L}$, $F_{G}$, $F_{P}$ and $F_{D}$ respectively, we can rewrite the equation of motion ($F=ma$) of a CME moving in the solar wind as:
\begin{eqnarray}
	\label{eqn:cmeforce}
	M_{*}\frac{dv_{i}}{dt}  & = & F_{L} + F_{G}+F_{P}+F_{D}  \nonumber \\
	& =  & F_{L}+F_{G}+F_{P}- \rho_{e}AC_{D}(v_{i} - v_{e})|v_{i}-v_{e}|
	\end{eqnarray}
	$	\textrm{with }F_{L}  =  \mathbf{J} \times \mathbf{B} \textrm{, } F_{G} = -GMm/r^{2} \textrm{ and } F_{P} = - \nabla p$, where A is the cross-sectional area of the CME, $C_{D}$ is the drag coefficient and subscript $i$ or $e$ refers to internal CME or external solar wind quantities respectively, and $M_{*}=M+M_{V}$ where $M_{V}$ is the virtual mass $M_{V} \approx \rho_{e} / 2\tau$, and $\tau$ is the CME volume. The gravitational force $F_{G}$, and pressure gradient $F_{P}$ far from the Sun are assumed to be negligible, and the Lorentz force $F_{L}$ has been shown to be negligible in full MHD simulations of the CME dynamics \citet{Cargill1996p4855}. Equation \ref{eqn:cmeforce} can be re-written in the following manner:
\begin{equation}
	\label{eqn:aerodrag}	\frac{dv}{dt}=\frac{F_{D}}{M_{*}}=-\gamma C_{D}(v_{i}-v_{e})|v_{i}-v_{e}| \textrm{, where } \gamma = \frac{\rho_{e}A}{\tau(\rho_{i}+\rho_{e}/2)}
\end{equation}
this is a first-order differential equation in terms of $v$, and can be numerically integrated given the initial conditions.
	
In order to arrive at the aerodynamic drag model, a number of implicit assumptions have been made. The first of these is that the CME can be treated as a solid body in a fluid flow, and that there is no feedback between the CME and solar wind; the second that the magnetic nature of the CME and solar wind plays no significant role in the interaction. Finally, that the assumption that the plasma of the solar wind can be represented as a typical fluid. Simulations of flux-ropes in plasma flows have shown that, in most cases, the feed-back between the flux-rope and solar wind is small, and they also show that $C_{D}\,\sim 1$. Note that the drag coefficient determined from MHD simulations will include some of the magnetic and plasma effects that would be left out of the aerodynamic equation.

It should be noted that in the above equations that $\rho_{e}$, $\rho_{i}$, $\tau$, $A$ and $v_{e}$ are all functions of $r$. If we wish to describe the total CME evolution then we would need to use the equation of state describing the internal pressure, temperature, etc. in balance with the external parameters, and let the CME evolve accordingly. This would yield the theoretical evolution of the parameters mentioned above, but as of yet none of the models have encompassed this much detail. However, statistical studies have been performed, and values extracted, by fitting a function of the form $<\gamma> = \alpha_{R}R^{-\beta_{R}}$ to velocity-time data and best-fit parameters were obtained \citep{Vrsnak2001p173}. In \figref{simdrag} we see a number of simulations of the deceleration a CME due to aerodynamic drag, which were obtained by numerical integration of Equation \ref{eqn:modeldrag} using a combination of the best-fit parameters from \citet{Vrsnak2001p173} and realistic estimates. Various simulations were carried out for different initial distances and velocities \figref{simdrag}
\begin{equation}
	\label{eqn:modeldrag}
	\frac{dv_{cme}}{dr}=R_{\odot} \alpha_{R} R^{-\beta_{R}} \left ( 1 - \frac{v_{cme}}{v_{sw}} \right )|v_{cme}-v_{sw}|
\end{equation}
\citet{Reiner1998p1923} pointed out that, at least for some fast CMEs, a quadratic form of drag was inconsistent with the observations. Specifically they found that the velocity profile derived from CME-driven shocks producing Type II radio bursts could not be reproduced by a quadratic drag model. Instead the authors suggested a linear model was more appropriate in reproducing the observed kinematics and \citet{Vrsnak2002p1019} and \citet{Vrsnak2006p431} also investigated a linear model. A generalised drag equation can be written which will take both possibilities into account:
\begin{equation}
	\label{eqn:gendrag}
	\frac{dv_{cme}}{dr}=\alpha_{R}  R^{-\beta_{R}} \frac{1}{v_{cme}}\left (v_{sw} - v_{cme} \right )^{\delta}
\end{equation}
where the sign of the force must be taken into account positive if $v_{sw}>v_{cme}$, negative otherwise. The $\alpha$ and $\beta$ values will be different depending on the form of the drag which is specified by setting $\delta=1$ for linear and $\delta=2$ for quadratic models.
 
\figuremacroW{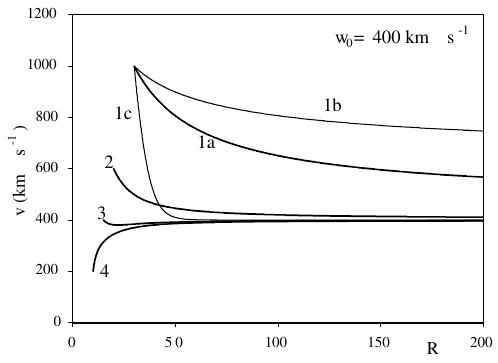}{Aerodynamic Drag Simulation}{Numerical integration of Equation \ref{eqn:modeldrag} for various parameters. All simulations use a model solar wind with asymptotic speed of 400 km\,s$^{-1}$, different distances for the driving force cut-off were assumed. Bold lines show deceleration for $\alpha_{R}= 2\, \times \, 10^{-3}$ and $\beta_{R} = 1.5$ for initial velocities $v_{0}=$1000, 600, 400 and 200 km\,s$^{-1}$ (curves 1a, 2, 3 and 4 respectively). The curves labelled 1b and 1c correspond to the deceleration with $\alpha_{R}= 10^{-3}$, $\beta_{R} = 1.5$ and $\alpha_{R}= 2\, \times \, 10^{-3}$, $\beta_{R} = 1$ respectively both with $v_{0}=$1000\,km\,s$^{-1}$ \citep{Vrsnak2002p1019}. The $x$-axis units are R$_{Sun}$.}{0.6}

\subsubsection{`Snow Plough' Model}	
The `snow plough' model is based on the conservation of momentum. As the CME sweeps up the solar wind ahead of it, theismaterial must be accelerated, thus momentum is transferred from the CME to the swept up material. This process gives rise to two coupled differential equations:
\begin{equation}
	\frac{dv}{dt}=-\frac{dM}{dt}\frac{(v_{i}-v_{e})}{M}
	\label{eqn:vel}
\end{equation}
\begin{equation}
	\label{eqn:mass}
	\frac{dM}{dt}=\rho_{e}A(v_{i}-v_{e})
\end{equation}
We can combine these two equations and substitute in the equation for mass to get:
\begin{equation}
	\label{eqn:snowplough}
	\frac{dv}{dt}=-\frac{\rho_{e}}{\tau \rho_i} A (v_{i}-v_{e}) (v_{i}-v_{e})
\end{equation}
which is a firstorder differential equation describing the motion of a CME due to a `snow plough' interaction. \figref{snowplough} shows a comparison of modelled CME propagation to the observations for the `snow plough' and aerodynamic drag models.

\figuremacroW{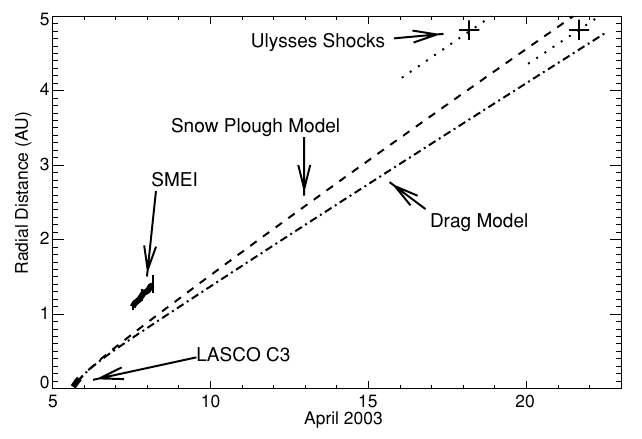}{Comparison of Aerodynamic Drag and `snow plough' Models to Observations}{Modelled height-time profiles for a transient using the `snow plough' (dashed curve) and aerodynamic drag (dash-dot curve) models compared with the LASCO, SMEI and Ulysses observations (heavy lines and + signs). The Ulysses shock speeds are indicated by the dotted line segments through their locations \citep{Tappin2006p233}.}{0.6}

By looking at \eqref{eqn:snowplough} and \eqref{eqn:aerodrag} we can see that while the physical mechanisms behind the snow plough and aerodynamic drag models are different, they have very similar mathematical forms. From \eqref{eqn:aerodrag} we can see that in the limit when $\rho_{e} \ll \rho_{i}$ we recover the simpler snow plough model.
	
\subsubsection{Flux Rope Model}
\cite{Chen1996p27499} has extended his model to include the interplanetary propagation of CMEs: close to the sun the forces on the flux rope are described by \ref{rq:chenflxrp}, and far from the Sun the forces due to gravity, coronal pressure and the solar magnetic field are not significant. The equation of motion is reduced to:
\begin{equation}
	F_{R} = \frac{I_{t}^{2}}{c^{2}R} \left [ \ln \left ( \frac{8R}{a} \right )  - \frac{1}{2}\frac{\overline{B_{t}}^{2}}{{B_{pa}}^{2}} - 1 + \frac{\xi_{i}}{2} \right ] + F_{d}
\end{equation}
far from the Sun. He assumed that the drag force was an aerodynamic type of the form:
\begin{align}
	F_{d} =C_{D}\rho_{a}m_{i}A(v_{sw}-v)|v_{sw}-v|
\end{align} where $\rho_{a}$ is the solar wind density and $m_{i}$ is the internal mass of the flux rope. \figref{chendrag} shows the evolution of CME in this model out into interplanetary space. The CME is accelerated by the Lorentz force over the first $\sim$24 hours before drag takes over and begins to equalise the CME's speed to the background solar wind. For the initial speed used in this simulation this corresponds to a distance of about 70$R_{\odot}$ or an elongation of $20^{\circ}$. While the exact source of long duration acceleration is not known it has been observed in a number of CMEs at large distances from the Sun \citep{Howard2007p610,Manoharan2001p1180,Manoharan2006p345,Tappin2006p233}.

\figuremacroW{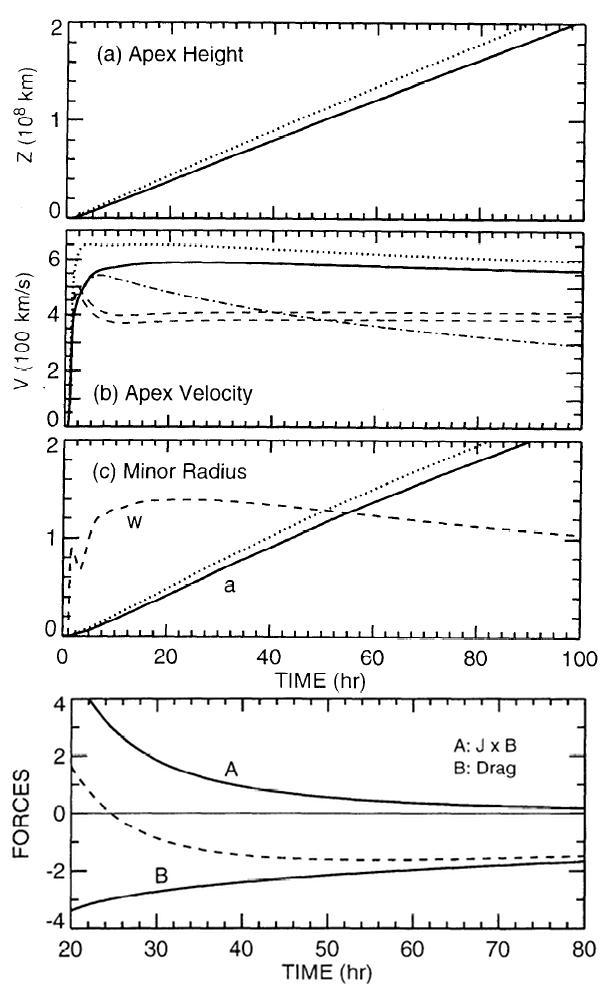}{Flux Rope CME Propagation Simulation}{Toroidal instability simulation of CME propagation \citep{Chen1996p27499}. (a) Apex height, (b) apex velocity and (c) minor radius dynamics $a(t)$ in $10^{7}$\,km and $w(t)$ in 50\,km\,s$^{-1}$. Dotted curve simulation with more faster flux injection. (botton) The Lorentz and drag forces acting on the CME, dashed curve is net force. The units are in $10^{17}$\,dyn.}{0.5}


\section{Shocks}
\label{shocks}
\figuremacroW{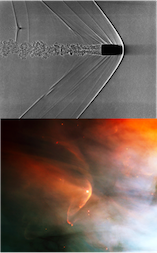}{Imaging Observations of Bow Shocks Formed by a Bullet and Star}{(top) A shadowgraph of a bullet traveling through air at about 1.5 times the speed of sound (or $M=1.5$) a bow shock can be seen ahead of the bullet and a turbulent wake behind it. (bottom) A Hubble Space Telescope observation of the bow shock around the very young star, LL Ori in the Orion nebula. Images courtesy of \href{http://www.nasa.gov/mission_pages/}{NASA.GOV}}{0.7}

A shock wave or simply a shock is a disturbance across which properties of the medium such as pressure, density, velocity, temperature change in a nearly discontinuous manner. Shock waves may be generated through a number of mechanisms: 1) abrupt changes in the properties of the medium, for example caused by explosions; 2) propagation at supercritical velocities such as supersonic aircraft; and 3) a non-linear wave steepening, for example waves in the ocean. Shocks are formed across many scales and in different conditions, from astrophysical shocks such as planetary bow shocks \mbox{\citep{Slavin1981p11401}}, or the shock at the edge of the Heliosphere \citep{vanBuren1995p2914,Decker2008p67} to shocks generated by the re-entry of the Apollo mission capsules \citep{Glass1977p269}. \figref{shockimgs} illustrates the bow shock caused by a supersonic projectile in a wind tunnel and from a star moving through the ISM. Shocks can be categorised in terms of the angle between the flow direction and the shock normal as, parallel (or normal), perpendicular, or oblique, and as moving or stationary shocks. Shocks caused by supersonic flows over bodies form attached to the body or, if the flow is deflected more than some critical angle a detached shock will form. The distance at which the detached shock forms called the stand-off distance is a complex function of both the shape of the object and properties of the medium. Bow shocks are detached shocks which occur when a blunt object moves relative to a medium at supersonic (supeaflv\'{e}nic) speeds \citep{rathakrishnan2010applied}. The difference between a moving and standing shock can be understood as follows: in a moving shock the shock moves through a medium supersonically, while in a standing shock the shock is at rest and the medium flows supersonically. The difference between them is only the frame of reference so the system is invariant under Galilean transforms.

In gas-dynamic shocks (collisional shocks), the important physical processes are the collisions between the molecules. They allow temperature and density perturbations to propagate, and the associated viscous forces lead to dissipation. Collisions also allow the temperature equalisation of different species of molecules. Gas-dynamic shocks are often understood in terms of the macroscopic fluid equations containing quantities such as density, temperature and bulk (rather than thermal) velocity. Space plasmas, on the other hand, are rarefied and as a result collisions are very infrequent. The lack of collisions means that different particles can have very different temperatures which can be non-Maxwellian, and in the presence of magnetic fields can even lead to anisotropic temperature distributions. The dissipation mechanisms are complex, involving the interaction between fields and particles. In these collision-less shocks the ensemble interaction of the particles and the fields is the important physical processes. MHD describes these interactions between the large scale field and particles in terms of macroscopic density, pressure and bulk velocity, similar to the gas-dynamic case.

In either gas dynamic or MHD cases, these continuum descriptions can not describe the shock itself, as they can not represent the kinetic microscopic processes that control the shock; particle collisions in the gas-dynamic case and wave-particle interactions in plasma shocks. However, the continuum descriptions are applicable on either side of the shock. Shocks are often considered to be an infinitely thin interface, this this is a good approximation when the shock is physically thin compared to the length scales involved, but thick with respect to the mean free path or the Debye and ion gyro-radius, in gas-dynamic and plasma shocks respectively. By applying the mass, momentum and energy conservation laws across the shock, the jump conditions or Rankine-Hugoniot equations can be derived. In the following sections, the notation $[X]=X_{u}-X_{d}$ is used to give the difference between quantity $X$ upstream and downstream of the shock.

\subsection{Gas-dynamic Shocks}
\label{sec:gdshocks}
The simplest description of a shock is in the shock rest frame, where gas moving supersonically (faster than information can be transmitted) flows into the shock from upstream. At the shock irreversible processes alter the speed, density and temperature of the medium and ,as a result the out flow downstream is subsonic. A shock is thus an entropy-increasing, or irreversible, wave that causes the transition from supersonic to subsonic flow. The Mach number ($M=v/v_{c}$) is defined as the ratio of the shock speed in the upstream medium to the sound speed. The Mach number is always greater than unity upstream, and less than unity downstream. In the rest frame of the shock, the application of the conservation of mass energy and momentum gives the Rankine-Hugoniot equations:
\begin{align}
	\label{gdmass}
	\left [ \rho u_{n} \right ] &= 0 \\
	\label{gdmon}
	\left [ \rho u_{n}^{2} + p \right ] &= 0 \\
	\label{gdmot}
	\left [ \rho u_{n}u_{t} \right ] &= 0 \\
	\label{gden}
	\left [ \left ( \frac{\rho u^{2}}{2} + \frac{\gamma}{\gamma -1}p \right )u_{n} \right ] &= 0
\end{align}
where $u$ is the flow speed, and $u_{n}$ ($u_{t}$) is the normal (tangential) component to the shock and the other symbols have their usual meaning. Combining \eqref{gdmass} and \eqref{gdmon} gives $[u_{t}]=0$, so the tangential component of the flow is continuous and thus we can consider a coordinate system moving along the shock with speed $u_{t}$. In this the normal incidence frame (see \figref{shockframe}~(left)) the flow speeds $u$ and $u_{n}$ are identical. Making a Galilean transform into the laboratory frame, the mass continuity equation \eqref{gdmass} can bewritten $[\rho (v_{s} - u_{n})]=0$. This can be rearranged in terms of the shock speed $v_{s}$:
\begin{align}
	v_{s}= \frac{\rho_{d}u_{n,d}- \rho_{u}u_{n,u}}{\rho_{d}- \rho_{u}}.
\end{align}
It should be noted that the shock speed alone is not a good indicator of the energetics involved, as small density change can result in a high shock speed but the total energy in terms of the compression ratio and mass motion may be low.

\figuremacroW{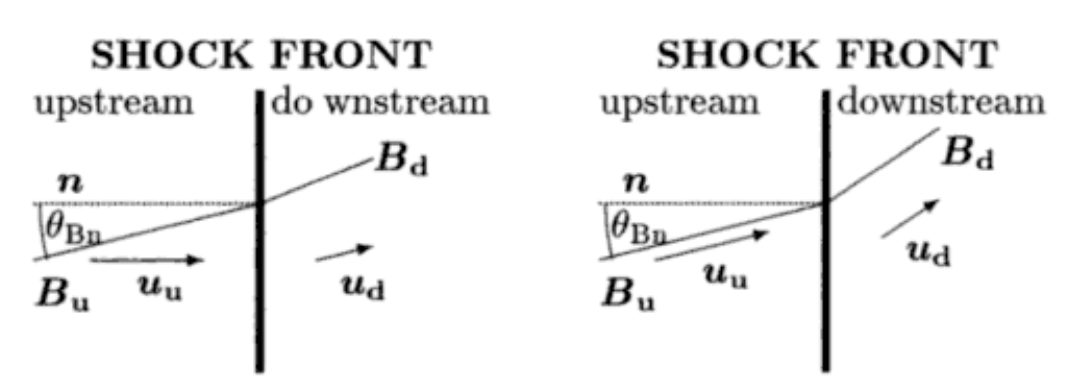}{Shock Frames of Reference}{(left) Normal incidence shock frame and (right) de Hoffmann-Teller frame \citep{kallenrode2004space}.}{0.5}

\subsection{Magnetohydrodynamic Shocks}
\label{sec:mhdshocks}
A crucial difference between the gas-dynamic and MHD shocks is the magnetic field. This must be accounted for in the conservation equations and also in terms of both the flow direction and magnetic field direction, as these do not have to be parallel. In the normal incidence frame \figref{shockframe}~(left) the upstream flow is normal to the shock and oblique to the magnetic field, downstream the flow is oblique to both the flow and shock normal. A transformation can be made under which the flow becomes parallel to the magnetic field. This is known as the de Hoffmann-Teller transformation or frame. This reference frame moves parallel to the shock at the de Hoffmann-Teller speed ($v_{HT} \times \mathbf{B}=-\mathbf{E}$). Thus, the induced field in the shock front vanishes. The Rankine-Hugoniot equations for an MHD shock are:
\begin{align}
	\label{mhdmass}
	\left [ \rho \mathbf{u}\cdot  \mathbf{n} \right ] &= 0 \\
	\label{mhdmon}
	\left [ \rho  \mathbf{u} ( \mathbf{u} \cdot  \mathbf{n}) + \left ( p + \frac{ \mathbf{B}^{2}}{2 \mu_{0}} \right )  \mathbf{n} - \frac{( \mathbf{B}\cdot  \mathbf{n}) \mathbf{B}}{\mu_{0}}  \right ] &= 0 \\
	\label{mhden}
	\left [  \mathbf{u} \cdot  \mathbf{n} \cdot \left (  \frac{\rho  \mathbf{u}}{2} + \frac{\gamma}{\gamma -1}p + \frac{ \mathbf{B}^{2}}{\mu_{0}}  \right ) - \frac{( \mathbf{B}\cdot\mathbf{n})(\mathbf{B}\cdot\mathbf{u})}{\mu_{0}}  \right ]&= 0 \\
	\label{mhdmx1}
	\left [ \mathbf{B}\cdot \mathbf{n} \right ] &= 0 \\
	\label{mhdmx2}
	\left [ \mathbf{n} \times (\mathbf{u}\times \mathbf{B}) \right ] &=0.
\end{align}
The normal component of the magnetic field ($B_{n}$) must be continuous from \eqref{mhdmx1}, and from \eqref{mhdmx2} the tangential component of the electric field must also be continuous. Thus, the Rankine-Hugoniot jump conditions are a set of five equations for the five unknowns $\rho$, $\mathbf{u}$, $p$, $B_{n}$ and $B_{t}$. The jump conditions allow the calculation of the downstream parameters from knowledge of the upstream conditions, or vice versa.

The solutions to the jump conditions actually describe a number of discontinuities which are not necessarily shocks. A contact discontinuity is formed when there is no flow across the discontinuity ($u_{n}=0$) and is associated with a density jump while all other parameters remain unchanged. The magnetic field has a component normal to the discontinuity ($B_{n}\ne 0$), so the two sides are not completely separate, but tied to move at the same tangential speed $u_{t}$. A tangential discontinuity completely separates two regions no flux crosses the boundary ($B_{n}=0$ and $u_{n}=0$), and the tangential components change ($[B_{t}] \ne 0$ and $[u_{t}]\ne0$). The plasma and field properties can change arbitrarily across the boundary, but static pressure balance is maintained, i.e. $[p+B^{2}/2\mu_{0}]=0$. A rotational discontinuity requires pressure equlibrium but flux flow across the boundary $u_{n}\ne0$ and $B_{n}\ne 0$. The normal flow speed is $u_{n}=B_{n}/\sqrt{\rho \mu_{0}}$ and the change in the tangential flow speed is related to the change in the tangential magnetic field $[u_{t}]=[B_{t}/\sqrt{\rho \mu_{0}}]$. These rotational discontinuities can be viewed as large amplitude waves and are related to the transport of magnetic signals through their dependance on the Alfv\'{e}n speed.

A shock differs from the discontinuities above as there is a flow across the boundary ($u_{n}\ne0$) with compression and changes to the flow speed. An important parameter for MDH shocks is the angle between the magnetic field and shock normal $\theta_{Bn}$ and this leads to classification as perpendicular ($\theta_{Bn}=90^{\circ}$), parallel ($\theta_{Bn}=0^{\circ}$) or oblique for intermediate values. Oblique shocks are often subdivided into quasi-parallel ($0^{\circ}<\theta_{Bn}<45^{\circ}$) and quasi-perpendicular ($45^{\circ}<\theta_{Bn}<90^{\circ}$) shocks. In a parallel shock $B_{t}=0$ and the magnetic field is unchanged by the shock and behaves like a gas-dynamic shock except the collective interactions are mediated by the field, and not through collisions. In a perpendicular shock $B_{n}=0$, and both the pressure and magnetic field strength change. For the MHD jump conditions we can write the shock speed as:
\begin{align}
	v_{s} = \frac{\rho_{d}\mathbf{u}_{d} - \rho_{u}\mathbf{u}_{u}}{\rho_{d}-\rho_{u}}\cdot \mathbf{n}.
\end{align}

In a gas-dynamic shock there is only one critical speed called the sound speed, $v_{c}$, while in the MHD case different wave modes have different critical speeds resulting in three critical speeds -- the slow, fast, and Alfv\'{e}n speeds. Only the slow and fast mode are compressive and form true shocks; the Alfv\'{e}n mode only forms shocks in an anisotropic plasma, while in an isotropic plasma it results in a rotational discontinuity. The phase speed of the slow and fast modes is given by:
\begin{align}
	2 v^{2}_{sl, fa} = ( v_{c}^{2} + v_{A}^{2}) \pm \sqrt{ ( v_{c}^{2} + v_{A}^{2})^{2} - 4v_{c}^{2}v_{A}^{2}\cos^{2}\theta}
\end{align} taking the positive value gives the fast mode,  and the negative gives the slow mode. Taking $\theta=90^{\circ}$ gives the fast magnetosonic speed as $v_{ms}=\sqrt{v_{c}^{2}+v_{A}^{2}}$, on the other hand taking $\theta=0^{\circ}$ gives two solutions $v_{A}$ and $v_{s}$. As in the gas-dynamic case we can define critical ratios in terms of the characteristic speed, the fast magnetosonic Mach number ($M_{ms}=v/v_{ms}$) and the Alfv\'{e}n Mach number ($M_{A}=v/v_{A}$). 

\chapter{CME Observations and Instrumentation} 
\label{chap:inst}


\ifpdf
    \graphicspath{/}
\else
    \graphicspath{{3/figures/EPS/}{3/figures/}}
\fi


\hrule height 1mm
\vspace{0.5mm}
\hrule height 0.4mm
\noindent
\\	{\it In this chapter the details of CME observations, instrumentation, the data reduction process, and conversion of data to physical coordinates are described. This begins with an introduction to the mechanism which allows CMEs to be imaged namely Thomson scattering and its effect on the  observations. Next, the design principles behind the first coronagraph are outlined. Following this detailed descriptions of the relevant instruments from the Solar and Heliospheric Observatory (SOHO) and Solar Terrestrial Relation Observatory (STEREO) missions are given. Finally, a discussion on the conversion of data coordinates to physical coordinates and the various coordinate systems is presented.} \\
\hrule height 0.4mm
\vspace{0.5mm}
\hrule height 1mm

\newpage


\section{Observations of Coronal Mass Ejections}
\label{cmeobs}
The majority of the radiation from a CME is due to the scattering of photospheric radiation by free electrons in the CME plasma. While some CMEs contain bright prominence material which primarily emits at H$\alpha$ (6563\,{\AA}), most CMEs are dominated by the scattered radiation. The scattering mechanism is Thomson scattering which is a special case of the general theory of the scattering of electromagnetic waves \citep{jackson1975classical}. Thomson scattering applies when: (1) the coherence length of the radiation is small compared with the separation of the particles or, in the case of an incoherent source such as the photosphere, the wavelength must be small compared to the typical separation of the particles; and (2) the rest mass energy of the scatterers greatly exceeds the photon energy. This is the case throughout the corona for white-light. In the case of radio waves condition (1) is not met and coherent theory must be used. 

\subsection{Thomson Scattering}
\label{s:thomscat}
\figuremacroW{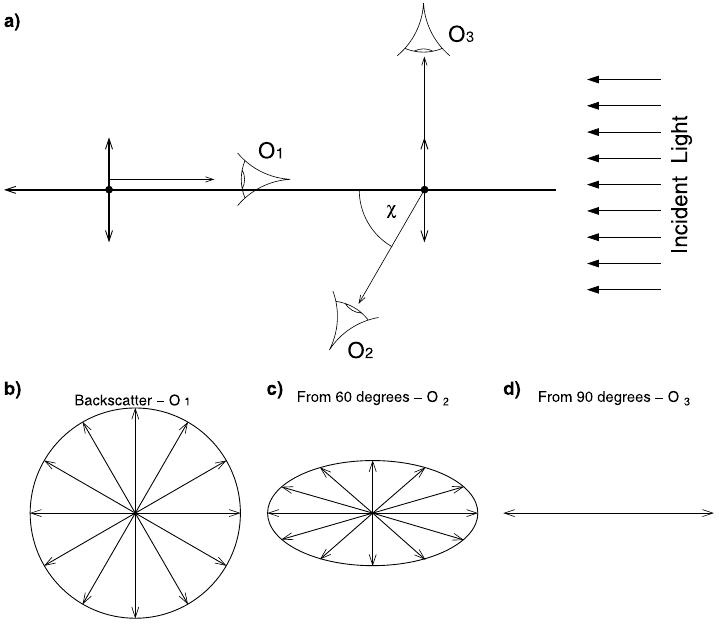}{Thomson Scattering Schematic}{Schematic demonstrating how the angular variation in Thomson scattering arises (a) the conceptual set up, the scattering angle $\chi$ is included for the oblique observer (O$_{2}$), (b)Ð(d) the scattered electric vectors as seen by observers at O$_{1}$ at $\chi =180^{\circ}$, O$_{2}$ at $\chi =60^{\circ}$ andO$_{3}$ at $\chi= 90^{\circ}$ respectively \citep{Howard2009p31}.}{0.8}

An unpolarised monochromatic plane electromagnetic (EM) wave incident on an electron will accelerate the electron which will then radiate symmetrically about the direction of the incident wave. Since the electric field of an EM wave is always perpendicular the direction of propagation, the acceleration of the electron will be confined to the plane perpendicular to the propagation direction. An observer at a scattering angle ($\chi$) of 0$^{\circ}$ or 180$^{\circ}$ would see unpolarised light, while one at $\chi=90^{\circ}$ would see only linearly polarised light, though both observe the same intensity (see \figref{thomscat}). The differential cross-section for Thomson scattering by an electron is:
\begin{align}
	\frac{d\sigma}{d\omega} = \frac{1}{2} \left ( \frac{e^{2}}{4 \pi \epsilon_{0}m_{e}c^{2}} \right )^{2} (1 +\cos^{2} \chi )
\end{align} where $\sigma$ is the cross-section, $d\omega$ is solid angle element at the scattering angle $\chi$, $e$ is the electron charge and $m_{e}$ is the mass of the electron. Integrating over all solid angles gives the total cross-section for scattering as:
\begin{align}
	\sigma_{t} = \frac{8 \pi }{3} \left ( \frac{e^{2}}{4 \pi \epsilon_{0}m_{e}c^{2} } \right )^{2} = \frac{8 \pi }{3} r_{e}^{2} \qquad[\text{m}^{2}]
\end{align}	
where $r_{e}$ is the classical electron radius. The differential cross-section for perpendicular scattering, $\sigma_{e}$, can be written as:
\begin{align}
	\sigma_{e}=  \left ( \frac{e^{2}}{4 \pi \epsilon_{0}m_{e}c^{2}} \right )^{2} = r_{e}^{2}\qquad[\text{m}^{2}\,\text{sr}^{-1}].
\end{align}	
Thus far, we have only considered the scattering of light from a point source by a single electron. The solar photosphere is neither a point source nor uniform, and so it is necessary to integrate the scattering component over the light from the whole visible disk. The intensity of radiation from the solar photosphere decreases towards the limb, this is known as limb darkening and can be characterised by:
\begin{align}
	I = I_{0}(I - u + u \cos \phi )
\end{align}
 where $u$ is the limb-darkening coefficient (a function of wavelength), and $\phi$ is the angle between the emitted radiation and the radius vector of the emitting point. The tangential ($I_{T}$) and radial ($I_{R}$) vector components of Thomson-scattered radiation from a single electron in the solar corona may be calculated in terms of a small number of measurable parameters \citep{Minnaert1930p209, vandeHulst1950p135,billings1966corona,Howard2009p31} as:
\begin{align}
	I_{T} = I_{0}\frac{\pi \sigma_{e}}{2z^{2}} \left [ (1-u)C  +uD \right ]
\end{align}
and
\begin{align}
	I_{P} = I_{0}\frac{\pi \sigma_{e}}{2z^{2}} \sin^{2} \chi \left [ (1-u)A  +uB \right ]
\end{align}
where
\begin{align}
A & = \cos \Omega \sin^{2} \Omega,  \\
B & = -\frac{1}{8} \left [ 1 - 3 \sin^{2} \Omega - \frac{\cos^{2} \Omega}{\sin \Omega} (1 + 3 \sin^{2} \Omega) \ln \left ( \frac{1 + \sin \Omega}{\cos \Omega}\right ) \right ], \\
C & = \frac{4}{3} - \cos \Omega - \frac{\cos ^{3} \Omega}{3}, \\
D &= \frac{1}{8} \left [ 5 + 3 \sin^{2} \Omega - \frac{\cos^{2} \Omega}{\sin \Omega} (5 - \sin^{2} \Omega) \ln \left ( \frac{1 + \sin \Omega}{\cos \Omega}\right ) \right ],
\end{align}
and $\Omega$ is the solid angle of the Sun as seen by the scatterer and $I_{P} = I_{T}-I_{R}$. The total scattered light intensity is therefore:
\begin{align} 
	I_{tot} = (I_{T}+I_{R})=2I_{T}-I_{P}.
\end{align}

The locus of all the points that maximise the scattered intensity form what is known the Thomson sphere (TS) as shown in \figref{thomsphere}~(left). \figref{thomsphere}~(right) shows the ratio of the correct to assumed plane-of-sky brightness and indicates it is a valid assumption out to about 70\,$R_{\odot}$m, after which the brightest feature may be far from the plane-of-sky. There is an often quoted misconception that scattering efficiency is maximised on the TS which it is not in fact the opposite occurs as the cross-section for scattering is {\it minimised} on the TS. However the TS also indicates the point of closest approach, thus the amount of incident radiation is maximised as is the density of scatterers, and hence the scattered intensity is maximised for the above constraint.

\figuremacro{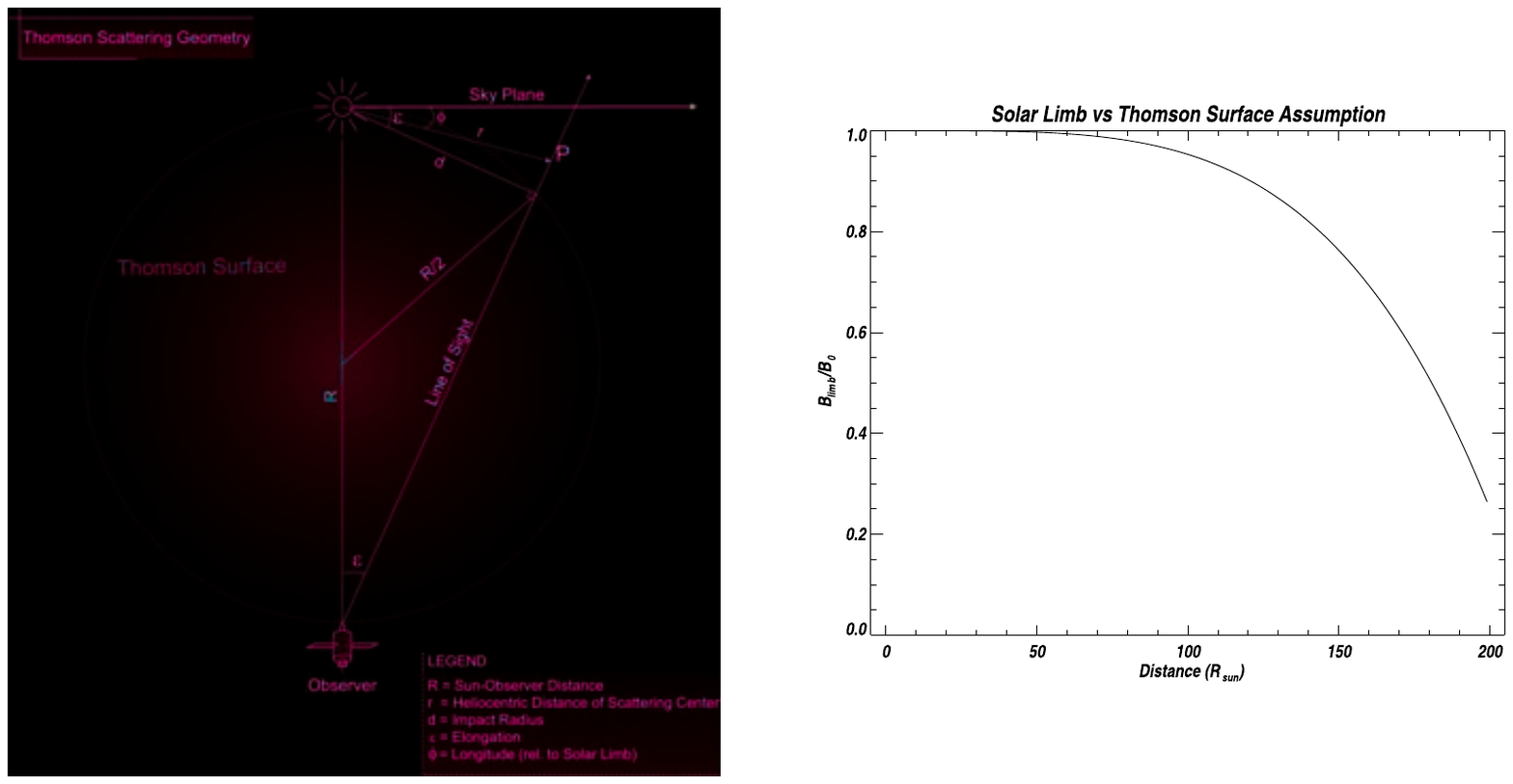}{Thomson Sphere and Scattered Intensity}{Thomson sphere geometry (left) and the range of validity of the plane-of-the-sky assumption (right).The ratio $B_{limb} /B_{0}$ is the ratio of the brightness calculated using the  plane-of-the-sky assumption over the brightness derived from the full treatment \citep{Vourlidas2006p1216}.}

While the effects of Thomson scattering on the plane-of-sky assumptions may be negligible close to the Sun they severely impede the ability to unambiguously identify CME structures in observations, as \figref{thommorph} shows. The observations (\figref{thommorph}~left) the show the graduated cylindrical shell (GCS) model \citep{Thernisien2006p763} almost edge-on and nearly side-on, but in the synthetic data (\figref{thommorph}~right) or observations it is nearly impossible to disguising between the two orientations and certainly not unambiguously. This highlights the problems in interpreting coronagraph observations and trying to compare them to theory.

\figuremacroW{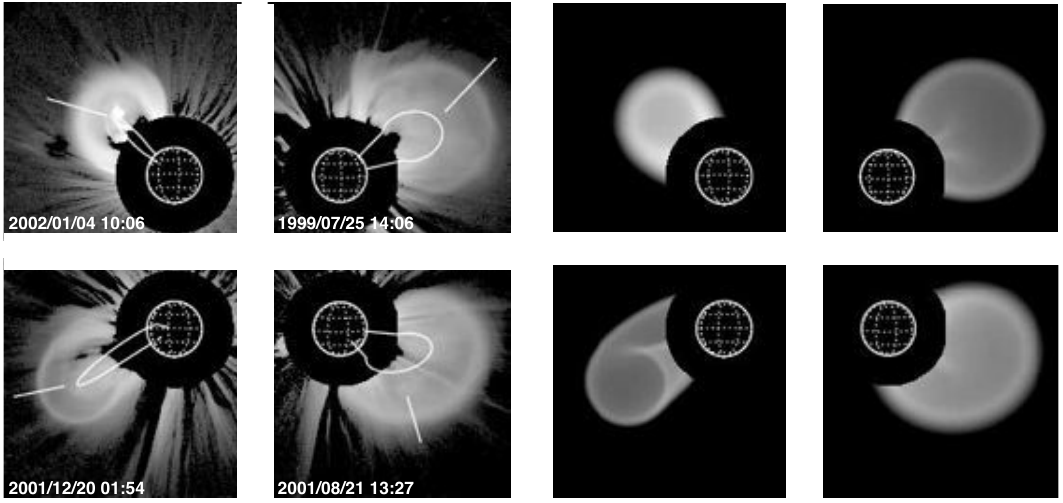}{Thomson Scattering Effect on Morphology}{LASCO observations (left) and the graduated cylindrical shell (GCS) model synthetic images (right) from \cite{Thernisien2006p763}. The wire frame on the LASO data indicates the orientation of the GCS axis.}{0.9}

\subsection{Projection Effects}
\label{sec:projeff}
Close to the Sun, the plane-of-sky assumption will only affect the magnitude of the derived kinematics, but not their profile shape. For example, any acceleration seen is real, but its magnitude is subject to possibly large uncertainties. Using on-disk features and assumptions about the propagation direction can allow accurate kinematics to be derived for selected events, especially those on the limb. Even then, care must be taken in their interpretation \citep{Byrne2009p4915}. However, far from the Sun the Thomson scattering and its effects on the plane-of-sky assumption become much more important. \figref{projecteffn} shows both bubble and shell models of a CME at different times in its propagation. The differences between the observed ($\times$), true ($\bigcirc$) and inferred ($+$) fronts are clear, and become extremely large as the CME reaches observer-like distances.
\figuremacroW{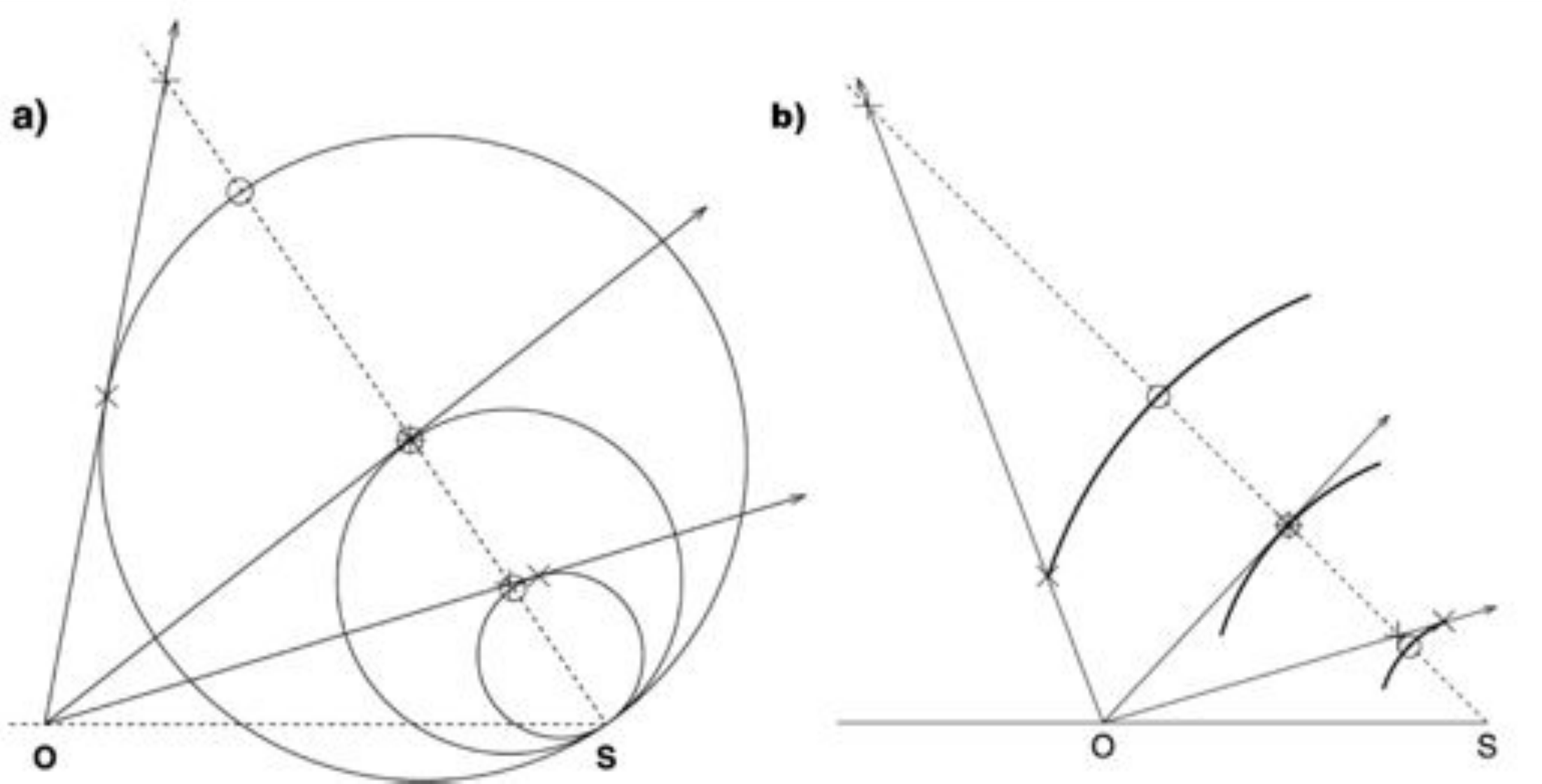}{Basic CME Structure and Observational Interpretations}{Basic CME structures. (a) The expanding bubble and (b) a simple shell with an angle of 30¡ at three different times during their propagation. The tangent drawn from the observer O across the CME surface shows the location of the relative leading edge. The $\times$ symbols represent the location of the leading edge seen by the observer the $\bigcirc$ symbols show the true location of the leading edge at the central location and the $+$ symbols the inferred location of the leading edge based on the central location \citep{Howard2009p31}.}{0.9}
The effects of making the plane-of-sky assumption when deriving kinematics are shown in \figref{kinspos} which shows simulations of the true kinematics (black), as well as those that would be observed from STEREO A (red) and B (green) like positions for a point-like CME propagating $12^{\circ}$ east of the Sun-Earth line. The CME is close to the plane-of-sky for STEREO B but far from A as seen in the inset (\figref{kinspos}~d). Not only are the kinematics significantly under estimated close to the Sun from and B, as the CME distance increases apparent acceleration, due to the plane-of-sky assumption begins to take place. It should be clear that the plane-of-sky assumption is unsuitable for studying CMEs at large distances from the Sun.
\figuremacroH{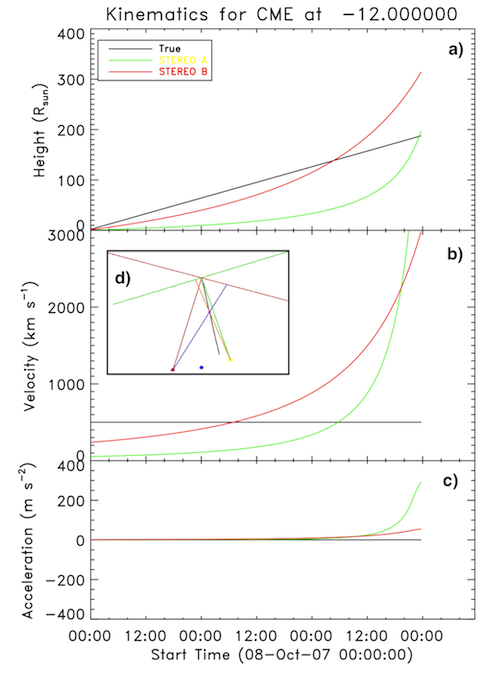}{Plane-of-Sky Projected Kinematic Representative of STEREO}{Plane-of-Sky projected kinematics representative of STEREO A (Red), STEREO B (green) and true (black) for a CME propagating 12$^{\circ}$ east of the Sun-Earth line. (a) Height, (b) velocity (c) acceleration and (d) geometry of simulation.}{0.6}
Other approaches have been developed such as the ``point-P'' method or the ``fixed-$\phi$'' method. The ``point-P'' method \citep{Howard20064105} assumes the CME is intrinsically a very broad, uniform, spherical front, centred on the Sun, with the relation ship between the observed elongation to height given by:
\begin{align}
r = d \sin \epsilon 
\end{align} where $d$ is the Sun-observer distance, and $\epsilon$ is the elongation angle. The ``fixed-$\phi$'' method \citep{Kahler2007p9103} assumes that the CME is a relatively narrow, compact structure travelling on a fixed, radial trajectory at an angle, $\phi$ relative to the observer's line-of-sight to the Sun and the derived height is given by:
\begin{align}
	r = \frac{d \sin \epsilon}{\sin \epsilon + \phi }.
\end{align}
More recently another method known as the ``harmonic mean'' has been developed \citep{Lugaz2009p479}. This is so called as it corresponds to the harmonic mean of values calculated from the other two methods the derived height is given by:
\begin{align}
	r = \frac{2d \sin \epsilon}{1 + \sin \epsilon + \phi}.
\end{align}

\section{Coronagraphs}
\figuremacro{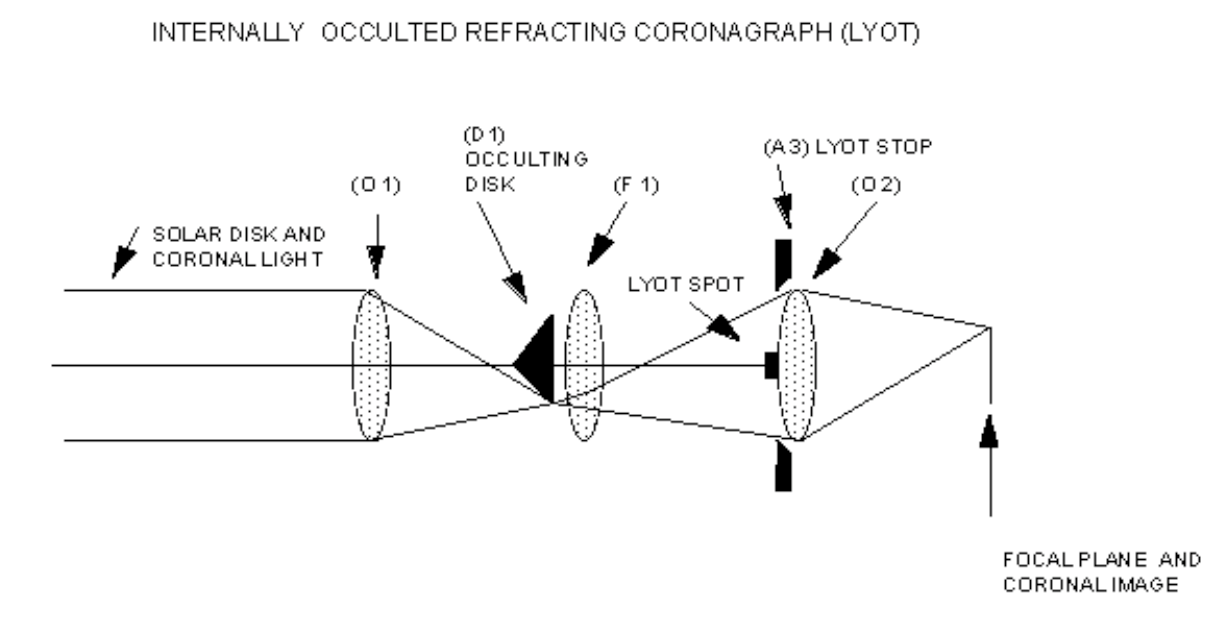}{Optical Layout of Lyot Coronagraph}{Optical layout of Lyot internally occulted coronagraph adapted from \citet{lyot1939p580}.}

A coronagraph is a telescope which aims to reproduce the spectacular images of the solar corona during eclipse observations. The two main problems with developing a coronagraph are: the scattered light in the telescope, and the sky brightness itself, both of which will overwhelm the weak corona. Bernard Lyot solved these problems and constructed the first coronagraph in 1930 \citep{lyot1939p580}. The optical layout of this coronagraph (see \figref{Lyot_coronagraph}) is designed to reduce internal scattering and blockoff the solar disk. The first element was an objective lens that was free from inclusions, and highly polished to reduce scattering and reflections. Directly behind the objective lens (O1), and in its focal plane, is the artificial moon or occulter which reflects away the bright solar disk image. The first field lens (F1) forms an image of the objective lens, and a screen placed here would show an image of the objective with a bright halo and central bright spot due to refraction, and double reflection, respectively. Lyot introduced what are now known as the Lyot spot and Lyot stop to remove these aberrations. Finally the second objective lens (O2) forms an image of the occulter (blocking the solar disk) and corona on to the detector plane.

There are two different configurations for coronagraphs, internally occulted (occulter is inside the first optical element, O1 in \figref{Lyot_coronagraph}), and externally occulted (occulter outside of the first optical element, O1 in \figref{Lyot_coronagraph}). Each configuration has its own unique advantages and disadvantages. The internally occulted coronagraph allows the very inner corona to be imaged, but as the front lens and aperture are directly illuminated by the solar disk, scattered light within the telescope is a problem. The externally occulted coronagraph cannot image the inner corona due to the diffraction limit (in space, the sizes of the telescopes are limited), but scattered light is not as  much of a problem as the aperture and objective lens are not directly illuminated by the solar disk. Modern coronagraphs use the same basic design, but have added baffles to reduce internal stray light further and polarisers to separate the signature of CMEs from that of the corona. The coronal light is a combination of polarised and unpolarised light, while CMEs contain only polarised light. Polarisers block some of the coronal light, allowing the CME to be more easily identified.

\begin{sidewaystable}[p]
	\centering
		\begin{tabular}{|c|c|c|c|c|c|c|} 
			\hline
			{\bf Mission} & {\bf Instrument} & {\bf FOV } & {\bf Image size} & {\bf Pixel size} & {\bf Cadence} & {\bf Filters/Polarisers} \\ 
			\hline\hline 

			\multirow{3}{*}{SOHO} & EIT & Full disk & 1024$\times$1024 & 2.6 arcsec & 12\,min &171{\AA}, 195{\AA}, 284{\AA}, 304{\AA}   \\
			\cline{2-7}
			& LASCO C2 & 1.5\rng6.0$\,R_{\odot}$ & 1024$\times$1024 & 11.4 arcsec & 30\,min & 0, $\pm60^{\circ}$  various filters 400\rng835\,nm \\ 
				\cline{2-7}		
			& LASCO C3 & 3.7\rng30.0$\,R_{\odot}$ & 1024$\times$1024 & 56 arcsec & 30\,min & 0, $\pm60^{\circ}$ various filters 400\rng1050\,nm \\  
\hline
			\multirow{5}{*}{STEREO} & EUVI & Full disk$<1.7\,R_{\odot}$ & 2048$\times$2048 & 1.6 arcsec & 5\,min & 171{\AA}, 195{\AA}, 284{\AA}, 304{\AA} \\
\cline{2-7}
			& COR1 &1.4\rng4.0$\,R_{\odot}$ & 1024$\times$1024 & 7.5 arcsec & 5\,min & 0, $\pm$60$^{\circ}$ at H-$\alpha$ (656.3\,nm)\\ 
\cline{2-7}
			& COR2 &2.5\rng15.0$\,R_{\odot}$ & 2048$\times$2048 & 14.7 arcsec & 15\,min & 0, $\pm$60$^{\circ}$ whitelight \\ 
\cline{2-7}
			& HI1 & 4$^{\circ}$\rng24.0$^{\circ}$ & 1024$\times$2048 & 70.0 arcsec & 40\,min & whitelight \\
\cline{2-7}			
			& HI2 &18.7$^{\circ}$\rng88.7$^{\circ}$ & 1024$\times$2048 & 2 arcmin & 2\,hours & whitelight \\ 
\hline
		\end{tabular}
	\caption[Summary of Instruments]{Summary of instruments and their properties}
	\label{instsum} 
\end{sidewaystable}

Coronagraphs are limited to observe only a small distance ($\leq$30\,$R_{\odot}$) from the Sun and cannot image CMEs as they propagate to the Earth at $\sim$215\,$R_{\odot}$, though this isone of the `holy grails' of solar physics.  Extended coronagraphs or heliospheric imagers aim to fulfil this goal and two instruments have already validated the ability to measure the electron scattered CME signal against the strong zodiacal light and stellar background. The Zodiacal Light Photometer \citep{pitz1976p19} on the Helios spacecraft launched in 1974, and the Solar Mass Ejection Imager (SMEI; \citealt{eyles2003p319}) instrument, on the Coriolis spacecraft launched in 2003, have demonstrated the ability to detect and track CMEs far from the Sun \citep{tappin2004v31}. The Heliospheric Imagers on-board STEREO image from $\sim$4\rng88$^{\circ}$ elongation (15\rng215\,$R_{\odot}$\symbolfootnote[1]{Depends strongly on the distance of the object from the plane-of-sky}) and thus, for the first time, allow CMEs to be tracked from the Sun to the Earth from two perspectives. The results in this thesis are largely based on STEREO SECCHI observations in particular simultaneous (A and B) observations of CMEs in COR1, COR2, and HI. A summary of the instruments and their important features and properties is listed in Table~\ref{instsum}.

\section{Solar and Heliospheric Observatory (SOHO)}
\label{sec:soho}
SOHO \citep{domingo1995p1} was a joint  ESA NASA mission which was launched on December 2, 1995 by an Atlas II rocket. SOHO was launched into an orbit around the first Lagrange point (L1), located along the Sun-Earth line, about 1\% of the distance to the Sun, thus allowing continual monitoring of the Sun. The main scientific goals of the mission were to: (i) study the solar interior using helioseismology techniques; (ii) study the heating of the solar corona; and, (iii) investigate the solar wind and its acceleration. There are twelve complementary instruments onboard: 3 devoted to helioseismology probing the inner structure of the Sun by observing solar oscillations; 3 measuring {\it in-situ} plasma properties (densities, velocities, magnetic field, composition, etc.) of the solar wind; and 6 telescopes with imagers or spectrometers studying the solar disk and atmosphere. A schematic of the SOHO spacecraft and its instruments is shown in \figref{sohoschm}.

\figuremacroW{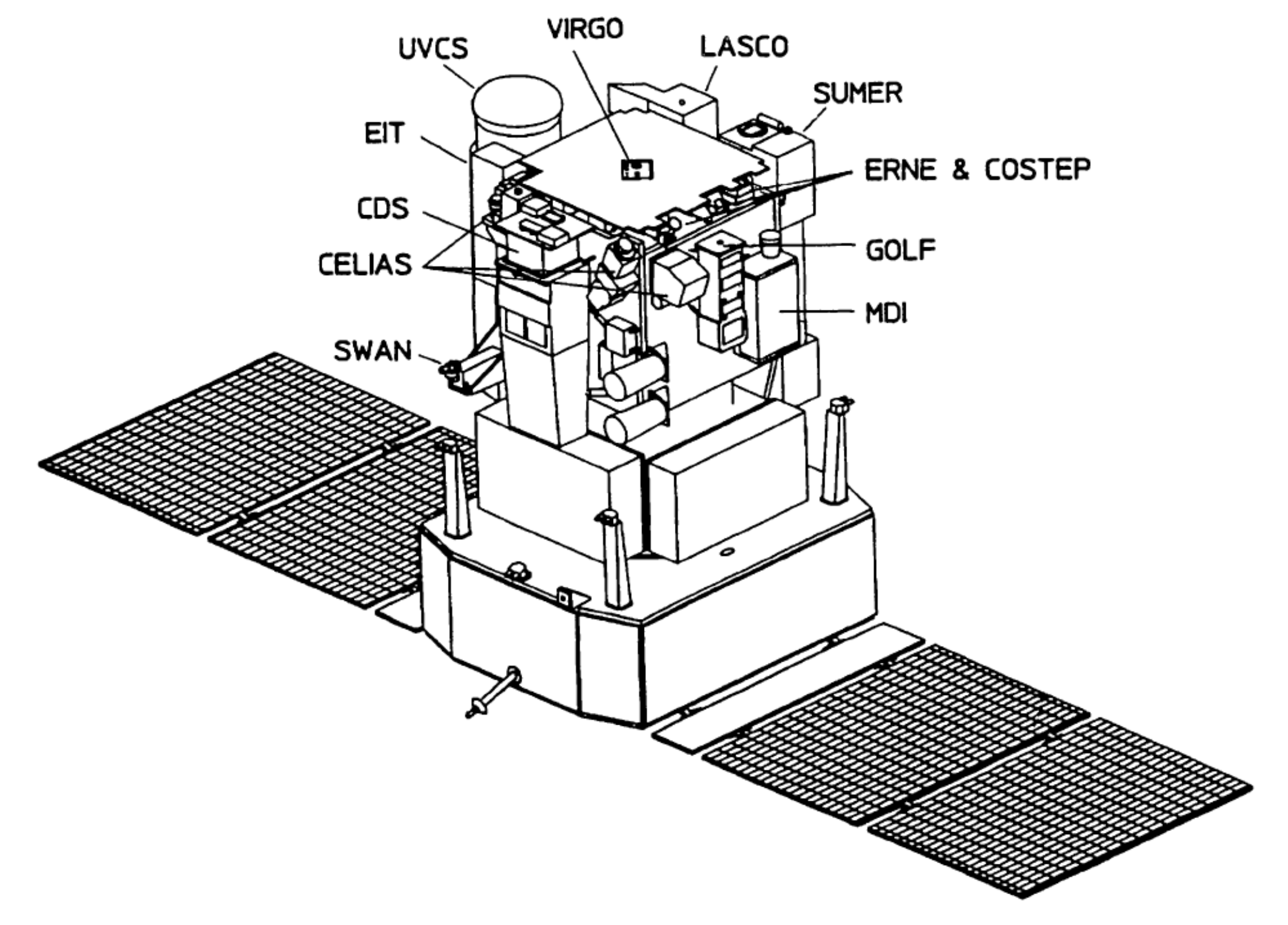}{Schematic of the SOHO Spacecraft}{A schematic of the SOHO spacecraft and instruments \citep{domingo1995p1}.}{0.8}

Both Extreme-Ultraviolet Imaging Telescope (EIT) and Large Angle Spectrometric Coronagraph (LASCO) use the same CCD detectors, which are STIe (formerly Tektronics) 1024$\times$1024, three-phase, multi-phased-pinned type with square pixels of 21\,$\mu$m. The full-well capacity of the CCDs is about 150,000 electrons and the quantum efficiency is about 0.3-0.5 in the optical range 500\rng700\,nm when using front side illumination. For EIT, the CCDs were thinned, back-side illuminated, and had anti-reflective coatings applied in order to keep the quantum efficiency high at 0.27\rng0.36 in the EUV 171.4\rng303.8\,{\AA} range. The CCDs have 40 non-imaging (over- or under-scan) pixels on each line which can be used for calibration and engineering purposes.

\subsection{Extreme-Ultraviolet Imaging Telescope (EIT)}
\label{sec:eit}
\figuremacroW{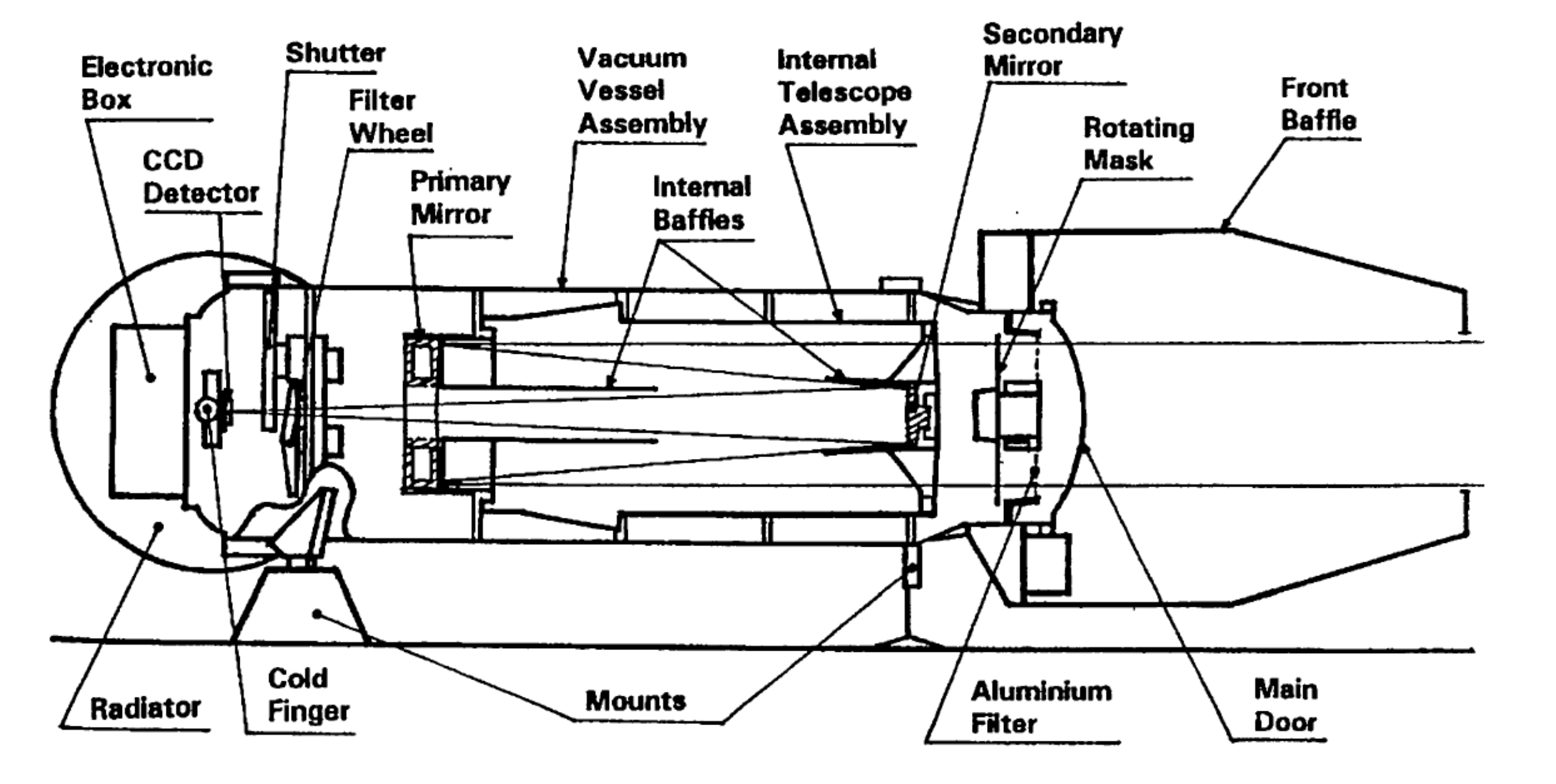}{Schematic of EIT}{Schematic of the EIT telescope \citep{delaboudiniere1995p291}.}{0.9}

\figuremacroW{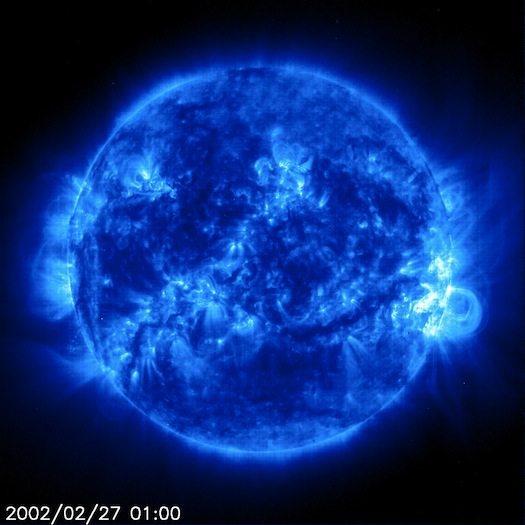}{Sample EIT {171\AA} Observation}{Observation showing active regions and magnetic loops as recorded by EIT in the Fe IX/X 171 line. The temperature of this material is about 1 million K in the lower corona. Image courtesy of the \href{http://sohowww.nascom.nasa.gov/gallery/images/20020227eit171looprt.html}{SOHO website}.}{0.9}

EIT \citep{delaboudiniere1995p291} is a Ritchey-Chr\'{e}tien telescope which images the corona and transition region on the solar disk, and up to 1.5\,$R_{\odot}$ above the solar limb. Light enters the telescope through an initial filter of 700\,{\AA} of cellulose sandwiched between two 1500\,\AA ~thick aluminium films, which reject visible and IR radiation, and then passes through a quadrant selector (see \figref{eit}). Each quadrant of the primary and secondary mirrors has multilayer coatings of molybdenum and silicon deposited on them, each was optimised to reflect in a different wavelength bands centred on 304\,\AA, 284\,\AA, 195\,\AA, and 171\,\AA. Probing the solar atmosphere at peak temperatures of $8.0\times10^{4}$\,K, $2.0\times10^{6}$\,K, $1.6\times10^{6}$\,K, and $1.3\times10^{6}$\,K respectively (assuming quiet sun conditions). Finally, a set of filters just in front of the focal plane block long wavelength radiation before imaging,  and a shutter controls the exposure time. An image is formed on the CCD and as EIT has a field-of-view 45\,arcmin square, each pixel is 2.6\,arcsec in size. A typical image from EIT from the {171\,\AA} channel dominated by Fe~{\sc ix} and {\sc x} lines is shown in \figref{eit171img}.

\subsection{Large Angle Spectrometric Coronagraph (LASCO)}
\label{sec:lasco}

\figuremacroH{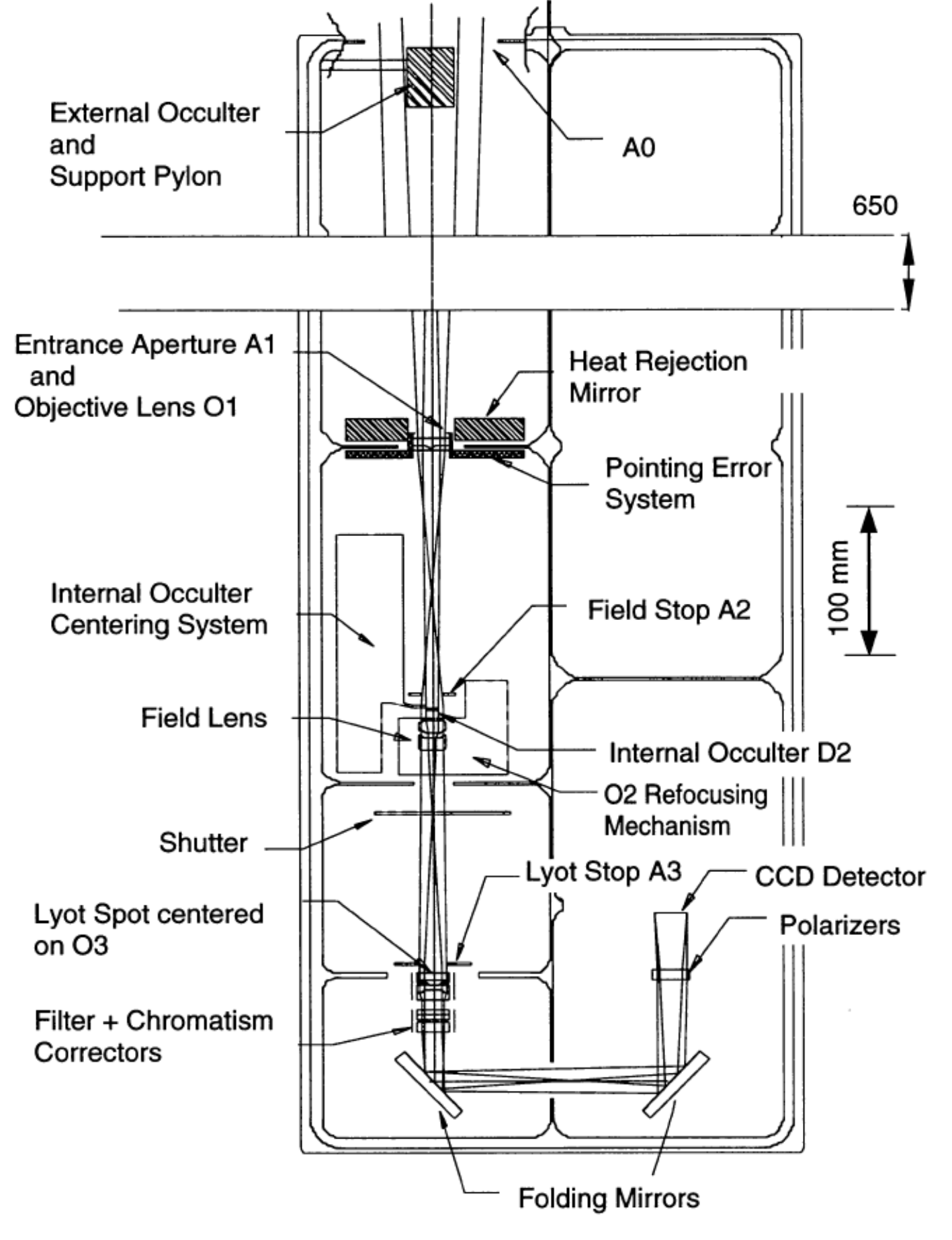}{Schematic of C2}{Optical layout of the C2 coronagraph \citep{brueckner1995p357}.}{0.5}

The LASCO \citep{brueckner1995p357} instrument consist of three coronagraphs which image the corona from 1.1\rng30\,R$_{\odot}$ (C1: 1.1\rng3\,$R_{\odot}$, C2: 1.5\rng6.0\,$R_{\odot}$, C3:  3.7\rng30\,$R_{\odot}$). The C1 coronagraph has not operated since the loss and subsequent recovery of the spacecraft in 1998. C2 is an externally occulted Lyot type coronagraph. Light enters the C2 coronagraph and encounters the external occulter, which was a new design a cone with sharp threads whose cone angle is slightly larger than the angle subtended by the Sun from L1 (\figref{c2}). This design achieves a light rejection level of 1.5$\times10^{-5}$, and at this level, light refracted by the next element in the path (the entrance aperture) becomes a concern. As a result, a polygonal serrated aperture was developed, with each side behaving as a knife edge, such that its direct diffraction avoids O1 (\figref{c2}). Light then passes through an oversized diaphragm A2, preventing propagation of light scattered off the edges of the aperture, and enters the objective O2. O2 is a two element design which apodises light diffracted by O1, and ghost images created by O1 are blocked by the Lyot spot (a metallic layer deposited onto O3). The final element, O3, is a four element design which magnifies and focuses an image of the corona seen by O2 onto the CCD detector, and results in an angular resolution of 11.4\,arcmin per pixel. Two plane mirrors fold the optical path to reduce the over all instrument length. The shutter and filter wheel are mounted as close to the final pupil as possible, and behind O3, respectively. The polariser is placed closer to the image plane, in front of the CCD detector. The pointing error system detects any imbalance in the penumbra created by O1 using four photodiodes behind four symmetric holes located about O1. The differential output of  paired diodes is amplified, coded, and telemetered to the ground where commands can be sent to move the pointing legs until satisfactory pointing is achieved.

C3 is also an externally occulted Lyot type coronagraph. Light enters the instrument through A0, where the occulting disk shadows the coronagraph entrance aperture, A1, from direct sunlight (\figref{c3}). The occulting disk is an assembly of three disk on a common spindle configured to minimise diffracted sunlight falling on A1 and the primary lens. The primary lens forms an image of the occulted corona and is followed by the internal occulter which blocks the image of the external occulter and halo. Behind this is the field lens, a multi-element anti-reflection coated lens which forms a collimated image of the corona on the relay lens. It also presents a sharp image of the instrument aperture A1 to the Lyot stop. The elements of the relay lens are all placed behind this stop to minimise diffracted light falling on it. The relay lens focuses an image of the corona on to the CCD detector, and it also contains the Lyot spot. The resulting pixel resolution is 56\,arcmin per pixel. Between the relay lens and CCD lies a triple mechanism containing the filter, polariser, and shutter. Centred baffles are placed along the instrument so that the field, objective and relay lenses see only the rear of the baffles, or the walls shadowed by the preceding baffle. All the surfaces are anodised and coated in a dense black paint.  

\figuremacroH{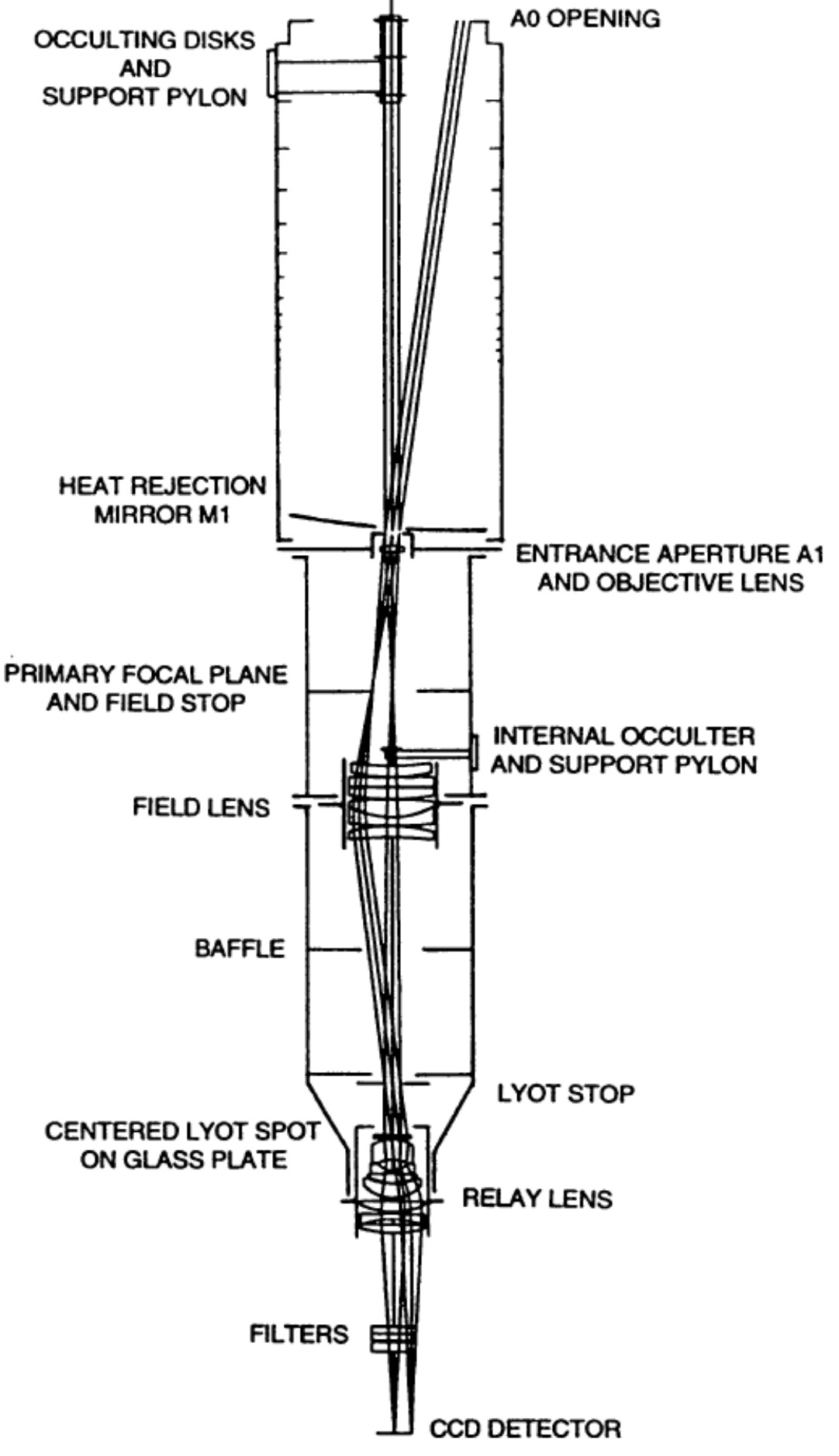}{Schematic of C3}{Optical layout of the C3 coronagraph \citep{brueckner1995p357}.}{0.6}

Both C2 and C3 have a number of filters in the range 400\rng850\,nm for C2,  400\rng1050\,nm for C3 and polarisers at 0$^{\circ}$, $\pm60^{\circ}$.  The polarisers are used to obtain total brightness $B$, or polarised brightness $pB$, images via a combination of observations taken at polariser positions $I_{a}=-60^{\circ}$, $I_{b}=0^{\circ}$, $I_{c}=+60^{\circ}$, combined using \citep{billings1966corona}:
\begin{align}
	B & = \frac{2}{3} \left ( I_{a} + I_{b} +I_{c} \right ) \\
	pB & = \frac{4}{3} \left[   \left( I_{a} + I_{b} +I_{c} \right) ^{2} - 3 \left( I_{a}I_{b}+I_{a}I_{c}+I_{b}I_{c} \right) \right] ^{1/2}.
\end{align}
A sample LASCO observation from C3 is shown in \figref{lascoc3}.

\figuremacro{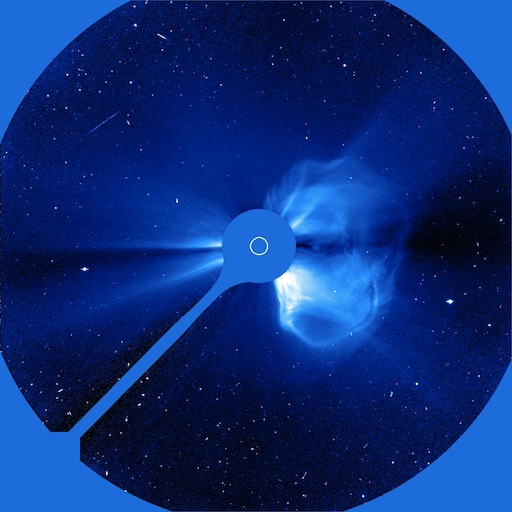}{Sample LASCO C3 Observation}{A typical C3 observation, the occulter, support pillar and non-imaging pixel have been set to zero. The white circle on the occulter indicates the Sun's size. A CME is visible off to the right of the occulter. Image courtesy of the  \href{http://sohowww.nascom.nasa.gov/gallery/SolarCorona/large/las021.jpg}{SOHO website}.}

\section{Solar Terrestrial Relation Observatory (STEREO)}
\label{sec:stereo}

\figuremacroW{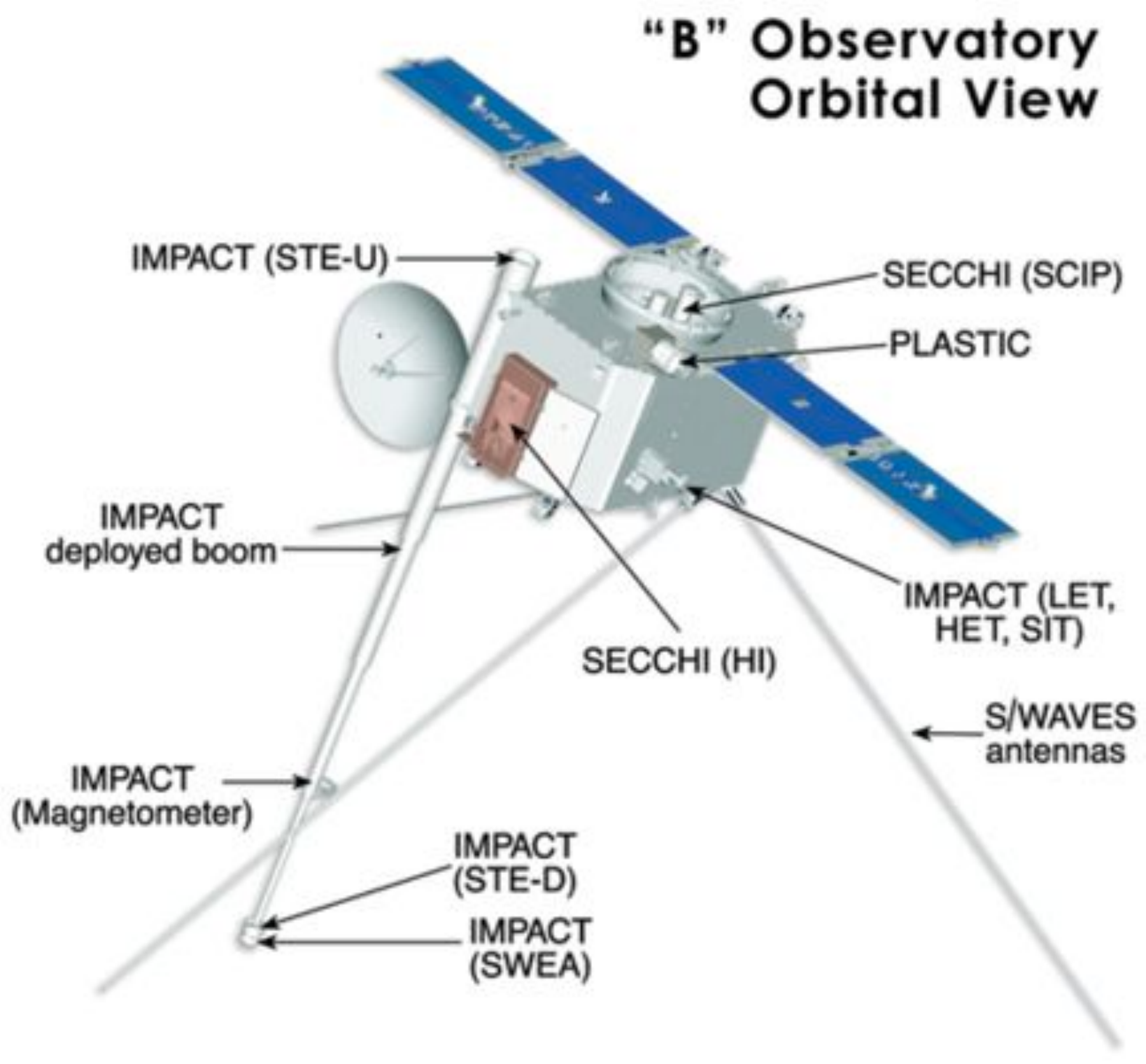}{Schematic of the STEREO B Spacecraft}{Schematic of the STEREO B spacecraft . The positions of the some of the instruments and their detectors are indicated \citep{Kaiser2008p5}.}{0.6}

The NASA STEREO \citep{Kaiser2008p5} mission was lunched on a Delta 2 rocket on October 26, 2006. The mission consists of two near-identical spacecraft which after a series highly eccentric Earth orbits and Moon sling shots, were placed in orbit around the Sun. The Ahead spacecraft (STEREO-A) travels ahead of the Earth, slightly closer to the Sun and the Behind spacecraft (STEREO-B) lags behind Earth, slightly further from the Sun. The spacecraft separate at about $\sim$\,22$^{\circ}$ per year from Earth, or $\sim$\,45$^{\circ}$ as viewed from the Sun. \figref{stereo_orbit} shows the positions of the spacecraft at three different times. The STEREO mission was designed to study the causes and mechanisms behind the initiation of CMEs, and then follow their propagation through the inner Heliosphere. STEREO will also be used to study the site of energetic acceleration and develop 3D models of solar wind properties such as magnetic topology, temperature, velocity, and density. As such, the STEREO spacecraft carry an almost identical payload consisting of optical, radio, as well as {\it in situ} particle and field instruments. These are divided into four suites: Sun-Earth Connection Coronal and Heliospheric Investigation (SECCHI; \citealt{howard2008p67}); In situ Measurements of PArticles and CME Transients (IMPACT; \citealt{luhmann2008p117}); PLAsma and SupraThermal Ion Composition (PLASTIC; \citealt{galvin2008p437}); and STEREO/WAVES (S/WAVES; \citealt{bougeret2008p487}). \figref{stereospc} shows one of the STEREO spacecraft and its instruments.

\figuremacroSW{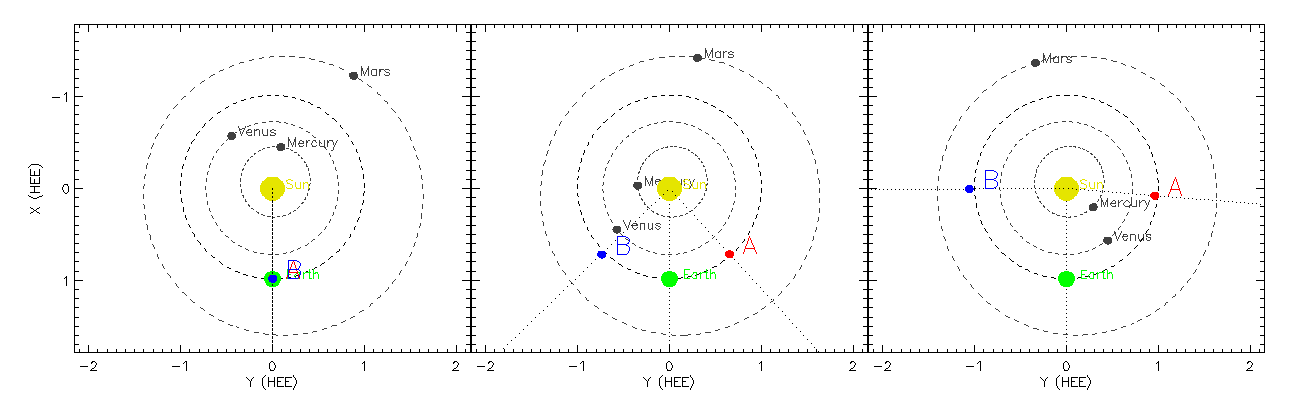}{STEREO Orbit Progression}{The positions of the STEREO spacecraft and the planets in the inner Heliosphere from left to right on 1 Jan 2007, 1 Jan 2009 and 1 Jan 2011. Images courtesy of \href{http://stereo-ssc.nascom.nasa.gov/where/make_where_gif.php}{Where is STEREO}.}{0.30}

\subsection{Sun Earth Connection Coronal and Heliospheric Investigation \mbox{(SECCHI)}}
The remote sensing optical suite, SECCHI, consist of an extreme ultraviolet imager (EUVI) which images the solar disk in four wavelengths 171\,\AA, 195\,\AA, 284\,\AA~and 304\,\AA, two coronagraphs (COR1 and COR2) which image the corona from 1.4\rng15\,$R_{\odot}$ in white-light, and the Heliospheric Imagers (HI) which image the inner Heliosphere ($\sim$4\rng88$^{\circ}$ elongation) in white-light. This combination of instruments can image from the solar surface to beyond 1\,AU from two perspectives. Combining the two perspectives allows for the possibility of 3D reconstructions of solar features such as, loops, prominences and CMEs.

Each of the scientific instruments on SECCHI uses a three-phase, back-illuminated, non-inverted mode (to ensure good full well capacity, 150k to 200k electrons) CCD model CCD42-40 manufactured by E2V in the United Kingdom. There are 2048$\times$2052 image pixels, each measuring 13.5\,$\mu$m on a side and providing a total imaging area of 27.6 mm square. The total readout is 2148$\times$2052 pixels, providing 100 columns of non-imaging over- and under-scan regions for calibration and engineering purposes. The CCDs used for visible light detection (COR-1, COR-2, HI-1 and HI-2) have an anti-reflective coating on their backside (illuminated side). The quantum efficiency (QE) of these devices varies from roughly 80\% at 500\,nm, to 34\% at 900 nm. The backside of the EUVI CCD has no coating in order to provide sensitivity at shorter wavelengths, with a quantum efficiency of 74\% at 17.1\,nm, and 70\% at 30.3\,nm.

\subsubsection{Extreme Ultraviolet Imager (EUVI)}
\label{sec:euvi}
EUVI \citep{wuelser2004p112} is of similar design to that of SOHO/EIT except it provides higher resolution (40\% better) and cadence capabilities. It is a Ritchey-Chr\'{e}tien telescope which images the solar disk and atmosphere up to 1.7\,$R_{{\odot}}$. It provides pixel limited resolution of 1.6\,arcsec per pixel across the entire field-of-view. Light enters the telescope through the entrance filter (150\,nm thick aluminium) which blocks undesired UV, optical and IR radiation, and passes through the aperture selector to one of the four quadrants. Each quadrant of the primary and secondary mirrors uses multilayer reflective coatings, optimised for four wavelength bands centred on 171\,\AA, 195\,\AA, 284\,\AA\  and 304\,\AA. The light continues through a filter wheel with redundant thin aluminium filters to remove the remainder of visible and IR radiation. The image is then formed on the CCD, and a rotating blade shutter controls exposure times. \figref{euvi-csec} shows the layout of EUVI. A typical EUVI observation in the {195 \AA} is shown in \figref{euvitypical}.

\figuremacroW{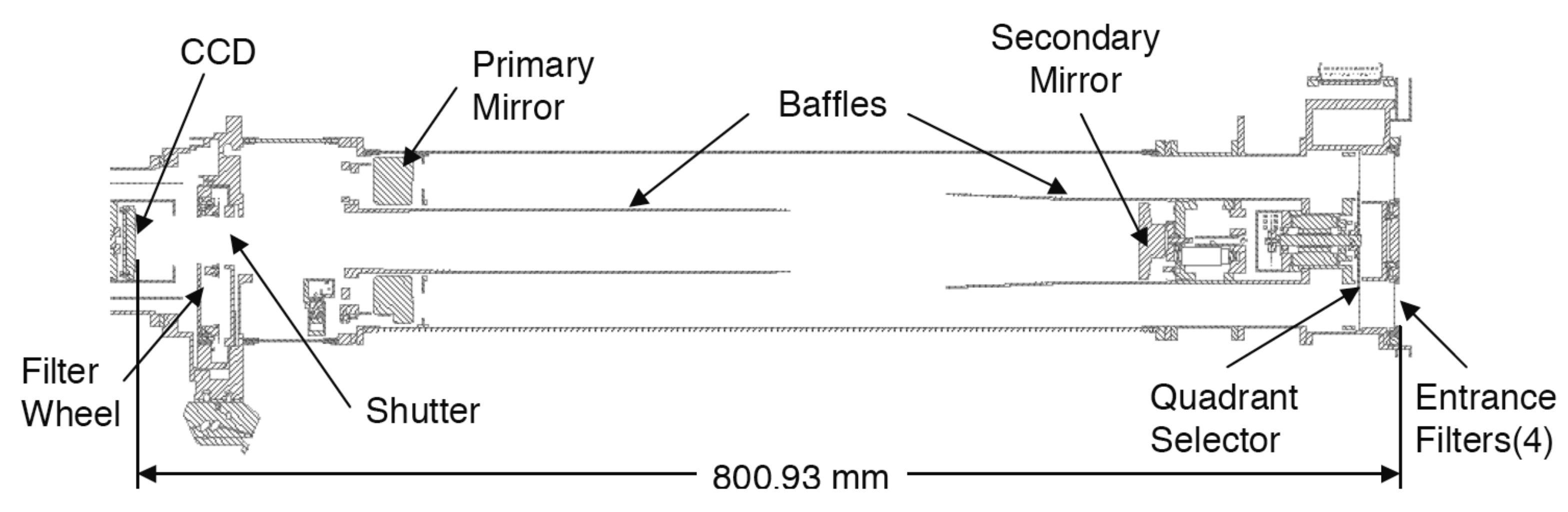}{Schematic of EUVI}{A cross section of the EUVI telescope \citep{wuelser2004p112}}{0.9}

\figuremacroW{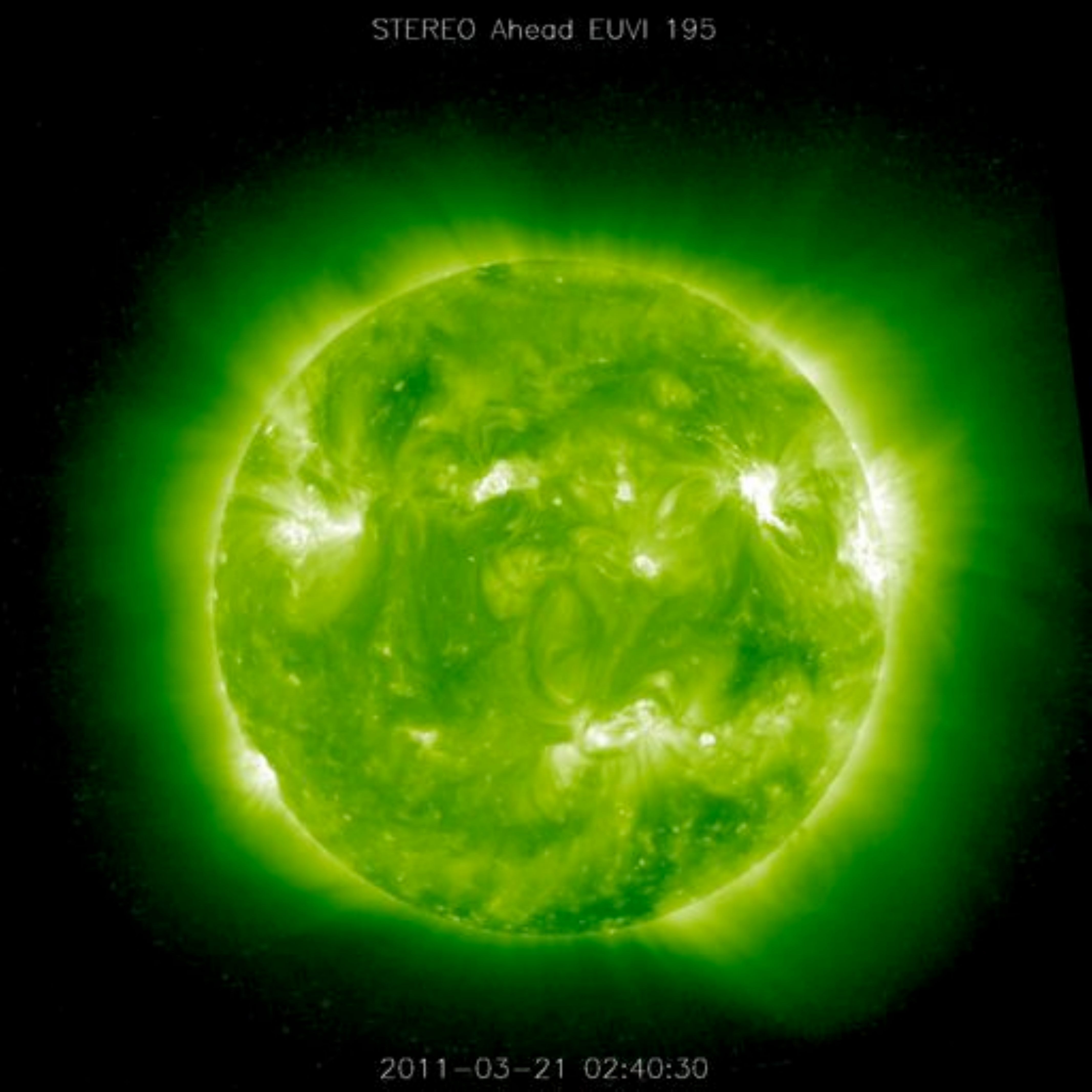}{Sample EUVI {195\AA} Observation}{Image of the Sun, taken by the SECCHI Extreme Ultraviolet Imager (EUVI) on the STEREO Ahead observatory on March 21, 2011 at 02:40:30 UT. The 195 Angstrom bandpass is sensitive to the Fe~{\sc xii} ionisation state of iron, at a characteristic temperature of about 1.4 million degrees Kelvin. Image courtesy of the \href{http://stereo.gsfc.nasa.gov/browse/2011/03/21/ahead/euvi/195/2048/20110321_024030_n4euA_195.jpg}{Stereo Science Centre}.}{0.9}

\subsubsection{COR1 and COR2}
\label{sec:cor1cor2}
The inner SECCHI coronagraph, COR1, is a Lyot internally occulted refractive telescope which images the corona from 1.4 to 4\,$R_{\odot}$ at 656\,nm (H$\alpha$) at 7.5\,arcsec per pixel (with onboard 2$\times$2 binning). It is the first space-borne internally occulted refractive telescope (in contrast to the internally reflective design of LASCO/C1) and this design enables better spatial resolution closer to the limb than an external design. COR1 is dominated by instrumentally scattered light which cannot be removed by Lyot principles, but as it is largely unpolarised it can be removed by talking polarised observations and calculating pB images.

\figuremacroW{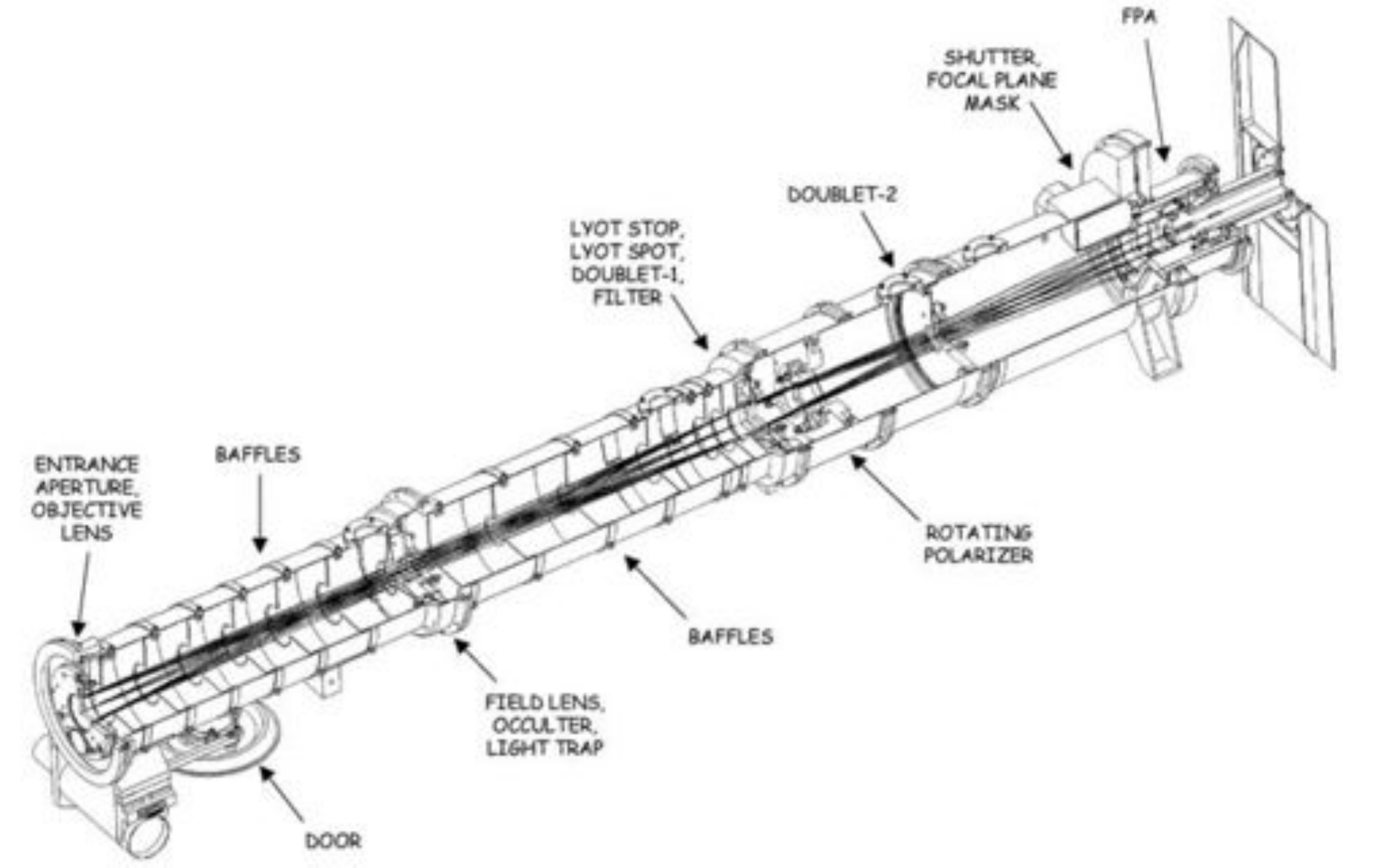}{Schematic of COR1}{A cut away view of the COR1 coronagraph \citep{howard2008p67}.}{0.9}

\figref{cor1inst} shows the opto-mechanical layout of the instrument. Light enters through the front aperture where the singlet objective lens focuses the solar image onto the occulter. The occulter is mounted on a stem at the centre of the field lens, and the tip of the occulter is cone-shaped to direct sunlight into the surrounding light trap. The light trap is wedge-shaped, and as a result the light must reflect many times before it can escape and is thus effectively absorbed. Two doublet lenses act as a telephotolens and focus the coronal image onto the CCD while maintaining diffraction-limited resolution (\figref{cor1inst}). A bandpass filter 22.5\,nm wide centred on the H$\alpha$ line at 656\,nm is placed just behind the first doublet. The Lyot stop, spot, first doublet, and bandpass filter all form a single linear optical assembly (\figref{cor1inst}). A linear polariser mounted on a hollow core motor rotational stage is located between the two doublets and a rotating blade shutter is located just in front of the focal plane detector assembly (\figref{cor1inst}).  

The objective lens only focuses the solar image accurately at one wavelength (H$\alpha$), due to chromatic aberration, so the internal occulter is designed to block all other the solar photospheric light between 350\rng1100\,nm. Diffracted light from the front aperture is focused onto a Lyot spot by the field lens, this removes the largest source of stray light. Placing baffles at various points between the front aperture and the Lyot stop removes additional stray light. To remove ghosting from the objective lens a Lyot spot is glued to the front surface of the doublet lens immediately behind the Lyot stop (\figref{cor1inst}). A focal plane mask located between the shutter and detector removes light diffracted from the edge of the occulter. At orbital perigee at the design wavelength the resulting solar image is completely occulted to 1.4\,$R_{\odot}$ and vignetted to 1.9\,$R_{\odot}$.

\figuremacroW{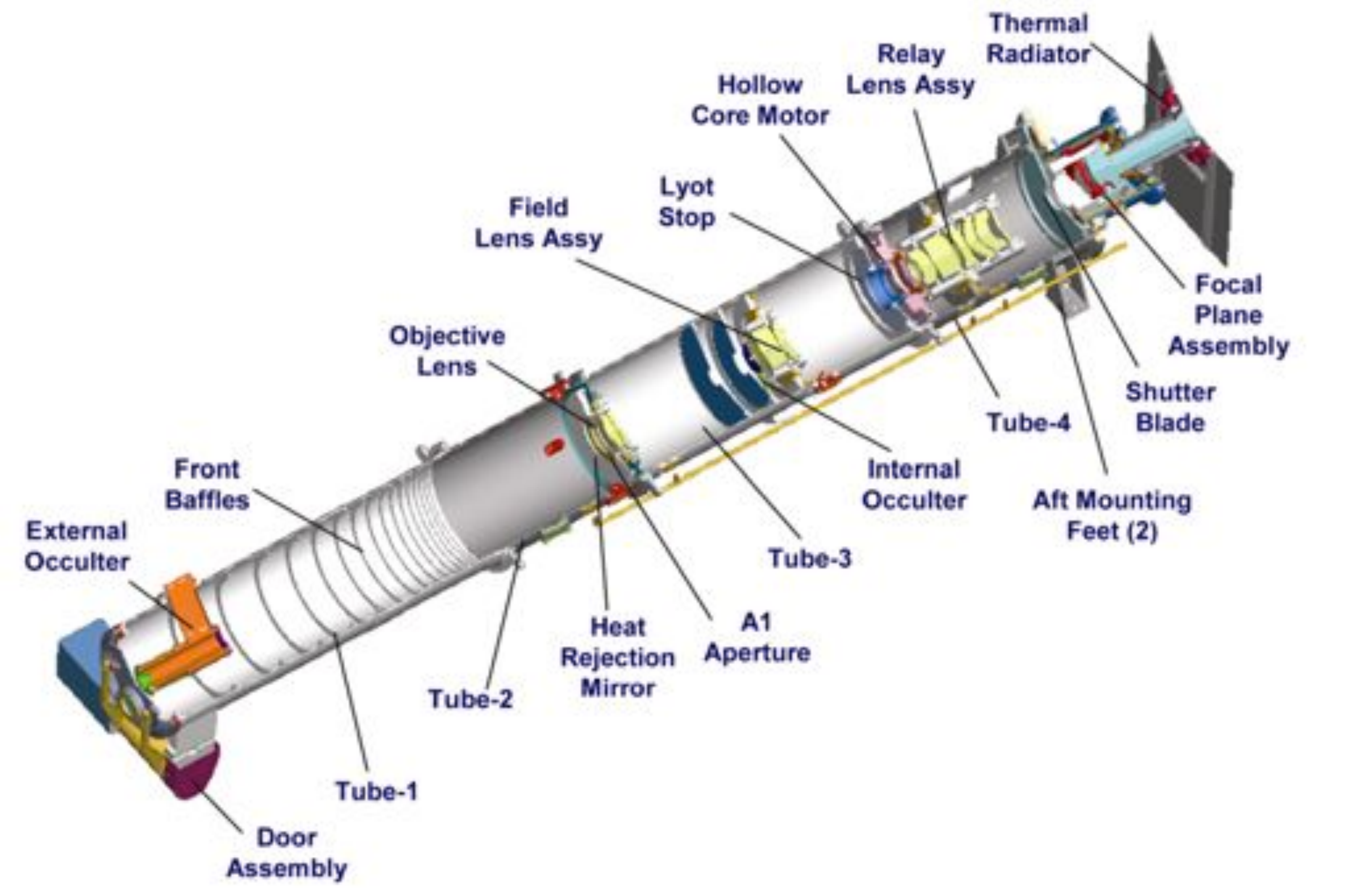}{Schematic of COR2}{A cut away view of the COR2 coronagraph \citep{howard2008p67}.}{0.9}

The outer coronagraph COR2 is an externally occulted Lyot coronagraph, of similar design to that of the SOHO/LASCO C2 and C3 coronagraphs, which images the corona in white light (650\rng750\,nm) from 2.5\rng15\,R$_{\odot}$ with 14.7\, arcsec per pixel (2048$\times$2048; \citealt{howard2008p67}). Solar radiation enters the instrument through the aperture A0 (see \figref{cor2inst}) where a three-disk occulter shades the objective lens from direct solar radiation and creates a deep shadow at the lens aperture A1. A mirror rejects heat back through the entrance aperture. The objective lens (O1) focuses an image of the external occulter onto the internal occulter, and the field lens (O2) focuses the A1 aperture onto the A3 aperture or Lyot stop. The objective lens also creates an image of A0 onto the field stop A2. The internal occulter and aperture stops block images of the brightly illuminated edges of the external occulter, A0, and A1 apertures. Between the O2 and O3 elements lies a linear polariser mounted in a hollow-core motor. The third lens group forms an image of the corona onto the focal plane where the detector is located. A bandpass filter transmits radiation from 650\rng750\,nm (FWHM) optimising throughput, and a rotating blade shutter controls exposure time. A typical COR2 observation is shown in \figref{cor2typical}

\figuremacroW{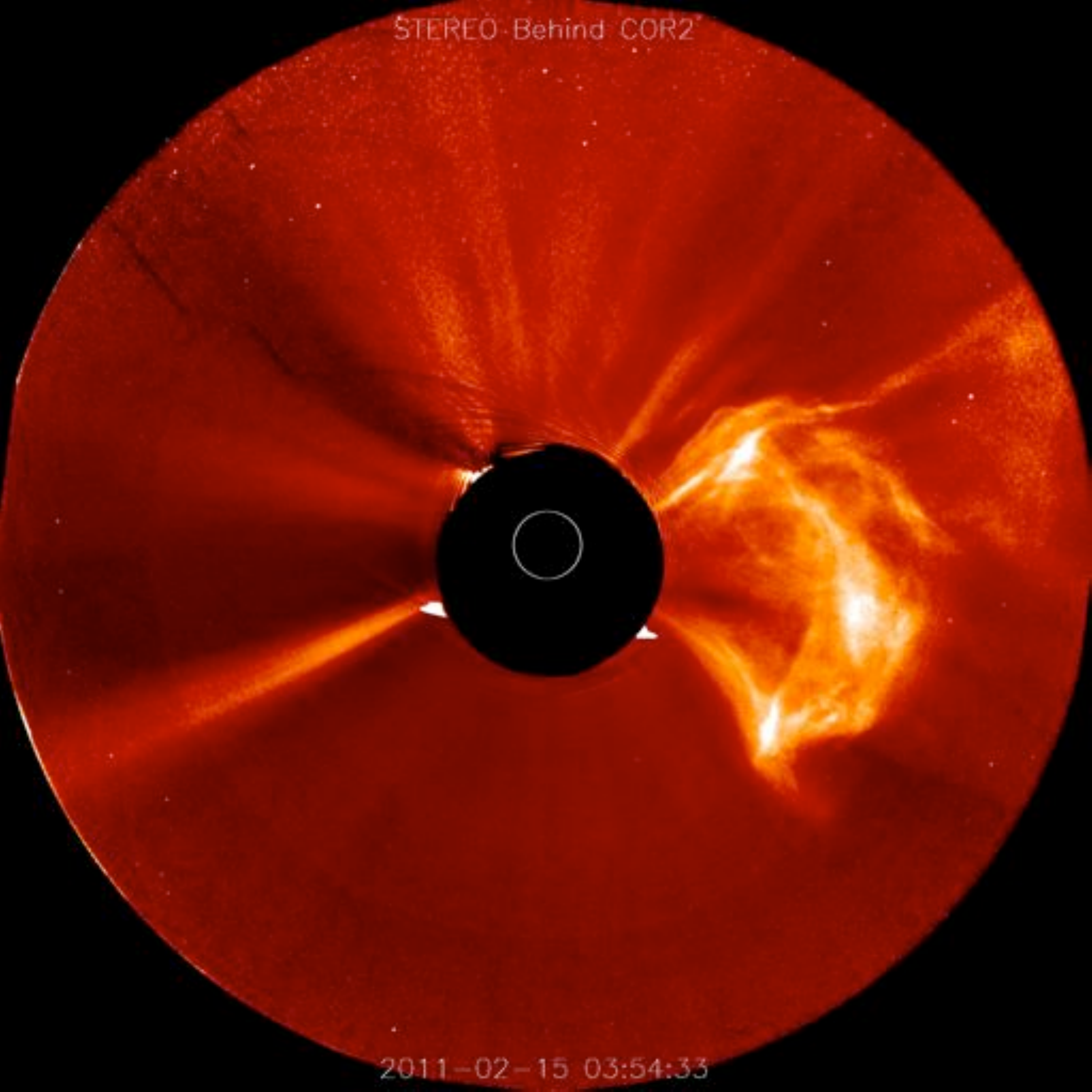}{Sample COR2 Observation}{Image of the solar corona, taken by the SECCHI outer coronagraph (COR2)
on the STEREO Behind observatory on February 15, 2011 at 03:54:33 UT. Image courtesy of the \href{http://stereo-ssc.nascom.nasa.gov/browse/2011/02/15/behind/cor2/2048/20110215_035400_d4c2B.jpg}{Stereo Science Centre}.}{0.9}

Both COR1 and COR2 take sequences of three polarised observations at $0^{\circ}$ and $\pm60^{\circ}$ for a complete polarisation sequence. This results in a cadence of 8 and 15 minutes respectively. These images are telemetered to the ground where they can be combined to give both total brightness (B) and polarised brightness (pB) images.

\subsubsection{Heliospheric Imager (HI)}
\label{sec:hi}
The HI instrument is a combination of two refractive optical telescopes (HI1 and HI2) with a multi-vane, multi-stage light rejection system which images the inner Heliosphere in white-light  (HI1: 630--730\,nm, HI2: 400--1000\,nm ). HI1 has a FOV of 20$^{\circ}$ degrees centred on an elongation of 13.28$^{\circ}$\,degrees with $70$\,arcsec pixel$^{-1}$ plate scale, while HI2 has a FOV of 70$^{\circ}$\,degrees centred on an elongation of 53.36$^{\circ}$ degrees from the Sun with $2$ arcmin pixel$^{-1}$ plate scale \citep{eyles2009p387}.

\figuremacroFP{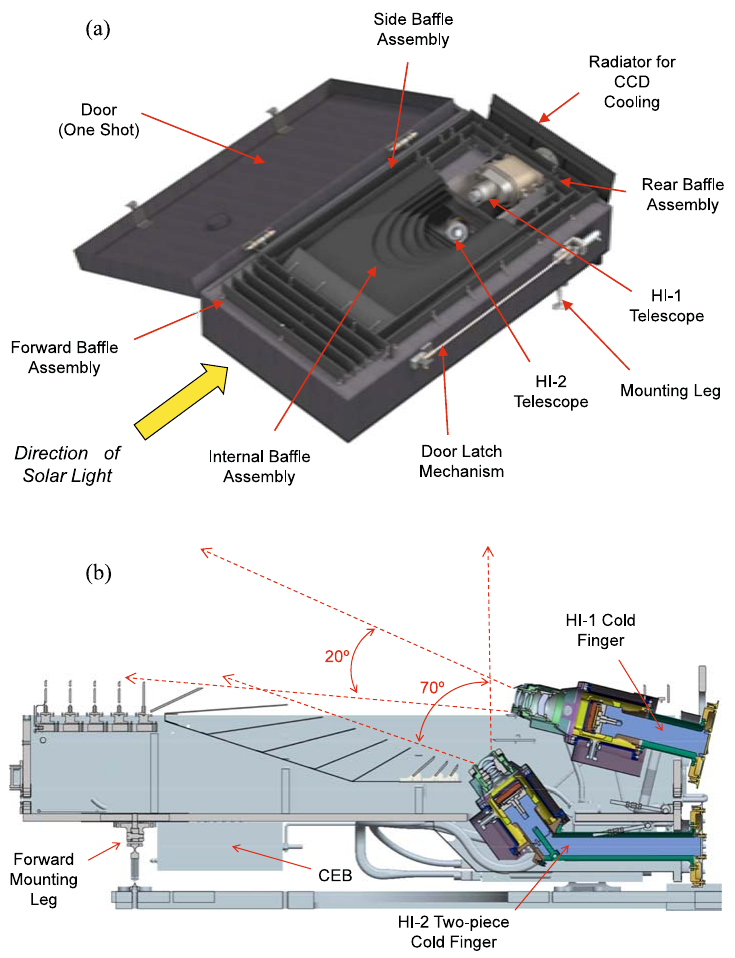}{HI Concept and Cross-section}{(a) Heliosphereic Imager concept. (b) Cross sectional view through the instrument, showing the fields-of-view of the two telescopes \citep{eyles2009p387}.}{0.8}

\begin{figure}[!t] 
   \centering
   \centerline{
   \includegraphics[width=0.45\textwidth]{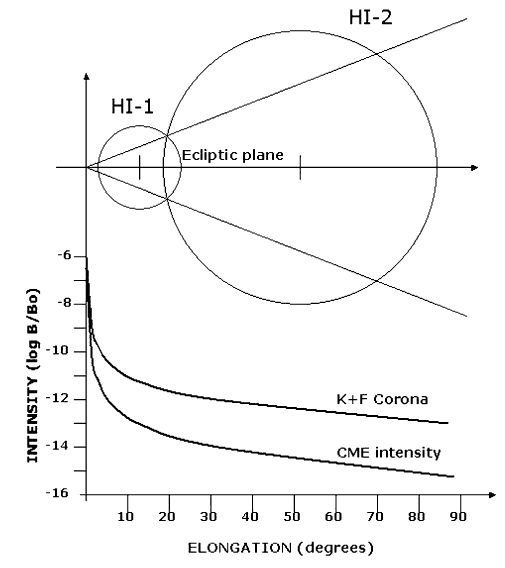}
   \includegraphics[width=0.55\textwidth]{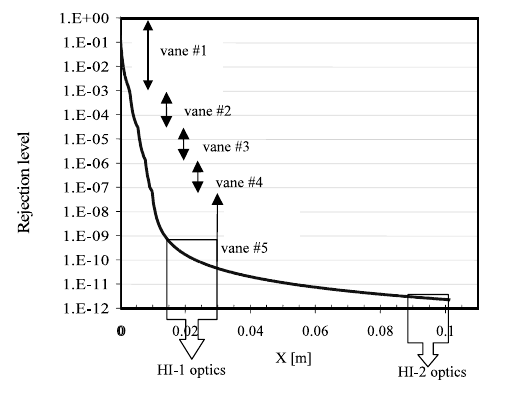}}
   \caption[HI fields-of-view and theoretical light rejection Levels]{Left) The fields-of-view and geometry of the HI telescopes and anticipated intensity of a typical CME and coronal background. Right) Theoretical light rejection levels for HI in typical conditions as a function of distance below the shadow of first vane. Also shown are the approximate positions of HI1 and HI2 optics \citep{eyles2009p387}.}
    \label{fig:lightlevels}
\end{figure}

The HI concept is based on the laboratory measurements of \citet{buffington1996p6669}, who determined the scattering rejection as functions of the number of occulters and their angle below the occulting edge. The baffle system consists of the forward, internal, and perimeter baffles (see \figref{hi_pic_diag}(a)). The forward baffle system uses Fresnel diffraction with the positions and heights of the vanes arranged in an arc such that the $(n + 1)^{th}$ vane lies below the shadow line of the $(n - 1)^{th}$ and $n^{th}$ vanes (see \figref{hibaffles}). The forward baffle is the primary source for light rejection in the instrument and provides the low levels of residual light needed of 10$^{-10}$--10$^{-12}$, as computed using Fresnel's second order approximation of the Fresnel-Kirchhoff diffraction integral for a semi-infinite half-screen see (\figref{fig:lightlevels} Right). These high levels of light rejection are vital to detect faint CMEs (see \figref{fig:lightlevels} Left)


\figuremacroW{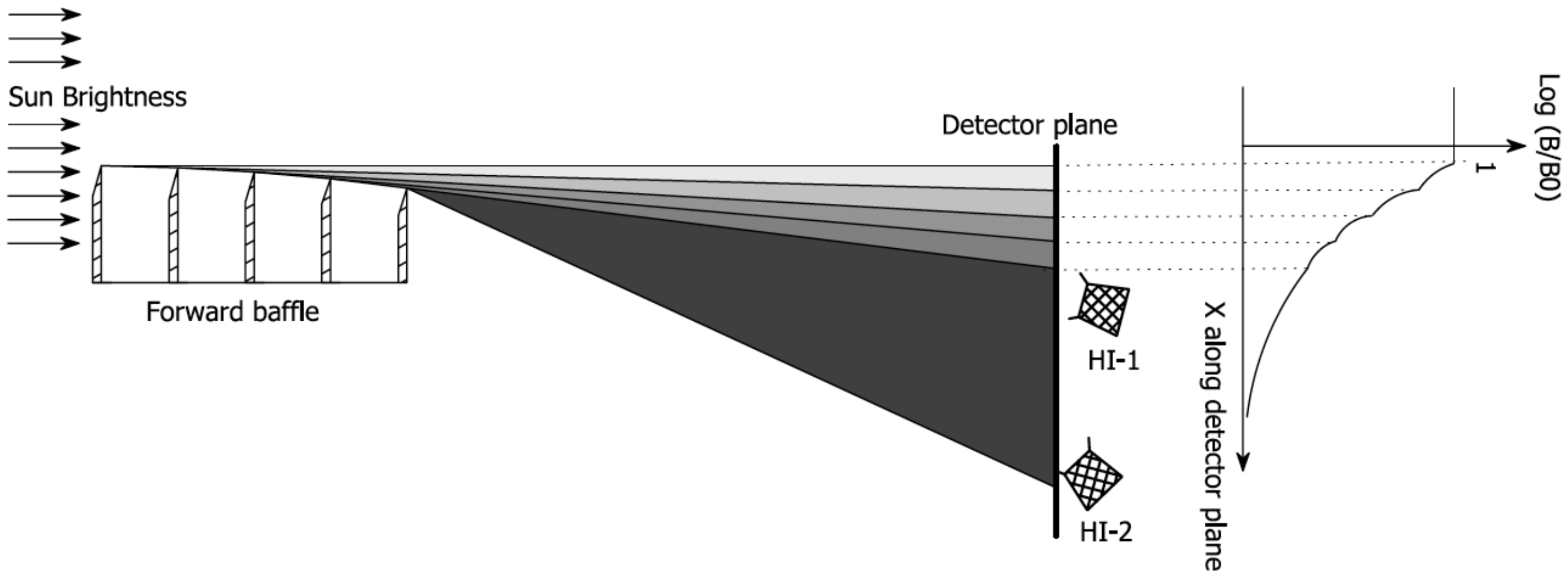}{HI Baffle System Concept}{Concept of the cascade of knife-edge diffraction system, also shown are the positions of the entrance apertures for HI1, HI2 and the form of the rejection function \citep{eyles2009p387}.}{0.9}

The internal baffle consists of five principal vanes and a number of secondary vanes (see \figref{hi_pic_diag}(a)). The system is designed to attenuate unwanted light entering the HI optical systems through multiple reflections in the baffles. Finally the  perimeter baffles (side and rear baffles) are designed to  shield the HI optical systems from reflections of photospheric light below the horizon defined by the baffles.

\figuremacroW{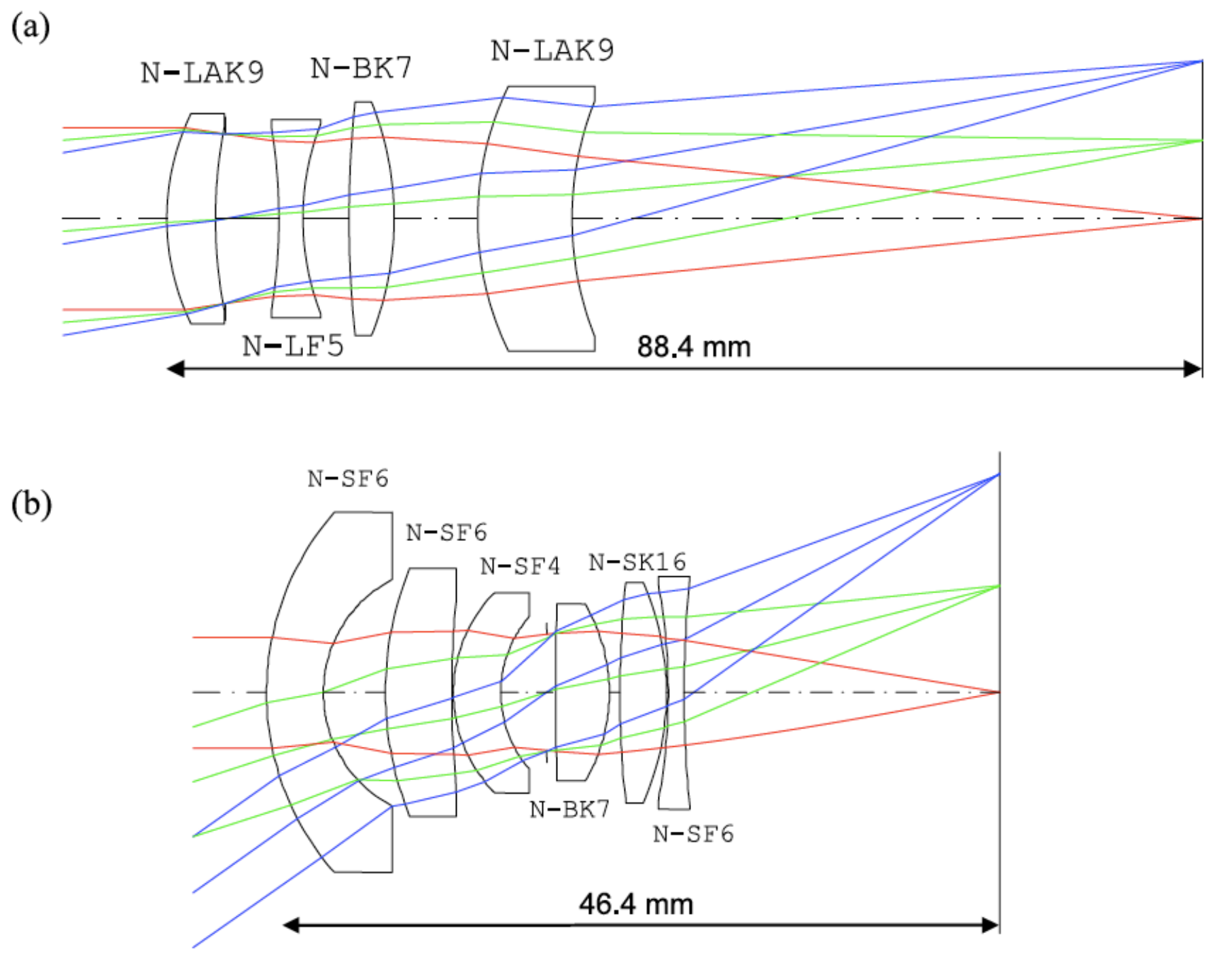}{Schematic of HI1 and HI2 Optics}{Optical layout of the HI1 (a) and HI2 (b)  telescopes. The Schott glass type is indicated beside each element \citep{eyles2009p387}.}{0.9}

The optical configurations of the two HI telescopes is shown in \figref{hioptics}. Neither of the telescopes is diffraction limited as such they are optimised to minimise the rms (root mean square) spot diameter over the operational temperature range. The design of the HI1 optics was mainly constrained by the location of the first lens which must lie in the appropriate position in the shadow pattern of the forward baffle and the lens barrel and housing must be below the shadow line cast by the first baffle, so direct sunlight cannot reach these components. The spectral response is controlled by means of multilayer coatings on two internal lenes which have the minimum angles of incidence. All the lenses are coated with anti-reflective (AR) coatings to minimise ghost images. The HI2 optics were designed to give a wide field-of-view, wide bandpass and minimise ghosting. Due to the wide bandpass the effectiveness of the AR coatings is minimised. To combat this effect the optics were designed to spread out ghost images over large regions of the focal plane. So the ghost images from bright objects contribute to the overall background rather than generating multiple images,  with their spacing dependant on the position of the object in the FOV. The lens barrels were optimised using ray-tracing software, with some cavities being added to act as light traps. All the internal surfaces and mountings were treated with either black copper oxide, or black chromium coatings. Due mainly to mechanical constraints neither of the HI telescopes have a shutter mechanism.

\figuremacroW{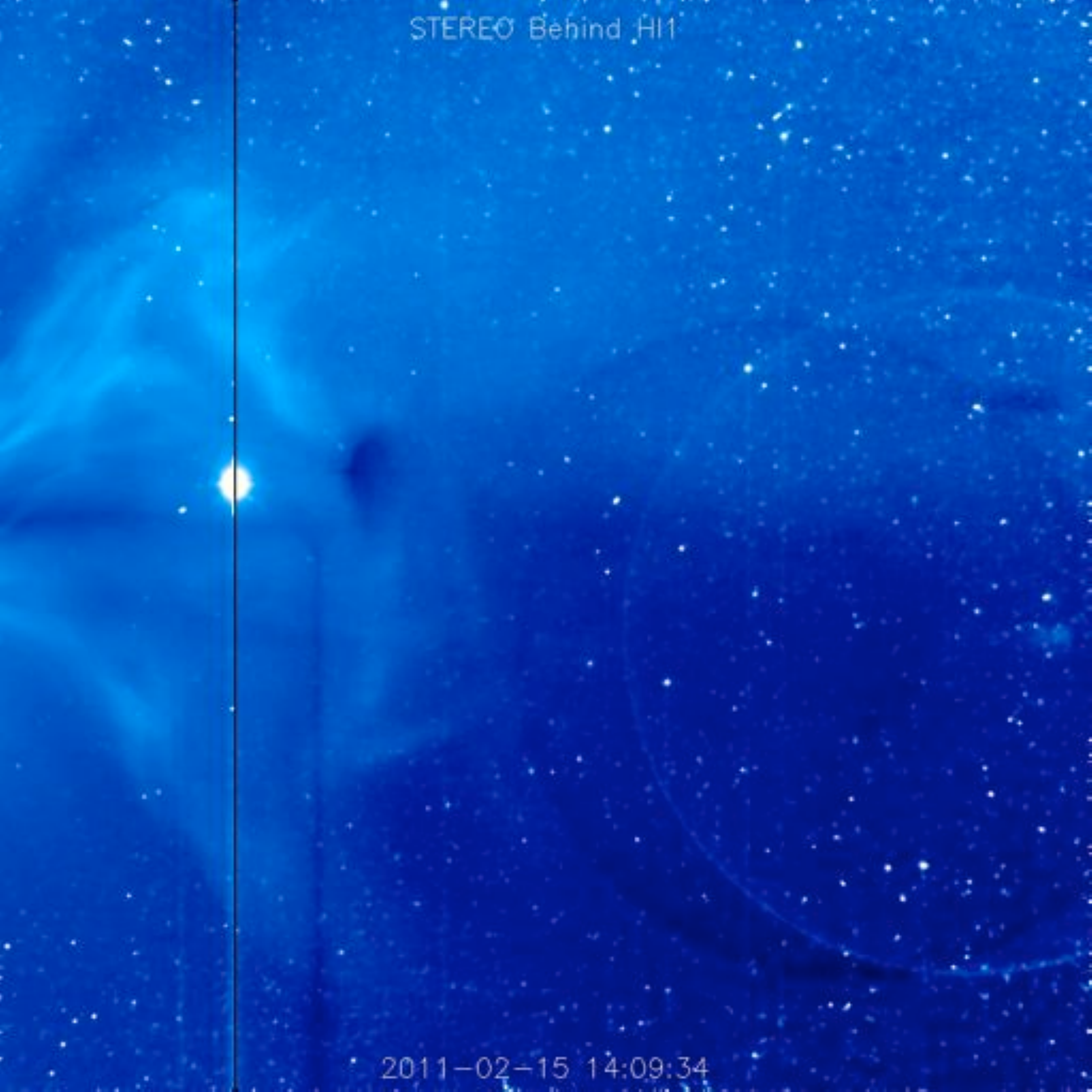}{Sample HI1 Observation}{
Image of the solar corona, taken by the SECCHI inner Heliospheric Imager (HI-1)
on the STEREO Behind observatory on February 15, 2011 at 14:09:34 UT. Image courtesy of the \href{http://stereo-ssc.nascom.nasa.gov/browse/2011/02/15/behind/hi1/1024/20110215_140901_s4h1B.jpg}{Stereo Science Center}.}{0.9}

\figuremacroSW{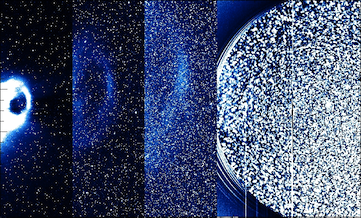}{CME Progression through HI1 and HI2 Fields-of-view}{A composite image showing the progression of a CME  through HI1 and HI2 Fields-of-view and its decreasing intensity as it does so. The image is made up of three HI1 observations taken at different times and a single HI2 image.}{0.60}

The images from both HI1 and HI2 are binned on-board to give 1024\,$\times$\,1024 pixels at stated plate scale. In order to attain the required signal-to-noise ratio needed to image the extremely faint CMEs, the HIs need long exposure time images. Due to limited dynamic range, and to avoid cosmic ray swamping, a series of shorter exposures are taken and summed on-board: in the case of HI1 30 images of 40\,s exposure each are summed (total exposure time 1200\,s) and the resulting cadence is 40 minutes; for HI2 99 images with a 50\,s exposure time each are summed (total exposure time 4950\,s) resulting in an image cadence of 2 hours. Neither HI1 nor HI2 have shutters and hence all images must be corrected for the smearing caused by shutterless readout (i.e., each row sees the scene from the pixel below it). Some of the objects in the HI FOV are bright enough to saturate the CCD, which was designed to bleed along columns allowing these to be identified and corrected later \citep{eyles2009p387}. A typical HI1 observation is shown in \figref{hi1typical}, and \figref{himimg} shows a CME as it progresses through the the HI1 field-of-view and into HI2.

\section{Data Reduction}
\label{sec:datareduct}
The data were reduced and calibrated using {\sc SolarSoft} library \citep{1998freelandp497}. Each of the instruments has its own data reduction software which corrects for various effects and produces science-grade observations. The data reduction can consist of the following:
\begin{itemize}

\item {\bf On-board processing}: numerical operations applied on board the spacecraft to keep the data within the valid range of the compression algorithm, which may include taking the square root of the image or dividing the image by 2, 3 or 4 multiple times, the operations are noted in the image header and corrected for when the data is reduced.

\item {\bf De-biasing} is the removal of the electronic offset applied to each pixel. CCD electronics are designed to produce a positive offset value to avoid issues with negative numbers. For most science grade CCDs the bias is derived from the overscan pixels on the detector chip, recorded in the image header and subtracted off when the data is reduced.

\item {\bf Flat-Fielding} is the correction of the image for non-uniform illumination (including vignetting) of the detector, and pixel-to-pixel variations (hot pixels) in the CCD response. Once a flat-field is derived all the images are divided by this to flatten the CCD response when that data is reduced.

\item {\bf Bleeding} occurs when the a signal exceeds the full well capacity of a pixel. The excess charge ``bleeds'' into adjacent pixels in the same column, causing columns of saturated pixels and in severe cases there can be horizontal bleeding as well. These regions can be flagged and removed during data reductions.

\item {\bf Distortions} the optics of telescopes produce distortions such as pin cushion or barrel distortion. The images can be geometrically corrected for these during date reduction.

\item {\bf Calibration} involves converting the digital numbers (DN) to a physical unit such as mean solar brightness (MSB).

\end{itemize}

For some instruments, backgrounds which aim to remove the static coronal signature can be subtracted during the reduction process (default for COR1) or after, to enhance the faint coronal signatures. In the case of the HIs which have no shutter, additional corrections need to be applied to account for pixel smearing during readout. Smearing is the effect of the extra exposure time a pixel will experience during readout and clear phase as a result of there being no shutter. In the case of HI this means a bright object (such as a star) will leave vertical trails above the position during read out. Also during the clear phase the pixels are exposed leaving vertical trails in the opposite direction. As the readout, exposure, and clear times are known this effect can be correct using 
\begin{equation}
\label{eqn:desmear}
	\mathbf{I}=\mathbf{T}^{-1}\mathbf{R}
\end{equation}
where $\mathbf{I}$ is the corrected image, $\mathbf{T}$ is a matrix with the exposure time along the diagonal, the readout time above the diagonal and the clear time below, $\mathbf{R}$ is the uncorrected image. Finally, the pointing information in the HI headers are updated with pointing information derived from the star fixes in the images, as dust impacts on the telecscopes can shift the optical alignment \citep{2009brownp185}.
 
 \section{Coordiante Systems}
 \label{sec:coords}
 
The derivation of phyical coordinates from pixel coordinates is a multi step procedure. In the first step pixel coordinates are transformed to intermediate coordinates i.e., they are converted into the relevant units (meters, degrees, arcsec), but they are not adjusted for the reference point of the observations or projection/geometrical effects. These intermediate coordinates are then transformed by the various possible projections and rotated into celestial coordinates. The conversion from pixel to intermediate coordinates is given by:
\begin{equation}
	x_{i} = s_{i}\sum_{j=1}^{N}m_{ij}\left (  p_{j} - r_{j}\right )
	\label{eqn:intercoords}
 \end{equation}where $p_{i}$ refers to pixel coordinates, $r_{j}$ is the reference pixel, $m_{ij}$ is a linear transformation matrix and $s_{i}$ is a scale function, with subscript $i$ referring to pixel axes and subscript $j$ referring to coordinate axis \citep{2002Calabrettap1077}. 

\figuremacro{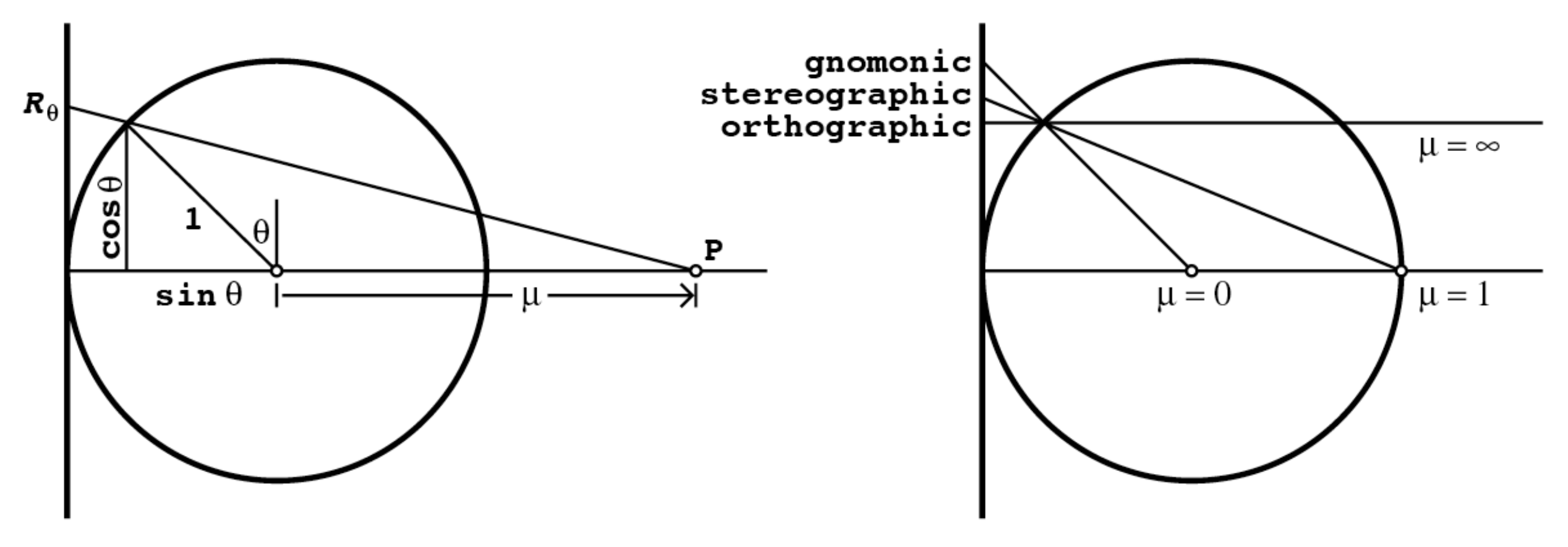}{Projection Geometries}{Geometry of the azimuthal (zenithal) perspective projections (Left), the point of projection at P is one spherical radius from the centre of the sphere; the three important special cases (right) from \cite{2002Calabrettap1077}.}

Traditionally almost all solar imaging telescopes use the gnomonic or `TAN' projection. In this projection a feature at an angle $\theta$ to the instrument's optical axis or bore-sight is projected onto the image plane at a radial distance $R_{\theta}$ given by:
\begin{align}
	R_{\theta} = F \tan(\theta)
\end{align} 
 where $F$ is the focal length of the optical system, assuming symmetry about the optical axis. The position of the feature in the image plane is given by:
 \begin{align}
 	x = R_{\theta} \cos(\phi) \\
	y = R_{\theta} \sin(\phi)
 \end{align}
 where $\phi$ is the angular position of the feature about the optical axis. However, for the wide-angle optics of HI, this type of projection is not adequate, and it was found that a Azimuthal Perspective (AZP) could accurately represent the behaviour of the HI optics. In this projection $R$ is given by: 
\begin{align}
\label{eq:azproj}
R_{\theta} = F_{p}\frac{(\mu +1)\sin(\alpha)}{\mu +\cos{\alpha}}
\end{align}
where $F_{p}$ is the paraxial focal length, and $\mu$ is the distortion parameter. Inspecting \eqref{eq:azproj} it should be clear that the gnomonic projection is a special case of the AZP projection with $\mu=0$. Figure \ref{projs} shows the AZP projection for a unit sphere and three special cases. This figure shows why, for most solar telescopes which have a narrow field-of-view a gnomonic projection is suitable. Once the coordinates have been de-pojected they can be converted to the Sun-centred physical coordinates.

\subsection{Heliographic Coordinates}
The well-known solar coordinate system expresses the position of a feature on the solar surface in terms of latitude, $\Theta$, and longitude, $\Phi$, and can be extended to 3D by adding the radial distance, $r$, from the centre of the Sun. There are two basic variations of the heliographic coordinate system called Stonyhurst and Carrington which both use the same rotation axis only differing in longitude. 

\subsubsection{Stonyhurst}
The origin of the Stonyhurst coordinate system is at the intersection of the solar equator and central meridian as seen from Earth. $\Theta$ and $\Phi$ are given in degrees increasing northwards and westwards, respectively, and $r$ is either physical the absolute physical distance in $m$ or relative to $R_{\odot}$ as shown in \figref{stonyhurst}.

\figuremacroW{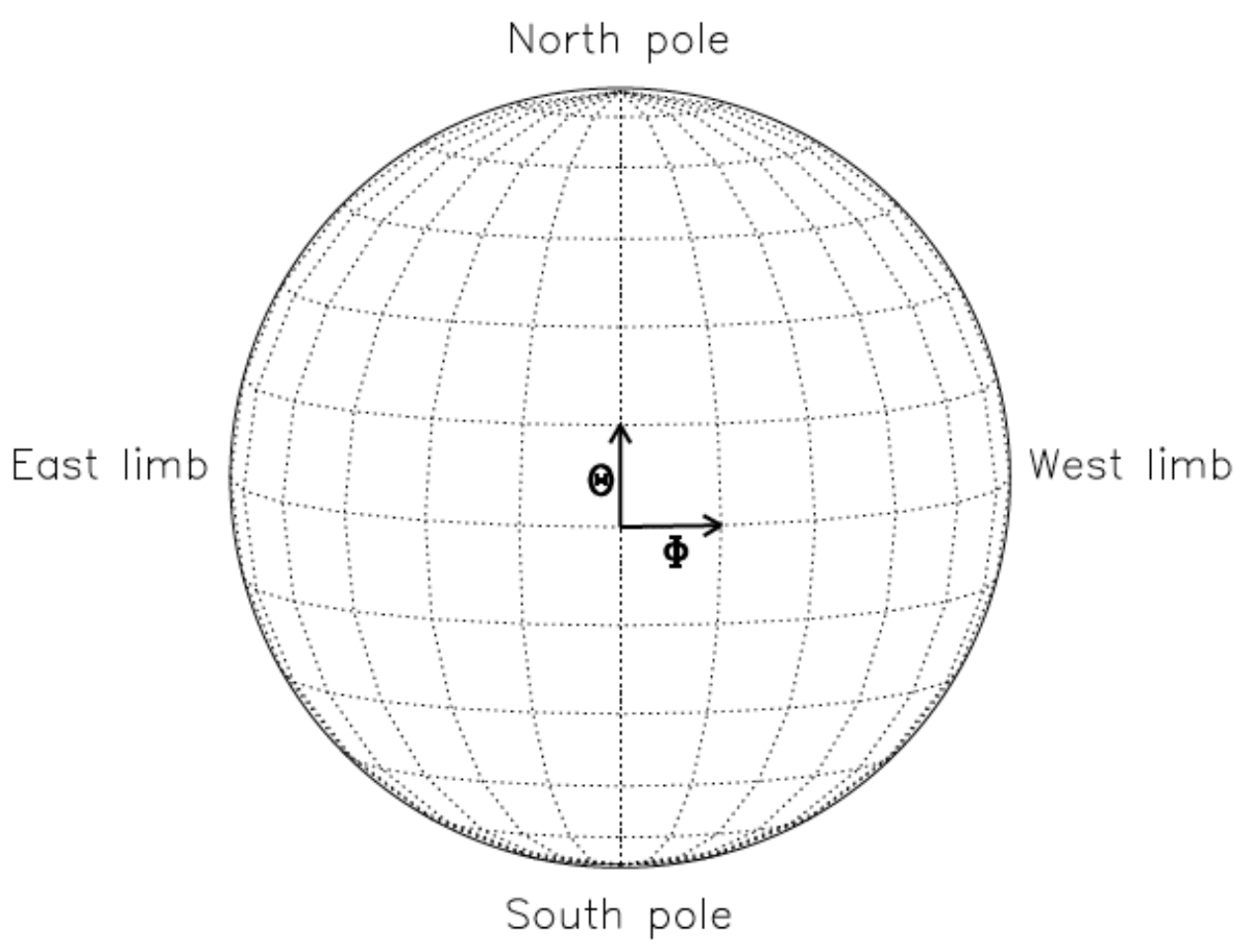}{Stonyhurst Coordinates}{A diagram of the Sun showing the Stonyhurst coordinate system \citep{2006Thompsonp791}.}{0.60}

\subsubsection{Carrington}
While Stonyhurst is fixed with respect to the Sun-Earth, line Carrington coordinates rotate at the mean solar rotation rate with the first rotation commencing on 9 November 1853 and being sequentially numbered thereafter. Both of these systems have limitations, especially in representing features far off the solar disk.

\subsection{Heliocentric Coordinates} 
Heliocentric coordinates express the true position of a feature in terms of physical units from the centre of the Sun. Heliocentric Aries Ecliptic (HAE), Heliocentric Earth Ecliptic (HEE) and Heliocentric Earth Equatorial (HEEQ) are all examples of heliocentric coordinate systems consisting of three mutually perpendicular axes $X$, $Y$ and $Z$. HEEQ is closely related to Stonyhurst heliographic by
\begin{align}
	\Theta = \tan^{-1}(Z_{HEEQ} / \sqrt{X^{2}_{HEEQ}+Y^{2}_{HEEQ}} ), \\
	\Phi = arg(X_{HEEQ}, Y_{HEEQ}) \nonumber 	\\
	\nonumber \\
	X_{HEEQ} = r \cos(\Theta) \cos(\Phi) \\
	Y_{HEEQ} = r \cos(\Theta) \sin(\Phi) \nonumber \\
	Z_{HEEQ} = r \sin(\Theta) \nonumber
\end{align}

No single perspective solar observation can be truly accurately transformed into heliocentric coordinates. Sun-Observer lines-of-sight are not truly parallel, so the physical position will depend somewhat on the distance along a line-of-sight, also no distinction is made between the possible projections. However even with these problems Heliocentric coordinates are useful, especially for representation of the results of 3D reconstruction and form the basis of the helioprojective coordinate system.

Both of the coordinate systems discussed below are observer specific i.e. non-terrestrial observers will measure different coordinates to those measured from Earth. As such, information about the position of the observer must be provided to properly define the coordinate system.

\subsubsection{Heliocentric-Cartesian}
Heliocentric-Cartesian coordinates are a true cartesian system with each of the axes being mutually perpendicular, and all lines of constant $x$ ($y$ or $z$) being parallel,  see \figref{heliocentric}. The $z$-axis is defined to be parallel to the observer-Sun line and pointing towards the observer. The $y$-axis is defined to be perpendicular to $z$ and in the plane containing the solar North pole axis and increasing northwards. The $x$-axis is perpendicular to both $x$ and $y$ and increases westward. The distance along each axis is expressed in a physical distance of meters, or relative to $R_{\odot}$

\figuremacroW{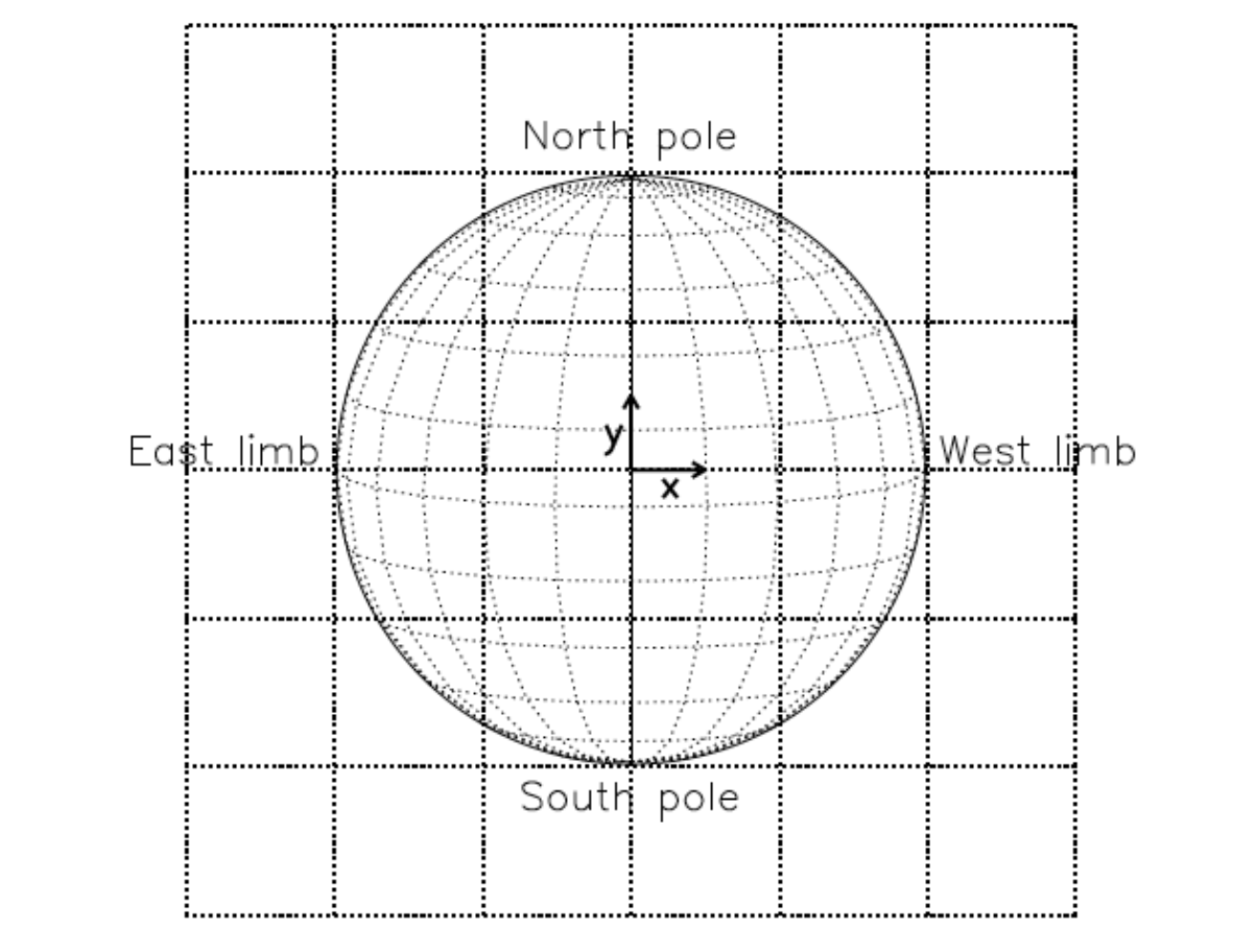}{Heliocentric Coordinates}{A diagram of the Sun with lines of constant Heliocentric cartesian ($x$, $y$) overlaid. The $z$ axis would point out of the page \citep{2006Thompsonp791}.}{0.60}

\subsubsection{Heliocentric-Radial}
Heliocentric-Radial coordinates share the same $z$ axis as the previous coordinate system, but $x$ and $y$ are replaced with $\rho$ and $\phi$. The impact parameter $\phi$ is the radial distance from the $z$-axis, again expressed in a physical unit or relative to $R_{\odot}$. The position angle $\phi$ is measured in degrees counterclockwise from solar north as viewed by the observer.

\subsection{Projected Coordinate Systems}
It has already been stated that observations from a single viewpoint can only approximate heliocentric coordinates. The observations are projected against the celestial sphere and this needs to be accounted for, in order to accurately describe the observations. Helioprojective coordinates mimic heliocentric coordinates but replace physical distances with angles. All angles are measured from Sun centre as seen by the observer. As the coordinates are spherical in nature they take into account exactly what projections are used.

\subsubsection{Helioprojective-Cartesian Coordinates}
\label{sssec:hpc}
This is the projected version of heliocentric cartesian coordinates where $x$ and $y$ have been replaced by $\theta_{x}$ and $\theta_{y}$. The helioprojective equivalent of $z$ is $\zeta = D_{\odot}-d$ surface of constant $\zeta$ represent spheres centred on the observer, with the sphere of $\zeta =0$ passing through the centre of the sun. The relationships between helioprojective-cartesian and heliocentric-carteian are
\begin{align}
	x &= d \cos(\theta_{x}) \sin(\theta_{y}), \\
	y &= d \sin(\theta_{y}), \nonumber \\
	z &= D_{\odot} - d \cos(\theta_{x}) \cos(\theta_{y}), \nonumber \\
	\nonumber \\
	d &= \sqrt{x^{2}+y^{2}+(D_{\odot}-z)^{2}}, \\
	\theta_{x} &= arg(D_{\odot}-z, x), \nonumber \\
	\theta_{y} &= \sin^{-1}(y/d). \nonumber
\end{align}
where $D_{\odot}$ is the observer-Sun distance, and $d$ is the observer feature distance. Close to the Sun, where the small angle approximation holds, the relationship reduces to 
\begin{align}
	x \approx d \left ( \frac{\pi}{180^{\circ}}\right ) \theta_{x} \approx D_{\odot} \left ( \frac{\pi}{180^{\circ}}\right ) \theta_{x}\\
	x \approx d \left ( \frac{\pi}{180^{\circ}}\right ) \theta_{y} \approx D_{\odot} \left ( \frac{\pi}{180^{\circ}}\right ) \theta_{y}. \nonumber \\	
\end{align}
These are the default coordinates for STEREO SECCHI observations.

\subsubsection{Helioprojective-Radial Coordinates}
The helioprojective version of helcentric-radial is where the impact parameter $\rho$ is replaced with the angle $\theta_{\rho}$.




\chapter{Data Analysis and Techniques} 


\ifpdf
    \graphicspath{/}
\else
    \graphicspath{{4/figures/EPS/}{4/figures/}}
\fi


\hrule height 1mm
\vspace{0.5mm}
\hrule height 0.4mm
\noindent
\\	{\it In this chapter the methods used to enhance and analyse the STEREO observations are described. CMEs can be extremely faint compared to the coronal background and in order to detect and track them image processing methods are applied to enhance the CMEs signature. Once the CMEs are identified, 3D reconstruction techniques are applied to the two perspective views to derive their true position. Using the true position of the CME derived from a number of observations, the kinematics can be calculated and compared to the various models. Due to the large number of ill-determined parameters in the drag models a rigorous statistical procedure was used to test if the drag models appropriately reproduce the observations. This chapter is based on methods described in \citeauthor*{Maloney2009p149}, Solar Physics, \citeyear{Maloney2009p149} and \citeauthor*{Byrne2010p74}, Nature Communications, \citeyear{Byrne2010p74}.
 } \\
\hrule height 0.4mm
\vspace{0.5mm}
\hrule height 1mm

\newpage

\label{chapter:methods}

\section{Image Processing}
\label{sec:imp}
In coronagraph observations, CMEs are observed as outwardly moving regions of stronger brightness intensity relative to the background corona. As CMEs can be extremely faint compared to the background, imaging processing techniques are applied to enhance and so enable their identification and tracking. Coronagraph observations are dominated by the K and F corona with an occasional star, comet or planet overcoming this signal. Thus the main goal in processing coronagraph images ia the removal of the static coronal signal, and the enhancement of the CME signal. This is accomplished through a number of techniques:

\begin{itemize}
	\item {\bf Background Subtraction:} Long term (days to weeks) background subtraction removes instrumental stray light and static coronal signals such as the F and K corona. The backgrounds are derived by taking the average, mean, minimum, or some combination of these operations applied to the series of images.  Short term (hours to days) variations can also be subtracted to remove further semi-transient features. 
	
	\item {\bf Running Difference:} Running difference images are created by subtracting the previous image from the current image and thus help in enhancing moving features. Formally this can be defined as:
	\begin{align}
		I'(x,y)_{i} = I(x,y)_{i} - I(x,y)_{i-1}
	\end{align}
where $I'_{i}(x,y)$ is the running difference intensity at pixel $(x,y)$, $I(x,y)_{i}$ is the current pixel intensity, and $ I(x,y)_{i-1}$ the previous pixel intensity.
	\item {\bf Base Difference:} Base difference images are formed by removing a base image from every subsequent image, usually the base image is a pre-event image.
	
	\item {\bf Normalised Radial Gradient filter:} It is well known that the intensity of radiation falls steeply falls moving away from the Sun, by a factor $\sim$10$^{4}$ between the limb and $\sim$3\,$R_{\odot}$. The normalising-radial-gradient filter (NRGF) \mbox{ \citep{Morgan2006p263}} attempts to account for this radial falloff using:
	\begin{align}
		I'(r,\phi) = \left [ I(r,\phi) - I(r)_{	\langle \phi \rangle} \right  ] / \sigma(r)_{\langle \phi \rangle}
	\end{align}
where $I'(r,\phi)$ is the NRGF intensity at height $r$ and position angle $\phi$, $I(r,\phi)$ is the original intensity, and $I(r)_{\langle \phi \rangle}$, $ \sigma(r)_{\langle \phi \rangle}$ are the mean and standard deviation of intensities calculated over all position angles at height $r$
\end{itemize}
The standard procedure used was to reduce the data as described in Section \ref{sec:datareduct} to produce total brightness ($B$) images from which backgrounds are then subtracted. The images were then spacially median filtered and smoothed, and at this point the various different types of image were created. The effects of processing COR1 observations are shown in \figref{cor1proc}. \figref{cor1proc}~(a) shows the reduced data, with instrumental stray light apparent even on the occulter, as is the dominant K and F corona. \figref{cor1proc}~(b) has had a background subtracted and a binary mask applied to remove all non-imaging pixels and some of the artefacts know as ``nails''. \figref{cor1proc}~(c) shows the effect of removal of an additional short term background from (b) while (d) shows a base difference image, (d) a running difference image and (f) NRGF applied to (c). \figref{cor2proc} shows the same but for COR2 observations.

\figuremacroFP{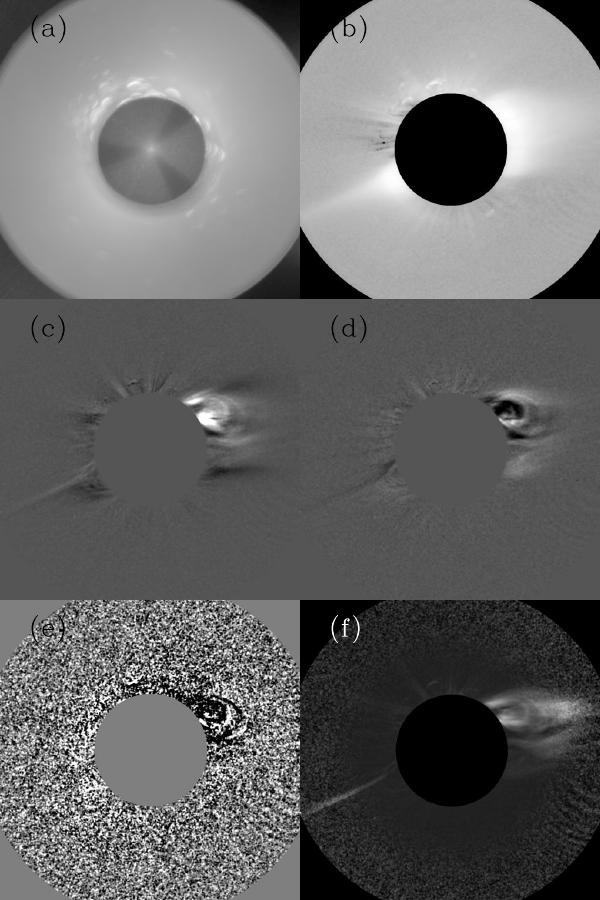}{COR1 Image Processing}{A COR1 observation from 08 November 2007 05:52UT. Different processing steps are shown: (a) reduced data, (b) processed and standard background subtraction,  (c) additional background subtraction, (d) base difference, (e) running difference, and (f) NRGF of (c).}{0.9}

\figuremacroFP{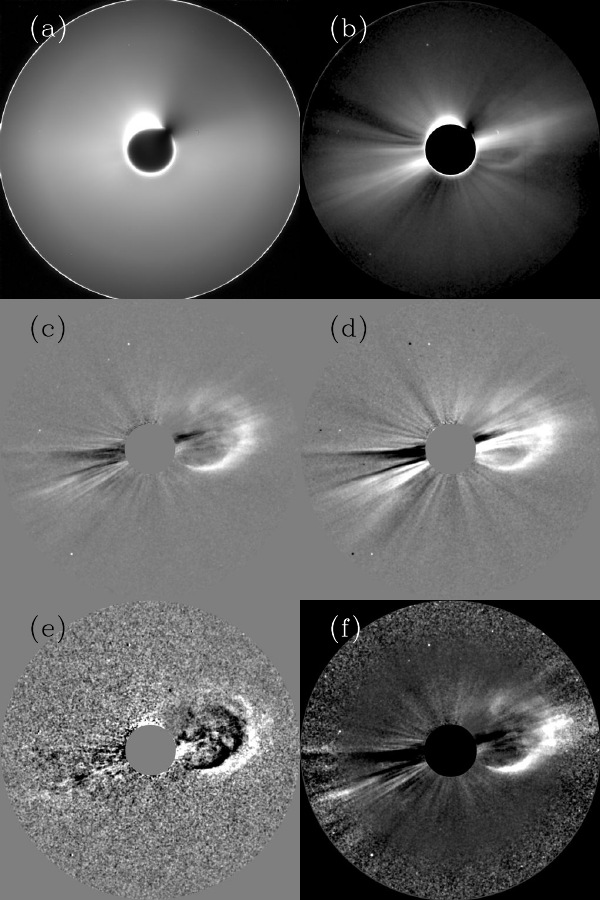}{COR2 Image Processing}{A COR2 observation from 09 November 2007 01:22UT. Different processing steps are shown: (a) reduced data, (b) processed and standard background subtraction,  (c) additional background subtraction, (d) base difference, (e) running difference, and (f) NRGF of (c).}{0.9}

\subsection{Image Processing for the Heliospheric Imager}
In the case of HI, the images are dominated by the static F corona, and upon removal of this stars, the HI can image down to a 13$^{\text{th}}$ magnitude star. The CME signal per pixel is often comparable to the background and far smaller than the contribution from stars. Thus the removal of the star field is critical for detecting CMEs far from the Sun. The exposure times for HI1, HI2 observations are 1,200\,s and 4,950\,s respectively (Section \ref{sec:hi}). The star field in the field-of-view of the instrument can be very well approximated as static over this time period, but there will be a shift ($\sim$1 pixel per image in HI1 and HI2) due to the motion of the satellite. A standard running difference image would then be contaminated with large signals due to this moving star field (see \figref{hi1proc}(c) and \figref{hi2proc}(c)). If the offset between two subsequent images can be found, then the images can be shifted and subtracted and only transient features will be left. Formally this can be defined as:
\begin{equation} 
	I \left (x, y \right ) _{i} = I \left (x, y \right )_{i} - I \left (x+\alpha, y+\beta \right )_{i-1} \\
	\label{Eq-aa}
\end{equation} where $\alpha$ and $\beta$ are the $x$ and $y$ offsets due to the motion of the satellite. However, there is a geometrical optical distortion in the HI images which makes this difficult. We developed a method to approximate this offset. A region (512x512) is extracted from both images and a local cross-correlation is performed in a running window (128x128) around these sub-images. The average offset from this procedure is used to shift the image. In a similar manner to the NRGF the HI images can be flattened to account for the radial falloff in intensity. The NRGF cannot be directly used as the HI images only sample a small range of $\phi$. However, by summing down the columns of a HI image, and then median filtering this with respect to time a pseudo NRGF can be derived which flattens the HI images as shown in \figref{hi1proc}~(d). \figref{hi2proc}~(d) shows the results of this modified running difference on HI1 and HI2 observations. The effects of the distortion are especially clear in the HI2 images.

\figuremacroFP{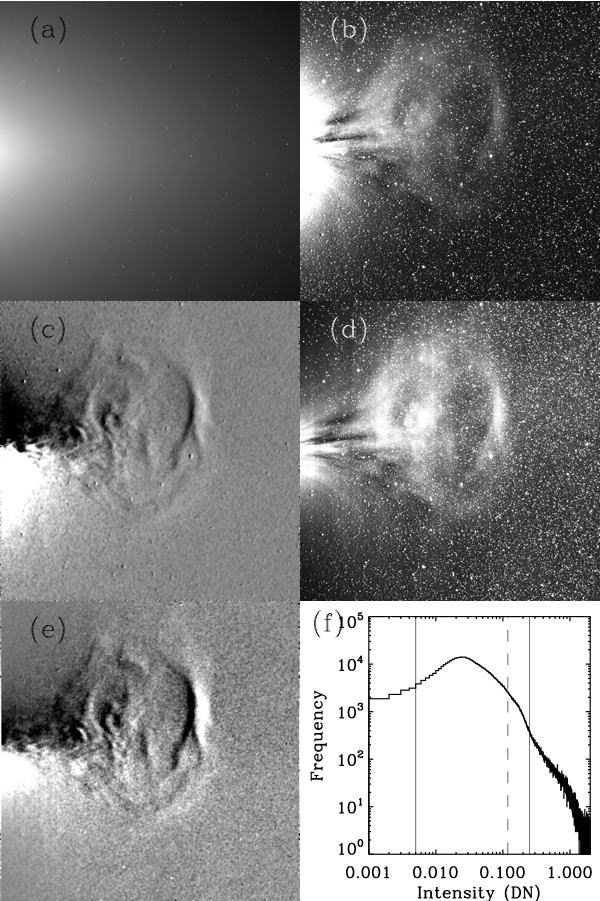}{HI1 Image Processing}{A HI1 observation from 10 November 2007 09:29UT. Different processing steps are shown: (a) reduced data, (b) standard background subtraction,  (c) running difference, (d) `radial' filtered, (e) modified running difference \eqref{Eq-aa}, and (f) intensity histogram of (b) the solid vertical lines show the thresholds used to display (b) and the dashed line is the mean of image (b).}{0.85}

\figuremacroH{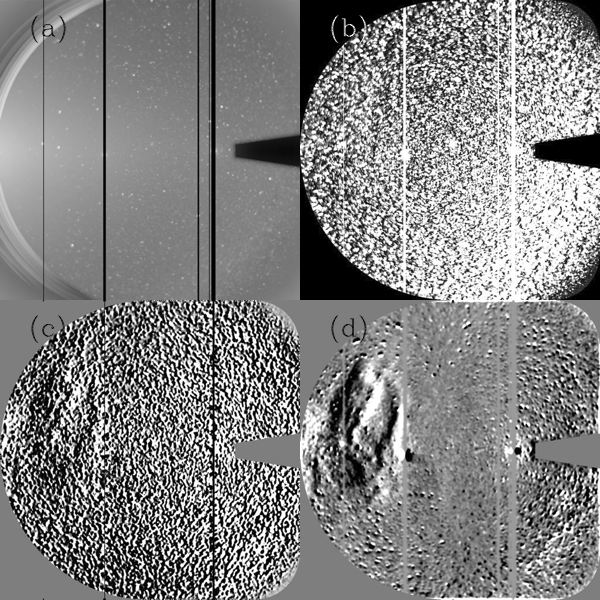}{HI2 Image Processing}{A HI2 observation from 12 November 2007 06:10UT. Different processing steps are shown: (a) reduced data, (b) standard background subtraction,  (c) running difference and (d) modified running difference \eqref{Eq-aa}.}{0.6}

\section{Three Dimensional Reconstruction} 
\label{section:3drecon}

\figuremacroW{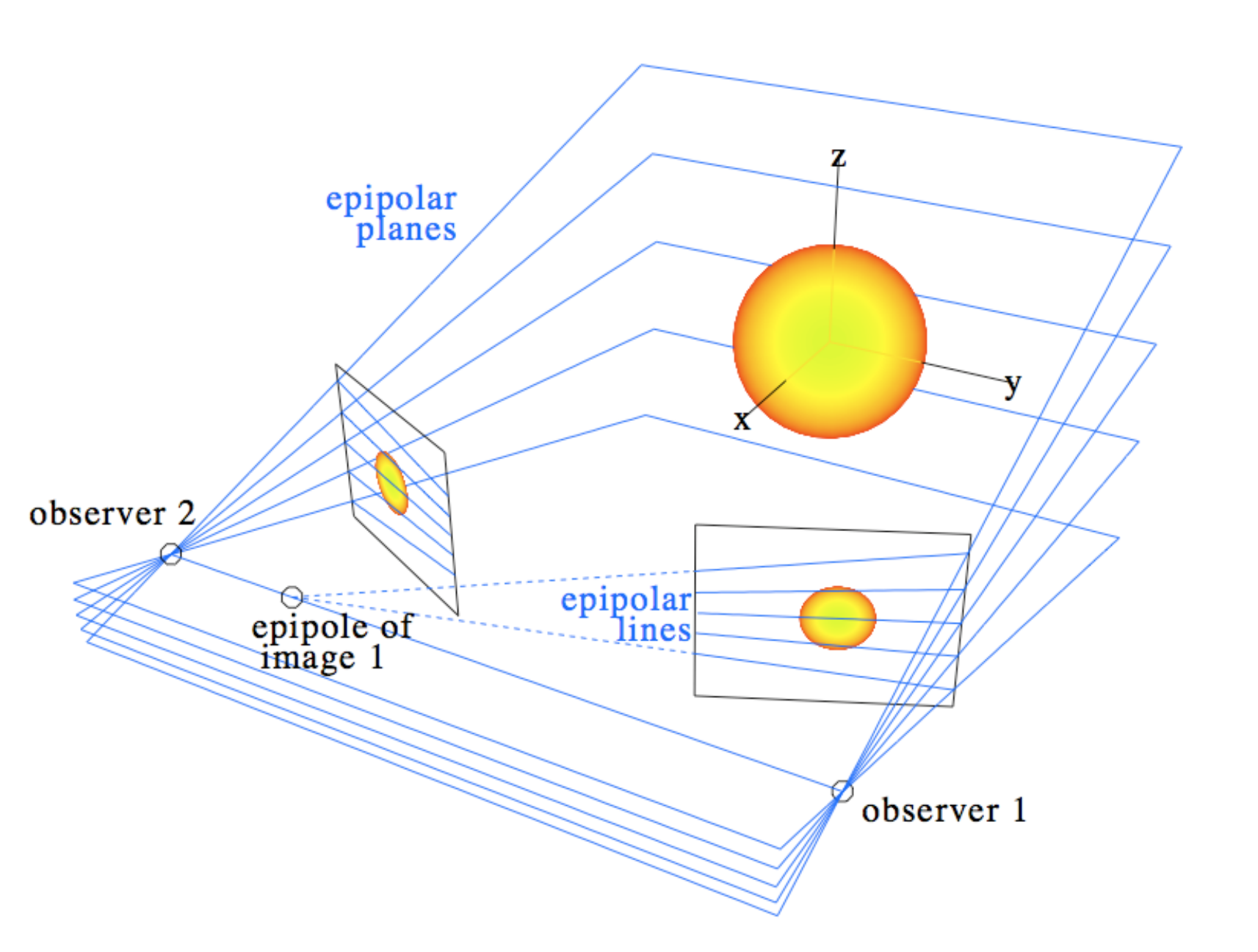}{Epipolar Geometery}{Orientation of epipolar planes in space and the respective epipolar lines in the stereo images for two observers looking at the Sun \citep{Inhester2006astroph}.}{0.9}

The concept of stereoscopy was first suggested by \citet{Wheatstone1839p625}. The reconstruction of three dimensional (3D) information from images of an object observed from two different perspectives is well studied and the general reconstruction methods are well developed. The main technique used in this thesis is known as `tie-pointing' and relies on the epipolar constraint. This states that any feature identified on an epipolar line in one STEREO image must lie along the same epipolar line in the other stereo image \citep{Inhester2006astroph}. An epipolar line is the projection of the epipolar plane (the plane containing the two observers and the point P) in the stereo images (see \figref{epipolargeo}).

\subsection{Tie-pointing}
\label{ssec:tp}
One possible implementation of this method of stereoscopy is outlined below. We implicitly assume that two simultaneous observations of a feature (the point P; see top panel of \figref{3drecon}) from two different perspectives are available.  If the feature is identified in one of the images (from A for example), the angles $\theta_{x}^{A}$ and $\theta_{y}^{A}$ (bottom panel \figref{3drecon}) can be derived, and the only unknown is the depth of the feature or the distance along the line-of-sight $d$ (see bottom panel of \figref{3drecon}). One can arbitrarily pick two values for $d$ ($d_{0}$, $d_{1}$) thereby giving two sets of 3D coordinates $\left[ \theta_{x}^{A}, \theta_{y}^{A}, d_{0} \right]$ and $\left[ \theta_{x}^{A}, \theta_{y}^{A}, d_{1} \right]$, which can then be transformed into Heliocentric-Cartesian as described in Section \ref{sssec:hpc}
   
\figuremacroFP{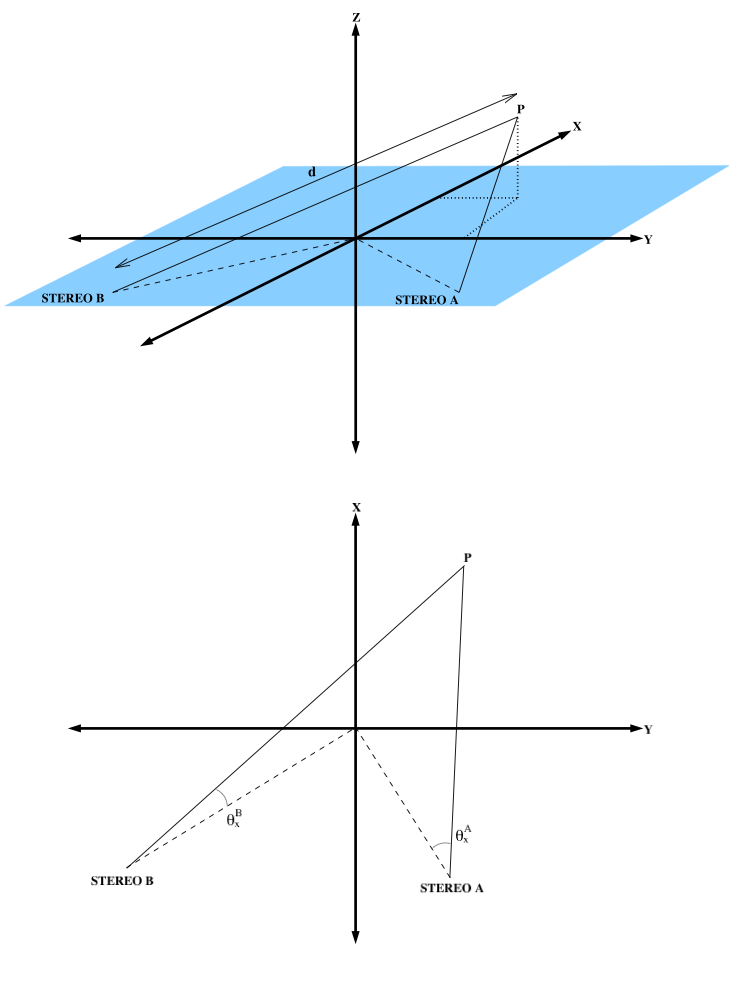}{3D Reconstruction Geometry}{(top) 3D representation of the geometry of the observations with the lines-of-sight (LOS) from the two spacecraft (solid), spacecraft Sun centre lines (dashed) and the point to be reconstructed P. (bottom) Projection in the $x-y$ plane showing the angles $\theta_{x}^{A}$ and $\theta_{x}^{B}$, where $x$ corresponds to image pixel axis. The point P is found by solving the equations of the LOS for their intercept. The same can be done in the $x-z$ plane for $\theta_{y}$.}{0.80}
   
These coordinates (of the line-of-sight), as viewed from the second perspective, can then be derived by reversing the transform, and the epipolar line can be over-plotted on the second image. If the position of the same feature along this epipolar line can be found, this gives all the information needed to complete the reconstruction. The same procedure is carried out for this second point, selecting two values of $d$ and transforming these into the reconstruction coordinate system. Now the problem is reduced to finding the point of intersection of two lines in 3D, which can be simplified to solving for the intersection point of two lines in 2D in two planes, for example the $x-y$ and $x-z$ plane (see bottom panel of \figref{3drecon}).

Due to the nature of the HI instruments, there are often only observations available from one spacecraft as a result, an additional constraint is  needed. In the HI FOV, the CMEs are at large distances from the Sun so we assume that the CME propagates pseudo-radially, continuing along the trajectory that it followed in the COR1/2 FOV. Based on this assumption, we use the best fit of the trajectory COR1/2 data in the $x-y$ (ecliptic) plane and constrain the CME to propagate along this fit. This fit line is treated exactly the same as the line-of-sight above. The equations are solved for the point of intersection between this fit line and the observed line-of-sight, yielding the $x$, $y$ position. The $z$ coordinate is then calculated by assuming two distances along the line-of-sight to find the equation of the line with respect to $x$ or $y$, and substituting in the relevant coordinate calculated yields the corresponding $z$ value.
\subsection{Elliptical Tie-pointing}
\label{ssec:etp}
The tie-pointing technique is best suited to study small features close to the Sun, such as coronal loops \citep{Aschwanden2008p827}. When applied to study CMEs a number of the assumptions do not hold. Firstly, as CMEs are large and curved the lines-of-sight will not intersect upon it but ahead of it, see \figref{surf3d} and \figref{ellipticaltiept}~(a). Also, due to the motion of the CME and the Thomson scattering geometry, the same part of the CME cannot be rigorously identified from both perspectives. Consequently CMEs cannot be fully reconstructed by tie-pointing alone, though their position in 3D space can be localised by the intersection of sight lines tangent to the CME front \citep{deKoning2009p167,Pizzo2004p21802}.

\figuremacro{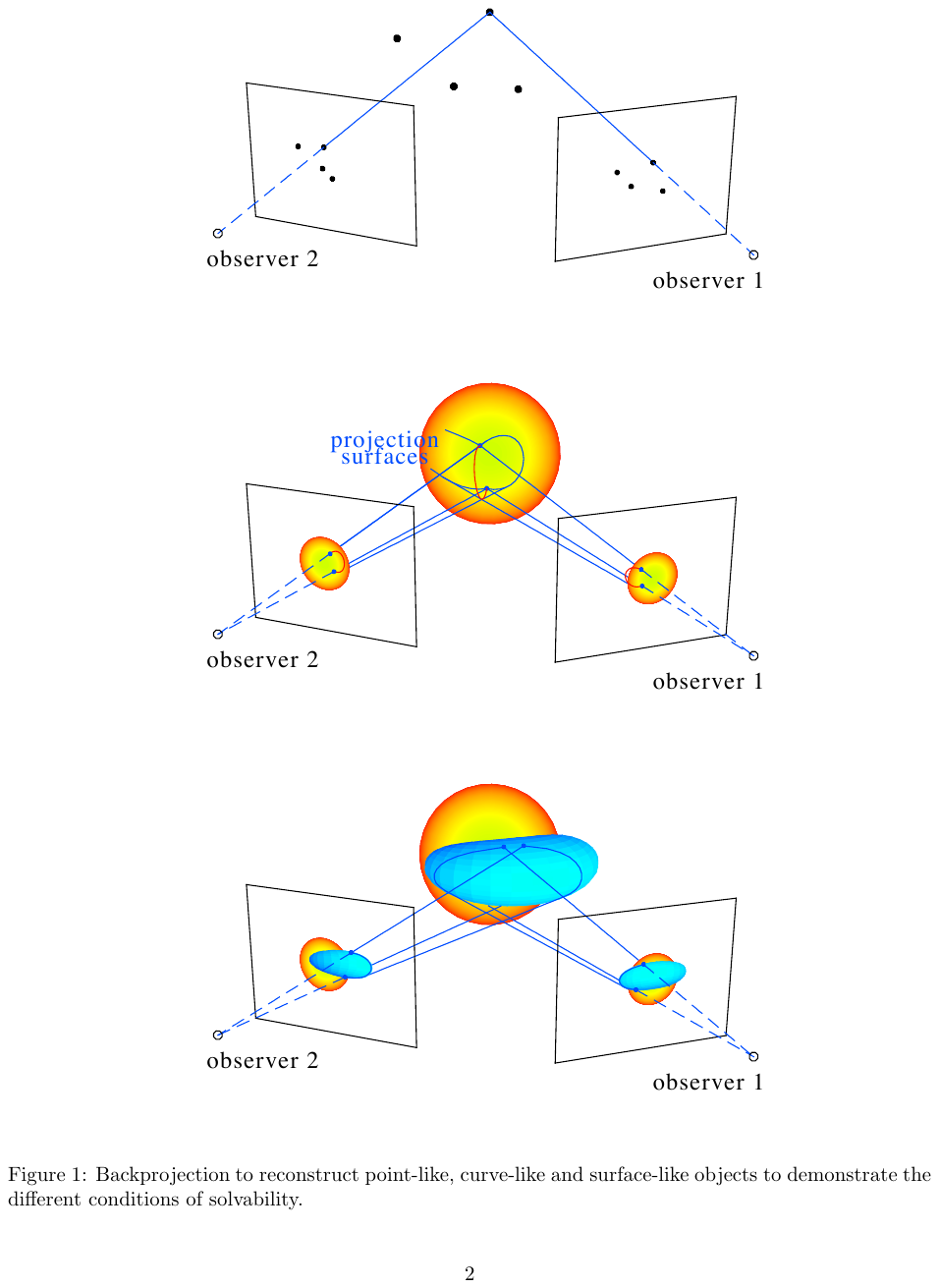}{Tie-Point of a Curved 3D Surface}{ A schematic of tie-pointing a curved 3D surface within the epipolar geometry. As the lines-of-sight are tangent to different surfaces they may not insect upon it limiting the accuracy of the reconstruction \citep{Inhester2006astroph}.}

Characterising the CME front as an ellipse from both view points allows this localisation to be carried out for a series of epipolar planes or slices. Each slice corresponds to a quadrilateral in 3D. As the CME is known to be a curved object, an ellipse is inscribed in the quadrilateral such that it is it tangent to each side (line-of-sight), giving an optimal reconstruction of the CME front in that slice \citep{Byrne2010p74}. By repeating this for a number of slices the entire CME can be reconstructed. Then, applying this method to series of observations, allows the 3D reconstruction to be studied as a function of time. For a more detailed discussion of this technique see the PhD thesis by \citet{byrnethesis2010}.

\figuremacroSW{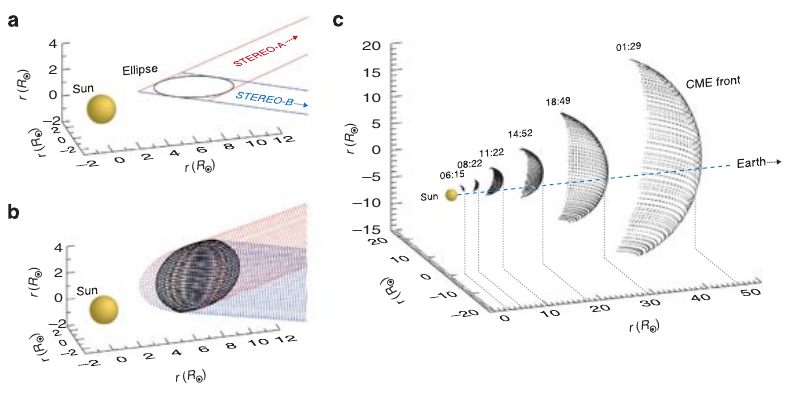}{Elliptical Tie-Pointing}{The elliptical tie-pointing technique shown for a Earth-directed CME which occurred on 12 December 2008. (a) For any given epipolar plane the CME localisation will give a quadrilateral in that plane. An ellipse can be inscribed within the quadrilateral such that is it tangent to each line-of-sight giving reconstruction of the CME front in the plane. (b) The full reconstruction is achieved by stacking multiple ellipses from a series of epipolar planes. (c) Applying this technique to a series of observations allows the CME to reconstruct as it propagates into the Heliosphere \citep{Byrne2010p74}.}{0.45}


 \section{Drag Modeling} 
In Section \ref{sec:cmepropagation} a number of possible drag models were shown and a generalised drag equation \eqref{eqn:gendrag} was written:
\begin{align}
	\label{eq:gendrag}
	\frac{dv_{cme}}{dr}=\alpha_{R}  R^{-\beta_{R}} \frac{1}{v_{cme}}\left (v_{sw} - v_{cme} \right )^{\delta}
\end{align}
where $v_{sw}=f(R,v_{sw}^{\text{ asymptotic}})$. There are a number of solar wind models which could be used in the drag model, for example the Parker solar wind as described in \sref{ssec:solwin} but the one chosen for this work was based on coronagraphic measurements of blobs in the solar wind \citep{Sheeley1997p472}:
\begin{align}
	v_{sw} = v_{sw}^{\text{ asymptotic}}  \sqrt{1 - e^{(R - 2.8)/8.1}}.
\end{align} It should be clear that this is a non-linear ordinary differential equation and has no analytical solutions. In oder to evaluate this equation, a numerical integration scheme was used, 4$^{\text{th}}$ order Runge-Kutta (RK4). This method is outlined below:
\begin{align}
	y' = f(t, y) \text{ where } y(t_{0})=y_{0} \qquad \left [y' = \frac{dy}{dt} \right ]
\end{align}
then the RK4 solution to this problem is:
\begin{align}
	y_{n+1} &= y_{n} + \frac{1}{6} \left (  k_{1} + 2k_{2} +2 k_{3} + k_{4}  \right ) \\
	t_{n+1} &= t_{n} + h
\end{align}
where $y_{n+1}$ is the RK4 approximation at $t_{n+1}$ and:
\begin{align}
	k_{1} & = h f(t_{n}, y_{n}), \\
	k_{2} & = h f(t_{n}+\frac{1}{2}h, y_{n}+\frac{1}{2}k_{1}), \\
	k_{3} & = h f(t_{n}+\frac{1}{2}h, y_{n}+\frac{1}{2}k_{2}), \\
	k_{4} & = h f(t_{n}+h, y_{n}+k_{3}). 
\end{align}
The $k$ are the changes in the $y$ values estimated from the slope at the begining ($k_{1}$), midpoint ($k_{2}$, $k_{2}$) and end ($k_{4}$), which are then averaged with greater weight given the estimates at the midpoint. The RK4 method is a 4$^{th}$ order method so the accumulated error is $O(h^{4})$ while the error per step is $O(h^{5})$. \figref{cmeaerodrag} shows the results of this numerical integration of \eqref{eq:gendrag} for a number of values of $\alpha$ and $\beta$ in the both the quadratic ($\delta=2$) and linear cases ($\delta=1$), as described in \sref{sec:cmepropagation}.

\figuremacroW{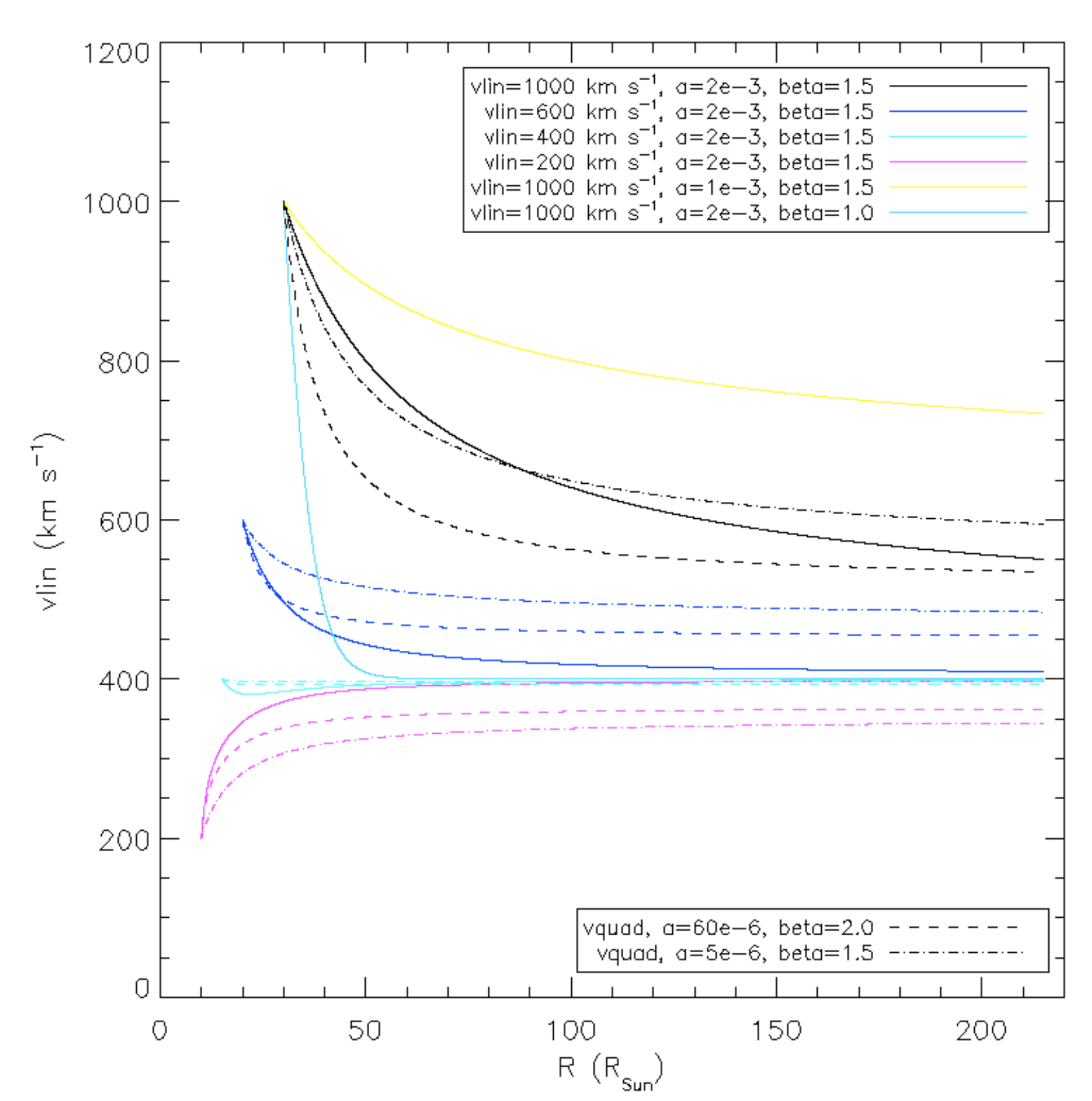}{CME Drag Simulation}{The results of numerically integrating the drag equation using input values for a number of possible scenarios.}{0.8}

\section{Bootstrap Technique}
\label{sec:bootst}
 The bootstrap technique \citep{efron1993} is part of a broader group of techniques knows as resampling methods. These methods allow a series of `numerical' experiments to be carried on data already gathered. The basic idea is to take original data, $F$, consisting of $n$ independent and identically distributed (i.i.d.) points and resample it to obtain $m$ sub samples and $m$ estimates of $\theta$, the parameter(s) of interest.
\begin{equation}
F \left ( X:x_{1}, x_{2},  \ldots, x_{n}\right), \theta \rightarrow
 \left\{ \begin{aligned}
 		 & F_{1}^{*}\left ( X:x_{i}, x_{i+1},  x_{j}\right) \theta_{1}^{*} \\
 		 & F_{2}^{*}\left ( X:x_{i}, x_{i+1},  x_{j}\right) \theta_{2}^{*} \\
 		& \vdots \\
 		& F_{m}^{*}\left ( X:x_{i}, x_{i+1},  x_{j}\right) \theta_{m}^{*}
       \end{aligned}
 \right.
\end{equation}
The simplest resampling method is known as the `jack-knife'. In this method, given a set of observations from some distribution $X=\left (x_{1},x_{2},x_{3},\ldots,x_{n} \right )$ each data point is in turn removed from the set and the measurements of interest the mean and standard deviation for example, are recalculated and stored. This allows $n-1$ resamples to be computed. The next resampling method is knows as the `m-out-of-n' or `deleted-D jackknife' where m out of n data points are removed this allows $\binom{n}{m}$ resamples to be carried out, however  caution must be taken to ensure $m$ is not made too large. The methods outlined above are generally applicable when the data is from a distribution, as the samples need to be i.i.d. However there are other methods which can be applied to non-i.i.d. data such as `cross validation'. In `cross-validation' the data is split up into roughly $k$ equal parts for each of the $k^{th}$ parts, fit all other parts and compute $\theta^{*}$. Or, if testing a model, the model can be fitted to the other $k-1$ parts and then the prediction compared to the $k^{th}$.

\subsection{The Bootstrap}
The bootstrap can be thought of as an extension to the jackknife though its mathematical foundation and insight into the data are very different. Instead of removing m data points we randomly sample, with replacement, from the original sample making a data set of the same size. The bootstrap technique is outlined below
\begin{enumerate}
	\item Collect $n$ data points to form $X={x_{1}, x_{2},\ldots,x_{n}}$
	\item Randomly resample with replacement $n$ from $X$ to form $X^{*}$ the bootstrap sample.
	\item Compute parameters of interest $\theta^{*}$
	\item Repeat steps 1 and 2 $N$ times
\end{enumerate}
N is chosen to be very large 1,000 to 10,000 depending on the requirements.

\subsection{Linear Regression}
In linear regression the problem is generally to find the correct values of some predictor to match the responses, for example the kinematic constants for a ball under parabolic motion. The data sets for linear regression consist of two parts: the predictor $c_{i}$ and the response $y_{i}$, and together they can be written
\begin{align}
	\mathbf{x}_{i} = (\mathbf{c}_{i}, y_{i}),
\end{align} 
where $\mathbf{c_{i}}$ is a $1 \times p$ vector $\mathbf{c}_{i} = (c_{i1}, c_{i2}, \ldots, c_{ip})$. For a given set $\mathbf{x}_{i}$, the conditional expectation given the predictor $\mathbf{c}_{i}$ is
\begin{align}
	\mu_{i} = E \left ( y_{i} | \mathbf{c}_{i} \right ) \qquad \left ( i = 1,2, \ldots ,n \right ).
\end{align} 
The term `linear' applies to the key assumption that the expectation value $\mu_{i}$ is a linear combination of the components of the predictor $\mathbf{c}_{i}$ or,
\begin{align}
	\mu_{i} = \mathbf{c}_{i} \boldsymbol{\beta} = \sum_{j=1}^{10} c_{ij} \beta_{j}.
\end{align}
$\boldsymbol{\beta}$ is the parameter vector or regression vector, $\boldsymbol{\beta} = (\beta_{1}, \beta_{2},\ldots,\beta_{p})^{T}$, and the goal of regression is to determine these values. Going back to a ball under parabolic motion, the regression parameter could have the form $\boldsymbol{\beta}=(h_{0}, v_{0},a_{0})$, and in this case the predictor would be $\mathbf{c}_{i} = (1, t, 1/2t^{2})$. This may seem to be nonlinear but again the linear term applies to the form of the expectation value (a linear combination of $\mathbf{c}_{i}$) and the fact that is it a quadratic function of $t$ doesn't matter.

Regression problems are usually expressed as: 
\begin{align}
	y_{i} = c_{i1}\beta_{1} +  c_{i2}\beta_{2} + \ldots c_{ip}\beta_{p}+\epsilon_{i} = \mathbf{c}_{i} \boldsymbol{\beta} +\epsilon_{i} \qquad \text{for } i = 1,2,\ldots,n 
\end{align} 
where $\epsilon_{i}$ are assumed to be errors from a random sample of a unknown distribution $F$ with a mean or expectation value of 0, 
\begin{align}
	F \rightarrow ( \epsilon_{1}, \epsilon_{2}, \ldots, \epsilon_{n}) = \boldsymbol{\epsilon} \text{ where }E_{F}(\epsilon)=0.
\end{align}
The goal is to estimate the parameter vector $\boldsymbol{\beta}$ from the pairs of $(\mathbf{c}_{1}, y_{1}), (\mathbf{c}_{1}, y_{1}), \ldots, (\mathbf{c}_{n}, y_{n})$. If we assume a trivial value of $\mathbf{b}$ for $\boldsymbol{\beta}$ the residual squared error (RSE) can be found
\begin{align}
	\text{RSE}(\mathbf{b}) = \sum_{i=1}^{n} (y_{i} - \mathbf{c}_{i}\mathbf{b})^{2}.
\end{align}
Finally the least-squares estimate of $\boldsymbol{\beta}$, or the value of $\mathbf{b}$ that minimises the RSE giving, $\boldsymbol{\beta}'$ is
\begin{align}
	\text{RSE}(\boldsymbol{\beta}') = \min_{\mathbf{b}} \left [ \sum_{i=1}^{n} (y_{i} - \mathbf{c}_{i}\mathbf{b})^{2} \right ].
\end{align}
An extremely usefully consequence of the linear regression formulation is that we can write it as a series of matrix operations. If we set a matrix $\mathbf{C}$ an $n \times p$ matrix with its $i^{th}$ row $\mathbf{c}_{i}$ (the design matrix), and let $\mathbf{y}$ be $(y_{i}, y_{i}, \ldots,y_{n})^{T}$, then we can write
\begin{align}
	\mathbf{C}^{T}\mathbf{C}\boldsymbol{\beta} = \mathbf{C}^{T}\mathbf{y},
\end{align}
which gives $\boldsymbol{\beta}'$ as
\begin{align}
	\boldsymbol{\beta}' = (\mathbf{C}^{T}\mathbf{C})^{-1}\mathbf{C}^{T}\mathbf{y}.
\end{align}

\subsection{Bootstrap Linear Regression}
The model for linear regression has two components: the parameter vector $\boldsymbol{\beta}$, and $F$ the probability distribution of the errors,
\begin{align}
	P = (\boldsymbol{\beta}, F)
\end{align}
The general bootstrap algorithm outlined above requires that we estimate $P$, we have least-squares estimates of $\boldsymbol{\beta}$, $\boldsymbol{\beta}'$ but no measure of $F$. If $\boldsymbol{\beta}$ was known we could directly calculate the errors (or residuals) $\epsilon_{i} = y_{i}-\mathbf{c}_{i}\boldsymbol{\beta}$ but we don't know $\boldsymbol{\beta}$. However we can get an approximate error distribution using $\boldsymbol{\beta}'$,
\begin{align}
	 \epsilon'_{i} = y_{i}-\mathbf{c}_{i}\boldsymbol{\beta}'.
\end{align}
The simplest approximation to $F$ is the empirical distribution of $\epsilon'_{i}$, so $F'$ is this distribution with a probability of $1/n$ and each residual.

With our estimate of the error distribution $F'$ we can now calculate $P'=(\boldsymbol{\beta}', F')$ or $P'\rightarrow \mathbf{x}^{*}$, which must be the same as $P \rightarrow  \mathbf{x}$. To generate the $\mathbf{x}^{*}$ we select a random sample of bootstrap error terms
\begin{align}
	F' \rightarrow ( \epsilon_{1}^{*},\epsilon_{2}^{*},\ldots,\epsilon_{n}^{*} ) = \boldsymbol{\epsilon}^{*},
\end{align}
where each is $\epsilon^{*}$ is a randomly drawn from $\epsilon'$. The bootstrap responses are then generated according to
\begin{align}
	y_{i}^{*} = \mathbf{c}_{i} \boldsymbol{\beta}' + \epsilon_{i}^{*}
\end{align}
notice that $\boldsymbol{\beta}'$ is a constant for all data points. Finally, the bootstrap least-squares estimate $\boldsymbol{\beta}'^{*}$ is the $\mathbf{b}$ that minimises of the RSE of the bootstrap data,
\begin{align}
	\sum_{i=1}^{n} ( y_{i}^{*} - \mathbf{c}_{i}\boldsymbol{\beta}' )
^{*}  = \min_{b} \left [ \sum_{i=1}^{n} ( y_{i}^{*} - \mathbf{c}_{i}\mathbf{b}) \right ].
\end{align}
Errors and confidence intervals may be constructed from the bootstrap distributions of the of $\boldsymbol{\beta}^{*}$.

It should be clear that there are two possible ways to use the bootstrap, with linear regression the $\mathbf{x}_{i} = (\mathbf{c}_{i}, y_{i})$ pair in which case the bootstrap sample is:
\begin{align}
	x^{*} = \left \{   (\mathbf{c}_{i_{1}}, y_{i_{1}}), (\mathbf{c}_{i_{2}}, y_{i_{2}}), \ldots, (\mathbf{c}_{i_{n}}, y_{i_{n}})  \right \}
\end{align} or bootstrap the residuals in which case the bootstrap sample is:
\begin{align}
	x^{*} = \left \{ (\mathbf{c}_{1},\mathbf{c}_{1}\boldsymbol{\beta}'+\epsilon'_{i_{1}} ), (\mathbf{c}_{1},\mathbf{c}_{1}\boldsymbol{\beta}'+\epsilon'_{i_{2}} ), \ldots, (\mathbf{c}_{1},\mathbf{c}_{1}\boldsymbol{\beta}'+\epsilon'_{i_{n}} )    \right \}
\end{align}
where $i_{1}, i_{2}, \ldots, i_{n}$ are random samples of the integers from 1 to $n$. The choice of which to use depends on the problem at hand, but in the limit as $n$, the number of samples grows large, both tend to the  same result.

\subsection{General Bootstrap fitting}
Bootstrapping the residuals can also be applied in a broader manner in the following way:
\begin{enumerate}
	\item Make an initial fit of a model $\mathbf{y}=f(\mathbf{x}, \boldsymbol{\theta})$ to the data $y_{i}$ (this need not be via least-squares), and obtain fit parameters $\theta$
	\item Construct an empirical residual distribution $\epsilon_{i}'  = y_{i} - f(\mathbf{x}, \boldsymbol{\theta}')$
	\item Construct abootstrap residual sample $\epsilon^{*}$ by randomly sampling from the equally weighted empirical distribution $\epsilon'$.
	\item Make a bootstrap data set $y_{i}^{*} = f(\mathbf{x}, \boldsymbol{\theta}') + \epsilon^{*}$
	\item Fit a model to the bootstrap data $\mathbf{y}^{*}=f(\mathbf{x}, \boldsymbol{\theta}^{*})$, and obtain bootstrap estimate of $\theta$, $\theta^{*}$
	\item Repeat steps 3-5 N times  
\end{enumerate}
Where the fitting of the model to the data can be carried out by any number of means. With computational power of modern computers N is often chose to be very large, e.g.  10,000\rng100,000.




\chapter{CME Trajectories in Three Dimensions} 


    \graphicspath{/}


\hrule height 1mm
\vspace{0.5mm}
\hrule height 0.4mm
\noindent
\\
{\it
Before STEREO, single view-point observations obscured the true trajectories of CMEs. The STEREO mission provides two perspective views of CMEs in the inner Heliosphere. This, for the first time, enables the reconstruction of the 3D trajectories of CMEs in a range of heliocentric elongations ($\sim$2\rng88 degrees). In this work, a number of CMEs, which were simultaneously observed by COR1 and COR2 from both STEREO satellites, were selected. These observations were then used to derive the CMEs trajectories in 3D out to $\sim$15\,R$_{\odot}$. The reconstructions using COR1 and 2 observations support a radial propagation model. Assuming pseudo-radial propagation at large distances from the Sun, the CMEs 3D positions were extrapolated into the HI field-of-view. This enables the 3D trajectories of CMEs to be reconstructed from $\sim$1.4\rng240\,R$_{\odot}$. It was demonstrated that CMEs undergo acceleration in the inner Heliosphere; CMEs slower that the solar wind are accelerated, while CMEs faster than the solar wind are decelerated, both tending to the solar wind velocity. This effect will have a significant impact on the forecasting of CME arrival times at 1\,AU and hence space weather. This chapter is based on work published in \citeauthor*{Maloney2009p149}, Solar Physics, \citeyear{Maloney2009p149}.} \\
\hrule height 0.4mm
\vspace{0.5mm}
\hrule height 1mm

\newpage

\section{Introduction}
\label{S-intro}
	
Prior to the launch of STEREO, CMEs could only be routinely observed in white light up to 32\,R$_{\odot}$ in the plane-of-sky using the SOHO LASCO (\sref{sec:lasco}), and rarely to Earth using the Solar Mass Ejection Imager (SMEI) instrument \citep{Jackson2004p177,Howard2007p610}. CME related shocks are observed in radio data, but rely on density models to relate the observations to heights in the corona, and provide no information on the direction of propagation. Some work has been done using radio observations from two observatories to reconstruct 3D positions from radio triangulation \citep{Reiner1998p1923,Reiner2007pa2}. Another possibility to track CMEs over large distances is during fortunate quadrature observations such as SOHO-Sun-Ulysses \citep{Bemporad2003p567}. CMEs have also been tracked out into the Heliosphere when they interact with various spacecraft such as Wind, ACE, Ulysses, Cassini, Voyager 1 and 2 \citep[][and references therein]{Lario2005p09S11,Richardson2005pL03S03,Gopalswamy2005pA09S15}.

The identification of events over interplanetary time/distance scales when these are not available requires the comparison of coronagraph observations with {\it in situ} data. The relationship between images and the {\it in situ} data is complex: coronagraphs image solar radiation which has been Thomson scattered by electrons in the CME and in-situ measurements provide the actual densities and magnetic fields of a track through the CME. All of these problems are compounded when multiple events occur in a short time frame. 

\figuremacroH{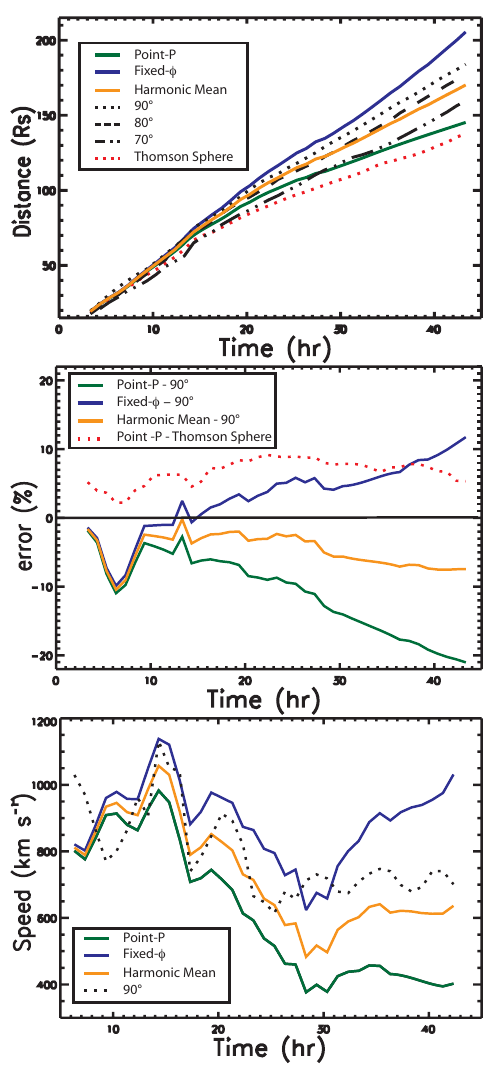}{Kinematics Derived form various Methods from Synthetic SECCHI Observations}{Position (top), error (center) and speed (bottom) of a simulated CME and as derived from the synthetic SECCHI images from a 3D MHD simulation with the different methods \citep{Lugaz2009p479}.}{0.6}

In an attempt to overcome some of these problems, statistical studies have been carried out to try to extract physics from the observations, however these types of study can miss vital event-to-event variations \citep{Gopalswamy2005p2289}. In a recent paper \citet{Owens2004p661}, compared a number of empirical models to predict the arrival time of CMEs at 1\,AU. They cited the main sources of error as projection effects, assuming that solar wind conditions are the same for all CMEs, and the determination of the point of intersection of the CME with the spacecraft with respect to {\it in situ} measurements.

Coronagraph observations are subject to projection effects (\sref{sec:projeff}) which are one of the leading sources of uncertainty in determining kinematics \citep{Howard2008p1104}. Other methods of deriving height from observations such as the ``point-P'', ``fixed-$\phi$'', and ``harmonic mean'' have been used, but a recently study of these methods on simulated data shows the problems associated with them. \figref{compmeth} shows the height, error in the height, and velocities derived from synthetic observations from the various methods, from which it is clear none of them accurately produce the CME kinematics.

Various techniques have been developed to overcome some of these limitations such as tomographic techniques which use the fact that the observations are taken in a rotating reference frame to reconstruct the 3D nature of the corona (at the loss of time resolution; \cite{Frazin2000p1026}). Another possibility is the polarimetric technique of \citet{Moran2004p66}. Other methods that use {\it a priori} knowledge of the system can also be use to infer the 3D position, such as pre-assuming the CME geometry. This type of analysis, known as forward modelling, has been applied both to single view point and stereoscopic observations. In the later case, the model parameters are modified in either a semi-automatic or manual fashion to simultaneously fit the observations from both perspectives \citep{Thernisien2009p111,Boursier2009p131}. 

Stereoscopy has been used in astrophysics for a long time, with one of the earliest applications being the determination of the distances to near-by stars using parallax. Using stereoscopy techniques, the reconstruction of the 3D coordinates of features identified in images from two vantage points is a well-defined linear problem \citep{Inhester2006astroph}. The biggest challenge in using this method is the identification and matching of features in the stereo image pairs. Stereoscopy has been successfully applied to on-disk features such as coronal loops \citep{Aschwanden2008p827} and extended to prominences and CMEs \citep{Liewer2009p57,Mierla2008p385,Temmer2008p95,Thompson2011p1138,Srivastava2009p213}. Also, assuming CMEs travel at constant velocity at large distances from the Sun, the trajectory can be derived by inverting the equation for the fixed-$\phi$ case \citep{Sheeley2008p853}. For a recent review of the methods which have been applied to coronagraph observations see \cite{Mierla2010p203}.

In order to better understand the acceleration and propagation of CMEs the true 3D trajectory needs to be known to facilitate comparison between observations and theory. From a space weather perspective, knowing the 3D trajectory will lead directly to better predictions and also indirectly through a better understanding of CME acceleration and propagation.

In this chapter the trajectories four CMEs are reconstructed in 3D. The observations are presented in \sref{S-Obsred}. The true 3D trajectories of the four events are presented  \sref{S-results}. \sref{S-disc} contains a discussion of the results, the conclusions drawn, and future work.

\section{Observations}
 \label{S-Obsred}

\figuremacroFP{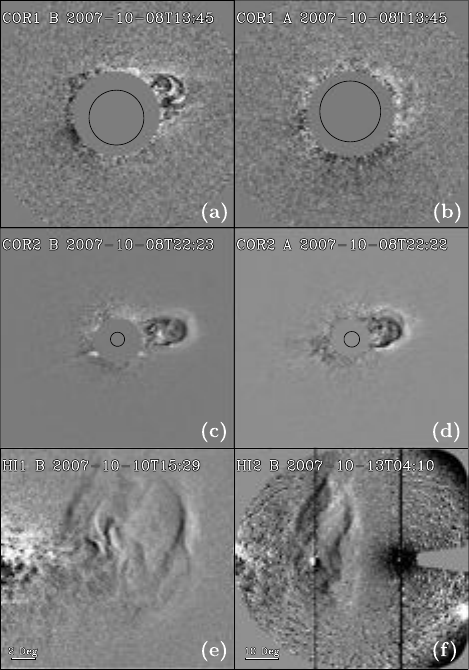}{Sample Observations from the 2007 October 08--13 Event}{Sample observations from 2007 October 08-13 event from: (a) COR1 Behind 13:45\,UT, (b) COR1 Ahead 13:45\,UT, (c) COR2 Behind 22:23\,UT, (d) COR2 Ahead 22:22\,UT, (e) HI1 Behind 2007-10-10 15:29\,UT, (f) HI2 Behind 2007-10-13 04:10\,UT.}{0.8}

\figuremacroFP{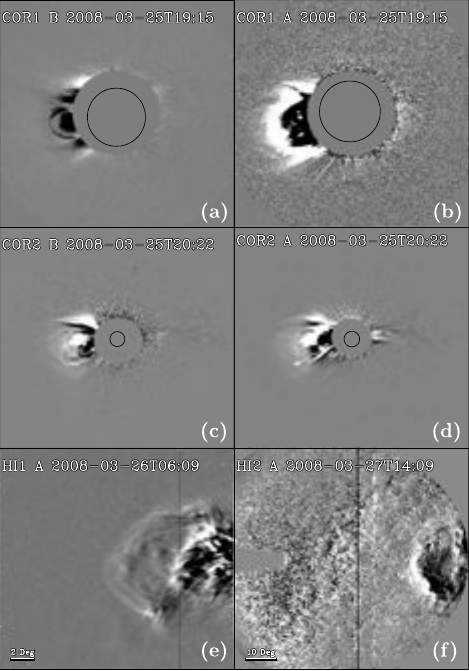}{Sample Observations from the 2008 March 25-28 Event}{Sample observations from 2008 March 25-28 event from: (a) COR1 Behind 19:15, (b) COR1 Ahead 19:15\,UT, (c) COR2 Behind 22:22\,UT, (d) COR2 Ahead 22:22\,UT, (e) HI1 Ahead 2008-03-26 06:09\,UT, (f) HI2 Ahead 2008-03-27 14:09\,UT.}{0.8}

\figuremacroFP{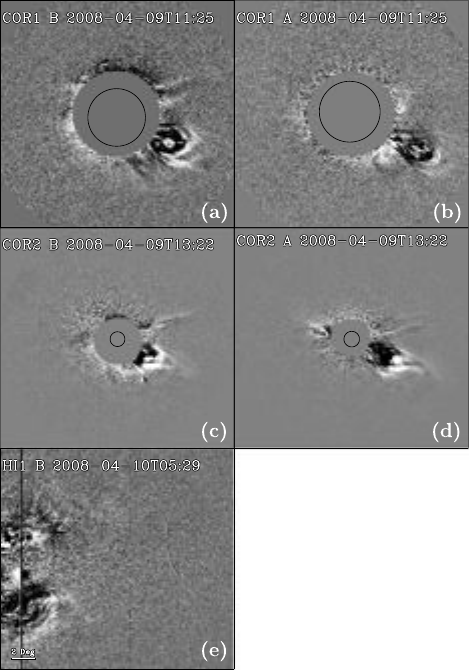}{Sample Observations from the 2008 April 09--10 Event}{Sample observations from 2008 April 09-10 event from: (a) COR1 Behind 11:25\,UT, (b) COR1 Ahead 11:25\,UT, (c) COR2 Behind 13:22, (d) COR2 Ahead 13:22\,UT, (e) HI1 Behind 2008-04-10 05:29\,UT.}{0.8}

\figuremacroFP{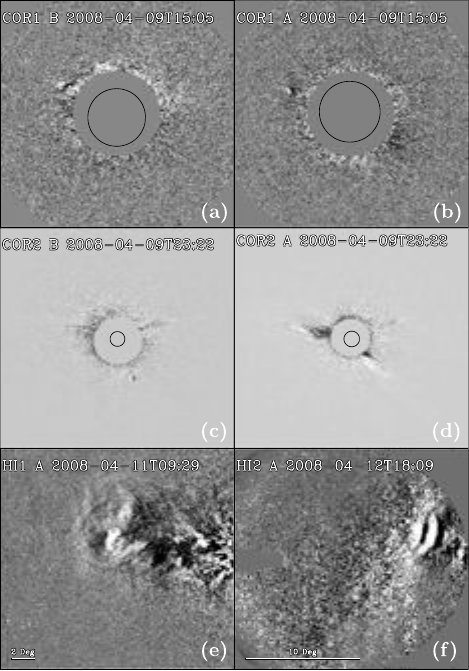}{Sample Observations form the 2008 April 09-13 Event}{Sample observations from 2008 April 09-13 event from: (a) COR1 Behind 15:05\,UT, (b) COR1 Ahead 15:05\,UT, (c) COR2 Behind 23:22\,UT, (d) COR2 Ahead 23:22\,UT, (e) HI1 Behind 2008-04-11 09:29\,UT, (f) HI2 Behind 2008-04-12 18:09\,UT.}{0.8}
	
The dates of the observations are 2007 November 18,  2008 March 25, 2008 April 9 and finally 2008 April 10. The CME catalogue hosted at the Coordinated Data Analysis Workshop (CDAW\footnote[1]{\href{http://cdaw.gsfc.nasa.gov/CME\_list}{http://cdaw.gsfc.nasa.gov/CME\_list}}; \cite{Gopalswamy2009p295}) provided heights derived from SOHO/LASCO observations which were deprojected and used to corroborate the accuracy of the 3D reconstruction. Figures \ref{2007oct8-13obs}, \ref{2008march25-28obs}, \ref{2008april9-10obs}, \ref{2008april9-13obs} show the reduced and processed observation from COR1, COR2, HI1 and HI2.

\subsection{Uncertainties in the Reconstructed Heights}
 \label{SS-Errors}

In order to compare our results to theory we need to have an estimate of the uncertainties in our reconstruction. We use a statistical approach to find the spread of the data which we can then use as an estimate of the uncertainty. The 3D height data is de-trended with a third order polynomial and the moments of the distribution of the residuals are then calculated. The standard deviation ($\sigma$) of the distribution is taken as a good measure of the uncertainty. As a check on the statistics, the number of data points greater than one $\sigma$ from the mean were calculated and compared to the 31.8\% which we expect if the data is normally distributed. If the number of points in the individual fields-of-view are too small this approach will not work. In \figref{errplot} we show data from the 2007 October 09 event for COR2 and HI2 as examples. The top plots show the height-time data as well as the best fit line using a 3\textsuperscript{rd} order polynomial. The bottom plots show the de-trended data, mean, and one standard deviation ($\sigma$) either side of the mean. The value of the 1$\sigma$ and the number of points greater than 1$\sigma$ from the mean are indicated on the plots.

\figuremacro{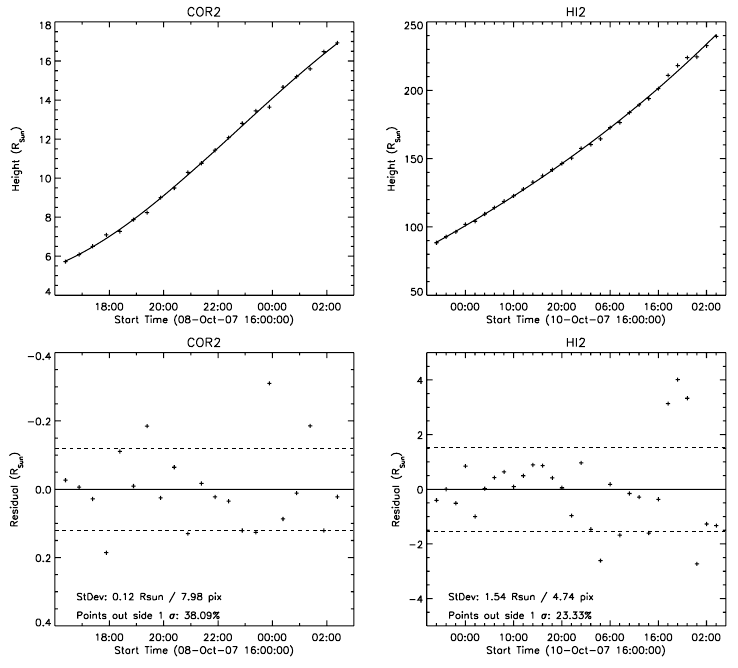}{3D Trajectory Error Analysis}{Height-time plots for the 2007 October 09 event in the COR2 and HI2 FOV, the line is the best fit to the data with a 3\textsuperscript{rd} order polynomial (top row). The de-trended data, standard deviation $\sigma$ and the percentage of points out side of a one $\sigma$ error (bottom row). The dashed lines are plotted at $\pm$\,1\,$\sigma$ from the mean \citep{Maloney2009p149}.}

\section{Results}
\label{S-results}
The observed apex of each CME was identified using point and click methods on the data. These points were then used to calculate the 3D coordinates and thereby plot the 3D trajectories. From these we were able to calculate the CMEs launch angles with respect to the ecliptic plane, and in the ecliptic plane with respect to the Sun-Earth line. To highlight the advantages these new data give, we estimated the velocity of the CME as viewed from A and B using a plane-of-sky assumption, as well as from the 3D reconstruction. These estimates were calculated using the first and last data point in the COR2 FOV and $v=\Delta\,h/\Delta\,t$ where $h$ and $t$ are the heights and times of the first and last observations. We also estimated the velocities on larger time/distance scales by using the first COR2 data point and the last data point (from HI1 or HI2) in the same manner.

Using the derived launch angles observations from other spacecraft can be de-projected and compared to the heights derived from the 3D reconstruction. \figref{lascodproj} shows one such comparison of the deprojected  heights derived from LASCO C2 and C3 taken from the CDAW catalogue and STEREO COR2 and HI1. Within errors (approximately the symbol size) the reconstructed and de-projected heights agree in the range of the range 5\rng18\,$R_{\odot}$. This demonstrates the CME reconstruction method is suitable and gives accurate CME height and launch angles.

\figuremacroW{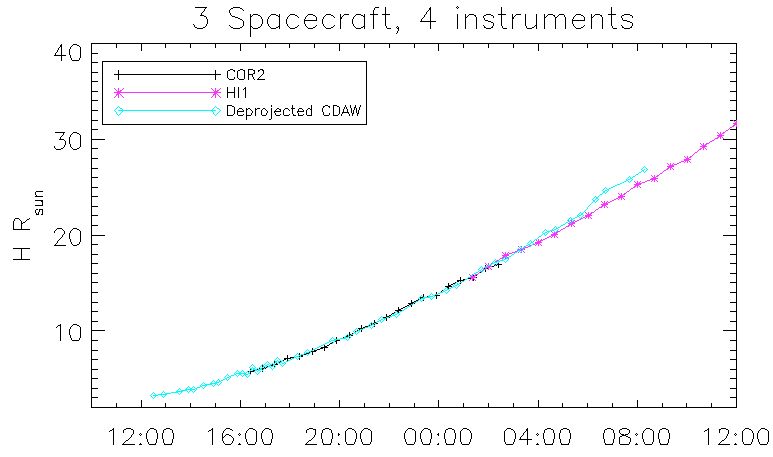}{STEREO 3D CME Height and LASCO CDAW De-projected Height}{STEREO 3D CME Height from COR2 and HI1 and LASCO CDAW deprojected height using the propagation angle derived (from the 3D reconstruction) for the CME. The x-axis is the time of day in hours.}{0.7}

\subsection{Event 1: 2007 October 08--13}
\label{SS-event1}
This event was first observed in COR1 B on 2007 October 8 at 10:15\,UT off the western limb, but was extremely faint in images from both spacecraft so the COR1 data were not used for the 3D reconstruction. The first data point used for the 3D reconstruction was observed at 15:52\,UT on the same day by COR2 A and B instruments, the last data point was observed at 04:01\,UT on 2007 October 13 by the HI 2 B instrument. From the 3D data, the launch angle in the $x$-$y$ plane was found to be $56^{\circ}$ (where $0^{\circ}$ corresponds to an Earth-directed CME, and $90^{\circ}$ to a CME off the western limb), thus this was a front side event (see \figref{trajplot20071009}). The CME also had an out of ecliptic component of $11^{\circ}$ (where $0^{\circ}$ corresponds to in-ecliptic while $90^{\circ}$ corresponds to Solar North). The velocity seen from A was found to be 190\,km\,s$^{-1}$, from B 157\,km\,s$^{-1}$, and from the 3D reconstruction 216\,km\,s$^{-1}$. The mean velocity over the entire event was found to be 430\,km\,s$^{-1}$. The CMEs true height was tracked from 5.7\rng239.4 R$_{\odot}$ (\figref{errplot}, top right) with an estimated error of 0.15 R$_{\odot}$, 0.25 R$_{\odot}$ (\figref{errplot}, bottom left) and 2\,R$_{\odot}$ (\figref{errplot}, bottom right) in COR2, HI1 and HI2 respectively. The CMEs morphology evolves from a nearly circular loop in COR1 B and COR2 A and B, to an elliptical structure with much larger extent in the vertical than the horizontal direction in HI1 (see \figref{2007oct8-13obs}). In HI2, the front evolves from nearly linear to a concave front where the wings/flank of the CME are ahead of what was the CME apex (see \figref{2007oct8-13obs}). There is some indication of a multiple loop CME in COR2 A, HI1 B, and HI 2 B (see \figref{2007oct8-13obs}).	
	
\figuremacroH{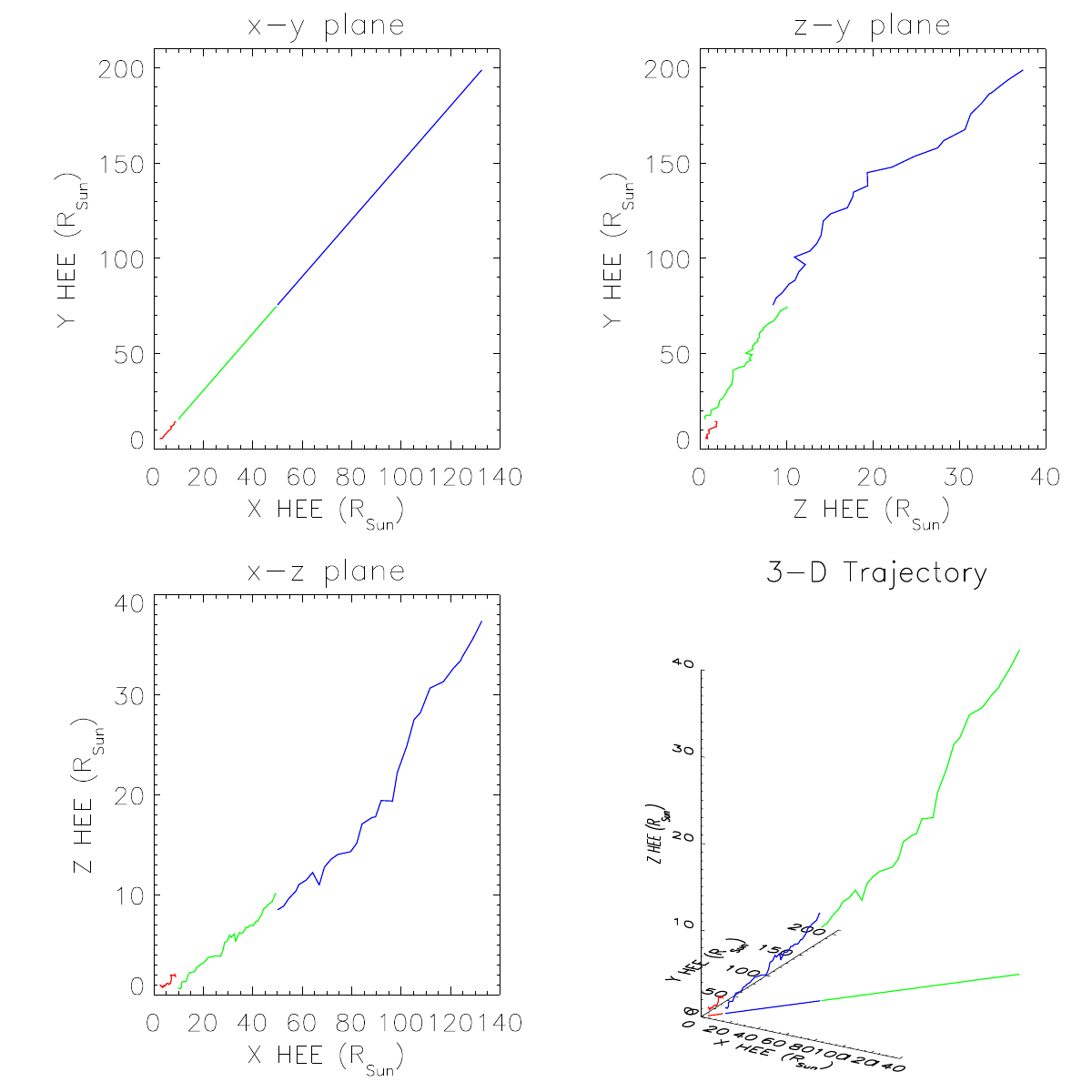}{3D CME Trajectory for 2007 October 08--13 Event}{CME trajectory for the 2007 October 08-13 event (\figref{2007oct8-13obs}): Cut in the $x$-$y$ plane (top left), cut in the $y$-$z$ plane (top right), cut in the $x$-$z$ plane (bottom left), 3D trajectory (bottom right) from \cite{Maloney2009p149}.}{0.60}	
		
\subsection{Event 2: 2008 March 25-27}
\label{SS-event2}
The second event was first observed on 2008 March 25 at 18:55\,UT in both COR1 A and B off the eastern limb. The last data point used was observed at 02:09\,UT on 2008 March 27. The CME was well observed in A and B COR1 but was somewhat faint and diffuse in COR2. The CME launch angle in the $x$-$y$ plane was found to be $-74^{\circ}$ which corresponds to a front-side event, while the angle with respect to the ecliptic plane was $-22^{\circ}$ (see \figref{trajplot20080325}). The velocity from A was found to be 863\,km\,s$^{-1}$, from B 1009\,km\,s$^{-1}$, and from the 3D reconstruction 1020\,km\,s$^{-1}$. The velocity over the entire event was found to be 494\,km\,s$^{-1}$. The CMEs height was tracked from 1.9\rng139.7\,R$_{\odot}$ with an estimated error of 0.3\,R$_{\odot}$ in HI1, and 0.6\,R$_{\odot}$ in HI2. There were not enough data points in the COR1/2 FOV to estimate the error. The morphologies from both perspectives are very similar a nearly circular profile consisting of one faint loop which propagates all the way out into the COR 2 FOV (see \figref{2008march25-28obs}). In the HI1 FOV there appears to be a second loop-like feature behind the main front which is expanding at a similar rate as the main front. The CME is tracked until about halfway across the HI2 FOV before it becomes too difficult to identify. The second loop-like structure could be associated with a second flux rope, or perhaps prominence material.

\figuremacroH{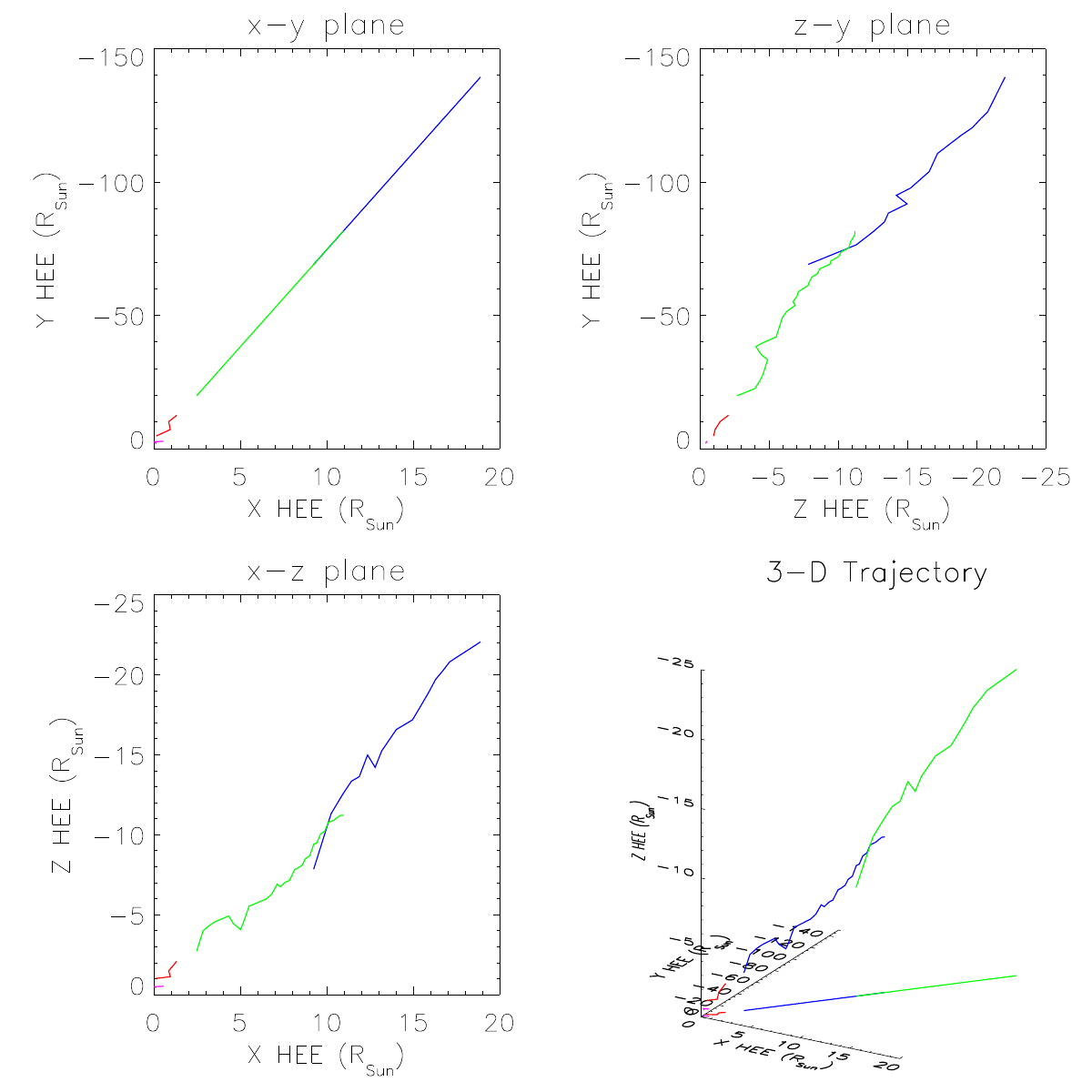}{3D CME Trajectory for 2008 March 25--27 Event}{CME trajectory for the 2008 March 25-27 event (\figref{2008march25-28obs}): cut in the $x$-$y$ plane (top left), cut in the $y$-$z$ plane (top right), cut in the $x$-$z$ plane (bottom left), 3D trajectory (bottom right) from \cite{Maloney2009p149}.}{0.60}

\subsection{Event 3: 2008 April 09--10}
\label{SS-event3}

This event was first observed on 2008 April 08 in both COR1 A and B at 10:05 off the western limb toward the South. The last data point was observed at 15:29\,UT on the 2008 April 10 in HI1 B. The CME is well observed in A and B images from both COR1 and COR2. Due to the large out-of-ecliptic angle the CME does not pass through the FOV of HI2 B. The CME launch angle in the $x$-$y$ plane was found to be $-71^{\circ}$ which corresponds to a front side event, while the out-of- ecliptic plane angle was $-60^{\circ}$. The velocity from A was found to be 504\,km\,s$^{-1}$, from B 513\,km\,s$^{-1}$, and from the 3D reconstruction 542\,km\,s$^{-1}$. The velocity over the entire event was found to be 354\,km\,s$^{-1}$.  The CMEs height was tracked from 1.9\rng50.4\,R$_{\odot}$ with an estimated error of 0.5\,R$_{\odot}$ in HI1. There were not enough data points in the COR1/2 to estimate the errors. The CME appears to be narrower in COR1 A than in COR1 B. When the CME enters the COR2 FOV it appears the same size in A and B (see Figure \ref{2008april9-10obs}). The CME has a more elliptical shape in HI1 and continues to propagate at a steep angle. There is no obvious trend of the trajectory back towards the ecliptic plane, which one may have expected (see \figref{trajplot20080409}).

\figuremacroH{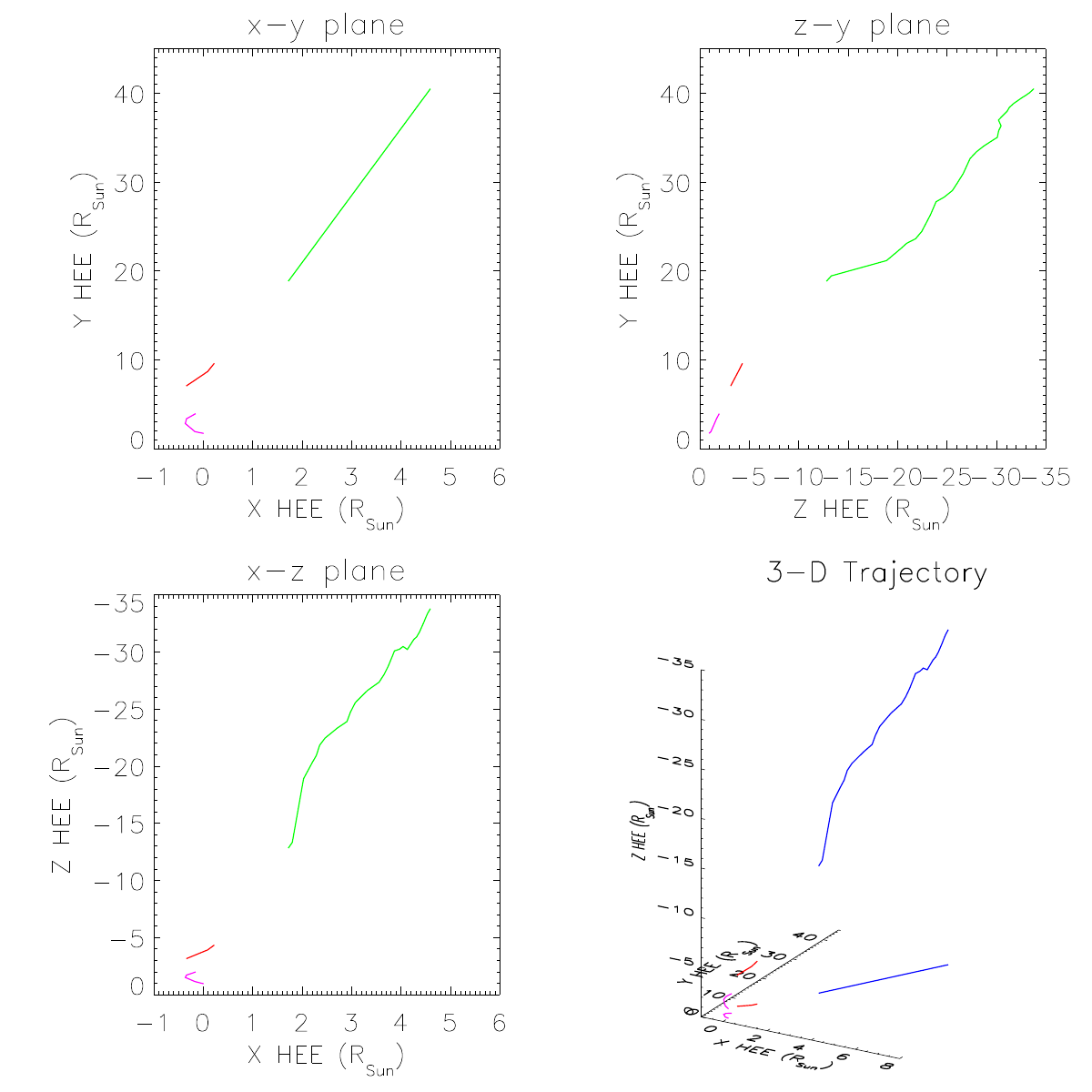}{3D CME Trajectory for 2008 April 09--10 Event}{CME trajectory for the 2008 April 09--10 event (\figref{2008april9-10obs}): cut in the $x$-$y$ plane (top left), cut in the $y$-$z$ plane (top right), cut in the $x$-$z$ plane (bottom left), 3D trajectory (bottom right) from \cite{Maloney2009p149}.}{0.60}
		
\subsection{Event 4: 2008 April 09--13}
\label{SS-event4}

This event was first observed off the eastern limb on the 2007 October 9 at 15:05\,UT by both COR1 instruments. The last data point was observed by HI2 A  at 00:09\,UT on 2008 April 13. The CME was extremely faint in both A and B COR1, but a small number of data points were able to be recorded. The CME was reasonably clear in COR2 A, but very faint in B. As a result the CME front was not visible and for this event a dark linear feature was tracked instead. The CME launch angle in the $x$-$y$ plane was found to be $-51^{\circ}$ which corresponds to a front-side event, while the out-of-ecliptic angle was $16^{\circ}$ (see \figref{trajplot20080410}). The velocity from A was found to be 109\,km\,s$^{-1}$, from B 217\,km\,s$^{-1}$, and from the 3D reconstruction 189\,km\,s$^{-1}$. The velocity over the entire event was found to be 338\,km\,s$^{-1}$. The CMEs height was tracked from 2.2\rng128.0\,R$_{\odot}$ with an estimated error of 0.8 R\,$_{\odot}$ in HI1 and 1.1\,R$_{\odot}$ in HI2 (there were not enough data points in the COR1/2 to estimate the errors). The CME appears very narrow in COR2 B images, and could be a loop structure with its axis nearly perpendicular to the plane of sky. The CME has an elliptical shape in HI1 with some material in the centre (see \figref{2008april9-13obs}). This shape is held into HI2 where the CME is lost as it crosses in front of the Milk Way, where the background subtraction technique does not work as well. 

\figuremacroH{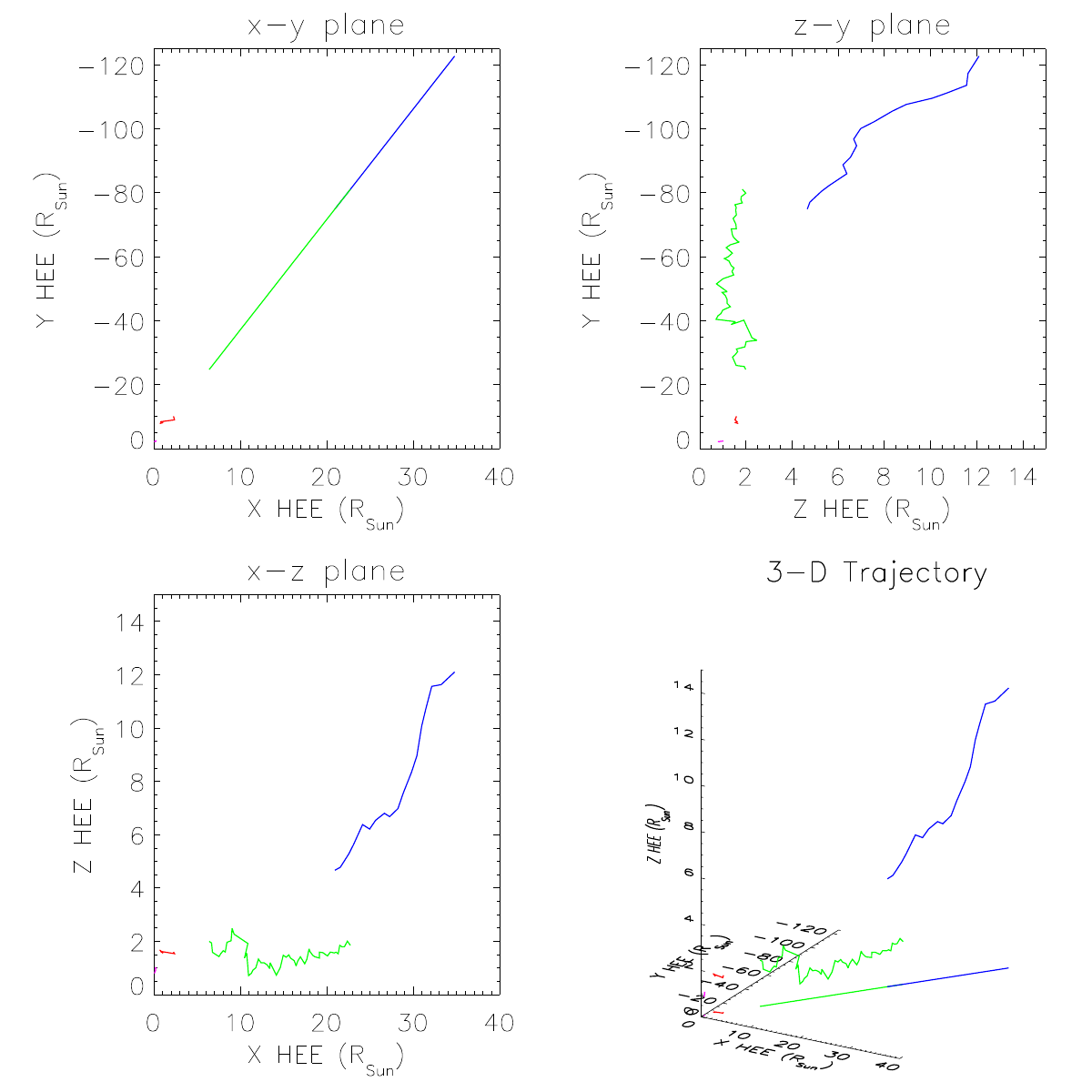}{3D CME Trajectory for 2008 April 09--13 Event}{CME trajectory for the 2008 April 09--13 event (\figref{2008april9-13obs}): cut in the $x$-$y$ plane (top left), cut in the $y$-$z$ plane (top right), cut in the $x$-$z$ plane (bottom left), 3D trajectory (bottom right).}{0.60}

\section{Discussion and Conclusions}
	\label{S-disc}

We have shown it is possible to reconstruct the 3D trajectories of CMEs in the inner Heliosphere using STEREO/SECCHI observations. These 3D reconstructions give the true position of the CMEs studied in the range between $\sim$2--240\,R$_{\odot}$ as a function of time. The COR1/2 data indicates that a radial propagation model is appropriate once the CME is above a few solar radii. This is in agreement with what we expect, as the forces acting on CMEs fall off steeply with increasing radius (especially forces which would act perpendicular to the propagation direction in the ecliptic). We have also shown that the velocities that would be derived from either of the spacecraft alone can be very different to that of the 3D velocity. Also we have shown that CMEs do undergo acceleration in the Heliosphere as the velocities in the COR2 FOV can be very different to those calculated using the first and last data points.
The reconstructions can also be of significant value for space weather forecasting. Using an empirical formula such as used in \cite{Gopalswamy2005p2289} we can try and estimate the arrival time of these CMEs at 1\,AU. The arrival time of a CME at one 1\,AU is given by:
\begin{equation}
\tau = \frac{-U+\sqrt{U^{2}-2ad}}{a}
\end{equation}
where
\begin{equation}
a = \frac{V(1\,\mbox{AU})-U}{\tau}
\end{equation} with $V(1\,\mbox{AU})$ the velocity at 1\,AU, $U$ the velocity in low corona, $\tau$ the time between the two measurements, and $d$ in this case 1\,AU. Now as we have no in-situ data at 1\,AU we use the fact that in all the events the CMEs have tended towards a typical solar wind value. This should be a good proxy for the velocity at 1\,AU. Using the 3D velocity and the velocity that differs the most, we can make estimates of the CME arrival time $\tau$. For the first event $\tau$ was calculated to be 5.39 days using the velocity derived from the B spacecraft, and 5.11 day from the 3D velocity. The difference between the two calculated arrival times for the events are $\Delta\tau=$0.28\,days, $\Delta\tau=$0.39\,days, $\Delta\tau=$0.17\,days, and $\Delta\tau=$0.10\,days for the four events respectively. Unfortunately, as none of the CMEs were detected in-situ near 1\,AU we can not tell if these differences correspond to improvements or not.  Another major advantage is the ability to predict if the nose or flank of the CME will interact with Earth --this was identified as one of the major sources of error in \cite{Owens2004p661}. The geo-effectiveness will be affected by which part of the CME interacts with the Earth, and also the arrival times will be affected as the CME flank should take longer to arrive at Earth.

There are a number of outstanding issues with the reconstructions. Some of these are the results of the reconstruction method, while others are related to the data used or its interpretation. The stereoscopy method requires that we can identify the same feature in both images and this is extremely difficult in some cases as the CME is very faint or has a very different appearance from each spacecraft. For example, COR1 observations for the first event were so faint the CME could not be identified in the A images (se Figure \ref{2007oct8-13obs}), and in both the third and fourth events the identification of the same feature was extremely difficult as the CME has a different appearance in each image (see Figures \ref{2008april9-10obs} and \ref{2008april9-13obs}). This is evident in the reconstructions as they are less contiguous, and there are large changes in the CME trajectory which do not make physical sense as there is no known source for such large forces out in the Heliosphere.

The different sensitivities and light rejection levels of the different instruments mean the CME may appear bigger or smaller just due to instrumental effects. It is believed that the discontinuities of the first event are due to this type of effect. Also as the front of the CME imaged is a line-of-sight integration through the CME (flux rope or magnetic bubble), the front seen from each spacecraft will be different. This is especially important in the HI FOV as CMEs can expand up to large sizes and we only have one point of view, but will also have an effect in COR1/2. This means the positions, and thus heights derived, will always be an upper limit on the actual values. If a model shape for CMEs is used  -- such as an ellipse or circle -- we can estimate the size of this effect. A CME which passes one of the spacecraft taking {\it in situ} measurements would allow a comparison of the reconstructed CME path and the actual CME path and provide a welcome test of the methods.
  
A number of interesting observations arise from these data, such as the possible multi-loop nature of two of the CMEs, the complex structure seen at the rear of the CMEs (especially in HI1), and finally the distortion of the CME front in Event 1. Firstly events 1 and 2 show possible multi-loop structures (see Figures \ref{2007oct8-13obs} and \ref{2008march25-28obs}): these could be attributed to a second flux rope or possibly due to prominence material. These features are only visible because of the unique features of the HI1 and HI2 instruments. Secondly, most of the events show complex structure at the rear of the CME. This is very apparent in Figures \ref{2007oct8-13obs}, \ref{2008april9-10obs} and \ref{2008april9-13obs}. This type of structure could be interpreted as the rear part of a flux rope if the flux rope axis was into, or out of, the image plane. Finally, the distortion of the CME front in event 1 as it progresses through the Heliosphere could be caused by a number of processes. This may be a Thomson scattering effect i.e., we are seeing different parts of the CME from different planes, and it only appears as though the CME front is distorted with the flanks ahead of the nose. The other possibility is that as the CME has expanded to extremely large size at this point and the nose and flanks are experiencing very different ambient solar wind conditions. As such, the CME flank may be in the high speed solar wind region (hence accelerated by aerodynamic drag), or the nose may be in a relative density-enhanced region due to the heliospheric current sheet and experiencing more drag than the CME flanks. This effect has previously suggested to occur and is known as `pancaking'.

These reconstructions are the first steps towards detailed study of CME kinematics in the heiloshpere. Using these trajectories, we can derive the full 3D CME kinematics which will not be subject to projection effects. When we combine these with CME mass estimates we will have all the information needed to study the forces acting on CMEs from $\sim$2\,$R_{\odot}$ to beyond 1\,AU. We plan to compare the derived forces with those predicted by the flux rope, `snow plough', and aerodynamic drag models \citep{Cargill2004p135,Vrsnak2006p431,Vrsnak2002p1019}. Having the 3D information will finally allow for a full comparison between the various theories and the data. In doing so we will be able to gain insight into some the interesting observations discussed here.




\chapter{CME Kinematics in Three Dimensions} 


    \graphicspath{/}
    \graphicspath{{6/figures/EPS/}{6/figures/}}


\hrule height 1mm
\vspace{0.5mm}
\hrule height 0.4mm
\noindent
\\
{\it 
The forces governing CME evolution during both their early acceleration, and later propagation are still unclear, although plane-of-sky coronagraph measurements have provided some insight into their kinematics near the Sun ($<$32\,R$_\odot$). The dual perspectives of the STEREO spacecraft are used to derive the 3D kinematics of CMEs over a range of heliocentric distances ($\sim$2\rng250\,$R_{\odot}$). Evidence for solar wind (SW) drag forces acting in interplanetary space, with a fast CME decelerated, and a slow CME accelerated, towards typical SW velocities was found. It was also found that the fast CME showed a linear dependence on the velocity difference between the CME and the SW, while the slow CME showed a quadratic dependence. The differing forms of drag for the two CMEs indicate the forces and thus mechanisms responsible for their acceleration may be different. Further, using an advanced 3D reconstruction technique and statistically rigorous fitting, we were able to show that a CME underwent aerodynamic drag at distances greater than 7\,$R_{\odot}$. This chapter is based on work published in \citeauthor*{Maloney2010p127}, Solar Physics, \citeyear{Maloney2010p127} and  \mbox{\citeauthor*{Byrne2010p74}}, Nature Communications, \citeyear{Byrne2010p74}.}\\
\hrule height 0.4mm
\vspace{0.5mm}
\hrule height 1mm

\newpage

\section{Introduction}
\label{Intro}
The kinematic evolution of CMEs can be broken into three phases: initiation, acceleration, and propagation \citep{Zhang2001p452}. During the acceleration phase magnetic forces such as the Lorentz force are thought to dominate and drive the eruption (\sref{sec:cmeforminit}). During the propagation phase, the initial acceleration has ceased, and the CME motion is dominated by the interaction between the SW and the CME (\sref{sec:cmepropagation}). The ``snow plough'', aerodynamic drag, and flux-rope models all aim to explain the motion of CMEs in the SW \citep{Tappin2006p233,Borgazzi2009p885,Vrsnak2010p43,Cargill2004p135,Chen1996p27499}. 

Before the launch of the STEREO mission, synoptic white-light CME observations were limited to 32\,$R_{\odot}$ using LASCO, while SMEI sometimes tracked CMEs to Earth ($\sim$215\,$R_{\odot}$). In radio observations, fast CMEs which drove shocks could be tracked to Earth \citep{Reiner2007pa2}. Interplanetary Scintillation (IPS) observations provided density and velocity measurements for both CMEs and the SW from 50\,$R_{\odot}$ to beyond 1\,AU, and using tomographic techniques, can give 3D information \citep{Manoharan2006p345,Manoharan2010p137}. CMEs are also observed in {\it in situ} measurements with WIND and ACE at L1 ($\sim$1\,AU), and occasionally CMEs can be tracked up to very large distances of up to 5\,AU using additional spacecraft \citep{Tappin2006p233}.

Numerical modelling has been used to study CME propagation with numerous approaches such as 1D Hydro simulations, 2.5D MHD simulations, and full 3D MHD simulations \citep{Gonzalez2003p1039,Cargill1996p4855,Cargill2002p879,Odstrcil1999p28225,Odstrcil200p2116O,Smith2009p12005S,Falkenberg2010Sp06004}. \cite{Cargill1996p4855}, \cite{Cargill2002p879}, and \cite{Cargill2004p135} used MHD simulations to show that aerodynamic drag was an appropriate description of a flux rope in a magnetohydrodynamic flow, and found that the drag coefficient was normally around unity, but can range between approximately $0.0$\rng$10.0$. He also found that there was a feedback interaction between the flow and flux rope which could significantly distort the flux rope depending on the strength of the magnetic field (see \figref{cargillsim}).
\figuremacroFP{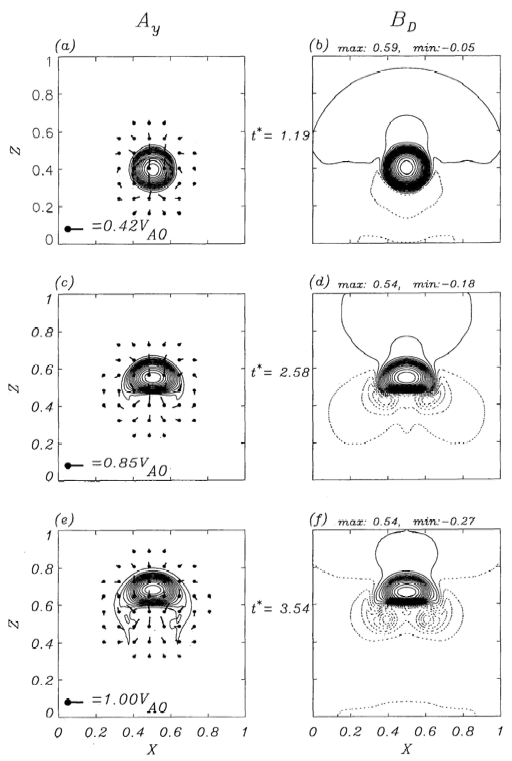}{Simulation of a Flux Rope in a Magnetohydrodynamic Flow}{Magnetic field lines projected in the $x$-$z$ plane (a), (c), and (e). Contours of constant $B_{D}=B_{y}-B_{y0}/B_{y0}$ where $B_{y0}$ is the initial external field strength (b), (d), and (f). Solid and dashed lines represent $B_{D}$ less than and greater than zero. Time is normalised to the Alfv{\'e}n wave crossing time of the flux rope $t_{0}=2a_{0}/v_{A0}$ \citep{Cargill1996p4855}.}{0.8}
 \citet{Smith2009p12005S} using Hakamada-Akasofu-Fry version 2 (HAFv2; \citealt{Fry2003pA4}), a modified kinematic code, simulated a series of events, predicting up to 89\% of the observed events. \citet{Falkenberg2010Sp06004} using the ENLIL \citep{Odstril1999p493,Odstril1999p483} code with a cone model, which is a 3D time-dependent MHD code, modelled an observed CME and compared the simulation to the measured {\it in situ} data, and found it showed good agreement given the correct input parameters. While ENLIL is a MHD code, the CMEs are simulated by time dependent velocity, density and temperature perturbations at the inner boundary, with no magnetic field component similar to HAVv2. The fact both ENLIL and HAVv2 do not model the CME magnetic field but still give accurate predictions may indicate that the magnetic field is not crucial in determining CME dynamics, while Cargill's results indicate it is for CME morphology.

Statistical studies comparing {\it in situ} with white light observations indicate a trend of CME velocity converging towards the SW velocity as they propagate to 1\,AU \citep{Gopalswamy2006p145}. Other studies based on white light observations have indicated that aerodynamic drag of some form may explain this trend \citep{Vrsnak2001p173,Shanmugaraju2009Sp351}.  Radio observations suggest that a linear form of aerodynamic drag may be most appropriate for fast CMEs \citep{Reiner2003p152}. \cite{Tappin2006p233} showed that acceleration can continue far out (5\,AU) into the Heliosphere. However, these studies are subject to the difficulties associated with the observations they are based on. For example, white light observations were limited to single, narrow, fixed, view-points meaning only observations of the inner Heliosphere could be made, and even these were subject to projection effects \citep{Howard2008p1104}. Also, linking features in imaging and {\it in situ} observations is complex and can be ambiguous, a problem exacerbated during periods of high activity. In the case of numerical simulations their complexity can make it hard to extract which effects are the most important, possibly obscuring the important underlying physics.

Using STEREO observations, a number of papers have been published which extract 3D information and study CMEs at extended heliocentric distances, over-coming some of the difficulties outlined above. \citet{Davis2009p08102D} identified a CME in HI1 and HI2,  and using a constant velocity assumption \citep{Sheeley2008p853} they derived the speed and trajectory of the CME. The predicated arrival time, based on the speed derived, agreed with the {\it in situ} observations. \citet{Wood2009p707} used the ``point-P'' and ``fixed-$\phi$'' methods to derive the height, speed, and direction from elongation measurements out to distances of $\sim$120\,$R_{\odot}$. A recent paper by \citet{Liu2010p82} tracked a CME to $\sim$150\,$R_{\odot}$ in 3D using J-maps from both spacecraft to triangulate the CMEs position in 3D. On the other hand, \citet{Maloney2009p149} tracked the trajectory of CME apexes in 3D using triangulation, some as far as 240 $R_{\odot}$. 

A general equation describing the motion of a CME may be written:
\begin{equation}
\label{eq1}
\rho \frac{Dv}{Dt}=\vec{j} \times \vec{B} - \nabla P + \rho \vec{g} + \vec{F}_{D}
\end{equation}
where the first term on the right is the Lorentz force, the second term is the force due to gas pressure, the third term is gravitational force, and the final term is the drag force. An equation describing the motion of a CME in the drag dominated regime may be written:
\begin{equation}
\label{eq2}
M \frac{dv}{dt}= - 1/2 C_{D} \rho_{sw} A_{cme} (v-v_{sw})|v-v_{sw}|
\end{equation}
where $C_{D}$ is the drag coefficient, $\rho_{sw}$ is the solar wind density, $A_{cme}$ is the CME area and $v_{sw}$ is the solar wind velocity, and $M$ is the CME mass. We use a parametric drag model similar to that of \cite{Vrsnak2002p1019} with the added parameter $\delta$, which determines if the drag is quadratic (aerodynamic) or linear (Stokes) (see \sref{sub:fluid}). This parametric form collapses the complex dependences of the CME area ($A_{cme}$), mass ($M$) and the solar wind density ($\rho_{sw}$) into a power-law which depends on heliospheric distance $R$. Thus \eqref{eq2} can be written:
\begin{equation}
\label{eq3}
	\frac{dv}{dt}= - \alpha R^{-\beta}(v-v_{sw})^{\delta}
\end{equation}
where $\alpha$, $\beta$, and $\delta$ are constants.

The CME trajectories were derived using the tie-point method (\sref{ssec:tp}) and from these the kinematics were calculated. These kinematics are then used to investigate the effects of drag on the CME.  We present the reconstructed CME kinematics for four events: one acceleration (CME 1), one decelerating (CME 2), and one with constant velocity (CME 3). For the final event (CME 4), the entire CME front is reconstructed, and we consider the midpoint kinematics of the reconstructed front this event showed a more complex acceleration profile. The observations are described in \sref{datobs}, and their analysis is described in \sref{kinmod}. The results for each event are presented in \sref{Res}. Finally, a discussion of the results followed by the conclusion is presented in \sref{DisConc}.

\section{Observations and Data Analysis}
\subsection{Observations}
\label{datobs}
The trajectories of four CMEs were reconstructed using observations from STEREO SECCHI observed during: 2007 October 8\rng13 (CME 1), 2008 March 25\rng27 (CME 2), 2008 April 9\rng12 (CME 3), and 2008 December 12\rng15 (CME 4).  The observations were reduced and processed as described in \sref{sec:datareduct} and \sref{sec:imp}. \figref{newfig2} shows reduced observations from the 2008 March 28 event where the CME is simultaneously observed in both COR1 and COR2 in both the Ahead and Behind spacecraft,  but only in from the Ahead spacecraft in HI. \figref{20081212obs} shows observation of the 2008 December 12 event.

\figuremacroSW{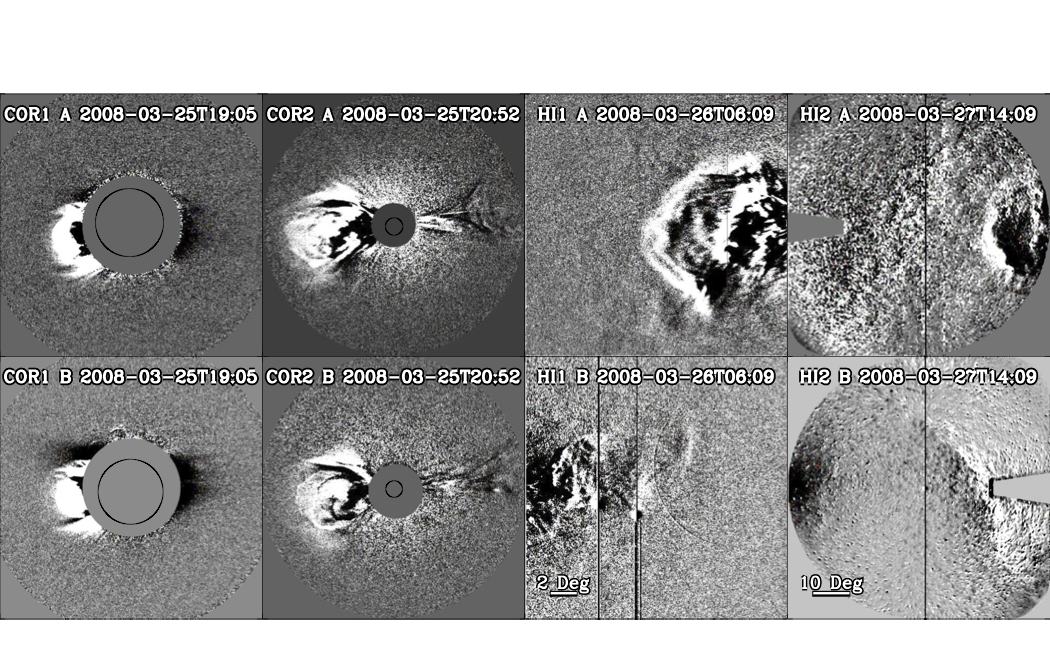}{Sample Observations from the 2008 March 25 Event}{Sample observations from the 2008 March 25 event showing the CME in COR1, COR2, HI1, and HI2 fields-of-view from STEREO-A (top row), STEREO-B (bottom row). Note the absence of a clear CME signature in the STEREO-B HI1 and HI2 images \citep{Maloney2010p127}.}{0.6}

\figuremacroFP{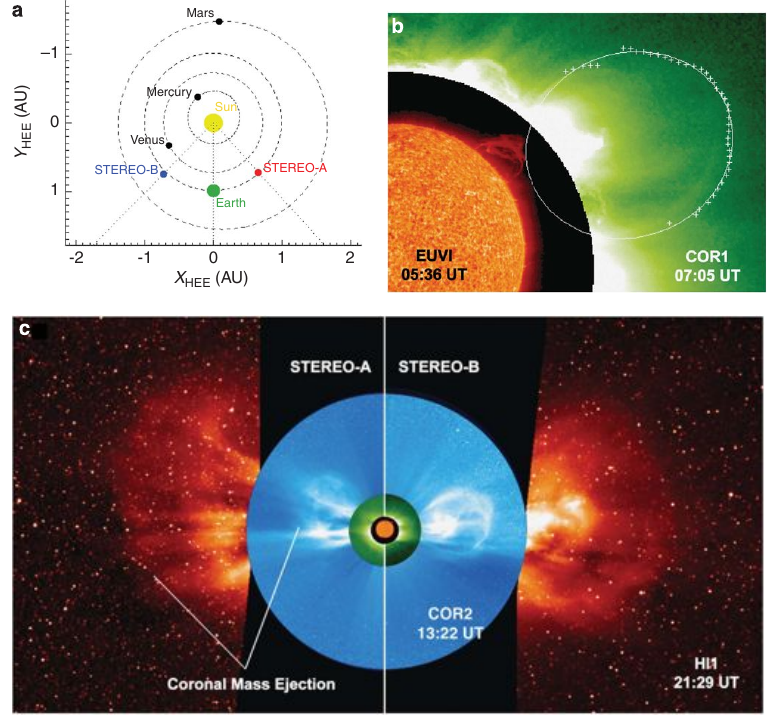}{Composite of STEREO-A and B Images from SECCHI of the CME on 12 December 2008}{(a) Indicates the STEREO spacecraft locations, separated by an angle of 86.7¡ at the time of the event. (b) Shows the prominence eruption observed in EUVI-B off the northwest limb from approximately 03:00 UT, which is considered to be the inner material of the CME. A multiscale edge detection and corresponding ellipse characterisation are over-plotted in COR1. (c) Shows the CME propagation towards Earth from STEREO-A and STEREO-B \citep{Byrne2010p74}.}{0.6}

CMEs 1\rng3 were observed in either the inner coronagraph (COR1) or outer coronagraph (COR2) simultaneously by both STEREO-A and STEREO-B. From these images the CME apex was localised via tie-pointing (\sref{ssec:tp}), and by tracking it through a series of images the trajectory was built up. In these events the CME was only observed in HI by one spacecraft, so an additional constraint is required to localise the CME apex: we assumed that the CME will continue along the same path, with respect to solar longitude, as it did in the COR1/2 field-of-view. CME 4 was observed by both A and B by all SECCHI instruments. For this event the entire CME front was reconstructed using the elliptical tie-pointing (\sref{ssec:etp}). From the 3D data the CMEs kinematics were derived.


\subsection{Kinematic Modelling}
\label{kinmod}
The kinematics were only fit during the time interval we believe that drag is at play and the observations are accurate. There was evidence for early acceleration which was not attributed to drag and as a result these data were not fit. Also, some observations of CMEs tracked far into HI1 or HI2 field-of-view were excluded from fitting as identification of the CME became ambiguous.

The kinematics of CMEs 1 and 2 were fitted, via a least-squares method, with a parametric model for the drag \eqref{eq3}. In order to test which form of drag is most suitable (linear or quadratic), we fitted \eqref{eq3} with $\delta$ set to 1, and then separately with $\delta$ equal to 2. A number of the model parameters can be fixed from the observations, such as the CME height and velocity. We assume that the CME tends to the SW speed, which was taken to be where the velocity plateaus. The model parameters obtained from the fitting were then compared with previous results from \cite{Vrsnak2002p1019}. From this comparison we infer which model best reproduces the kinematics, and hence is the most appropriate. Both the fast and slow CMEs (CME 1 and CME 2) were analysed using this method. The intermediate CME (CME 3) was fit with a constant acceleration model as this seemed most apt since there was no apparent acceleration.

In the case of CME 4 the kinematics were analysed using a bootstrapping method (\sref{sec:bootst}). An initial fit of \eqref{eq3} was carried out and used to calculate the residuals of the fit. These were then used to create the bootstrap samples by randomly sampling them and adding them back on to original fit, creating a bootstrap sample. All of the model parameters were allowed to vary within physically appropriate limits for a bootstrap run of 10,000 iterations. The observationally known parameters such as CME initial height and speed were compared to the resulting bootstrap values as a verification of the method.

\section{Results}
\label{Res}
CME 1 and CME 2 were fitted in three ways with $\alpha$, $\beta$, and $\delta$ all allowed to vary (black line), with fixed $\delta$ of two (magenta line), and one (orange line). In both cases the free fitting returned values that were not comparable to previous studies \citep{Vrsnak2001p173} showing the need to apply so form of physical constraint. The constrained fit parameters for the events are given in Table \ref{fitparms}. CME 3 was fit with a constant acceleration model (black line). CME 4 was fit using the bootstrap method with $h_{0}$, $v_{0}$, $v_{sw}$, $\alpha$, $\beta$, and $\delta$ allowed to vary.

\figuremacroFP{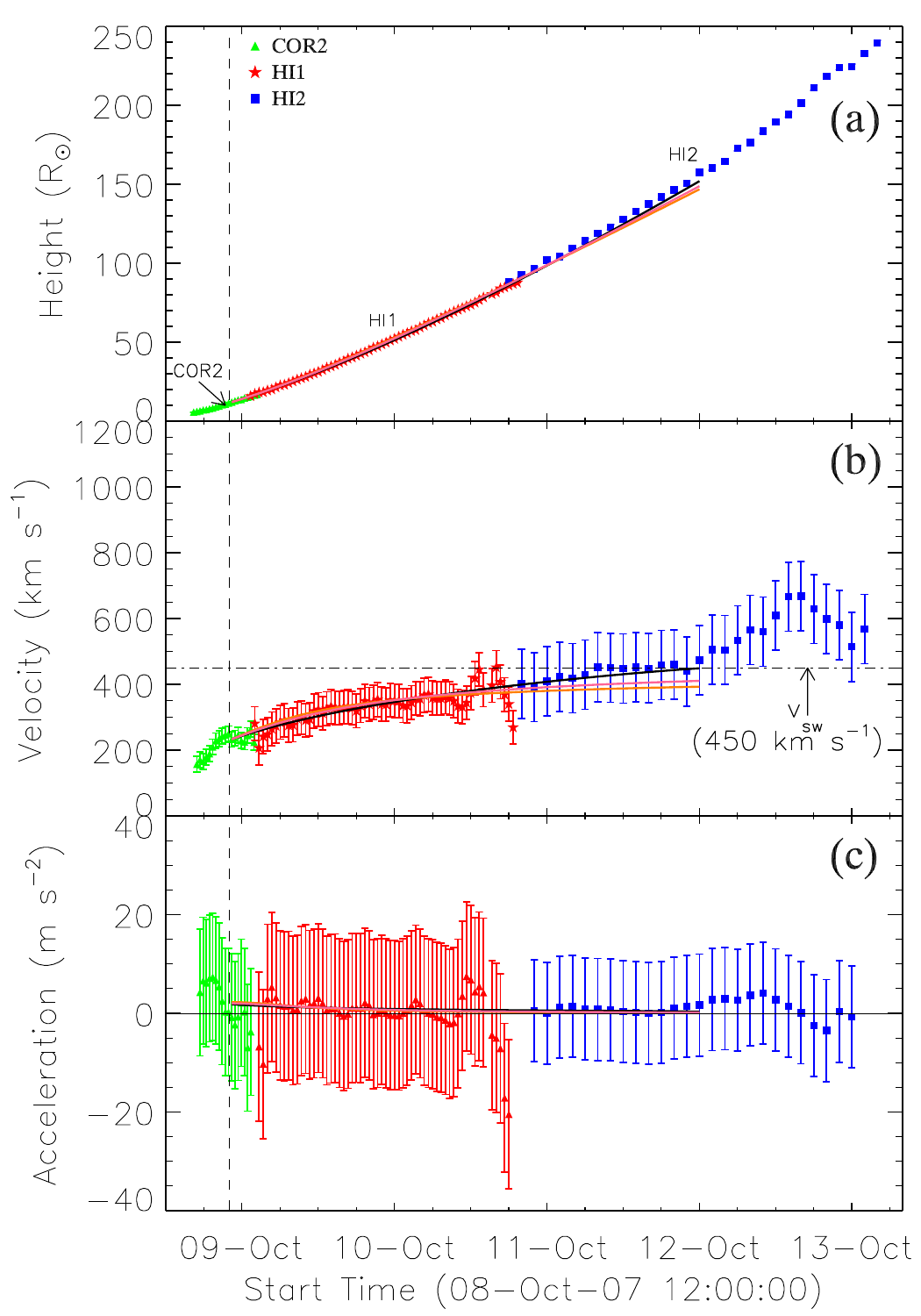}{Derived Kinematics for the 2007 October 8 Event}{Kinematics of the 2007 October 8 event, (a) height, (b) velocity, and (c) acceleration.  Vertical dashed line indicates separation between early and late phase acceleration. Horizontal dot-dash line indicates the assumed SW velocity. The black line corresponds to a fit varying $\alpha$, $\beta$ and $\delta$, the magenta fit has $\delta = 2$ while the orange fit has $\delta=1$}{0.6}

\subsection{CME 1 (2007 October 8--13)}
 \figref{fig3a}(a)-(c) shows the kinematics for the accelerating CME. This CME was first observed at 15:05~UT on 2007 October 8 off the west limb, and was found to be propagating at an angle of 56$^{\circ}$ from the Sun-Earth line. \figref{fig3a}(a) shows the height of the CME. \figref{fig3a}(b) shows the velocity profile which clearly shows the CME is undergoing acceleration, initial velocity of $\sim$150~km~s$^{-1}$, and final velocity of $\sim$450~km~s$^{-1}$. There may be two acceleration regimes; an early increased acceleration phase (before 18:00~UT on the October 8), followed by a drag acceleration. The early acceleration can be attributed to a magnetic driving force, and so was not fitted with the drag model. Later, when the CME reached the centre of the HI2 field-of-view, determining the front position became difficult, so this region was not fitted. \figref{fig3a}(c) shows the acceleration profile of the event. The $\delta=2$ (orange) fit gives the lowest chi-squared value.

\subsection{CME 2 (2008 March 25--27)}
The kinematics from the decelerating CME are shown in \figref{fig3b}(d)-(f). This CME was first observed at 18:55~UT on 2008 March 25 off the east limb and was found to be propagating at an angle of -82$^{\circ}$ from the Sun-Earth line. \figref{fig3b}(d) and (e) show the height and velocity profiles, and the velocity profile clearly demonstrates the CME is undergoing deceleration. The CME had an initial velocity (HI1) $\sim$\,800\,km\,s$^{-1}$ and final velocity of $\sim$\,375\,km\,s$^{-1}$. Due to the high speed of this CME, it was only observed in a small number of frames in COR1 and COR2. As a result, the kinematics were difficult to quantify in these instruments. However, there appears to have been an early acceleration feature. The deceleration in the HI1 and HI2 field-of-view continued until the CME reaches a near-constant velocity, and travels at this velocity throughout the rest of the field-of-view. \figref{fig3b}(d) shows the acceleration profile of the event. The $\delta=1$ (magenta) fit gives the lowest chi-squared value.

\figuremacroFP{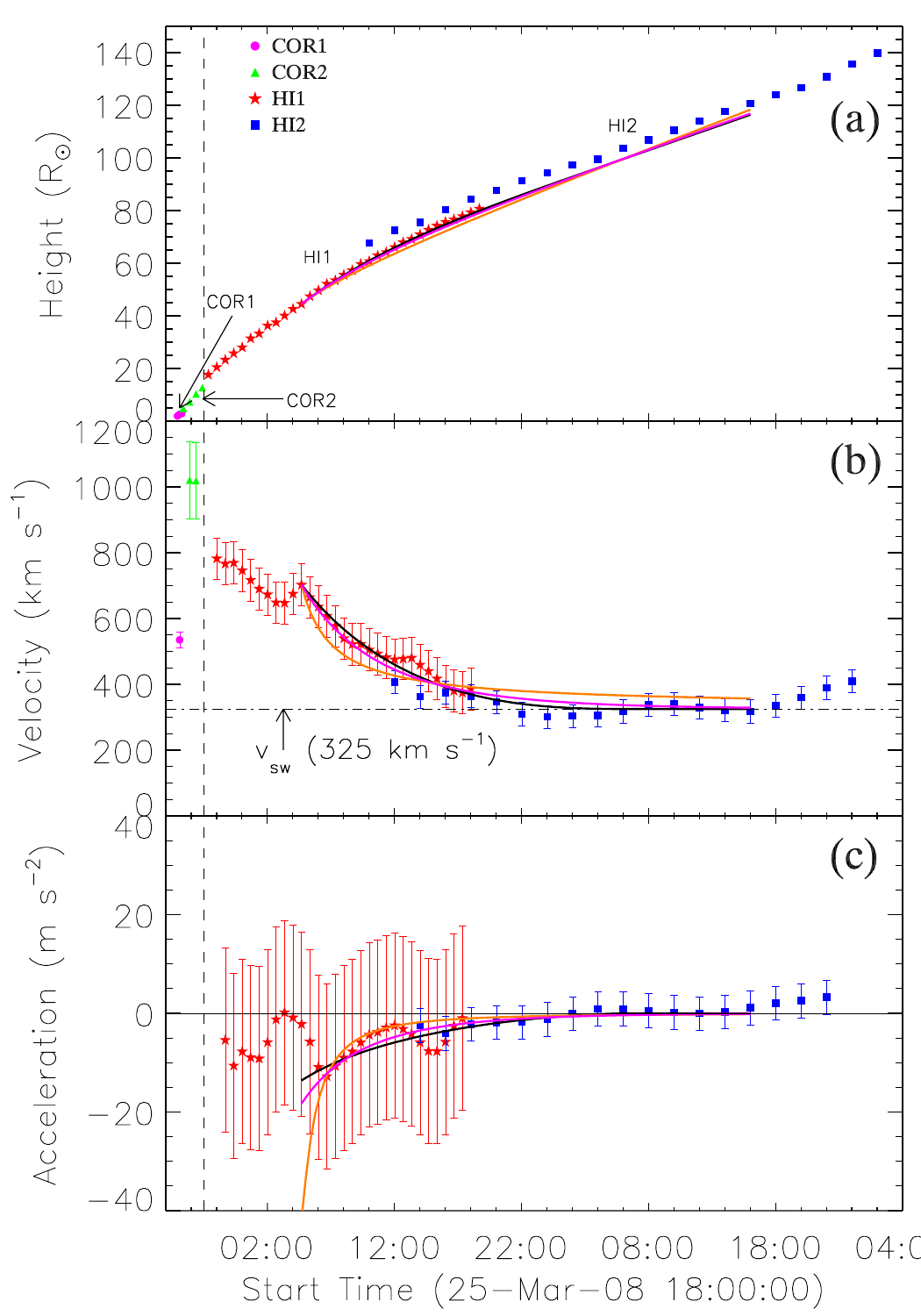}{Derived Kinematics for the 2008 March 25 Event}{Kinematics of the 2008 March 25 event, (a) height, (b) velocity, and (c) acceleration.  Vertical dashed line indicates separation between early and late phase acceleration. Horizontal dot-dash line indicates the assumed SW velocity. The black line corresponds to a fit varying $\alpha$, $\beta$ and $\delta$, the magenta fit has $\delta = 2$ while the orange fit has $\delta=1$}{0.6}

\figuremacroFP{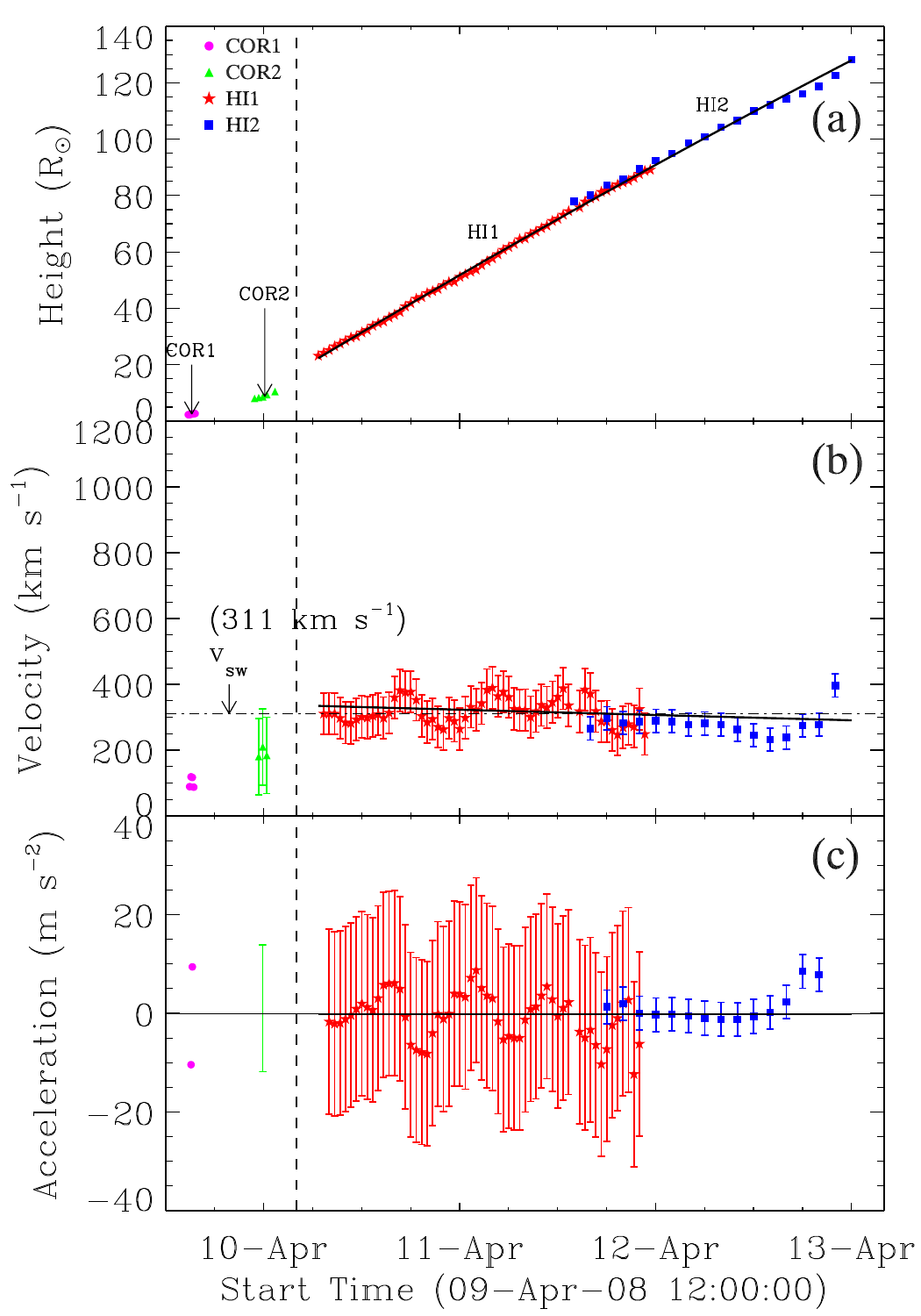}{Derived Kinematics of the 2008 April 09 Event}{Kinematics of the 2008 April 09 event (a) height (b) velocity (c) acceleration. This event shows an early acceleration (to left if dashed line) which levels of to a scatter about typical solar wind speeds. This event was fit with a constant acceleration the resulting fit parameters are $h_{0}$\,=\,22\,R$_{Sun}$,  $v_{0}$\,=\,334\,km\,s$^{-1}$, and a\,=\,$-0.18$\,m\,s$^{-2}$. The assumed solar wind value is indicated by the horizontal dot-dash line.}{0.6}

\subsection{CME 3 (2008 April 9--12)}
In \figref{fig4} we show the kinematics of the constant velocity CME. This CME was first observed at 15:05~UT on 2008 April 09 off the east limb and was found to be propagating at an angle of -73$^{\circ}$ from the Sun-Earth line. \figref{fig4}(a) shows the height of the CME.  \figref{fig4}(b) shows the velocity profile which has a scatter about $\sim$\,300\,km\,s$^{-1}$. Again, there may be some evidence in the COR1/2 observations for an early acceleration phase, but due the event's poorly observable features at this early stage, it is hard to quantify this. The departure from the fit after April 12 20:00~UT is thought to be due to error in the reconstruction,n as the CME apex becomes too faint to identify.  As this event shows no obvious acceleration it was not fitted with the drag model but with a constant acceleration model $h(t) = h_{0}+ v_{0}t + 1/2 at^2$ (thin black line). \figref{fig4}(c) shows the acceleration profile, the fit values ($h_{0}$\,=\,22\,R$_{Sun}$,  $v_{0}$\,=\,334\,km\,s$^{-1}$, and a\,=\,$-0.18$\,m\,s$^{-2}$) are consistent with negligible acceleration throughout the field-of-view.

%

\begin{table}
    \centering
    \begin{tabular}{lcccc}
    
        ~                               & $\alpha$  & $\beta$ & $\delta$ & $\chi^{2}$ \\ \hline
        CME 1 (2007 October 8--13)      & ~       & ~     & ~      & ~        \\ 
        \phantom{aaa}Linear (magenta)    & $1.61\times10^{-5}$ & -0.5  & 1.0    & 8.27     \\ 
        \phantom{aaa}Quadratic (orange) & $1.28\times10^{-7}$ & -0.5  & 2.0    & 6.74     \\
        & ~ & ~ & ~ & ~ \\ 
        CME 2 (2008 March 25--27)         & ~       & ~     & ~      & ~        \\ 
        \phantom{aaa}Linear (magenta)   & $1.02\times10^{-4}$ & -0.5  & 1.0    & 3.71     \\ 
        \phantom{aaa}Quadratic (orange) & $6.38\times10^{-7}$ & -0.5  & 2.0    & 17.63    \\

    \end{tabular}
    \caption[Summary of Fit Parameters]{Fit parameters for the accelerating events. CME 1, $v_{sw}=450$\,km\,s$^{-1}$, $v_{cme}=233$\,km\,s$^{-1}$ , $h_{0}=12$\,$R_{\odot}$. CME 2, $v_{sw}=325$\,km\,s$^{-1}$, $v_{cme}=702$\,km\,s$^{-1}$ , $h_{0}=44$\,$R_{\odot}$.}
    \label{fitparms}
\end{table}

\subsection{CME 4 (2008 December 12--15)}

\figuremacroFP{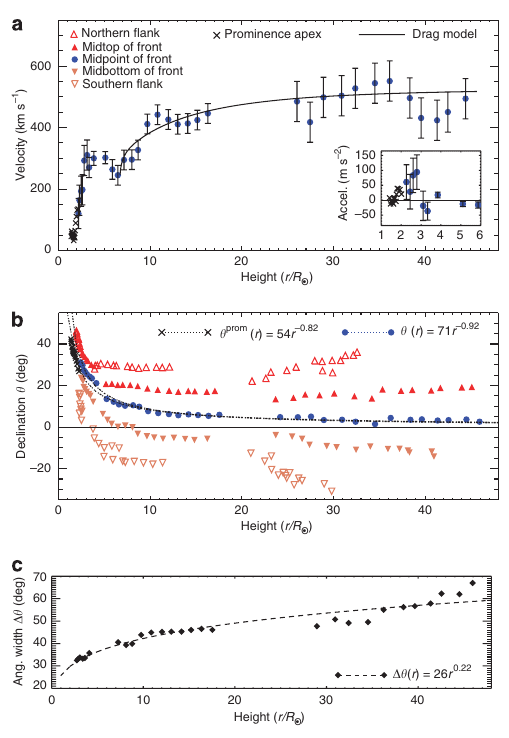}{Derived Kinematics of the 2008 December 12 Event}{Kinematics of the 2008 December 12 event, (a) velocity, (b) declination, and (c) angular width. The velocity in (a) is derived from the CME midpoint and the fit os an aerodynamic drag model \citep{Byrne2010p74}.}{0.8}

The prominence associated with this CME was observed at 50\rng55$^{\circ}$ north from 03:00~UT in EUVI-B images, obtained in the 304\,{\AA} passband, in the northeast from the perspective of STEREO-A, and off the northwest limb from STEREO-B. The prominence is considered to be the inner material of the CME, which was first observed in COR1-B at 05:35~UT (\figref{20081212obs}). This CME was Earth-directed thus $\sim$0$^{\circ}$ from the Sun-Earth line. The CME was initially propagating at 40$^{\circ}$ north from the ecliptic, but due non-radial propagation this tended towards zero as the CME attained a height of $\sim$50\,$R_{\odot}$ as shown in \figref{20081212kins}(b). This non-radial motion was found to be well characterised by a power law. The CME was also found to expand, its angular width increasing from approximately 33$^{\circ}$ to nearly 70$^{\circ}$ (\figref{20081212kins}(c)). Its expansion was also well characterised by a power law. \figref{20081212kins}(a) shows the velocity profile for the event, the CME was initially rapidly accelerated from approximately 100\rng300\,km\,s$^{-1}$ between 2\rng5\,$R_{\odot}$, before gradually rising to a scatter around a value of 550\,km\,s$^{-1}$. The acceleration peaks at approximately 100\,m\,s$^{-2}$ at a height of $\sim$3\,$R_{\odot}$ and then decreases. The early acceleration is attributed to a Lorentz force, while the subsequent increase in velocity, at heights above 7\,$R_{\odot}$, is predicted by theory to result from the effects of drag. From inspection of the Wang-Sheeley-Arge model \citep{Wang1990p355,Arge2000p10465} it appears that the CME is in a high speed solar wind stream of $\sim$550\,km\,s$^{-1}$.

\figuremacroFP{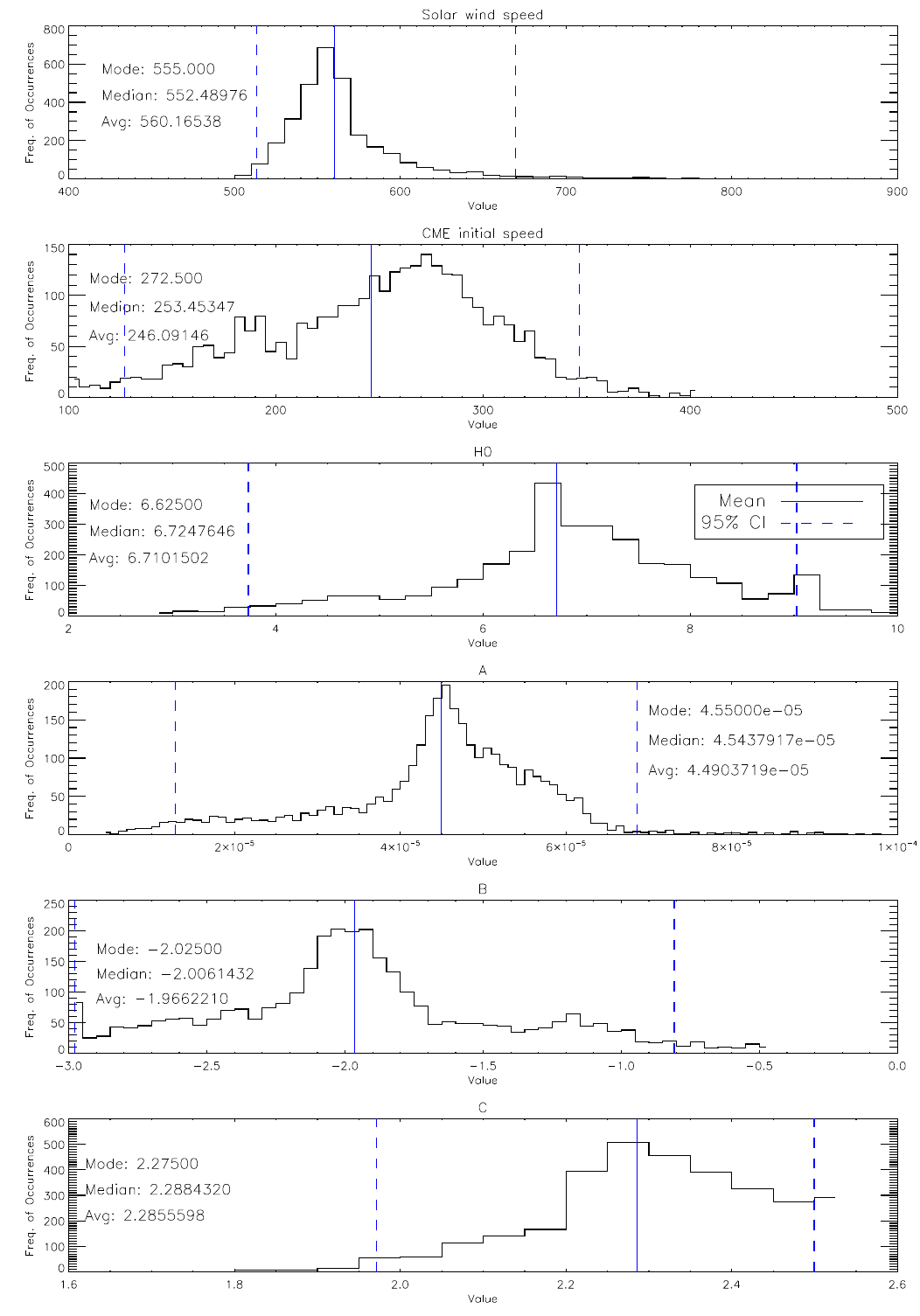}{Bootstrap Distributions for the 2008 December 12 Event}{Bootstrap distributions from a 10,000 iteration run for the parameters of \eqref{eq3} namely: asymptotic solar wind speed, CME initial speed, CME initial height, $\alpha$, $\beta$, and $\delta$.}{.8}

The initial CME height, CME velocity, asymptotic solar wind speed, $\alpha$, $\beta$ and $\delta$ are obtained from a bootstrap distributions for those parameters (\figref{bootstats}). These provide the best fit to the observations, as well as confidence intervals of the parameters. Best-fit values for $\alpha$ and $\beta$ were found to be $(4.55^{+2.30}_{-3.27} )\times10^{-5}$ and $-2.02^{+1.21}_{-0.95}$, which agree with values found in previous modelling work \cite{Vrsnak2001p173}. The best-fit value for the exponent of the velocity difference between CME and the solar wind, $\delta$, was found to be $2.27^{+0.23}_{-0.30}$, which is clear evidence that aerodynamic drag ($\delta$=2) functions during the propagation of the CME in interplanetary space.

The drag model provides an asymptotic CME velocity of $555^{+114}_{-42}$\,km\,s$^{-1}$ when extrapolated to 1\,AU, which predicts the CME to arrive $\sim$ 1 day before the Advanced Composition Explorer (ACE) or WIND spacecraft detected it at the L1 point. We investigate this discrepancy by using our 3D reconstruction to simulate the continued propagation of the CME from the Alfv{\'e}n radius ($\sim$\,21.5\,$R_{\odot}$) to Earth, using the ENLIL with Cone Model at NASA's Community Coordinated Modeling Center\footnote[1]{\href{http://ccmc.gsfc.nasa.gov/}{http://ccmc.gsfc.nasa.gov/}}. The height, velocity, and width from the 3D reconstruction were used as initial conditions for the simulation, it was found that the CME was actually slowed to $\sim$\,342\,km\,s$^{-1}$ at 1\,AU. This is a result of its interaction with an upstream, slow-speed, solar wind flow at distances beyond 50\,$R_{\odot}$. This CME velocity is consistent with {\it in situ} measurements of solar wind speed ($\sim$\,330\,km\,s$^{-1}$) from the ACE and WIND spacecraft at L1. Tracking the peak density of the CME front from the ENLIL simulation gives an arrival time at L1 of 08:09\,UT on 16 December 2008. Accounting for the offset in CME front heights between our 3D reconstruction and ENLIL simulation at distances of 21.5\rng46\,$R_{\odot}$ gives an arrival time in the range of 08:09\rng13:20\,UT on 16 December 2008. This prediction interval agrees well with the earliest derived arrival times of the CME front \citep{Davis2009p08102D,Liu2010p82}. This was identified as plasma pileup ahead of the magnetic cloud flux rope in the in-situ data of both ACE and WIND (\figref{mc}). The CME subsequently impacted the Earth.

\section{Discussion \& Conclusions}
\label{DisConc}
We have shown that it is possible to derive the 3D kinematics of features (the CME apex in this case) in the inner Heliosphere ($\sim$2\rng250\,$R_{\odot}$) using STEREO observations. The 3D kinematics are free from the projection effects of traditional 2D kinematics but may contain artifacts from the 3D reconstruction method (e.g, \citealt{Maloney2009p149}) and other sources. Both CME 1 and CME 2 showed two regimes in their velocity profiles, a low down ($<$\,15\,$R_{\odot}$) early rapid acceleration (in comparison to later values), followed by a gradual acceleration far from the Sun ($>$\,30\,$R_{\odot}$). The early acceleration is thought to be due to a magnetic driving force, as the solar wind velocity low in the corona $(v_{sw}(\le10$\,R$_{\odot})\le$\,268\,km\,s$^{-1}$, \citealt{Sheeley1997p472}) is lower than the velocity already attained by the CMEs in both cases. Here we assume that the later acceleration is due to the interaction between the SW and the CME, as in each case the CME attains a final velocity close to typical values for the solar wind.

Considering CME 2 in \figref{fig3b}(d)-(f), it can clearly be seen that the velocity levels off to a constant value typical of the solar wind. We interpret this as the CME reaching the local solar wind speed, and as a result, the force acting on the CME going to zero. For CME 1 \figref{fig3a}(a)-(c), the velocity initially increases, however, there is a plateau towards the end after April 11 6:00 UT which occurs at SW like speeds. The height measurements towards the end are very scattered and show a rapid increase. This is most likely due to losing the front to the background noise and triangulating a different feature. CME 3 propagates at a roughly constant velocity, which is consistent with the drag interpretation. The CME appears to have already attained the local SW speed and therefore is not accelerated. The fitting results show that a linear dependence produces a better fit for the fast event (CME 2), while a quadratic dependence better fits the slow event (CME 1). The differing range of the interaction CME 1 $\sim$\,120\,$R_{\odot}$ and CME 2 $\sim$\,80\,$R_{\odot}$ may be explained by the suggestion that wide, low mass, CMEs are more affected by drag than narrow massive CMEs \citep{Vrsnak2010p43}.

\cite{Reiner2003p152} suggest that for fast events, a linear model of drag better reproduces the kinematics, which agrees with our findings. \cite{Vrsnak2001p173} also suggested that a linear dependence might be appropriate. From a theoretical perspective, a quadratic dependence corresponds to aerodynamic drag, while a linear dependence suggests Stokes' or creeping drag. It is not currently clear which model is more physically correct. The fit parameters obtained do not agree with those found by \cite{Vrsnak2001p173} and while our values are not unphysical, it is not clear why they differ so much from the previous studies.

The mechanism behind the apparent differing forms of drag, linear ($\delta=1$) and quadratic ($\delta=2$), for the slow and fast event are unclear. The application of any hydrodynamic theory such as drag, to a CME, may be missing vital physics. Could the magnetic properties play a role modifying the form of the drag (reconnection, suppression of turbulence, wave energy transport)? For example, \cite{Cargill1996p4855} showed that, depending on the orientation of the flux rope and background magnetic field (aligned or non-aligned), the drag coefficient can vary between 0.0\rng10.0. They also found that the magnetic field of the flux rope is important for its survival as it propagates. Further, which form of drag is correct for a CME in the SW, the low Reynolds number viscous dominated Stokes'  drag, or the high Reynolds number turbulence dominated  aerodynamic drag? In order to address these questions a larger sample study is needed in order to verify that these effects are a recurring and observable phenomena and also to build up the statistics.

In one case, that of CME 4, a bootstrap analysis allowed us to conclude that the CME acceleration was consistent with aerodynamic drag. However using this model, or a ballistic, model to extrapolate an arrival time at Earth to compare to the {\it in situ} data, gave an arrival time about a day too early. An ENLIL simulation of the CME indicated that the CME was decelerated by a slow speed wind stream ahead of the CME, at distances greater than 50\,$R_{\odot}$. The importance of the dynamic interaction between CMEs and the solar wind is highlighted by the initial acceleration and subsequent deceleration of the CME.

We have shown that it is possible to derive the true 3D kinematics for a number of CMEs in the inner Heliosphere. Based on this, we have been able to conclusively show that CMEs undergo acceleration in the inner Heliosphere, and more specifically, that due to its range and strength, this acceleration is believed to be the result of some form of drag. In one case we were able to show this acceleration was due to aerodynamic drag. Drag acceleration has important implications for space weather predictions, and for the analysis techniques which assume CMEs travel at constant velocity through the Heliosphere. The HI observations of CMEs in the Heliosphere provide a unique and limited opportunity to study the propagation of CMEs, and to understand the coupling between the solar wind and CMEs.




\chapter{Direct Imaging of a CME Driven Shock} 


\ifpdf
    \graphicspath{/}
\else
    \graphicspath{{7/figures/EPS/}{7/figures/}}
\fi


\hrule height 1mm
\vspace{0.5mm}
\hrule height 0.4mm
\noindent
\\	{\it Fast CMEs generate standing, or bow shocks, as they propagate through the corona and solar wind. Although CME-driven shocks have previously been detected indirectly via their emission at radio frequencies and measured directly {\it in situ}, direct imaging has remained elusive due to their low contrast at optical wavelengths. Here we report the first images of a CME-driven shock as it propagates through interplanetary space from 8\,R\,$_{\odot}$ to 120\,R\,$_{\odot}$ (0.5\,AU), using observations from the {\it STEREO} HI. The CME was measured to have a velocity of $\sim$1000\,km\,s$^{-1}$ and a Mach number of 4.1$\pm$1.2, while the shock front stand-off distance ($\Delta$) was found to increase linearly to $\sim$20~$R_\odot$ at 0.5~AU. The normalised standoff distance ($\Delta/D_{O}$) showed reasonable agreement with semi-empirical relations, where $D_O$ is the CME radius. However, when normalised using the radius of curvature,  $\Delta/R_{O}$ did not agree well with theory, implying that $R_{O}$ was under-estimated by a factor of $\approx$3\rng8. This is most likely due to the difficulty in estimating the larger radius of curvature along the CME axis from the observations, which provide only a cross-sectional view of the CME. This chapter is based on work published in \citeauthor*{Maloney2011p5}, The Astrophysical Journal Letters, \citeyear{Maloney2011p5}.} \\
\hrule height 0.4mm
\vspace{0.5mm}
\hrule height 1mm

\newpage


\section{Introduction} 
\label{s_intro}
Bow shocks occur when a blunt object moves relative to a medium at supersonic speeds \citep{rathakrishnan2010applied}. These shocks are formed across many scales and in different conditions: from astrophysical shocks such as planetary bow shocks \citep{Slavin1981p11401}, or the shock at the edge of the Heliosphere \citep{vanBuren1995p2914}, to shocks generated by the re-entry of the Apollo mission capsules \citep{Glass1977p269}. CMEs which travel faster than the local fast magnetosonic velocity (with respect to the solar wind velocity) produce such standing shocks in the frame of the CME \citep{Stewart1974p203,Stewart1974p219}. Interplanetary (IP) CME-driven shocks have previously been detected in radio observations as Type II bursts and using {\it in-situ} measurements. Direct imaging of shocks, on the other hand, has remained elusive, primarily due their low contrast \citep{Vourlidas2009p139,Gopalswamy2008p560}.

\figuremacroW{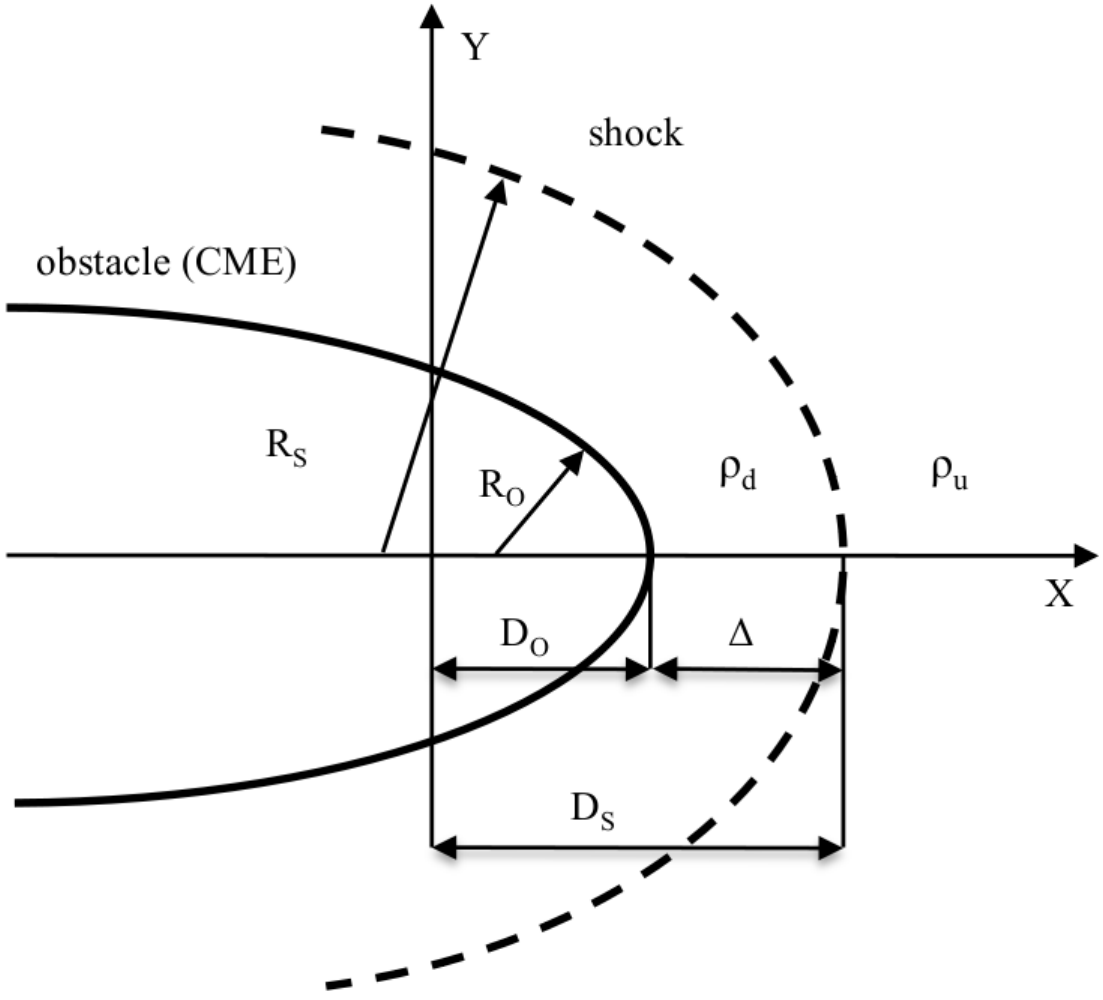}{CME and Shock Diagram}{Diagram of the various quantities used to describe the shock and CME.}{0.6}

In principle the jump conditions in combination with the MHD or fluid equations allow any shock to be modelled and understood (\sref{shocks}). However, the simulations are very complex, and it is often necessary and some times more insightful to work with analytical solutions or semi-emprical relations. There may also be unknown properties and assumptions have to be made about the underlying system such as it geometry. This is more so in the case of MHD shocks as the addition to the shock and flow properties, the magnetic fields must be known.

The shape, size, and standoff distance of a shock are controlled by several factors: the shape and size of the obstacle;  the velocity difference between the obstacle and the medium with respect to the sonic speed (i.e., the Mach number); and the properties of the medium, such as the ratio of specific heats ($\gamma$) and the magnetic field. Imaging observations of a shock give no information about the magnetic field strengths or directions, or the flows speeds or directions. As a result the jump conditions are not useful in this case. However, imaging observations allow the speed of the CME to be measured, and, using a model, its relative speed to the background medium and Mach number can be derived. The standoff distance, CME shock and shock shape can also be readily measured from the imaging observations. 

Relationships between the shock standoff distance and the compression ratio have been derived by a number of different authors. The well known semi-empirical relationship of \cite{Sieff1962p19} has the form:
\begin{equation}
	\label{eq:sieff}
	\frac{\Delta}{D_{O}}=0.78 \frac{\rho_{u}}{\rho_{d}},
\end{equation}
which was derived for a spherical object, where $\Delta$ is the shock standoff distance, $D_{O}$ is distance from the centre to the nose of the obstacle (in this case the radius), and $\rho_{u}$, $\rho_{d}$ are the densities upstream and downstream of the shock respectively. Using gas-dynamic theory, \cite{Spreiter1966p8946} demonstrated that $\rho_{u} / \rho_{d}$ could be written in terms of the upstream sonic Mach number, $M_{s}$, and the ratio of specific heats $\gamma$:
\begin{eqnarray} 
	\label{eq:spreiter}
	\frac{\Delta}{D_{O}}= 1.1\frac{(\gamma-1)M_{s}^{2}+2}{(\gamma+1)M_{s}^{2}}.
\end{eqnarray}
The increase in the coefficient in the standoff relations from 0.78 to 1.1 is due to the fact that the object under consideration (Earth's magnetosphere) in \eqref{eq:spreiter} is more blunt than a sphere; specifically, it is an elongated ellipse. Neither \eqref{eq:sieff} or \eqref{eq:spreiter} behave as expected at low Mach numbers, where the shock should move to a large standoff distance. A modification which corrects for this enables \eqref{eq:spreiter} to be written in the form:
\begin{equation}
	\label{eq:farris1}
	\frac{\Delta}{D_{O}}=1.1\frac{(\gamma-1)M_{s}^{2}+2}{(\gamma+1)(M_{s}^{2}-1)},
\end{equation} where the additional term in the denominator ensures the shock moves to a large distance as the Mach number approaches unity \citep{Farris1994p9013}. They also suggested that using the obstacle radius of curvature rather than radius would be more suitable as it accounts for the shape of the obstacle, resulting in:
\begin{equation}
	\label{eq:farris2} \frac{\Delta}{R_{O}}=0.81\frac{(\gamma - 1)M_{s}^{2}+2}{(\gamma+1)(M_{s}					^{2}-1)},
\end{equation} where $R_{O}$
is the obstacle radius of curvature.

In general, a conic section can be represented by $y(x)^{2} = 2R(D-x)+b(D-x)^2$ where $b$ is the bluntness ($b<-1$: blunt elliptic; $b = -1$: spherical; $-1 < b < 0$: elongated ecliptic; $b = 0 $: parabolic; $b > 0$: hyperbolic). The shape of the shock fronts are known to be represented by a modified conic section. One such parameterisation of the shock front, from \cite{Verigin2003p9014}, is:
\begin{eqnarray}
	\label{Eq6}
	y^{2}(x)=2R_{S}(D_{S}-x)+\frac{(D_{S}-x)^{2}}{M_{s}^{2}-1} \nonumber \\
			\cdot (1 + \frac{b_{S}(M_{s}^{2}-1)-1}{1 + d_{S}(D_{S}-x)/R_{S}}),
\end{eqnarray} 
where $b_{S}$ is the bluntness of the shock, and $d_{S}$ is related to the asymptotic downstream slope or Mach cone (see \figref{f1}).

\figuremacroSW{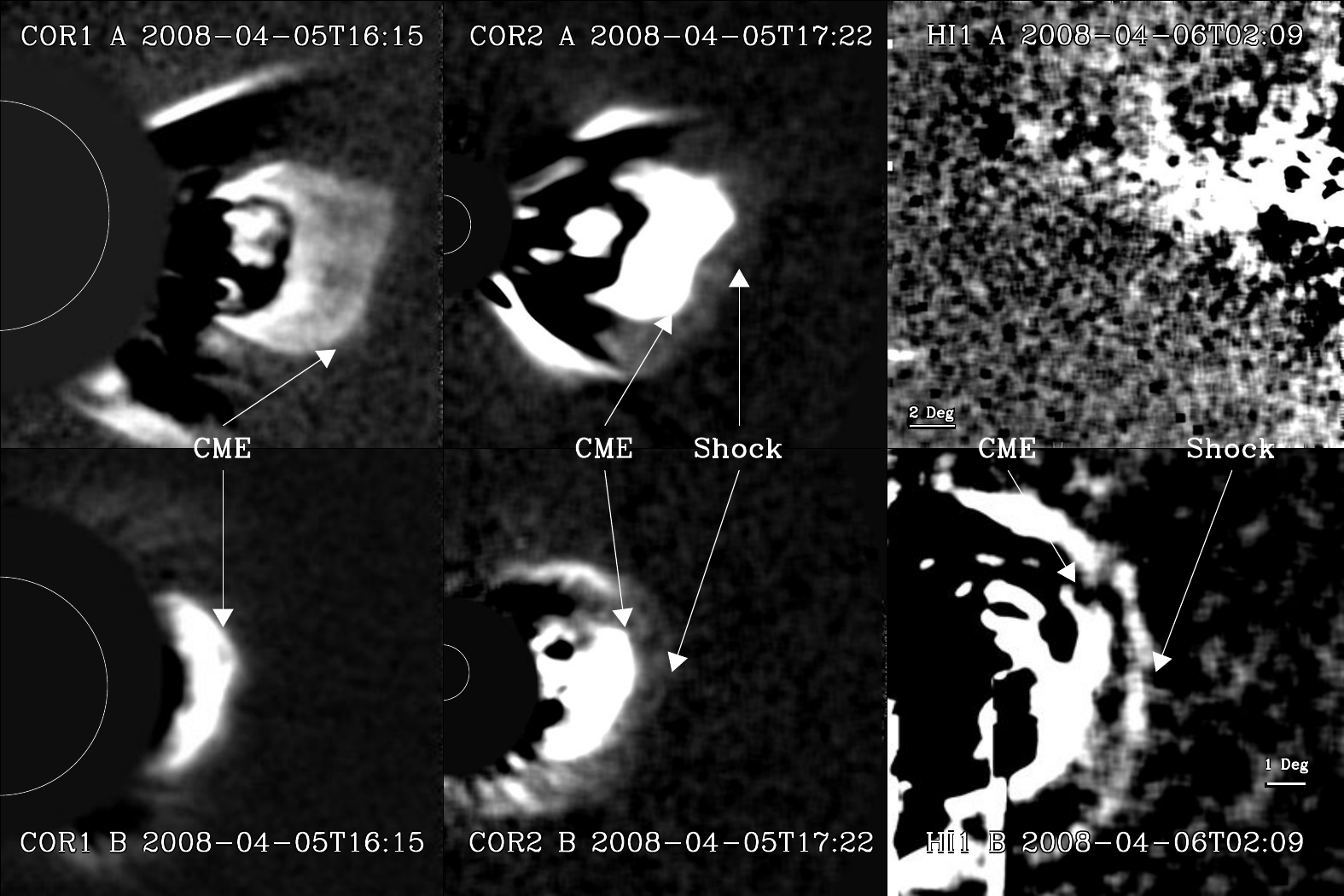}{CME and Shock Observations}{Simultaneous STEREO observations of the CME and shock front from Ahead (top row) and Behind (bottom row). The CME and shock are indicated on the individual images where applicable. No CME or shock is visible in the HI 1 A observation (the Sun is off the right hand edge of the HI 1 A image). These images have been severely clipped and smoothed to make the shock more discernible \citep{Maloney2011p5}.}{0.50}

The relationships between the standoff distance and the Mach number have been investigated from a number of perspectives, including numerical modelling, analytical relations, laboratory experiments and {\it in-situ} measurements of planetary bow shocks \citep{Spreiter1995p433,Spreiter1980p7715}. These have shown that, in general, the semi-empirical relations provide an adequate description of shocks, with the low Mach regime being an exception \citep{Verigin2003p1323}. Depending on the physical context the sonic Mach number ($M_{s}$) can be replaced with the magnetosonic Mach number ($M_{ms}$), when dealing with plasmas such as the solar wind and CMEs. It has been shown that using gasdynamic relations works well when dealing with magnetised plasmas when the MHD mach numbers are high. It also provides a good approximation when the Alfv\'{e}n ($M_{A}$) or fast magnetosonic ($M_{ms}$) Mach numbers are low and these Mach numbers are substituted for the gasdynamic ($M_{s}$) Mach numbers \citep{Fairfield2001p25361}. The reason this makes a good approximation can be seen by looking at the Rankine-Hugoniot equations or the jump conditions (\sref{shocks}): in the case of both gas-dynamic and MHD shocks the mass continuity relation is the same:
\begin{align}
	\left [ \rho u_{n} \right ] &= 0.
\end{align}
This is the most important quantity in determining the standoff distance. The standoff decreases until the mass flux flowing around the body matches the mass flux crossing the shock. 

Standoff distances of CME-driven shocks have been investigated from an {\it in-situ} perspective by many authors \citep[e.g.,][]{Russell2002p527,Lepping2008p125,Odstrcil2005p02106}. \cite{Russell2002p527} found the shock standoff distance ($\Delta$; thickness of magnetosheath)  was of the order of 21\,R$_{\odot}$ at 1\,AU. \cite{Lepping2008p125} derived an average $\Delta$ of about 8\,R$_{\odot}$ at 1\,AU. However, when considering the CME radius (flux rope radius) as $D_{O}$, the typical $\Delta$ expected from \eqref{eq:farris1} is about 5\,R$_{\odot}$ at 1\,AU. \cite{Russell2002p527} proposed that \eqref{eq:farris2} may be more suited as it accounts for the fact the CME front may not be circular and that the radius of curvature at the nose is a dominant factor in determining the standoff distance. However, they found that \eqref{eq:farris2} did not fit the observations either, and speculated this may be due to observational effect of only measuring one of the radii of curvature of the CME. The underlying structure of a CME is believed to be a flux rope: which has two characteristic curvatures; a smaller one due the curvature perpendicular to its axis (the radius when viewed as a cross-section), and the larger curvature along the axis.

We investigate if the shock relations hold for a CME-driven IP shock. Specifically, we use direct observations of a CME-driven shock observed in COR2 and HI1 instruments of SECCHI on STEREO. In Section \ref{s_obs}, we present SECCHI observations of the CME and resulting shock, and describe the analysis technique. The results of our analysis are presented in Section \ref{s_res}. We discuss our results and state our conclusions in Section \ref{s_disc}

\section{Observations and Data Analysis}
\label{s_obs}
The CME analysed first appeared in the COR1 (\sref{sec:cor1cor2}) coronagraph images from STEREO-B at 15:55\,UT on 2008 April 5. It was most likely associated with a B-class flare from NOAA active region 10987, which was just behind the west limb as viewed from Earth. \figref{f1} shows the CME as it propagates out from the Sun into the different instruments' fields-of-view from 8\,R$_{\odot}$ to 120\,R$_{\odot}$. The CME was visible in both A and B spacecraft in the inner and outer coronagraphs, but was only visible in HI1 (\sref{sec:hi}) from STEREO B. The CME propagation direction was found to be $\sim$106$^{\circ}$ west of the Sun-Earth line, the spacecraft were at a separation angle of 48$^{\circ}$ degrees (from each other). The shock is visible as a curved brightness enhancement in both the COR2 observations in \figref{f1} and also the HI1 B observations. The assumption was made that the curved front is a shock, but there were no radio or {\it in situ} data available to corroborate. However, due to the CME's velocity ($\sim$1000\,km\,s$^{-1}$) and the smoothness and position of the feature ahead of the CME, it can be argued that this is a legitimate assumption \citep{Bemporad2010p130,Ontiveros2009p2670}.


\figuremacro{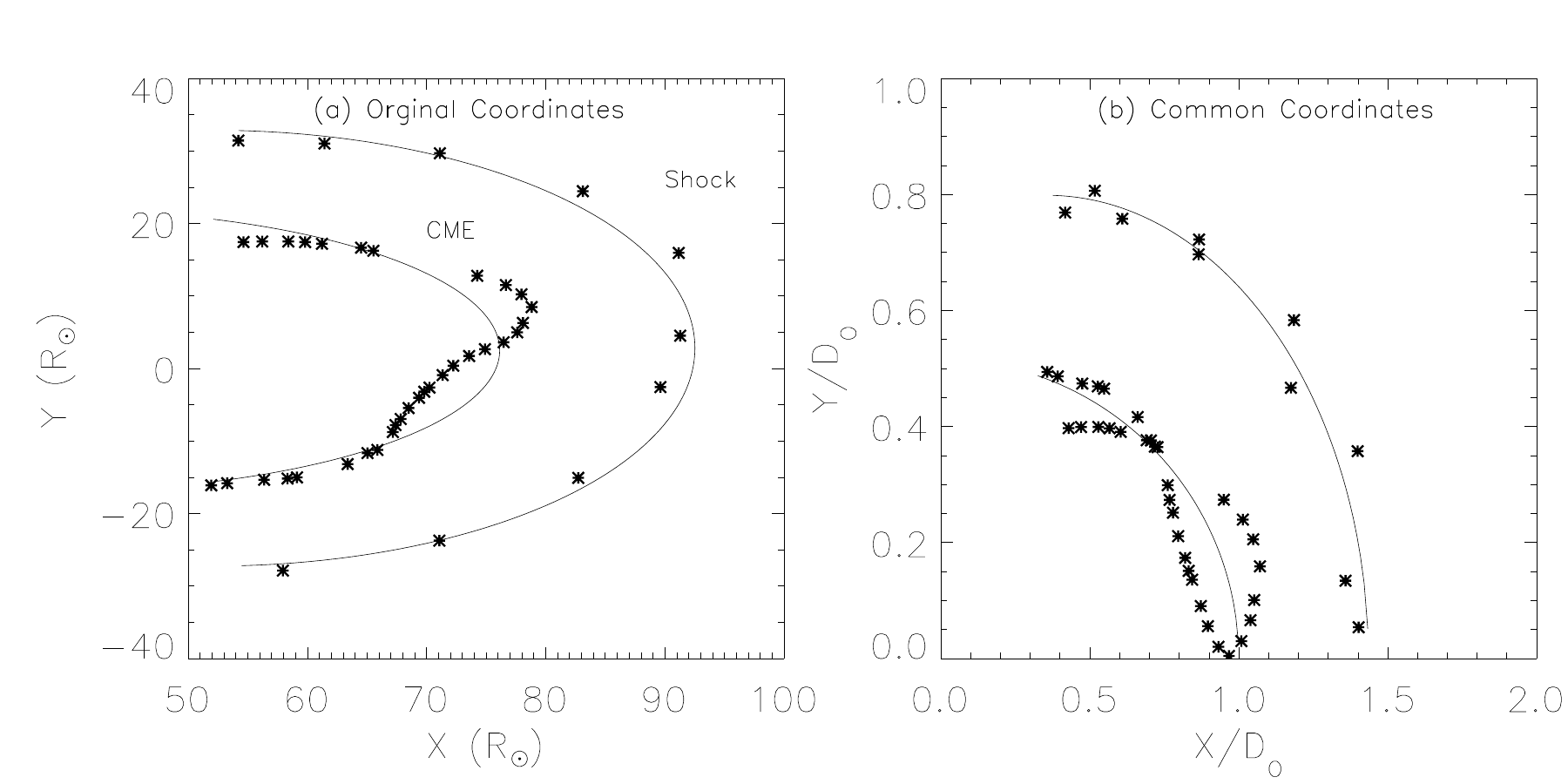}{Projection of the 3D Reconstructions of CME and Shock as well as Shock Front Fitting}{(a) 3D reconstruction of the CME and shock front viewed perpendicular to the propagation direction with the initial ellipse fits overplotted. (b) Data transformed into a coordinate system centred on the initial ellipse fit. Overplotted is the subsequent fit to the shock front using \eqref{Eq6}. Units have been normalised with respect to $D_{O}$ on the right \citep{Maloney2011p5}.}

For each observation in which the CME or shock was visible the front was identified. A number of points along this front were then manually chosen. For the observations where the CME or shock was observed from both spacecraft, the front was localised in three dimensions using the tie-point method (\sref{section:3drecon}). As the CME or shock was only observed by one spacecraft in the HI field-of-view, we used the additional assumption of pseudo-radial propagation, based on the direction derived from COR1 and COR2 to localise the front (\sref{section:3drecon}). The resulting data consisted of a series of points in 3D for the CME and shock for each observation time. \figref{f3}(a) shows the 3D reconstruction of both the shock front and CME front, viewed perpendicularly to the direction of propagation (assumed to be a cross-section). The techniques for deriving the 3D coordinates of features in the COR1/2, and especially the HI field-of-view are not without error. In the case of the event studied here, the CME was close ($<$10$^{\circ}$) to the plane-of-sky of STEREO A. As a result, errors in position should be small.

In order to compare with relationships in \sref{s_intro}, the data were transformed into a coordinate system centred on the CME. To accomplish this, each CME front was fitted with an ellipse. The centre coordinates of these fits were then used to collapse all the data on to a common coordinate system centred on the CME. The shock front was fitted with \eqref{Eq6}, which gave the shock properties such as the shock standoff distance $\Delta$,  the Mach number $M$, and the radius of curvature at the nose of the obstacle $R_{O}$. \figref{f3}(a) shows data and the initial fit, \figref{f3}(b) shows the shifted data and the shock fit using \eqref{Eq6}. The fast magnetosonic Mach number was calculated using $M_{ms} = ({v_{cme}-v_{sw}})/{v_{ms}}$, where $v_{cme}$ is the CME velocity, $v_{sw}$ is the solar wind velocity and $v_{ms}$ is the fast magnetosonic speed. Since $v_{sw}$ and $v_{ms}$ were not known at the position of the CME, a model corona was used to evaluate them. This was based on the Parker solar wind solution with a simple dipolar magnetic field of the form $B(r)=B_{0}(R_{\odot}/r)^{3}$, where {$B_{0}$} was 2.2\,G at the solar surface \citep{Mann2003p329}. For each of the paired CME and shock observations the standoff distances $\Delta$ (=$D_{S}-D_{O}$) were obtained by three different means: (i) using the 3D coordinates of the furthest point ($\textrm{max}(h)$, where $h=\sqrt{x^2+y^2+z^2}$) on the shock and the CME as $h_{shk}$ and $h_{cme}$ respectively; (ii) the previous method applied to the data in the common coordinate system giving $D_{O}$ and $D_{S}$; and (iii) the front fitting procedure also applied to produce standoff distances. However the results of method (i) cannot be used with the relations from Section \ref{s_intro} as they are not in a CME/obstacle centred coordinate system, though the results from method (ii) and (iii) can be compared to \eqref{eq:spreiter}, \eqref{eq:farris1} and \eqref{eq:farris2}.

\section{Results}
\label{s_res}

\figuremacroH{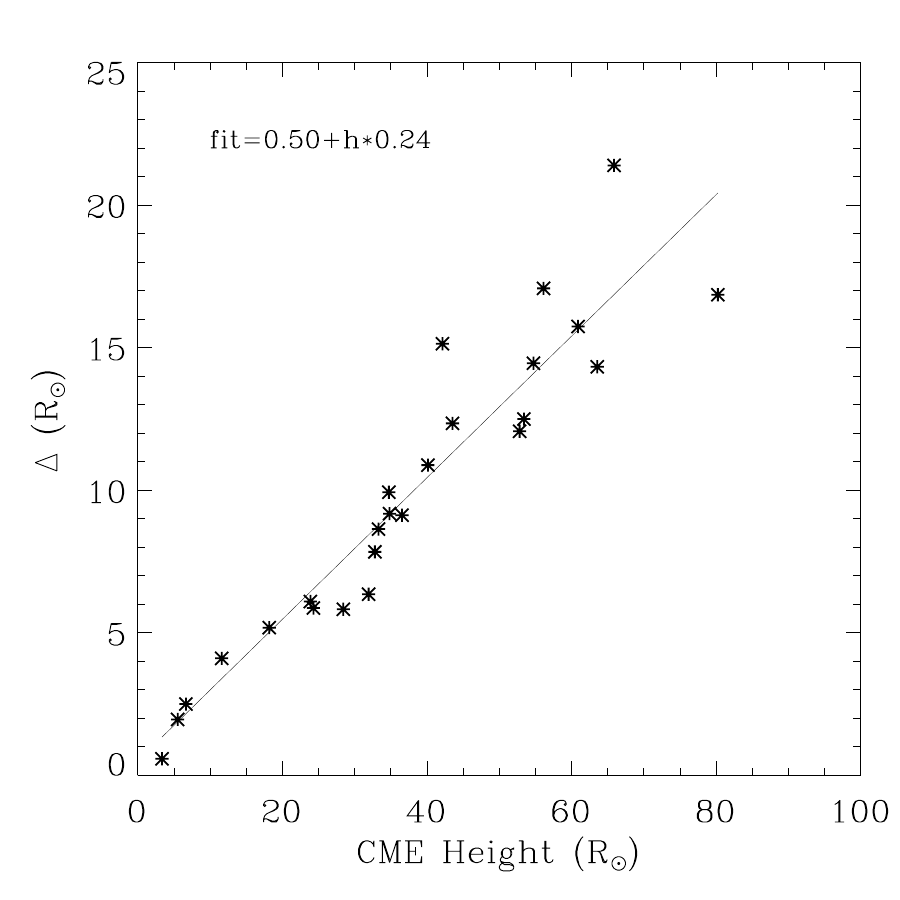}{Plot of the Shock Stand-Off Distance against CME Height}{Shock stand-off distance ($\Delta$) against CME Height and a linear fit to the data.}{0.5}

A summary of the shock properties derived from the observations as a function of time is shown in \figref{f4}(a)-(f). With the exception of the CME ($h_{cme}$) and shock heights ($h_{shk}$), all the properties have been derived from the data, collapsed on to a common coordinate system with respect to the CME. The gap between the first three data points and the others is a result of both the CME and shock leaving the COR2 field-of-view and entering the HI1 field-of-view. The contrast between shock and background in the first three and last three observations is extremely low, making identification of the shock difficult. As a result, these points are not reliable, and should be neglected. \figref{f4}(a) shows the derived heights of the CME and shock as they were tracked from 8\,R\,$_{\odot}$ to 120\,R\,$_{\odot}$ (0.5\,AU). Using a linear fit to $h_{shk}-h_{cme}$ (=$\Delta$) versus $h_{cme}$ (\figref{f6}), the extrapolated standoff distance at Earth was found to be $\sim$54\,R$_{\odot}$. \figref{f4}(b) shows the distance to nose of the CME ($D_{O}$) and shock ($D_{S}$) front directly measured (filled symbols); also shown are the values derived from fits to the shock and CME front (hollow symbols). The increasing offset between the two is due to the differing centres of their coordinate systems, as one is elliptic and the other is parabolic. \figref{f4}(c) shows the standoff distance $\Delta$ derived using $D_{O}$ and $D_{S}$ (filled symbols) and from the fits to the fronts (hollow symbols). Both are in general agreement and show an increase with time. The standoff distance normalised using $D_{O}$ is shown in \figref{f4}(d). The normalised standoff distance is roughly constant with a mean value of $0.37\pm0.09$\,${D_{O}}^{-1}$. The standoff distance normalised to the radius of curvature at the nose of the CME ($R_{O}$) is shown in \figref{f4}(e). The curvature could only be derived from the front fitting, and as such only hollow data points are shown. \figref{f4}(f) then shows the magnetosonic Mach number ($M_{ms}$) derived using: (i) the CME speed in conjunction with the coronal model (filled symbols); and (ii) the shock front fitted using \eqref{Eq6} (hollow symbols). The mean Mach number from the coronal model was 3.8$\pm$0.6, while a value of 4.4$\pm$1.6 was found using the front fitting method. The mean Mach number from both methods was 4.1$\pm$1.2.

\figuremacroFP{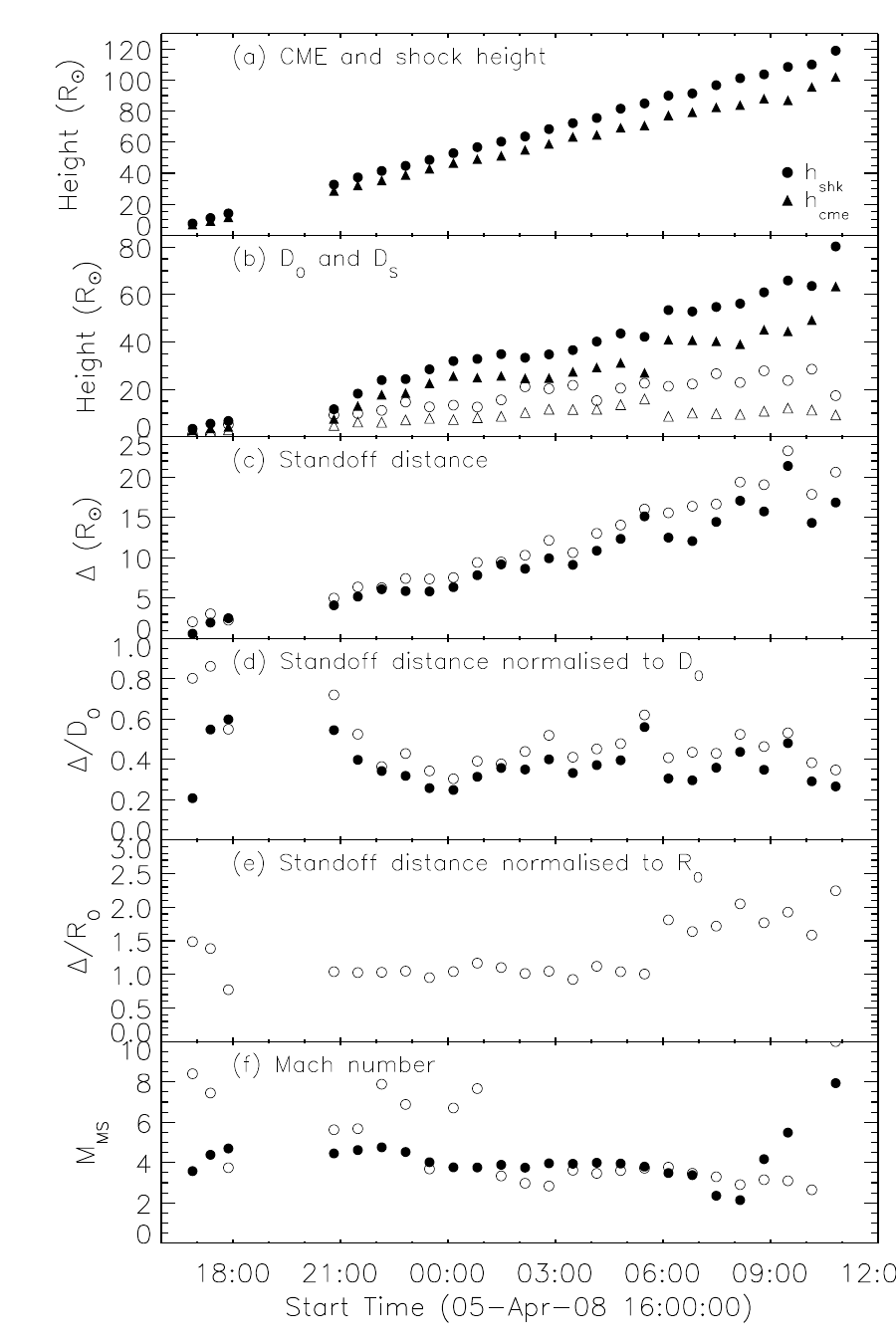}{Plot of Derived Shock Properties}{Shock properties derived directly from the observations (filled symbols) and from fits to the shock and CME (hollow symbols) as a function of time. (a) The maximum height of the CME front (triangles) and shock front (circles). (b) The distance to front of CME ($D_{O}$) and shock ($D_{S}$) in CME centred coordinate system. (c) The shock standoff distance $\Delta$. (d) The normalised standoff distance (${\Delta}/{D_{O}}$). (e) Standoff distance ($\Delta$) normalised by the radius of curvature of the CME ($R_{O}$). (f) The Mach number ($M$) derived the CME velocity and model for the corona (filled circles) and from the fits to the shock front (hollow circles) from \citet{Maloney2011p5}.}{0.8}

\figref{f5}(a) shows the relationship between the normalised standoff distance ($\Delta$/$D_{O}$) and Mach number ($M_{ms}$) for a number of models. The Mach numbers were calculated using the coronal model (filled symbols), and front fitting (hollow symbols). The normalised standoff distances were calculated using measured values of $D_{O}$ and $D_{S}$ (filled symbols), and fits to the CME and shock fronts (hollow symbols). Both show good general agreement between our observations and the models ($<$20$\%$). The model of \citet{Sieff1962p19} shows the poorest agreement, although this is not unexpected as it was derived for a circular obstacle and the CME is quite blunt compared to a circle. \figref{f5}(b) shows the relationship between the standoff distance normalised by the radius of curvature of the CME ($\Delta/R_{O}$) and Mach number for a number of models. In this case, $R_{O}$ can only be derived from the front fitting. These values are then plotted as a function of the Mach numbers derived using both methods described above (hence, each value of $\Delta/D_{O}$ appears twice). Our results do not agree with the expected relationship \eqref{eq:farris2} and indicate that the radius of curvature $R_{O}$ is underestimated by a factor of $\approx$3\rng8. One possible reason for this is that we have not considered the effect of the magnetic field of the CME, and solar wind effects on the shock. However, one would expect if this had a significant effect it would also affect the other relationship. It should be noted that the fast magnetosonic velocity and sonic velocity calculated from our model differ by less than 7\% after excluding the first three data points, as mentioned earlier. This also suggests that the magnetic field should not play a major role. A more likely reason is due to an observational effect similar to that suggested by \cite{Russell2002p527}, where only one radius of curvature of the CME is observed. The observations provide a cross-sectional view of the CME along one of its axes. As a result, we have no information on the curvature along other CME axes.

\figuremacroW{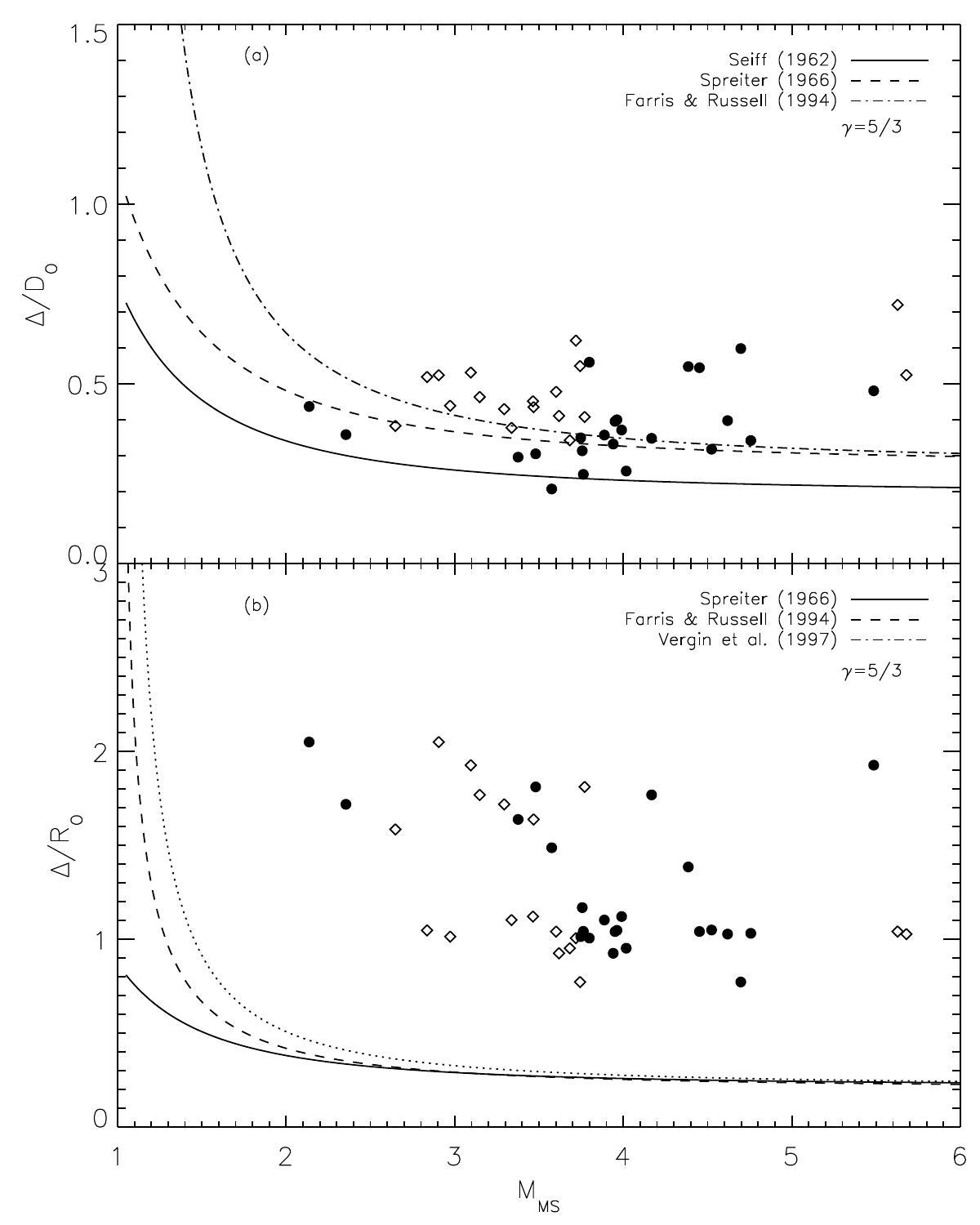}{Comparison of Normalised Standoff Distance to Models}{(a) Shock standoff distance normalised to $D_{O}$ as a function of Mach number. (b) Shock standoff normalised by $R_{O}$ as a function of Mach number. Also shown are the results of a number of semi-empirical models. Filled symbols indicate values derived using a coronal model, while hollow symbols indicate values derived using front fitting \citep{Maloney2011p5}.}{0.7}

\section{Discussion and Conclusions}
\label{s_disc}
For the first time, we have imaged a CME-driven shock in white light at large distances from the Sun. The shock was tracked from 8\,R$_{\odot}$ to 120\,R$_{\odot}$ (0.5\,AU) before it became too faint to be unambiguously identifed. The CME was measured to have a velocity of $\sim$1000\,km\,s$^{-1}$ and a Mach number of 4.1$\pm$1.2, while the shock front stand-off distance ($\Delta$) was found to increase  linearly to $\sim$20~$R_\odot$ at 0.5~AU. The normalised standoff distance (${\Delta}/{D_{O}}$) was found to be roughly constant with a mean of 0.37$\pm$0.09\,${D_{O}}^{-1}$. The normalised standoff distance derived using $D_{O}$ and $D_{S}$, and its relation to the Mach number ($M_{ms}$), were compared to previous relationships and showed reasonable agreement. The normalised standoff distance (${\Delta}/{D_{O}}$) and Mach number were also derived by fitting the CME and shock front, which agreed well with theory and our other method of estimation. The fitting also allowed us to find the CME radius of curvature ($R_{O}$), enabling us to test the relationship between ${\Delta}/{R_{O}}$ and the Mach number. In this case the derived ratios did not agree with the theoretical predictions and showed a significant deviation.

The faint nature of the shock front made its identification challenging, and thus the front location and characterisation showed some scatter (\figref{f4}). For example, the Mach numbers in \figref{f4}(f), show a large amount of variability especially from the front fitting. The standoff distances in \figref{f4}(c) show the same trend, and the two different methods give similar results. It should be noted that the first three, and last three, data points show large deviations from the rest of the data for a number of derived properties. These correspond to very low contrast observations, and hence should be ignored. The Mach number derived from our coronal model and the CME position and speed agrees with that derived from the shock front fitting method. This is a good indication that our methods accurately describe the shock even in the presence of large uncertainties.

Both sets of data for the normalised shock standoff distance $\Delta/D_{O}$ versus Mach number ($M_{MS}$) derived directly, and from front fitting, show good general agreement (Figure \ref{f5}(a)). The standoff distance normalised by the CME radius of curvature ($\Delta/R_{O}$) versus Mach number ($M_{MS}$) from either the fits, or derived directly, do not agree with any of the relationships (\figref{f5}(b)). Assuming that a CME can be modeled as a flux rope, it should have two radii of curvature. Our observations are a measure of a combination of these, which depends on the orientation of the flux rope. This observational affect implies that we may only be measuring the smaller of the two, and this leads to the underestimation of $R_{O}$. Finally, the general agreement between the Mach number derived from our model and the Mach number derived from the fits suggest that the fitting is not the source of the problem. Using a mean Mach number of 4 to derive a value of 0.26 from the semi-empical ratio $\Delta/R_{O}$, and the standoff distance at Earth calculated above (54\,R$_{\odot}$), we can estimate the radius of curvature of this CME at Earth to be $\sim$\,207\,R$_{\odot}$ ($\sim$\,0.95\,AU). This value could be tested using multipoint {\it in situ} measurements.

Imaging observations of CME-driven shocks opens up a new avenue for studying their fundamental properties. This type of observation will be highly complimentary to radio and {\it in situ} measurements. A complete picture of the shock could then be constructed, and the derived properties from the different observations could be compared and contrasted.  Furthermore, the analysis presented here will be applicable to future observations of shocks.




\chapter{Discussion \& Future Work} 


\ifpdf
    \graphicspath{/}
\else
    \graphicspath{{8/figures/EPS/}{8/figures/}}
\fi


\hrule height 1mm
\vspace{0.5mm}
\hrule height 0.4mm
\noindent
\\
{\it
The goal of my thesis research was to increase our understanding of CME evolution in the inner Heliosphere. To achieve this goal, a 3D reconstruction technique was developed and applied to CMEs far from the Sun and shown to be effective. From these reconstructions, the true CME kinematics could be extracted and studied. These kinematics were compared to theoretical models and conclusions on their applicability drawn. A CME-driven shock was imaged and its properties compared to semi-empirical models and shown to agree well. This chapter presents the main results and conclusions of this thesis and outlines possible future work.
}\\
\hrule height 0.4mm
\vspace{0.5mm}
\hrule height 1mm

\newpage

\section{Principal Results}

The purpose of this thesis research was to exploit the unique observations of the STEREO mission to further our understanding of CME evolution, focusing especially on their propagation at large distances from the Sun. This was achieved by extending and applying 3D reconstruction techniques to CMEs imaged by the SECCHI suite of instruments from both STEREO A and B. From these 3D reconstructions the true CME ki!d, free from the limiting uncertainties of projection effects. Investigating these kinematics revealed CMEs do undergo acceleration far from the Sun. The acceleration is consistent with some form of drag, and in one case could be shown to be due to aerodynamic drag. The advanced image processing methods used, allowed the detection and tracking of a CME-driven shock at large distances from the Sun. Analysing this, it was found that the shock stand-off distance agreed well with semi-empirical models.

The principal results from these studies may be summarised as follows:
\subsection{3D CME trajectories}
\begin{itemize}
	\item The STEREO observations allowed CMEs to be tracked from the solar surface to beyond 1\,AU. The modified running difference technique allows the extremely faint CME signature, in HI fields-of-view, to be identified and tracked. This, for the first time, enabled CMEs to be tracked in white light from the Sun to  the Earth. 
	\item The dual vantage point observations of STEREO were used to reconstruct the 3D trajectories of CMEs in the COR1 and COR2 field of view using a triangulation technique, namely tie-pointing. These reconstructions supported a pseudo-radial propagation model for CMEs at large distances from the Sun. This enabled the 3D reconstruction to be extended into the HI field of view, as often CMEs are only observed by either HI A or HI B due to their propagation direction. The assumption of pseudo-radial propagation provided the additional constraint necessary to triangulate the CME in the HI field-of-view. Thus, CME trajectories were reconstructed in the range between 2\rng240\,$R_{\odot}$.
	\item	 From examining the CME trajectories, initial estimates of the CME kinematics were made. These demonstrated that CMEs were being accelerated, with fast CMEs decelerated, slow CMEs accelerated, and both tending towards the solar wind speed. Previously, this acceleration could only be inferred, however using STEREO and the 3D reconstruction methods it was directly shown.
	\item The morphology of CMEs far from the Sun was revealed and showed their complex structure -- especially at the rear of some CMEs -- and the possible multi-loop structure of others. As CMEs propagate, their morphology evolves and, in one case, there was clear evidence for CME `pancaking'.
\end{itemize}

\subsection{3D CME Kinematics and Drag Modelling}
\begin{itemize}
	\item Using the 3D trajectories, the true kinematics of CME apexes were derived in the inner heliosphere, $<250\,R_{\odot}$, which were free from projection effects of traditional observations.
	\item The detailed analysis of the CME kinematics revealed that CMEs often contain multiple phases of acceleration: low down ($<$15\,$R_{\odot}$) rapid acceleration, and a more gradual acceleration far from the Sun ($>$15\,$R_{\odot}$). The rapid acceleration was attributed to magnetic forces, as the speed attained by the CMEs was already in excess of the solar wind at the same positions. The later acceleration was posited to be due to drag as the CMEs tended towards solar wind like speeds.
	\item In order to test if this acceleration was due to drag the CME kinematics were fitted with drag models. In the case of a fast CME the kinematics were best reproduced by a linear drag model, while in the case of a slow CME a quadratic drag model was deemed best. The differing forms of drag for that the two CMEs indicated the forces responsible for their acceleration might have been different.
	\item An entire CME front was reconstructed for the 2008 December 12 event using the elliptical tie-pointing technique. The kinematics of the midpoint of the CME front were extracted and studied. This CME was deflected into a non-radial trajectory which was well described by a power law. The angular width of the CME was found to increase super-radially and was also well described by a power law. The CME kinematics showed an early acceleration phase below 5\,$R_{\odot}$, with the CMEs speed increasing from approximately 100 to 300\,km\,s$^{-1}$, the acceleration peaked at ~100\,m\,s$^{-1}$ at $\sim$\,3\,$R_{\odot}$. This was followed by a more gradual acceleration up to a scatter around 550\,km\,s$^{-1}$. The kinematics beyond 7\,$R_{\odot}$ were fitted with a parameterised drag equation. Using a bootstrapping technique it was shown that this CMEs acceleration was due to aerodynamic drag. Using the final CME speed gave an estimated arrival time at L1 a day too early. This was investigated by running the ENLIL with cone model simulation, constrained by the CME properties derived from the 3D reconstruction. The simulation showed the CME was decelerated by a slow speed stream ahead of it, slowing the CME to approximately 330\,km\,s$^{-1}$ -- thus it predicted an arrival time within  hours of the actual arrival time.
	\item The dynamic interaction between CMEs and the solar wind was shown to contain multiple phases as the CME propagated through different solar wind regions. This is an important unknown factor in determining CME propagation. Accurate forecasting of CME arrival times without knowledge of the solar wind conditions the CME would encounter was shown to be difficult.  	
\end{itemize}

\subsection{CME-driven Shock}
\begin{itemize}
	\item A CME-driven shock was identified and tracked from 8\,$R_{\odot}$ to 120\,$R_{\odot}$ (0.5\,AU). Shocks have been imaged close to the Sun previously, and routinely studied indirectly by the radio emission they generate, however due to their low emission at optical wavelengths this was the first time a shock was directly imaged and analysed at such large distances from the Sun.
	\item The Mach number for the shock was found to be $ 4.1\pm1.2$ and the shock stand-off distance was found to increase linearly to 20$R_{\odot}$ at 0.5\,AU.
	\item  The normalised stand-off distance was found to be roughly constant with a mean value of $0.37\pm0.09$\,${D_{O}^{-1}}$. The shock stand-off distance normalised by the CME radius agrees with the semi-emprical model ($<$\,\%20). However the shock stand-off distance normalised by the CME radius of curvature differed significantly from the semi-emprical relations. This indicated that the CME radius of curvature was under estimated by a factor of $\sim$\,3\rng8. This was found to be most likely due an observational effect, as only a cross-section of the CME is observed, thus we cannot estimate the most likely larger radius of curvature along the flux rope axis.
	\item The radius of curvature of the CME along the flux rope axis was estimated to be $\sim$\,205\,$R_{\odot}$ (0.95\,AU) at 1\,AU, while the stand-off distance was estimated to be approximately 54\,$R_{\odot}$. These values can be compared to multi point {\it in situ} measurements to gauge their validity.  
\end{itemize}

\section{Future Work}
\figuremacroSW{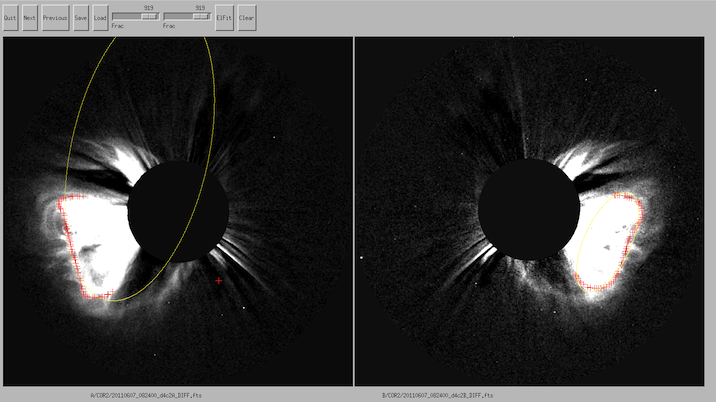}{xImager Software Screen Shot}{A screen shot of the xImager software. Controls are placed above the two STEREO images (A on the left, B on the right). The red $+$ symbols indicate data points selected along the font, the yellow curve is an ellipse characterisation of the data which has failed on the A image. At the bottom of each image the file location and filename is displayed.}{0.50} 

The methods developed and implemented in this thesis are the first to produce true CME kinematics at large distances from the Sun. Only a small number of events have been analysed, and in order to rigorously answer the open questions on CME propagation a large number of events need to be analysed and statistical measures computed. In order to achieve this an event catalog needs to be compiled and studied. This is a very time consuming task which is subject to large user bias. Using advanced imaging processing it may be possible to automate or semi-automate this process. This would firstly remove the user bias, and also allow a much larger number of events to be analysed. Numerous methods have been used to extract 3D information from the STEREO observation. An inter-comparison of their accuracy, applicability (or lack thereof) in studying CMEs at large distance from the Sun would be of great value. More realistic modelling of the ambient solar wind would allow a better comparison or the drag models to the derived kinematics.

\subsection{Event Catalogue}
Analysing a small number of events has led to some interesting results. In order to rigorously make statistically significantly conclusions about what forces govern CMEs during their later propagation a larger number of events must be analysed in an identical manner. To facilitate this large scale analysis, software was written which, using only on start and end times for each instrument, could download, reduce and process the observations producing standardised quicklook movies and data (base difference, running difference, etc). Further software was written (xImager) which allows the A and B observations to be analysed at the same time as shown in \figref{ximager}. This software also allows the scaling of the data, ellipse characterisation, and recording and editing of data points along the CME front. Two implementations of the tie-pointing algorithm have been written -- one which uses the ellipse characterisations, and the other which interpolates between the data. The two methods are necessary as the front of the CME is not always well characterised by an ellipse, as shown in \figref{ximager}. These methods take the output from the xImager code and automatically extract the 3D trajectories and kinematics for the event, and store them for later use.

Currently, a catalogue of 87 events between November 2007 and June 2011 has been compiled. Of these events, 15 have had a preliminary analysis performed. 
The remaining events have yet to be analysed. Once this is completed, analysis on the statistics of the kinematics can begin. Interesting questions like how many CMEs are accelerated, and by how much, could then be answered. Also, each event could be analysed using the bootstrap method to see if all slow and fast events show a different dependance on the difference between the CME and solar wind speed linear versus quadratic or if there is a distribution of values. A number of improvements could be made to the software,  including better handing of outliers in the kinematics, more robust 3D reconstruction, and better displaying and scaling methods to ease identification.

\subsection{Image Processing and Automation}
\figuremacro{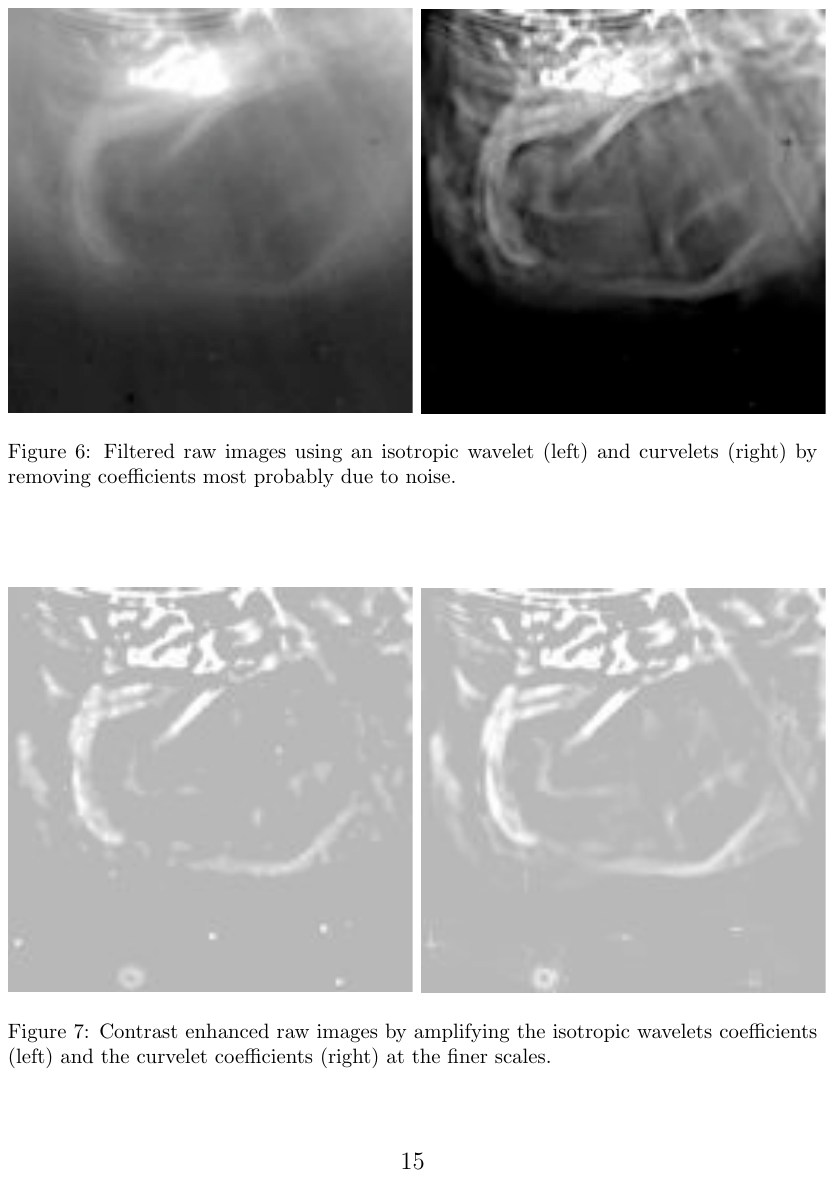}{Corongraph Images Filtered Using an Isotopic Wavelet and Curvelet}{Corongraph Images Filter using an Isotopic Wavelet (left) and a Curvelet right \citep{Gallagher2011p2118}.}
The faint nature of most CMEs is one of the key obstacles to studying their kinematics. The modified running difference technique was developed to enhance the CME signal. Various other image processing methods such as curvelets \citep{Starck2002p670,Cands1999p2495} have also been investigated. The idea of using curvelets is to represent a features as a superposition of functions of various lengths and widths obeying the $width\approx length^{2}$ scaling law. The first curvelets achieved by first decomposing the image into sub-bands or scales, and on each scale a local ridgelet \cite{Cands1999p2495} transform is performed giving the curvelet decomposition. More recent curvelet implementations give the decomposition coefficients in continous frequency as:
\begin{align}
	c(j, l ,k) = \iint \hat{I}(\nu)\hat{\phi}(R_{\theta_{l}}\nu)e^{ix^{j,l}_{k}\cdot \nu}d\nu,
\end{align}
at a scale $2^{-j}$, orientation $l$ and position $x^{j,l}_{k}=R_{\theta_{l}}^{-1}(2^{-j}k_{1},2^{-j/2}k_{2})$. The curvelet is obtain by rotation and translation of a mother curvelet $\psi_{j}$:
\begin{align}
	\psi_{j,l,k}(x) = \psi_{j}(R_{\theta_{l}}(x - x^{j,l}_{k})),
\end{align}
where $R_{\theta_{l}}$ is a rotation by $\theta_{l}$ radians and $\theta_{l}=2\pi2^{-|j/2|}l$. The waveform of $\psi_{j}$ is defined in terms of its Fourier transform in polar coordinates:
\begin{align}
	\hat{\phi}_{j}(r, \theta) = 2^{-3j/4}\hat{w}(2^{-j}r)\hat{\nu}(\frac{2^{[j/2]}\theta}{2\pi}).
\end{align}
This type of multi-scale analysis has already been performed on coronagraph images \citep{Gallagher2011p2118,Byrne2009p4915} as shown in \figref{cormultisc}. \figref{curvelet2} shows a curvelet decomposition of an HI1 image, in which each image is a reconstruction using one scale. The curvelet transform gives much more information than just scale, such as direction and angle. By combining filtering across scales and properties it should be possible to enhance the signal of CME compared to the background corona and stellar field.

\figuremacro{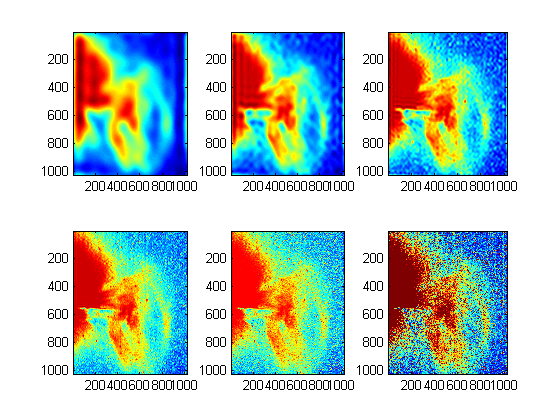}{Curvelet Decomposition of a HI1 Image}{Curvelet decomposition of a HI1 image showing different decomposition scales. The units are image pixels.}

Inspecting the HI images shows that there are clearly two size and intensity distributions present: small circular bright features (stars), and large some what amorphous blobs of very low intensity (CMEs) see \figref{cmedetect} (top row). Separating these two distributions seemed to lend itself to a two step process: firstly thresholding to perform the intensity separation, and secondly, morphological operators to provide the size/shape  separation. An initial threshold in intensity space gives a binary mask where all points below the threshold have been set to zero, and all points above set to unity. A morphological operator which draws conclusions on how a predefined shape (structuring element or kernel) fits or misses the shapes in the image is then applied to detect the desired property. Application of the morphological erode operation with a small circular kernel removes all small isolated features. The erode operation for binary data can be written:
\begin{align}
	C = A \ominus B = \underset{{b \in B}}{\cap} (A)_{-b}
\end{align}
where $A \ominus B$ is the erosion of image $A$ by structure element $B$, and $(A)_{-b}$ represents a translation of $A$ by $b$. In the case of grayscale values the minimum difference of the set is used. Following this, a morphological dilate operator with a slightly larger circular kernel serves to form contiguous blocks and leaves only the CME and largest stars (\figref{cmedetect}; bottom left). The dilate operation for binary data can be written:
\begin{align}
	C = A \oplus B = \underset{{b \in B}}{\cup} (A)_{b}
\end{align}
where $A \oplus B$ is the dilation of image $A$ by structure element $B$, and $(A)_{b}$ represents a translation of $A$ by $b$. In the case of grayscale values the maximum difference of the set is used. A contour operation allows the pixels corresponding to the front to be extracted, these can be over-plotted and the original image (\figref{cmedetect}; bottom right). This technique proved to work relatively well, but the same threshold can not be used on an entire image sequence, and attempts to find a metric that gives a threshold automatically have thus far failed. However a variable threshold interactively changed by the user, in addition to selection of the appropriate contour, could provided an aid to the analysis. This type of semi-automation could be included in future versions of the xImager software.

\figuremacro{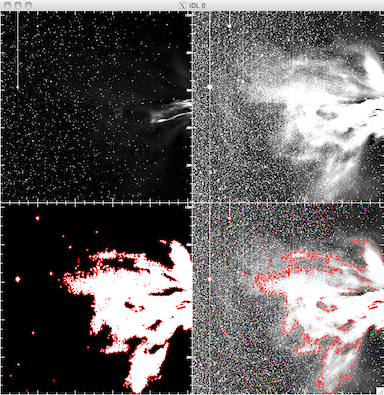}{CME Detection using Thresholding and Morphological Operators}{Original HI1 image scale to 95\% of the maximum intensity (top left) and the same image scaled to 0.5\% of the maximum intensity (top right). Resulting image after application of a morphological open and close operators with different size kernels (bottom left) the detected CME front over plotted (bottom right).}

\subsection{3D Reconstruction}
Numerous methods have been developed to derive the CME positions from STEREO observations. A detailed inter-comparison between all the available methods has not been done as yet. An especially interesting comparison would be between the two triangulation methods, tie-pointing (TP) and elliptical tie-pointing (ETP). It is known that in the TP method the reconstructed point will not lie on the CME, but either ahead of, or behind it, while for the ETP method the point should lie on the CME front. The difference between the two methods has not yet been either quantitatively or qualitatively evaluated. Studying the difference between the two methods as function of CME propagation angle and distance could be used to estimate these differences. Ideally, robust software to apply all the different methods to the extracted CME front would be developed, allowing an inter-comparison for any event studied. Understanding the effects of the various methods on the CME position, and hence kinematics, would enable better constraints on the errors used in the fitting process, thereby increasing confidence in the fit results.

\figuremacro{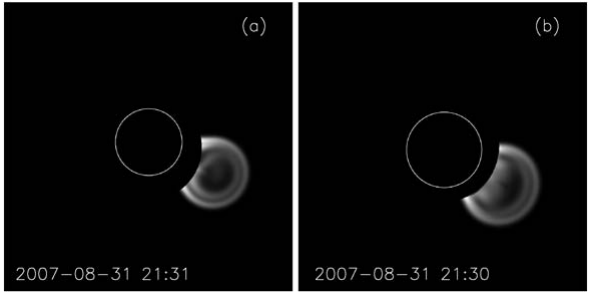}{Synthetic Graduated Cylindrical Shell STEREO Observations}{Synthetic Graduated Cylindrical Shell STEREO Observations from the perspective of A and B \citep{Mierla2009p123}.}

In the future a combination of forward modelling combined, with the 3D reconstructions, will lead to a more complete picture of the CME observations in terms of the CMEs structure and its kinematics. It was shown in \sref{s:thomscat} that using single viewpoint observations in combination with forward modelling was ambiguous. That is, the orientation of the model axis could not be uniquely determined, and, in most cases, two perpendicular orientations could both reproduce the data adequately. Using two viewpoints of STEREO should remove this ambiguity. Some work has already been done in this area, \figref{syntab} shows synthetic coronagraph images of a graduated cylindrical shell (GSC) model as observed by A and B \citep{Mierla2009p123}. Simultaneously constraining the fit from A and B observations should provide a unique best fit. There are many benefits of using a forward model like the GSC, for example better interpretation of the observed structures and more detailed kinematics can be extracted. It is well known that the velocity of a CME front is a combination of the CME velocity and its expansion velocity: using a model like the GSC the centre of mass velocity and expansion velocity can be derived. When comparing the CME kinematics to models the centre of mass velocity is the preferable measure.

\subsection{Kinematic Modelling}
The first obstacle in kinematic modelling is identifying and tracking the CME, the next problem is the accurate determination of its kinematics. Given height-time data, the first and second derivatives with respect to time need to be calculated, giving the velocity and acceleration respectively. As the data is not continuous, numerical approximations to the derivative operator must be used. Traditional methods for determining numerical derivatives such as forward, reverse, and centred difference have serious limitations. One serious limitation is the calculation of errors on the derivatives which depend inversely on the step size -- linearly for forward and reversed differences and quadratically for the centred difference -- double the candance and the error is quadrupled. This is extremely counter intuitive: as the number of data points is increased the errors also increase, even though more information is being added. More advanced methods like 3-Point Lagrangian interpolation suffer from the same problem, an error that depends inversely on step size. The bootstrap technique provides an accurate and robust way to test if the kinematics fit a profile or model, but requires that these be known beforehand. Some methods exist which may allow better kinematics to be obtained using only the data itself, such as the Kalman filter and inversion methods. These methods should be investigated in the future to see if they perform better than the current methods.

Once accurate kinematics have been derived, the next step is to compare them to theoretical models such as the drag model. In the drag model a number of the unknown parameters were absorbed into a power law parameterisation. Any knowledge of these parameters would allow a more accurate test of the drag model. Observation based models such as ENLIL and HAFv2 give the ambient solar wind conditions in the inner Heliosphere. \figref{enlilrun} is a snap shot in time from an ENLIL simulation showing solar wind density and velocity at various cuts through the simulation domain. Once the CME trajectory is known the solar wind properties such as density and speed can be extracted along the CME path. These can then be input into the drag models, which would be a great improvement to the current analytic solar wind models used. The ambient solar wind conditions have been shown to play a vital role in determining CME kinematics. Better constrains on these conditions would allow a better comparison between the observed kinematics and drag model predictions.

Also ENLIL and HAFv2 allow CMEs to be simulated. The CME are simulated by time dependent changes to boundary condition (density, temperature, velocity). Comparing the 1D drag models to these complex 3D simulations would allow a better understanding of which forces play the most important role and where they do so. This could give insight into which type of drag force is acting and the model best suited to study it.

\figuremacroW{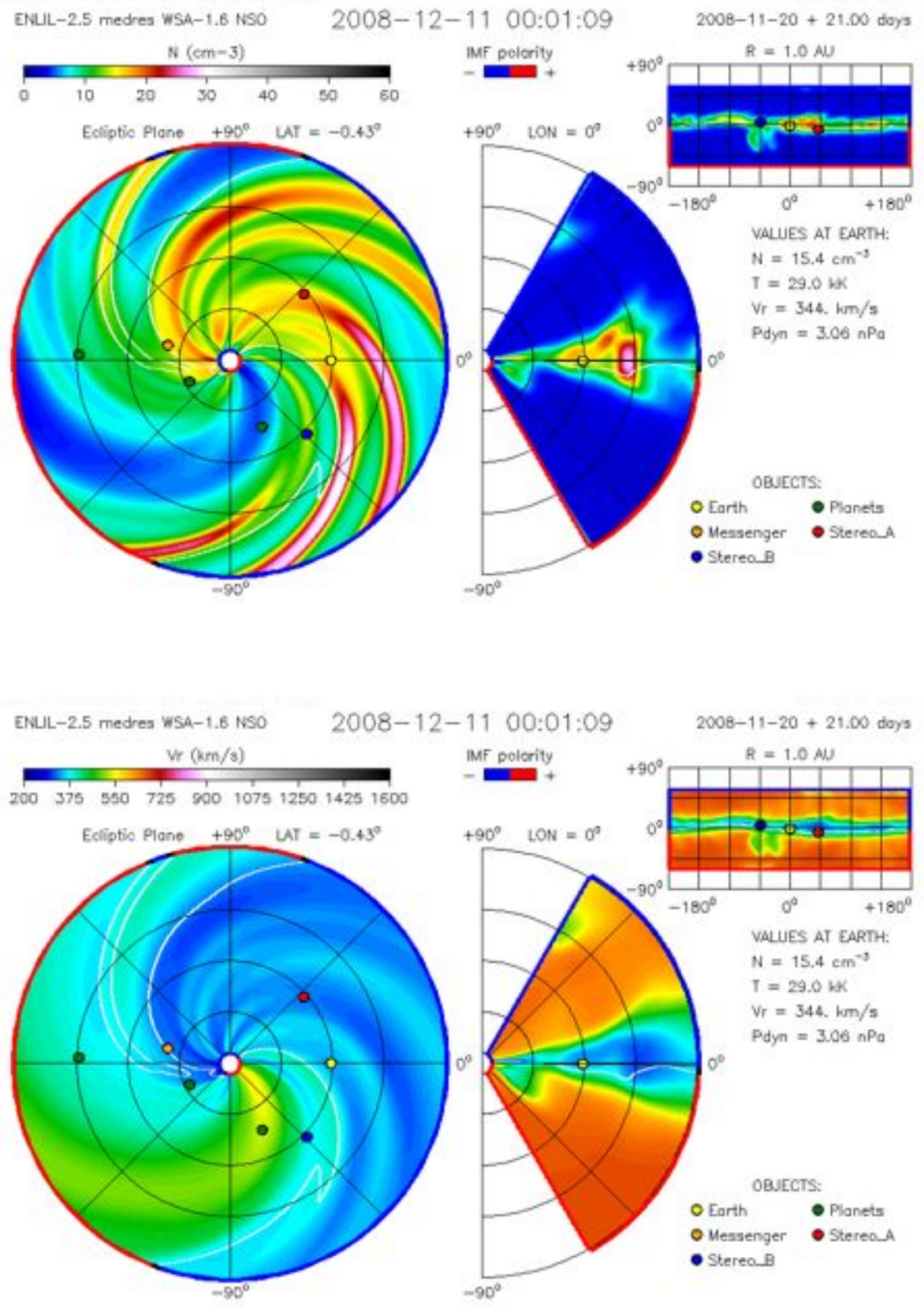}{ENLIL Simulation Results}{ENLIL simulation results showing the velocity and density from NASA's Community Coordinated Modeling Center \href{http://ccmc.gsfc.nasa.gov/}{(CCMC)}.}{0.95}







\begin{footnotesize} 

\bibliographystyle{jmb}
\renewcommand{\bibname}{References} 

\bibliography{thesisrefs} 

\end{footnotesize}

\newgeometry{}
\singlespace

\begin{abstractseparate}

Solar Coronal mass ejections (CMEs) are large-scale ejections of plasma and magnetic field from the corona, which propagate through interplanetary space. CMEs are the most significant drivers of adverse space weather on Earth, but the physics governing their propagation through the Heliosphere is not well understood. This is mainly due to the limited fields-of-view and plane-of-sky projected nature of previous observations. The Solar Terrestrial Relations Observatory (STEREO) mission launched in October 2006, was designed to overcome these limitations.

In this thesis, a method for the full three dimensional (3D) reconstruction of the trajectories of CMEs using STEREO was developed. Observations of CMEs close to the Sun  ($<$\,15\,$R_{\odot}$)  were used to derive the CMEs trajectories in 3D. These reconstructions supported a pseudo-radial propagation model. Assuming pseudo-radial propagation, the CME trajectories were extrapolated to large distances from the Sun (15\rng240\,$R_{\odot}$). It was found that CMEs slower than the solar wind were accelerated, while CMEs faster than the solar wind were decelerated, with both tending to the solar wind velocity.

Using the 3D trajectories, the true kinematics were derived, which were free from projection effects. Evidence for solar wind (SW) drag forces acting in interplanetary space were found, with a fast CME decelerated and a slow CME accelerated toward typical SW velocities. It was also found that the fast CME showed a linear dependence on the velocity difference between the CME and the SW, while the slow CME showed a quadratic dependence. The differing forms of drag for the two CMEs indicated the forces responsible for their acceleration may have been different. Also, using a new elliptical tie-pointing technique the entire front of a CME  was reconstructed  in 3D. This enabled the quantification of its  deflected trajectory, increasing angular width, and propagation from 2 to 46\,$R_{\odot}$ (0.2 AU). Beyond 7\,$R_{\odot}$, its motion was shown to be determined by aerodynamic drag. Using the reconstruction as an input for a 3D magnetohydrodynamic simulation, an accurate arrival time at the L1  Lagrangian point near Earth was determined.

CMEs are known to generate bow shocks as they propagate through the corona and SW. Although CME-driven shocks have previously been detected indirectly via their emission at radio frequencies, direct imaging has remained elusive due to their low contrast at optical wavelengths. Using STEREO observations, the first images of a CME-driven shock as it propagates through interplanetary space from 8\,$R_{\odot}$ to 120\,$R_{\odot}$ (0.5 AU) were captured. The CME was measured to have a velocity of $\sim$\,1000\,km\,s$^{-1}$ and a Mach number of $4.1\pm 1.2$, while the shock front standoff distance ($\Delta$) was found to increase linearly to $\sim$\,20\,$R_{\odot}$ at 0.5 AU. The normalised standoff distance ($\Delta/D_{O}$) showed reasonable agreement with semi-empirical relations, where $D_{O}$ is the CME radius. However, when normalised using the radius of curvature ($\Delta/R_{O}$), the standoff distance did not agree well with theory, implying that $R_{O}$ was underestimated by a factor of $\sim$\,3\rng8. This is most likely due to the difficulty in estimating the larger radius of curvature along the CME axis from the observations, which provide only a cross-sectional view of the CME. The radius of curvature of the CME at 1\,AU was estimated to be $\sim$\,0.95\,AU

\end{abstractseparate}










\end{document}